\documentclass[pra,twocolumn,eqsecnum,superscriptaddress,showpacs,amsmath,amstex,amssymb,citeautoscript]{revtex4-1}

\newcommand{\vect}[1]{\boldsymbol{#1}}

\usepackage[T1]{fontenc}
\usepackage[utf8]{inputenc}
\usepackage{lipsum}
\usepackage{amsmath}
\usepackage{amssymb}
\usepackage{bbm}
\usepackage{braket}
\usepackage{xcolor}
\usepackage{pifont}
\newcommand{\cmark}{\ding{51}}%
\newcommand{\xmark}{\ding{55}}%
\usepackage[mathscr]{euscript}
\usepackage[shortlabels]{enumitem}
\usepackage[justification=justified]{subcaption}
\usepackage{graphicx}
\usepackage{lipsum}
\allowdisplaybreaks
\usepackage{float}
\usepackage{graphicx}
\usepackage{dsfont}
\usepackage{comment}
\usepackage[colorlinks=true]{hyperref}  
\hypersetup{
    bookmarks=true,         
    unicode=false,          
    pdftoolbar=true,        
    pdfmenubar=true,        
    pdffitwindow=false,     
    pdfstartview={FitH},    
    pdftitle={Correlations in twisted trilayer graphene},    
    pdfauthor={Christos, Sachdev, Scheurer},     
    pdfsubject={},   
    pdfcreator={},   
    pdfproducer={}, 
    pdfkeywords={} {} {}, 
    pdfnewwindow=true,      
    colorlinks=true,       
    linkcolor=blue, 
    citecolor=blue,        
    filecolor=magenta,      
    urlcolor=blue           
}

\newcommand{\equref}[1]{Eq.~(\ref{#1})}
\newcommand{\equsref}[2]{Eqs.~(\ref{#1}) and (\ref{#2})}
\newcommand{\secref}[1]{Sec.~\ref{#1}}
\newcommand{\figref}[1]{Fig.~\ref{#1}}
\newcommand{\refcite}[1]{Ref.~\onlinecite{#1}}

\newcommand{\tableref}[1]{Table~\ref{#1}}
\newcommand{\appref}[1]{Appendix~\ref{#1}}

\newcommand{\pdagger}{{\phantom{\dagger}}}

\renewcommand{\approx}{\simeq}

\renewcommand{\vec}[1]{\boldsymbol{#1}}

\definecolor{wrongultramarine}{rgb}{1,0.5,0}
\newcommand{\change}[1]{}

\begin{document}

\title{Correlated insulators, semimetals, and superconductivity in twisted trilayer graphene}

\author{Maine Christos}
\affiliation{Department of Physics, Harvard University, Cambridge MA 02138, USA}

\author{Subir Sachdev}
\affiliation{Department of Physics, Harvard University, Cambridge MA 02138, USA}

\author{Mathias S.~Scheurer}
\affiliation{Institute for Theoretical Physics, University of Innsbruck, Innsbruck A-6020, Austria}

\begin{abstract}
Motivated by recent experiments indicating strong superconductivity and intricate correlated insulating and flavor-polarized physics in mirror-symmetric twisted trilayer graphene, we study the effects of interactions in this system close to the magic angle, using a combination of analytical and numerical methods. We identify asymptotically exact correlated many-body ground states at all integer filling fractions $\nu$ of the flat bands. To determine their fate when moving away from these fine-tuned points, we apply self-consistent Hartree-Fock numerics and analytic perturbation theory, with good agreement between the two approaches. This allows us to construct a phase diagram for the system as a function of $\nu$ and the displacement field, the crucial experimental tuning parameter of the system, and study the spectra of the different phases. The phase diagram is dominated by a correlated semimetallic intervalley coherent state and an insulating sublattice-polarized phase around charge neutrality, $\nu=0$, with additional spin-polarization being present at quarter ($\nu=-2$) or three quarter ($\nu=+2$) fillings of the quasi-flat bands. We further study the superconducting instabilities emerging from these correlated states, both in the absence and in the additional presence of electron-phonon coupling, also taking into account possible Wess-Zumino-Witten terms. In the experimentally relevant regime, we find triplet pairing to dominate, possibly explaining the observed violation of the Pauli limit. Our results have several consequences for experiments as well as future theoretical work and illustrate the rich physics resulting from the interplay of almost flat bands and dispersive Dirac cones in twisted trilayer graphene.
\end{abstract}

\maketitle
\tableofcontents

\section{Introduction}
Graphene-based moir\'e superlattice systems have attracted considerable interest in the last few years, motivated by the strongly correlated physics they display \cite{macdonald2019bilayer,andrei2020graphene,kennes2020moir,balents2020superconductivity,STMReview,ZaletelJournalClub}. While originally driven by the experimental realization of near-magic-angle twisted bilayer graphene (TBG) \cite{2018Natur.556...80C,SuperconductivityTBG}, related strongly-correlated moir\'e superlattices have emerged, such as twisted double-bilayer graphene \cite{2019arXiv190306952S,ExperimentKim,PabllosExperiment,burg2019correlated} and $\mathrm{ABC}$-trilayer graphene on hexagonal boron nitride \cite{2019NatPh..15..237C,SuperconductivityInTrilayer,chen2020tunable}. All of these systems exhibit low-energy bands which are energetically separated from the rest of the spectrum and can be tuned to be flat \cite{dos2007graphene,bistritzer2011moire,dos2012continuum}, enhancing the impact of correlations.   
The resulting correlated phenomena include interaction-induced insulating states \cite{2018Natur.556...80C,2019arXiv190306952S,ExperimentKim,PabllosExperiment,2019NatPh..15..237C}, superconductivity \cite{SuperconductivityTBG,2019arXiv190306952S,ExperimentKim,SuperconductivityInTrilayer}, and nematic order \cite{PasupathySTM,STMReview,Cao2020_nematics,rubioverdu2020universal,10.1088/2053-1583/abfcd6}, which are also integral parts of the phase diagrams of paradigmatic strongly correlated materials such as the cuprates. Furthermore, TBG has demonstrated that graphene moir\'e systems can also exhibit additional cascades of transitions that ``reset'' the band structure in an entire range of electron filling fractions \cite{2019arXiv191206150Z,2019arXiv191206145W}. This is likely related to the polarization of certain combinations of the internal ``flavor'' quantum numbers of the electrons in the quasi-flat bands \cite{Christos_2020,2021arXiv210401145K}, with interesting consequences for superconductivity \cite{PhysRevResearch.2.033062,Christos_2020}.

In addition to the twist angle between the layers as a tuning knob in TBG, the spectrum of twisted double-bilayer graphene can be efficiently tuned by applying a perpendicular electric displacement field $D_0$. Notwithstanding the interesting consequences for the correlated physics in the system \cite{2019arXiv190306952S,ExperimentKim,PabllosExperiment,burg2019correlated,rubioverdu2020universal,zhang2020visualizing,STMTDBGYazdani}, including the possibility of electrical control of the nematic director \cite{10.1088/2053-1583/abfcd6}, its superconducting properties \cite{2019arXiv190306952S,ExperimentKim} have been found to be more fragile than in TBG. Fortunately, mirror symmetric twisted trilayer graphene (MSTG), which consists of three layers of graphene with alternating relative twist angles, see \figref{TrilayerBands}(a), has very recently been realized experimentally \cite{Park_2021,Hao_2021,2021arXiv210312083C} and combines the best of both worlds: it can be tuned significantly by applying a perpendicular displacement field $D_0$ while exhibiting strong and reproducible superconductivity. These experiments indicate that MSTG also exhibits both interaction-induced resistive states around integer filling fractions $\nu$ as well as the aforementioned spontaneous band resetting in an extended region of $\nu$ with $2 \lesssim |\nu| \lesssim 3$, akin to the cascades in TBG \cite{2019arXiv191206150Z,2019arXiv191206145W}. As a function of $D_0$ and $\nu$, the largest superconducting region is found to emerge out of this reconstructed normal state. Remarkably, \refcite{2021arXiv210312083C} finds that the superconductor can sustain an in-plane magnetic field much larger than simple estimates of the Pauli limit.   
One additional crucial difference between MSTG and all of the other graphene moir\'e systems mentioned above is that MSTG not only exhibits quasi-flat bands but, at the same time, dispersive Dirac cones \cite{Khalaf_2019,Carr_2020,PhysRevLett.123.026402,Park_2021,Hao_2021,2021PhRvB.103s5411C,shin2021stacking,lei2021mirror}. The interplay between these different types of bands and interactions likely gives rise to rich physics but also provides theoretical challenges.

So far, only very few theoretical studies of interactions in MSTG exist  \cite{2021arXiv210203312R,2021arXiv210316132C,2021arXiv210413920L,2021arXiv210414026Q,2021arXiv210410176F,chou2021correlationinduced} and a systematic understanding of the nature and origin of possible particle-hole and superconducting instabilities as a function of $\nu$ and $D_0$ is still missing. The goal of this work is to help fill this gap by providing a detailed theoretical study of electron-electron interactions in the normal state of MSTG, allowing for a large class of possible instabilities, which is then used to analyze the order parameter and origin of the different superconducting phases of the system.  
In order to tackle the challenges associated with the simultaneous presence of flat and dispersive bands, we use a combination of analytical and numerical approaches.

More specifically, we start from the limit without displacement field, $D_0=0$, where the \textit{non-interacting} band structure of the system is just that of TBG and single-layer graphene \cite{Khalaf_2019,2021PhRvB.103s5411C,Carr_2020,shin2021stacking,PhysRevLett.123.026402,lei2021mirror}; these two ``subsystems'' are, however, coupled by the Coulomb interaction. In the limit where the TBG bands are perfectly flat, we construct exact eigenstates of the interacting Hamiltonian of MSTG at all integer $\nu$, which are also shown to be groundstates for a finite range of the strength of the interaction between the two subsystems. This analysis shows that, in certain limits, also the \textit{interacting} groundstates of MSTG are given by those of TBG and single-layer graphene. This is consistent with experiment \cite{Park_2021,Hao_2021}, observing a graphene Dirac cone at small $D_0$. 

To be able to address $D_0\neq 0$ and more realistic system parameters, we use Hartree-Fock (HF) numerics and analytic perturbation theory. Our HF approach is motivated by the success of this approach in TBG \cite{Xie_2020,bultinck2019ground,Liu_2021,Liao_2021,Bultinck_2020,Zhang_2020} and the fact that the exact groundstates at $D_0=0$ are Slater-determinant states (in the TBG sector). In the analytic perturbation theory, we start from the exact groundstates and study the $D_0$-induced deformations of the interaction matrix elements and band structure, the possible ordering in the graphene bands, the mixing between the graphene and TBG sectors for $D_0\neq 0$, as well as the finite bandwidth of the TBG bands as perturbations. This complements our HF numerics, as it allows us to pinpoint the novel energetic contributions in MSTG as compared to TBG and serves as an important validation of our numerics---in particular, concerning the fate of the graphene Dirac cones, which provides a challenge to any numerical study as they are energetically degenerate with the TBG bands only in a small fraction of the moir\'e Brillouin zone (MBZ). In our analytics, we take advantage of this fraction being small and use it as an expansion parameter. 
Overall, we find good agreement between the HF numerics and the perturbation theory, both confirming the form of the exact eigenstates at $D_0=0$ (but with realistic system parameters); discrepancies between the approaches are traced back to the impact on the energetics coming from additional remote bands only taken into account in the numerics.

Having established the correlated nature and possible particle-hole instabilities in the normal state of MSTG as a function of $D_0$ and $\nu$, we analyze the consequences for superconductivity. We study superconductivity both in the additional presence and absence of flavor polarization. Motivated by experiments \cite{StepanovTuning,Liu1261,SaitoTuning} on TBG, which indicate that electron-phonon coupling is important for pairing, we follow \refcite{PhysRevB.102.064501} and assume that electron-phonon coupling stabilizes superconductivity but leaves singlet and triplet almost degenerate; informed by results for particle-hole instabilities in MSTG, we can then investigate which of the two will be favored due to additional particle-hole fluctuations. We also discuss purely electronic pairing and comment on the relevance of Dirac cones and associated Wess-Zumino-Witten (WZW) terms, as studied previously in TBG \cite{Khalafeabf5299,Christos_2020}.

\begin{figure}[t]
    \centering
    \includegraphics[width=\linewidth]{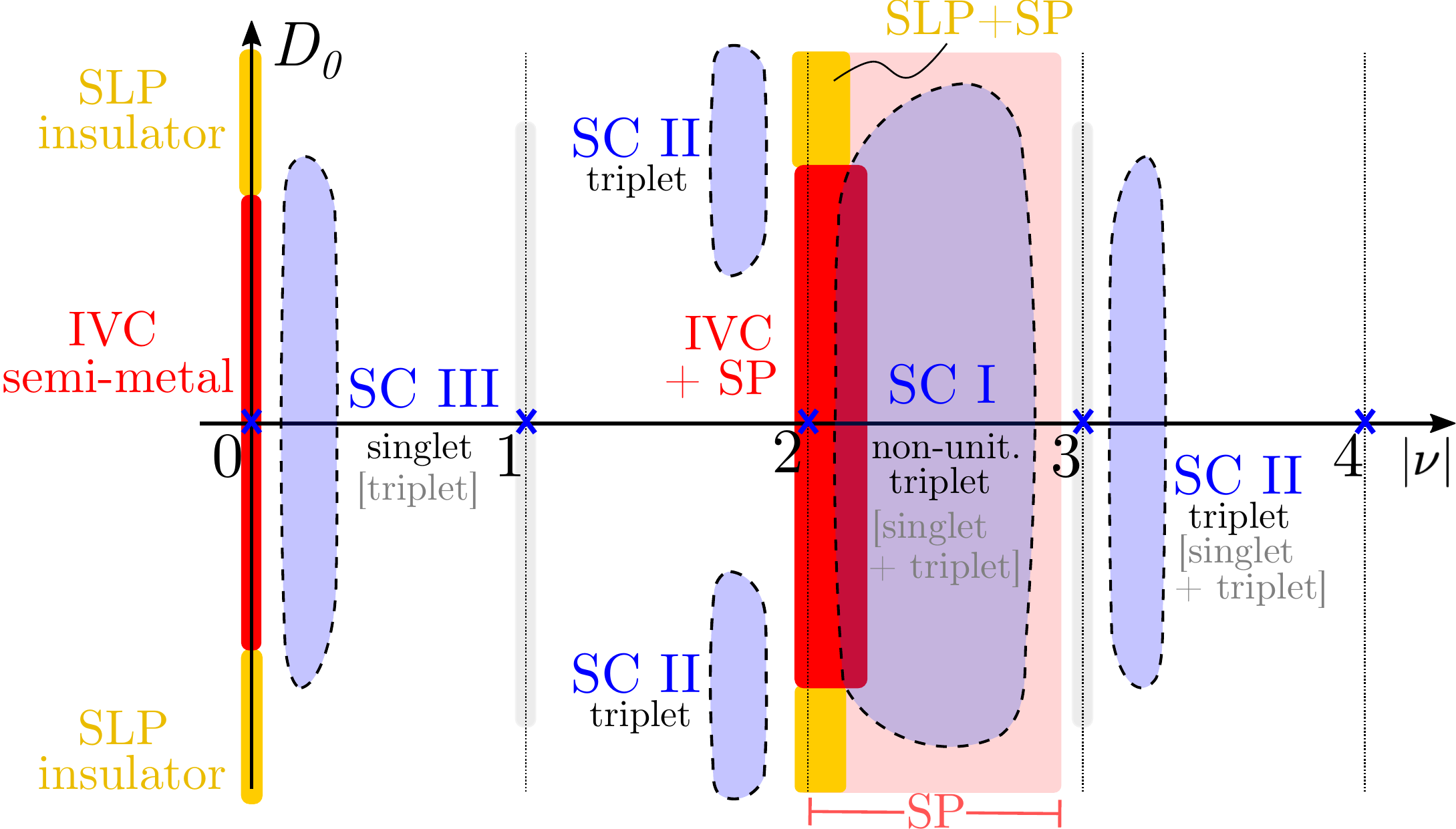}
    \caption{Schematic phase diagram for MSTG based on our analysis, as a function of displacement field $D_0$ and filling fraction $\nu$. Due to particle-hole symmetry of our model, the phase diagram only depends on $|\nu|$. We show the form of the dominant particle-hole instabilities, the intervalley coherent (IVC) semimetallic phase and a sublattice polarized (SLP) insulator at $\nu=0$ (charge neutrality); the IVC and SLP state coexist with spin-polarization (SP) around $|\nu|=2$. We discuss three different types of superconducting regimes, labelled SC~I--III. We indicate in black their respective spin structure for the expected sign of the intervalley Hund's coupling $J_H$ resulting from Coulomb interactions alone [with opposite sign in gray].}
    \label{SchematicPhaseDiagram}
\end{figure}

\subsection{Connection to experiment and phase diagram}
To concisely summarize our main findings, we here briefly discuss their consequences for the phase diagram of MSTG, see \figref{SchematicPhaseDiagram}, and their relation to experiment \cite{Park_2021,Hao_2021,2021arXiv210312083C}. Based on our exact groundstates, perturbation theory, and HF numerics, the (for $\nu\neq 0$ slightly doped) semimetallic nature of the graphene sector is expected to be stable against interactions for $D_0=0$. At charge neutrality, $\nu=0$, the leading instability is found to be an 
intervalley coherent state (IVC) in the TBG sector. While the two sectors start to mix when $D_0\neq 0$, the semimetallic character of the bands is retained; the Landau fan sequence of this state, $\pm 2, \pm 6, \pm 10, \dots$, is consistent with that observed in \refcite{Hao_2021}. Once $D_0$ increases beyond a certain critical value $D_0^c$ [which depends on the relaxation parameter $w_0/w_1$ in \equref{InterlayerCouplingMatrixElements}], the IVC state transitions into a sublattice-polarized (SLP) phase, where all Dirac cones are gapped out, leading to an insulator. For realistic parameters, we expect $D_0^c$ to be of the order of or potentially slightly larger than 
the range of the experimentally applied fields \cite{Park_2021,Hao_2021,2021arXiv210312083C}.

At $|\nu|=2$, we find self-consistent HF solutions with spin polarization, which co-exists with additional particle-hole instabilities. At small $D_0$, the leading additional symmetry-breaking state is again an IVC phase. Here, the TBG bands are completely spin polarized while the graphene cones' polarization is only partial and approaches zero as $D_0\rightarrow 0$. The spin polarization is our proposed mechanism for the experimentally observed band reset and reduced flavor number \cite{Park_2021,Hao_2021}.  At larger $D_0$, the IVC transitions into an SLP state. 
In \secref{Nu2ConnectionToExperiment}, we discuss how the bandstructures we find can give rise to the experimentally observed \cite{Park_2021} additional sign change of the Hall density at sufficiently large $D_0$.
Our obtained behavior at $|\nu|=0,2$ also agrees with another experimental feature: the conductance is suppressed \cite{Park_2021,Hao_2021} at $\nu=0$ for both $D_0=0$ and $D_0 > 0$ while the suppression at $|\nu|=2$ only sets in above a finite critical value of $D_0$, which is consistent with the IVC (SLP) being semimetallic (insulating) at $\nu=0$ and metallic (semimetallitc) at $|\nu|=2$.

As discussed in detail in \secref{BreakingSU2xSU2}, the relative spin-orientation of these orders in the two valleys depends on the sign of the intervalley Hund's coupling $J_H$ (see \equref{HundsCouplingInteraction}), which is not part of the Hamiltonian for our Hartree-Fock computations; the value or sign of $J_H$ is not determined in our computations. For $J_H <0$, which is expected for Coulomb interactions, both the SLP and SP have the same spin polarization in the two valleys, while the IVC order parameter does not carry any spin.

As indicated in blue in \figref{SchematicPhaseDiagram}, we investigate three different regimes of superconducting phases. For superconductivity (SC I) in the presence of spin polarization with or without additional IVC order, the order parameter will be a non-unitary triplet for $J_H <0$ and an admixed singlet-triplet phase \cite{PhysRevResearch.2.033062} for $J_H>0$. For superconducting phases (SC II) close to or in the range $2 \lesssim |\nu| \lesssim 3$ but not coexisting with additional particle-hole instabilities, we expect triplet (a singlet-triplet admixed phase) to dominate for $J_H <0$ ($J_H>0$). All of these superconducting states are consistent \cite{PhysRevResearch.2.033062} with the observed \cite{2021arXiv210312083C} behavior in low to intermediate magnetic fields. Finally, our analysis indicates that a superconducting phase (if present) close charge neutrality (SC III) should be in a singlet state for $J_H <0$.

\subsection{Organization of the paper}
The remainder of the paper is organized as follows. In \secref{ModelAndSymmetries}, we introduce the model for MSTG we study, its symmetries, and establish the basic notation used in this work. We begin our discussion of correlated physics in the limit of vanishing displacement field in \secref{ExactGroundStates}, where we construct exact groundstates. These build the starting point for our HF numerics, see \secref{HFNumericsSection} for $\nu=0$ and \secref{nu2} for $|\nu|=2$, and our analytical perturbation theory, presented in \secref{AnalyticalPerturbationTheory}. Furthermore, superconducting instabilities will be analyzed in \secref{Superconductivity}. Finally, \secref{SummaryOutlook} contains a short summary and discussion of results. Details of our analysis are provided in a set of appendices, labelled Appendix \ref{app:model} through \ref{AppendixOnWZW}.

\section{Model and symmetries}\label{ModelAndSymmetries}
In this section, we introduce the interacting model we consider in this work, which consists of a continuum model to describe the moir\'e bands supplemented by Coulomb repulsion. We further discuss its symmetries and define the notation used in the remainder of the paper.

\subsection{Continuum model and symmetries}
MSTG is constructed from three parallel sheets of graphene where the top ($l=1$) and bottom ($l=3$) graphene layers are aligned with one another and the middle ($l=2$) layer of graphene is twisted at a relative angle $\theta$ with respect to the top and bottom layers, as shown in \figref{TrilayerBands}(a). 
To compute the non-interacting band structure, we employ a continuum-model description, which is just a three-layer extension of the frequently applied continuum model for TBG \cite{dos2007graphene,bistritzer2011moire,dos2012continuum}. Denoting the electronic creation operator for an electron at position $\vec{r}\in\mathbbm{R}^2$, in sublattice $\rho=1,2$, layer $l=1,2,3$, valley $\eta=\pm$, and of spin $s=\uparrow,\downarrow$ by $c^\dagger_{\vec{r};\rho,l,\eta,s}$ the Hamiltonian reads in the absence of a displacement field as \cite{Khalaf_2019,Carr_2020,PhysRevLett.123.026402,2021PhRvB.103s5411C}
\begin{equation}
\begin{split}
    H^{\text{Full}}_{0,1}=&\int_{\vec{r}} c^\dagger_{\vec{r};\rho,l,\eta,s}\left(-iv_F (\vec{\rho}^\eta_{\theta_l})^\pdagger_{\rho,\rho'}\cdot\vec{\nabla}\right) c^\pdagger_{\vec{r};\rho',l,\eta,s}\\
    &+\int_{\vec{r}} \left[c^\dagger_{\vec{r};\rho,l,\eta,s}(T_{\eta,\vec{r}})_{\rho,\rho'} c^\pdagger_{\vec{r};\rho',l+1,\eta,s}+\text{H.c.}\right],  \label{NoninteractingModelWithoutDisplacement}
\end{split}
\end{equation}
where summation over repeated indices is implied, $\theta_l = (-1)^l \theta/2$, $(\vec{\rho}^{\eta=+}_{\theta})_{j} = e^{i\theta \rho_z/2} \rho_j e^{-i\theta \rho_z/2}$, with $\rho_j$ representing Pauli matrices in sublattice space (throughout we will use the same symbol for Pauli matrices and their indices, i.e., $s_j$, $\eta_j$ are Pauli matrices in spin and valley space), and $\vec{\rho}^{\eta=-}_{\theta} = -(\vec{\rho}^{\eta=+}_{\theta})^*$. While the first term in \equref{NoninteractingModelWithoutDisplacement} describes the Dirac cones of the two valleys, $\eta=\pm$, of each individual graphene layer in the continuum expansion around their respective K$_g$ and K$'_g$ points, the second line captures the tunneling between adjacent graphene layers parameterized by $T_{\eta,\vec{r}}$ (the direct hopping process between the outer two layers is neglected). As a consequence of the moir\'e superlattice, these tunneling matrix elements are modulated spatially. As is common, we focus on the lowest moir\'e-lattice harmonics for which symmetry allows for only two independent parameters, $w_0$ and $w_1$, in $T_{\eta,\vec{r}}$. It can be written as \cite{bistritzer2011moire}
\begin{equation}
    T_{+,\vec{r}} = e^{-i\vec{q}_1\vec{r}}\left[ \mathcal{T}_1 + \mathcal{T}_2 e^{-i\vec{G}_1^{\text{M}}\vec{r}} + \mathcal{T}_3 e^{-i(\vec{G}_1^{\text{M}}+\vec{G}_2^{\text{M}})\vec{r}} \right] \label{FormOfHoppingBetweenLayers}
\end{equation}
and $T_{-,\vec{r}}=(T_{+,\vec{r}})^*$.
Here $\vec{q}_1 = k_{\theta} (0,-1)$, $k_\theta = 2|\text{K}_g|\sin(\theta/2)$, connects the K and K$'$ points of the moir\'e lattice, $\vec{G}_1^{\text{M}} = -\sqrt{3}k_\theta (1,\sqrt{3})^T/2$ and $\vec{G}_2^{\text{M}} = \sqrt{3}k_\theta(1,0)^T$ are the basis vectors of the reciprocal lattice (RL) of the moir\'e lattice, $\text{RL} :=\{\sum_{j=1,2} n_j \vec{G}_j^{\text{M}}, \, n_j \in\mathbbm{Z} \}$. The matrices $\mathcal{T}_j$ in sublattice space in \equref{FormOfHoppingBetweenLayers} are given by
\begin{equation}
\mathcal{T}_j =  w_0 \rho_0 + w_1 \begin{pmatrix} 0 & \omega^{j-1} \\  \omega^{-(j-1)} & 0 \end{pmatrix}, \quad \omega = e^{-i \frac{2\pi}{3}}. \label{InterlayerCouplingMatrixElements}
\end{equation}
While rigidly rotated graphene layers correspond to $w_0=w_1$, lattice relaxation \cite{PhysRevB.96.075311,2019PhRvR...1a3001C} leads to $w_0\neq w_1$. We will here consider $w_0$ as a free parameter and study the physics as a function of it. 

As a result of the moir\'e modulations in \equref{FormOfHoppingBetweenLayers}, the tunneling matrix elements between the layers couple momenta related by RL vectors, which reconstructs the graphene cones of the first line in \equref{NoninteractingModelWithoutDisplacement}, leading to a (technically infinite) set of bands; the band energies will be denoted by $\epsilon_{n,\eta}(\vec{k})$, $n\in\mathbbm{Z}$, where $\vec{k}$ is in the first MBZ and we already used that \equref{NoninteractingModelWithoutDisplacement} is diagonal in the valley, $\eta$, and trivial in the spin index, $s$; this makes the band energies independent of spin and allows to label them by their valley index.

MSTG is distinguished from other moir\'e systems, such as TBG or twisted double-bilayer graphene, by a reflection symmetry $\sigma_h$ under exchange of the top and bottom layers of MSTG, with action $\sigma_h:\, c_{\vec{r};\rho,l,\eta,s}\rightarrow c_{\vec{r};\rho,\sigma_h(l),\eta,s}$ where $\sigma_h(2) = 2$, $\sigma_h(1) = 3$, and $\sigma_h(3) = 1$. Since it acts trivially in all internal indices ($\rho, \eta, s$) and $\vec{r}$, the Hamiltonian (\ref{NoninteractingModelWithoutDisplacement}) can be decomposed into sectors with different mirror eigenvalues, $\sigma_h = \pm 1$, by performing a unitary transformation, $V \in \text{U}(3)$, in layer space only, $c_{\vec{r};\rho,l,\eta,s} = V_{l,\ell} \psi_{\vec{r};\rho,\ell,\eta,s}$. As was pointed out before \cite{Khalaf_2019} and detailed in \appref{ContinuumModel}, the Hamiltonian in the mirror-even sector ($\sigma_h=+1$, $\ell=1,2$) turns out to be that of TBG with an interlayer hopping renormalized by a factor of $\sqrt{2}$. In the mirror-odd sector ($\sigma_h=-1$, $\ell=3$) the Hamiltonian is that of single-layer graphene (without any moir\'e modulation). This can be clearly seen in the spectrum shown in \figref{TrilayerBands}(b), which exhibits two almost-flat TBG-bands (shown in red) per spin and valley and unreconstructed Dirac cones at K (K$'$) associated with the graphene sector of valley $\eta=+$ ($\eta=-$). As such, there are three Dirac cones at K (and three at K$'$), one belonging to the graphene sector of MSTG in a single valley only and two Dirac crossings belonging to the TBG sector of both valleys. As can be seen, there are additional remote bands of both subspaces due to back-folding into the MBZ.

In addition to reflection symmetry, the Hamiltonian in \equref{NoninteractingModelWithoutDisplacement} has more point symmetries, forming the group $C_{6h}$: the model is invariant under three-fold rotational symmetry $C_{3z}$, with action, $C_{3z}:\, c_{\vec{r}} \rightarrow e^{i\frac{2\pi}{3}\rho_z\eta_z} c_{C_{3z}\vec{r}}$, which is also an exact symmetry of the moir\'e lattice in \figref{TrilayerBands}(a). While not an exact lattice symmetry, two-fold rotation perpendicular to the plane, $C_{2z}$, is a good approximate symmetry for small twist angles; in fact, it is an exact symmetry of the continuum in \equref{NoninteractingModelWithoutDisplacement} which can be verified by applying its action, $C_{2z}:\, c_{\vec{r}} \rightarrow \eta_x\rho_x c_{-\vec{r}}$. 

Besides these point symmetries, the model also exhibits the following exact internal symmetries: since there is no coupling between the valleys, it is invariant under a valley U(1) transformation, $\text{U}(1)_v: \, c_{\vec{r}} \rightarrow e^{i\eta_z\varphi} c_{\vec{r}}$. In combination with the absence of spin-orbit coupling, it is further invariant under the separate spin-rotation in each valley, forming the group $\text{SU}(2)_+ \times \text{SU}(2)_-$ with $\text{SU}(2)_{\pm}$ acting as $c_{\vec{r}} \rightarrow e^{i \vec{\varphi}\cdot \vec{s} (\eta_0\pm\eta_z)/2}c_{\vec{r}}$.
Furthermore, there is time-reversal symmetry, which is associated with the anti-unitary operator $\Theta$ with $\Theta c_{\vec{r}} \Theta^\dagger = \eta_x c_{\vec{r}}$; unless stated otherwise, we will always refer to this form of (spinless) time-reversal symmetry throughout the text, although combinations with spin-rotations (spinful time-reversal), $\Theta_s$, and with $\text{U}(1)_v$ rotations, $\widetilde{\Theta}$, will play a role further below.
Since the mirror symmetry protects any mixing between the graphene and TBG subspaces and $\text{U}(1)_v$ any mixing between the two valleys, the aforementioned Dirac crossings at K and K$'$ are protected by the combination $C_{2z}\Theta$, exactly as in graphene and TBG.

Focusing on specific limits, there are also additional internal symmetries, similar to TBG \cite{Tarnopolsky_2019,bultinck2019ground,bernevig2020tbg}: when $w_0=0$, $H^{\text{Full}}_{0,1}$ in \equref{NoninteractingModelWithoutDisplacement} changes its sign, $H^{\text{Full}}_{0,1}\rightarrow -H^{\text{Full}}_{0,1}$, if we apply the chiral symmetry operator $C: \, c_{\vec{r}} \rightarrow \rho_z L_C c_{\vec{r}}$ where $L_C$ is a matrix in layer space given by $L_C = V \text{diag}(1,1,-1) V^\dagger$. Therefore, $w_0=0$ will be referred to as the \textit{chiral limit} \cite{Tarnopolsky_2019}. Among other consequences for the interaction terms to be discussed below, the band structure has to be symmetric about zero energy at every given momentum $\vec{k}$ when $w_0=0$, i.e., $\text{spec}_{\vec{k},\eta} = -\text{spec}_{\vec{k},\eta}$, with $\text{spec}_{\vec{k},\eta} = \{\epsilon_{n,\eta}(\vec{k}),n\in\mathbb{Z}\}$.

Finally, when the additional rotation of the sublattice matrices in the first line of \equref{NoninteractingModelWithoutDisplacement} is neglected, $\vec{\rho}^\eta_{\theta_l} \rightarrow \vec{\rho}^\eta_{\theta=0}$, the ``unitary particle-hole symmetry'', previously discussed in TBG \cite{bultinck2019ground,bernevig2020tbg}, can be extended to MSTG: defining the unitary operator $P$ which acts as $P\psi_{\vec{r};l=1}P^\dagger = \eta_z\psi_{-\vec{r};l=2}$, $P\psi_{\vec{r};l=2}P^\dagger = - \eta_z\psi_{-\vec{r};l=1}$, and $P\psi_{\vec{r};l=3}P^\dagger = i \rho_y\eta_y \psi_{-\vec{r};l=3}$, one finds 
$PH^{\text{Full}}_{0,1}P^\dagger=-H^{\text{Full}}_{0,1}$, showing that the spectrum 
must obey $\text{spec}_{\vec{k}} = -\text{spec}_{-\vec{k}}$ where $\text{spec}_{\vec{k}} = \cup_\eta \text{spec}_{\vec{k},\eta}$. Since we will be focusing on small twist angles $\theta$ below, we will always assume that $\vec{\rho}^\eta_{\theta_l}$ has been replaced by $\vec{\rho}^\eta_{\theta=0}$ in \equref{NoninteractingModelWithoutDisplacement} when analyzing symmetries. We will also later set $\theta=0$ in our numerics.

For the detailed form of the model and the action of the symmetries in momentum space we refer the reader to \appref{ContinuumModel} and \ref{DiscussionOfSymmetries}, respectively. A brief list of the symmetries of $H_0$ and when they apply can be found in \tableref{ActionOfSymmetriesMainText}.

The primary interaction term we shall consider, in \equref{GeneralFormOfDensityDensityInteraction}, will preserve the symmetries of the continuum model. However, we will also consider the consequences of a Hund's coupling in \equref{HundsCouplingInteraction}, which will break the $\text{SU}(2)_+ \times \text{SU}(2)_-$ symmetry down to a more realistic $\text{SU}(2)_s$ (the group of simultaneous spin rotations in the two valleys).

\subsection{Adding a Displacement Field}
Motivated by recent experimental discoveries of electric-field-tunable correlated physics in MSTG \cite{Park_2021,Hao_2021,2021arXiv210312083C},
we now extend the model for MSTG in \equref{NoninteractingModelWithoutDisplacement} and include a perpendicular displacement field $D_0$. Finite $D_0$ results in a potential difference between the top and bottom layers of graphene. Suppressing all indices except layer $l$, a displacement field is represented in the continuum model as
\begin{equation}
    H^{\text{Full}}_{0,2}=D_0\sum_{\vec{r}}\left[c^\dagger_{\vec{r};l=1}c^\pdagger_{\vec{r};l=1}-c^\dagger_{\vec{r};l=3}c^\pdagger_{\vec{r};l=3}\right]
    \label{DisplacementField}
\end{equation}
and the full Hamiltonian becomes $H^{\text{Full}}_{0} = H^{\text{Full}}_{0,1} + H^{\text{Full}}_{0,2}$.
The displacement field $D_0$ breaks the mirror symmetry $\sigma_h$, resulting in hybridization between bands in the TBG and graphene sectors; this hybridization can only take place between graphene and TBG bands of the same valley and spin flavor, as $\text{U}_v(1)$ and spin-rotation symmetry [in fact the full $\text{SU}(2)_+ \times \text{SU}(2)_-$] are preserved by $H^{\text{Full}}_{0,2}$.

\begin{figure*}[tb]
    \centering
    \begin{subfigure}{.27\textwidth}
    \includegraphics[width=.98\textwidth]{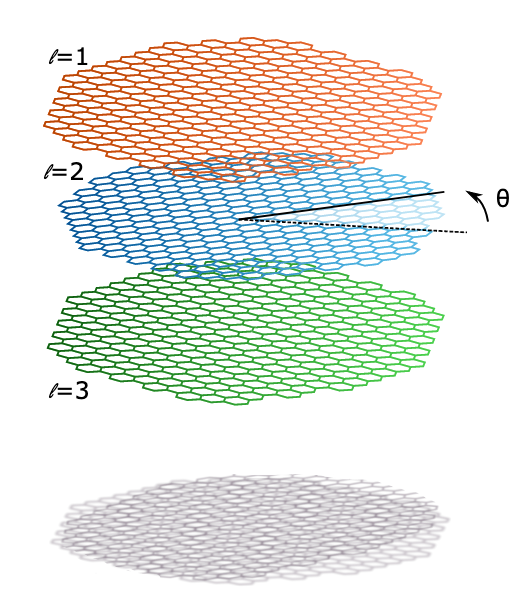}
    \caption{}
    \end{subfigure}
    \hfill
    \begin{subfigure}{.565\textwidth}
    \includegraphics[width=1\textwidth]{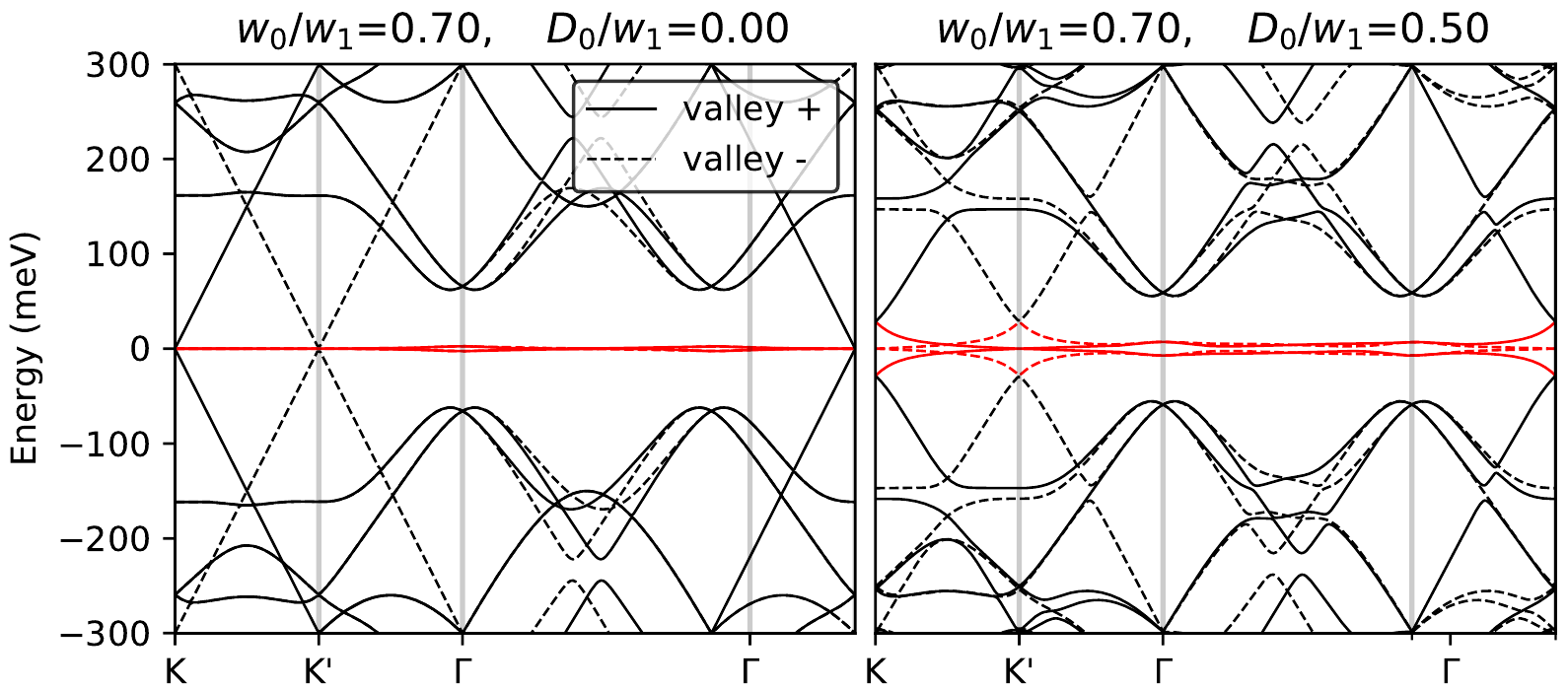}
    \caption{}
    \end{subfigure}
    \hfill
    \begin{subfigure}{.15\textwidth}
    \includegraphics[width=.98\textwidth]{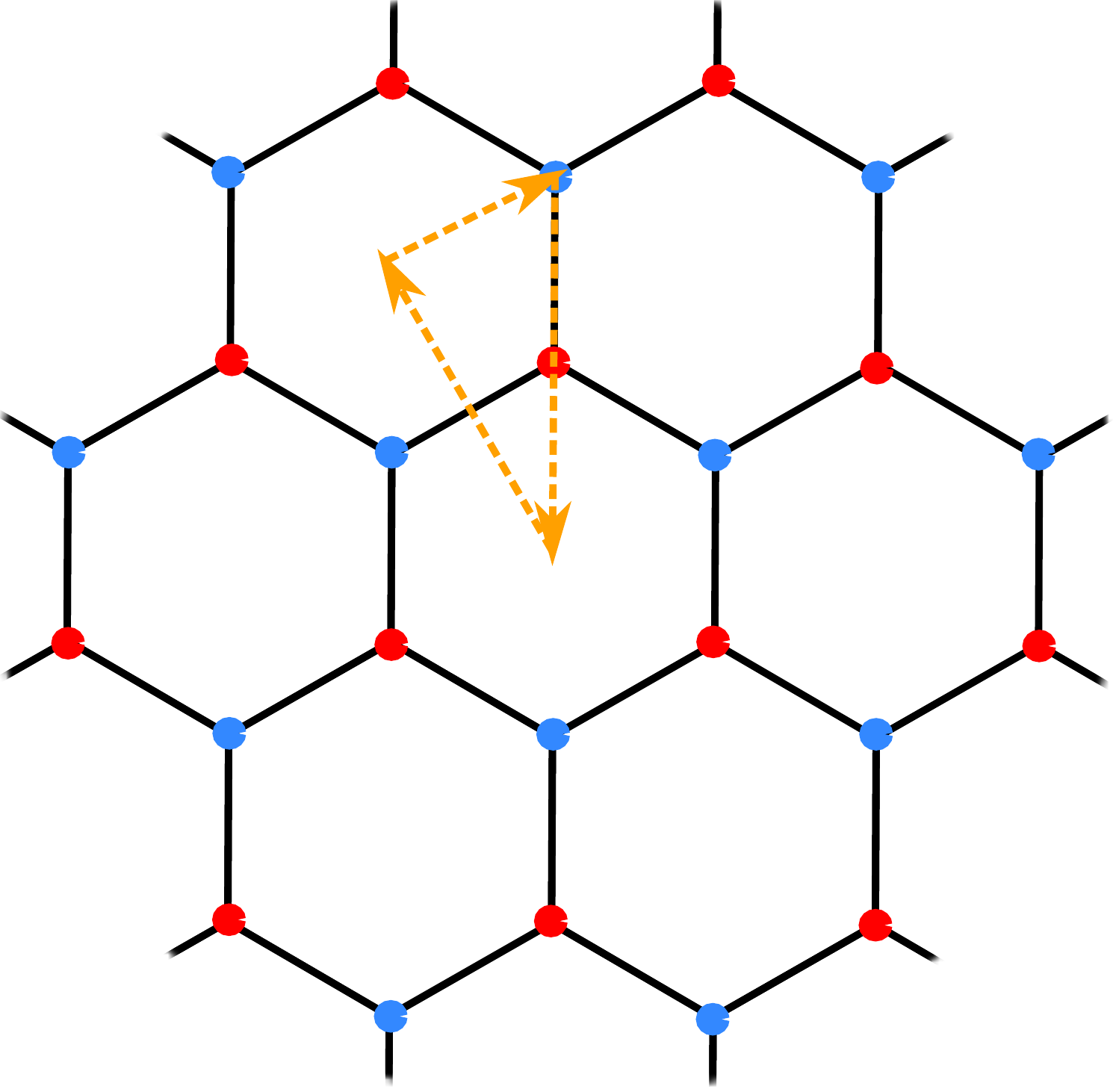}
    \caption{}
    \end{subfigure}
    
    \caption{ (a) Schematic of the geometry of MSTG. The top and bottom layers shown in red and green respectively are aligned, while the middle layer (blue) is twisted at relative angle $\theta$ to the top and bottom layers. (b) Band structure of MSTG at $D_0=0$ (left) and $D_0>0$ (right) for $w_1=124$ meV and twist angle $\theta=1.53^\circ$ along the one-dimensional cut through the MBZ shown in (c). We represent the spin-degenerate bands in the two different valleys $\eta=\pm$ with solid ($\eta=+$) and dotted ($\eta=-$) lines.
    At $D_0=0$, the band structure is given by that of TBG and single-layer graphene and we have colored the (two per spin and valley) quasi-flat TBG bands red. For $D_0>0$, the bands of these two subsystems mix but we continue to label the two bands per spin and valley closest to the Fermi as ``TBG-like'' bands, indicated in red.}
    \label{TrilayerBands}
\end{figure*}

Furthermore, when $D_0$ is finite, the system retains $C_{2z}$, $C_{3z}$, leading to the point group $C_{6}$. Time-reversal symmetry, $\Theta$, and, in the limit $w_0=0$, chiral symmetry, $C$, also persist. However, the unitary particle hole symmetry $P$ is no longer an exact symmetry for finite $D_0$.

At the K (K$'$) point of the MBZ, the Dirac crossings which belong to the valley $\eta=+$ ($\eta=-$) will hybridize as $D_0$ increases with the graphene bands of the same valley which also exhibit a Dirac cone at the Fermi level; this results in the Dirac points, which were pinned at the Fermi level at $D_0=0$, being pushed away from the Fermi level. As $C_{2z}\Theta$ symmetry is preserved for $D_0\neq 0$, these Dirac crossings cannot be gapped out. 
As a consequence of the absence of a graphene cone to hybrize with, the Dirac point of TBG in the other valley, $\eta=-$ ($\eta=+$), remains pinned at the Fermi level as $D_0$ increases. These features can be clearly observed in the band structure of MSTG with finite displacement field shown in \figref{TrilayerBands}(b).

\subsection{Projected low-energy model}\label{ProjectedLowEnergyBands}
In order to make the analytical and numerical study of interactions feasible in the system, we restrict our Hamiltonian to a finite set of bands in the vicinity of the Fermi level. 
Going forward, we will denote the two bands closest to the Fermi level per spin and valley as \textit{TBG-like bands} and the next two closest bands per spin and valley as \textit{graphene-like bands}. All other bands at energies further away from the Fermi level will be referred to as \textit{remote bands}.
When $D_0=0$, the TBG-like bands are labeled by mirror symmetry eigenvalue $\sigma_h=+1$ and can be identified exactly with the two flat bands of TBG. The graphene-like bands at $D_0=0$ are identified exactly with the bands of the continuum model of graphene in the vicinity of the K and K$'$ points. For finite $D_0$, the bands with different eigenvalues hybridize but we will retain our naming convention for the four bands closest to the Fermi level, i.e., the bands indicated in red in \figref{TrilayerBands} are TBG-like bands.

Note that away from the K points, the graphene-like bands are technically identified with the first remote bands of TBG with $\sigma_h=+1$ for $D_0=0$ within this convention. 
In our analytical calculations below, however, we will restrict the analysis to the TBG-like bands in the full MBZ and to the graphene-like bands only in the vicinity of the K and K$'$ points where they retain their graphene-like identity. Denoting the associated creation operators in these two sets of bands by $b^\dagger_{\vec{k};p,\eta,s}$ and $g^\dagger_{\vec{k};p,\eta,s}$, where $\vec{k}\in\text{MBZ}$ is the momentum, $p=+$ ($p=-$) labels the upper (lower) band in each sector, $\eta$ is the valley, and $s$ the spin quantum number of the electrons, the non-interacting Hamiltonian projected to these bands becomes
\begin{align}\begin{split}
    H_0 &= \sum_{\vec{k}\in\text{MBZ}}  W_{\text{TBG}}\epsilon_{(b,p),\eta}(\vec{k}) b^\dagger_{\vec{k};p,\eta,s}b^\pdagger_{\vec{k};p,\eta,s} \\
    & + \sum_{\vec{k}\in\text{MBZ}'} \epsilon_{(g,p),\eta}(\vec{k}) g^\dagger_{\vec{k};p,\eta,s}g^\pdagger_{\vec{k};p,\eta,s}. \label{FormOfNoninteractingHam}  \end{split}
\end{align}
Here the extra prime in MBZ$'$ in the second line indicates that we restrict the graphene-like degrees of freedom to the vicinity of the K and K$'$ points in our analytical calculations. Note that we further introduced the dimensionless parameter $W_{\text{TBG}} \in [0,1]$ that will allow us to organize the perturbation theory of \secref{PerturbationTheoryInBandwidth} in the TBG-like bandwidth. While the physical system corresponds to $W_{\text{TBG}}=1$, we define the \textit{flat limit} as $W_{\text{TBG}}=0$, which will play an important role below. Note that the Hamiltonian in \equref{NoninteractingModelWithoutDisplacement} realizes the flat limit exactly, $\epsilon_{(b,p),\eta}(\vec{k})=0$, only when $w_0=0$, $\theta$ is at the magic angle, and when $D_0=0$.  

In our numerics, we will keep all momenta of both sets of (in total $16$) bands in \equref{FormOfNoninteractingHam} and continue to use $g^\dagger_{\vec{k};p,\eta,s}$ as creation operators for the second lowest set of bands throughout the MBZ.
To check convergence, we will also study the impact of adding additional remote bands, as described in \appref{VaryingParameterInNumerics}.

In order to fix the phases of the wavefunctions of the TBG-like and graphene-like bands, we specify how they transform under the action of the various discrete symmetries discussed above or, equivalently how they act on the electronic operators $b^\dagger_{\vec{k};p,\eta,s}$ and $g^\dagger_{\vec{k};p,\eta,s}$. While a thorough discussion can be found in \appref{DiscussionOfSymmetries}, these representations are summarized in \tableref{ActionOfSymmetriesMainText}.

\begin{table}[tb]
\begin{center}
\caption{Summary of discrete single-particle symmetries (denoted $S$) of our model and when they apply (last column). For convenience of the reader and future reference, we show a redundant set of symmetries. We also indicate whether they are unitary or anti-unitary symmetries (second column) and whether they commute $+$ or anti-commute $-$ (third column) with the non-interacting Hamiltonian $H^{\text{Full}}_0$. Finally, we specify their action on the low-energy field operators in order to fix the phase of the Bloch states, using $\sigma_j$ to denote Pauli matrices acting on the upper/lower bands subspace.}
\label{ActionOfSymmetriesMainText}
\begin{ruledtabular}
 \begin{tabular} {ccccccc} 
$S$ & unitary & $[\cdot,\cdot]_{\pm}$ &  $S b_{\vec{k}}S^\dagger$ & $S g_{\vec{k}}S^\dagger$ & condition \\ \hline
$C_{2z}$ & \cmark  & $+$    & $\eta_1\sigma_0b_{-\vec{k}}$ &  $\eta_1\sigma_0g_{-\vec{k}}$ & ---  \\
$C_{3z}$ & \cmark  & $+$   & $\eta_0\sigma_0b_{C_{3z}\vec{k}}$ &  $\eta_0\sigma_0g_{C_{3z}\vec{k}}$ & ---  \\
$\sigma_{h}$ & \cmark  & $+$   & $\eta_0\sigma_0b_{\vec{k}}$ &  $-\eta_0\sigma_0g_{\vec{k}}$ & $D_0=0$  \\ \hline
$\Theta$ & \xmark  & $+$   & $\eta_1\sigma_0b_{-\vec{k}}$ &  $\eta_1\sigma_0g_{-\vec{k}}$ & ---  \\
$\Theta_s$ & \xmark  & $+$   & $s_2\eta_1\sigma_0b_{-\vec{k}}$ &  $s_2\eta_1\sigma_0g_{-\vec{k}}$ & ---  \\
$\widetilde{\Theta}$ & \xmark  & $+$   & $\eta_2\sigma_0b_{-\vec{k}}$ &  $\eta_2\sigma_0g_{-\vec{k}}$ & ---  \\
$C$ & \cmark  & $-$  & $\eta_3\sigma_2 b_{\vec{k}}$ &  $\eta_3\sigma_2g_{\vec{k}}$ & $w_0=0$  \\
$P$ & \cmark  & $-$  & $-i \eta_3\sigma_2 b_{-\vec{k}}$ &  $ i\eta_1\sigma_2 g_{-\vec{k}}$ & $D_0=0$, $\vec{\rho}_{\theta\rightarrow 0}$  \\ \hline
$C_{2z}\Theta$ & \xmark  & $+$   & $\eta_0\sigma_0b_{\vec{k}}$ &  $\eta_0\sigma_0g_{\vec{k}}$ & ---  \\
$C_{2z}P$ & \cmark  & $-$   & $\eta_2\sigma_2 b_{\vec{k}}$ &  $ i\sigma_2 g_{\vec{k}}$ & $D_0=0$, $\vec{\rho}_{\theta\rightarrow 0}$  \\
 \end{tabular}
\end{ruledtabular}
\end{center}
\end{table}

Based on these symmetry representations, one finds that the dispersions are constrained to have the form 
\begin{align}\begin{split}
    \epsilon_{(b,p),\eta}(\vec{k}) &= p\, \bar{E}^b_{0}(\vec{k}) + \eta \, \bar{E}^b_{1}(\vec{k})\,w_0 \\ &\quad + p\,\eta\, \bar{E}^b_{2}(\vec{k})\,D_0 + \bar{E}^b_{3}(\vec{k})\,w_0 D_0  
\label{TBGLikeDispersion}\end{split}\end{align}
and
\begin{equation}
    \epsilon_{(g,p),\eta}(\vec{k}) = p\, \bar{E}^g_{0}(\vec{k};\eta) +  \bar{E}^g_{1}(\vec{k};\eta) \, w_0 D_0. \label{GrapheneLikeDispersion}
\end{equation}
To illustrate the changes of the form of the band structures when symmetries are broken by $w_0\neq 0$ and $D_0\neq 0$, we introduced $\bar{E}^b_{j}(\vec{k})$ and $\bar{E}^g_{j}(\vec{k};\eta)$ which are functions of $w_0$ and $D_0$ that stay finite when $w_0, D_0\rightarrow 0$. 

As can be seen, the two terms in the first row of \equref{TBGLikeDispersion} just correspond to the TBG band structure (with, $w_0 = 0$, or without, $w_0\neq 0$, chiral symmetry). The second line shows that the hybridization with the graphene bands for $D_0\neq 0$ distorts the TBG band structure in a way not present in TBG, which is related to the $D_0$-induced breaking of $C_{2z} P$ symmetry. One of these additional terms leads to superexchange processes in the TBG-like bands that are not present in TBG and will be discussed in \secref{PerturbationTheoryInBandwidth}.
In \equref{GrapheneLikeDispersion}, the first term, $\bar{E}^g_{0}(\vec{k};\eta)$, simply becomes the graphene Dirac dispersion for $D_0\rightarrow 0$. We further see that the property $\epsilon_{(g,+),\eta}(\vec{k}) = -\epsilon_{(g,-),\eta}(\vec{k})$ is only violated if $w_0$ and $D_0$ are simultaneously non-zero. This is expected, as the graphene-like bands can only ``feel'' the broken chiral symmetry if they hybridize with the TBG-like bands. For the parameters in \figref{TrilayerBands}(b), this is a rather weak effect around the Dirac cones.

\subsection{Interactions and form factors}
As our goal is to study the interacting behavior of MSTG, we next add interaction terms to the Hamiltonian. We assume that these interactions in the full continuum model are of the density-density form. Our full interaction Hamiltonian thus reads as
\begin{equation}
    H_1^{\text{Full}} = \frac{1}{2N}\sum_{\vec{q}} V(\vec{q})  : \rho^{\text{Full}}_{\vec{q}}   \rho^{\text{Full}}_{-\vec{q}} : \label{GeneralFormOfDensityDensityInteraction}
\end{equation}
where $: \ldots :$ denotes normal ordering, $N$ is the number of moir\'e sites, and $\rho^{\text{Full}}_{\vec{q}}$ is the density operator of momentum $\vec{q}$ of the continuum electron operators $c_{\vec{r}}$ in \equref{NoninteractingModelWithoutDisplacement}. Note $\vec{q}\in\mathbbm{R}$ in the sum in \equref{GeneralFormOfDensityDensityInteraction} which is \textit{not} restricted to the MBZ. For our analytical discussions below, we will not have to specify the explicit form of $V(\vec{q})$ but only use $V(\vec{q}) > 0$ and $V(\vec{q}) = V(-\vec{q})$; consequently, our analytical results will be valid regardless of the details of the screening processes at high energies and of nearby gates and/or substrates, that affect the detailed form $V(\vec{q})$. For our numerics, we will use the single-gate-screened Coulomb potential:
\begin{equation}
    V(\vec{q})=\frac{1-e^{-2|\vec{q}|d_{s}}}{2A_{\text{moir\'e}}\epsilon_0\epsilon|\vec{q}|}, \label{ExplicitFormOfVOfq}
\end{equation}
where we have normalized by the real space area of the moir\'e unit cell $A_{\text{moir\'e}}$.
We vary the screening length $d_s$ in our numerical calculations from $d_s=10$ to $80$ nm and find relatively little dependence of the relative energies between phases and no qualitative change in the ground states. We will therefore take $d_s=40$ nm as our default value. We also vary the dielectric constant $\epsilon$ from $\epsilon=4$ to $\epsilon=15$. We find the phase boundaries depend on $\epsilon$, though no new phases emerge as $\epsilon$ is varied [the effect of varying interaction parameters is further discussed in \appref{VaryingParameterInNumerics}, especially \figref{VaryEpsilon}]. Unless otherwise specified, we will take $\epsilon=7$.

We note that the above interaction term has all the symmetries of the continuum model introduced in \secref{ModelAndSymmetries}, including the $\text{SU}(2)_+ \times \text{SU}(2)_-$ symmetry.
In the realistic system, there is a Hund's coupling, $J_H\neq0$, which breaks $\text{SU}(2)_+ \times \text{SU}(2)_-$ down to $\text{SU}(2)_s$. We discuss the form of the Hund's coupling in \appref{HundsCouplingAppendix} and \equref{HundsCouplingInteraction}, and will note its consequences in \secref{CandidateStatesConstruction}.

Neglecting normal ordering and replacing $\rho^{\text{Full}}_{\vec{q}} \rightarrow \tilde{\rho}^{\text{Full}}_{\vec{q}} = \rho^{\text{Full}}_{\vec{q}} - \text{const.}\times\delta_{\vec{q},0}$ in \equref{GeneralFormOfDensityDensityInteraction} leads to a new form of the interaction, $\widetilde{H}_1^{\text{Full}}$, which, however, can be rewritten as $H_1^{\text{Full}}$ by a redefinition of the chemical potential (and energy reference point). This is not the case anymore after projecting $\rho^{\text{Full}}_{\vec{q}}$ and $\tilde{\rho}^{\text{Full}}_{\vec{q}}$ in $H_1^{\text{Full}}$ and $\widetilde{H}_1^{\text{Full}}$, respectively, to a finite set of bands of $H_0$. As described in more detail in \appref{FormFactorsAndInteraction}, we follow \cite{bultinck2019ground,bernevig2020tbg} and rewrite the interaction before projection such that it exhibits particle-hole symmetry with respect to  the charge neutrality point before and after projection. Denoting the electronic creation operators for band $n$, of valley $\eta$, spin $s$, and with momentum $\vec{k}\in\text{MBZ}$ by $f^\dagger_{\vec{k};n,\eta,s}$, the projected interaction becomes 
\begin{subequations}
\begin{equation}
        \widetilde{H}_1 = \frac{1}{2N}\sum_{\vec{q}} V(\vec{q})  \delta\rho_{\vec{q}}   \delta\rho_{-\vec{q}}, 
\end{equation}\label{Hinteraction}
with the symmetrized density operators
\begin{align}
    \delta \rho_{\vec{q}} = \hspace{-0.6em} \sum_{\vec{k}\in\text{MBZ}} \left[ f^\dagger_{\text{MBZ}(\vec{k}+\vec{q})} F_{\vec{k},\vec{q}} f^\pdagger_{\vec{k}} - \frac{1}{2} \sum_{\vec{G}\in\text{RL}} \delta_{\vec{q},\vec{G}}\text{tr}(F_{\vec{k},\vec{G}})\right]. \label{ExpressionJustDeltaRho}
\end{align}\label{FormOfProjectedInteractionUsed}\end{subequations}
Here $\text{MBZ}(\vec{k}) := \vec{k} - \vec{G}_{\vec{k}} \in \text{MBZ}$ for the unique reciprocal lattice vector $\vec{G}_{\vec{k}} \in \text{RL}$. We suppressed all indices of $f_{\vec{k}}$ and $f^\dagger_{\vec{k}}$, which should be viewed as column and row vectors in band, valley, and spin space and introduced the matrix-valued \textit{form factors} $F_{\vec{k},\vec{q}}$, which contain all the microscopic details of the wavefunctions of the bands of $H_0$ [see \equref{DefinitionOfFormFactors} for a formal definition]. Note that \equref{FormOfProjectedInteractionUsed} holds for any subset of bands that we want to keep. As already discussed in \secref{ProjectedLowEnergyBands} above, we will only keep the TBG-like bands and graphene-like bands (around the K/K' points) in the analytics, while we will allow for more bands in the HF numerics.

We refer the interested reader to \appref{FormFactorsAndInteraction}, where a detailed discussion of constraints on $F_{\vec{k},\vec{q}}$ resulting from symmetries, Hermiticity, and the structure of $H_0$ can be found, and here only state a few properties of the form factors that we will explicitly refer back to in the main text. First, as a consequence of $\text{U}(1)_v$ and spin rotation symmetry, the form factors can only have non-diagonal matrix structure in band space,
\begin{equation}
    \left(F_{\vec{k},\vec{q}}\right)_{(n,\eta,s),(n',\eta,s)} = \delta_{s,s'}\delta_{\eta,\eta'} \left(F^{\eta}_{\vec{k},\vec{q}}\right)_{n,n'}.
\end{equation}
In accordance with our notation in \equref{FormOfNoninteractingHam}, we will use the multi-index notation $n=(t,p)$ where the ``type'' $t$ distinguishes between the graphene-like, $t=g$, and TBG-like, $t=b$, bands and $p=+$ ($p=-$) labels the upper (lower) band; for instance, for the electronic operators, it holds $f_{\vec{k};(b,p),\eta,s} = b_{\vec{k};p,\eta,s}$ and $f_{\vec{k};(g,p),\eta,s} = g_{\vec{k};p,\eta,s}$. As a consequence of the $\sigma_h$ symmetry for $D_0=0$, the form factors become block-diagonal in the sectors with different $\sigma_h$ eigenvalue. As such, $F_{\vec{k},\vec{q}}^{tt'}$ defined via 
\begin{equation}
    \left(F^{tt'}_{\vec{k},\vec{q}}\right)_{(p,\eta,s),(p',\eta',s')}= \delta_{s,s'}\delta_{\eta,\eta'} \left(F^{\eta}_{\vec{k},\vec{q}}\right)_{(t,p)(t',p')} \label{ThettpNotationOfF}
\end{equation}
obeys
\begin{equation}
    F^{tt'}_{\vec{k},\vec{q}} \propto \delta_{t,t'}, \quad \text{for } D_0=0.
\end{equation}
Consequently, the entire interacting Hamiltonian $H_0 + \widetilde{H}_1$ preserves the charge in the graphene and TBG system separately. This leads to simplifications, which will be exploited in \secref{ExactGroundStates} below. For $D_0\neq 0$, this is not the case anymore, as the form factors in \equref{ThettpNotationOfF} and, hence, the density operators in the interaction will \textit{not} be diagonal in the $t$ index and scatter electrons between the two different types of bands.

In the limit $D_0=0$, one can also compute the form factors in the graphene sector analytically [see \equref{FormOfGrapheneMatrixElements} for the full expression]. Most importantly for our purposes here, one finds that
\begin{equation}
    F^{gg}_{\vec{k},\vec{G}} = \delta_{\vec{G},0} \sigma_0 \eta_0 s_0, \quad \vec{G} \in\text{RL}, \quad \text{for } D_0=0, \label{SimplePropertyOfGrapheneFormFactorsMainText}
\end{equation}
where $\sigma_0$ are Pauli matrices in the ``band space'' (with indices $p=\pm$). Here only the $\vec{G}=0$ component is finite, which is related to the fact that the graphene bands do not ``feel'' the moir\'e superlattice for $D_0 = 0$.

\section{Exact groundstates at \texorpdfstring{$D_0=0$}{Lg}}\label{ExactGroundStates}
We begin our discussion of the interacting physics in the \textit{decoupled limit}, defined as $D_0=0$, where all bands can be labeled by their mirror eigenvalue $\sigma_h=\pm 1$, and the low-energy bands are those of TBG and single-layer graphene. 
While one might intuitively expect that the presence of the additional graphene Dirac cones, which have a much lower density of states than the (almost) flat bands close to the magic angle of TBG, is not strong enough to change the symmetry of the correlated insulating phase of TBG, it is \textit{a priori} not clear whether the density-density coupling between the two subsystems can also induce the same symmetry-breaking order (and potentially gap out) the graphene Dirac cones. It is further not clear whether \textit{exact} interacting ground states of MSTG can be identified in certain limits, similar to TBG \cite{2020arXiv200913530L,PhysRevLett.122.246401,bultinck2019ground}. These aspects will be addressed in this section.

\subsection{Hamiltonian and construction of eigenstates} 
To this end, let us focus on the flat limit, $W_{\text{TBG}}=0$, and postpone the perturbative treatment of the finite TBG bandwidth to \secref{PerturbationTheoryInBandwidth}. From our discussion in \secref{ModelAndSymmetries}, we can read off that the low-energy Hamiltonian of MSTG in the flat-decoupled limit is given by
\begin{equation}
    H_{\text{FD}} = H^{\text{g}} + H^{\text{b}} + \lambda H^{\text{gb}}, \label{ThreeTermsInHFD}
\end{equation}
consisting of three terms: $H^{\text{g}}$ is the Hamiltonian of (both valleys of) single-layer graphene with Coulomb repulsion [see \equref{HGraphene}], $H^{\text{b}}$ the interacting Hamiltonian of TBG in the flat limit, given by 
\begin{equation}
    H^{\text{b}} = \frac{1}{2N}\sum_{\vec{q}} V(\vec{q})  \delta\rho^{\text{b}}_{\vec{q}}   \delta\rho^{\text{b}}_{-\vec{q}} \label{FlatBandTBGHamiltonian}
\end{equation}
where we defined the projected subsystem density operators
\begin{equation}
    \delta\rho^{t}_{\vec{q}} = \hspace{-0.6em} \sum_{\vec{k}\in\text{MBZ}}\left[ t^\dagger_{\text{MBZ}(\vec{k}+\vec{q})} \left(F^{tt}_{\vec{k},\vec{q}}\right) t^\pdagger_{\vec{k}} - \frac{1}{2} \sum_{\vec{G}\in\text{RL}}\hspace{-0.1em} \delta_{\vec{q},\vec{G}}\text{tr}(F^{tt}_{\vec{k},\vec{G}}) \right]
\end{equation}
where $t=g,b$. Finally, the last term in \equref{ThreeTermsInHFD} describes the coupling between the two subsystems via a density-density interaction,
\begin{equation}
        \lambda H^{\text{gb}} = \frac{\lambda }{N}\sum_{\vec{q}} V(\vec{q})  \delta\rho^{\text{g}}_{-\vec{q}}   \delta\rho^{\text{b}}_{\vec{q}}, \label{FormOfCouplingBetweenTheTwoSubSys}
\end{equation}
where the additional prefactor $\lambda \in [0,1]$ has been introduced to adiabatically turn on the $H^{\text{gb}}$ interaction (the physical system corresponds to $\lambda=1$). 

For $\lambda=0$, the Hamiltonian is just the sum $H^{\text{g}} + H^{\text{b}}$ of the two commuting subsystem Hamiltonians. So its eigenstates are just given by (all combinations) of the individual eigenstates of graphene and TBG with the correct particle number: let us fix a certain integer filling $\nu_{\text{b}}=0,\pm 1, \pm 2, \pm 3, \pm 4$ of the TBG system. Then the graphene system will be at a corresponding filling, which we write formally as $\nu_g=\nu_g(\nu_{\text{b}})$ and its ground state will be a semimetal with ($\nu_g=0$) or without ($\nu_g\neq 0$) doping, that exhibits correlations but does not spontaneously break any symmetries; these properties are well established (theoretically and experimentally) for graphene \cite{RevModPhys.81.109}. 
Let us denote the ground-state of the graphene system at filling $\nu_g=\nu_g(\nu_{\text{b}})$ by $\ket{\Psi^{\text{g}}_{0}(\nu_{\text{b}})}$ and its (gapless) excited states with the same particle number by $\ket{\Psi^{\text{g}}_{j}(\nu_{\text{b}})}$, $j>0$. In the hypothetical absence of any correlations in the graphene subspace, $\ket{\Psi^{\text{g}}_{0}(\nu_{\text{b}})}$ ($\ket{\Psi^{\text{g}}_{j}(\nu_{\text{b}})}$) would just be Slater-determinant state(s) with the Dirac cones filled up to to the chemical potential (and some additional particle-hole excitations).

Exact ground states of the flat-band TBG Hamiltonian in \equref{FlatBandTBGHamiltonian} have been discussed previously \cite{2020arXiv200913530L}, which we will very briefly review here using our notation, in order to set the stage for the extension to MSTG. Upon defining new operators according to [referred to as ``chiral basis'' in \cite{bernevig2020tbg}; in this basis the chiral form factors, see \equref{FormFactorsToDiagonalize}, are diagonal]
\begin{equation}
    \tilde{b}^\pdagger_{\vec{k};c,\eta,s} = U_{c,p} b^\pdagger_{\vec{k};p,\eta,s}, \quad U = \frac{1}{\sqrt{2}} \begin{pmatrix} 1 & -i \\ 1 & i \end{pmatrix}, \label{TransformationInChiralBasis}
\end{equation}
consider the set of states
\begin{equation}
    \ket{\Psi_0^{\text{b}}(\nu_b)} = \prod_{\vec{k}\in\text{MBZ}} \prod_{c=\pm} \prod_{j_c=1}^{\nu_c} \tilde{b}^\dagger_{\vec{k};c,\eta^c_{j_c},s^c_{j_c}}  \ket{0^{\text{b}}} \label{ExactGroundStatesMainText}
\end{equation}
with arbitrary combinations of occupied flavors $\{\eta^{\pm}_j,s^{\pm}_j\}$ such that $\nu_++\nu_- = 4+\nu_{\text{b}}$. It was shown in \cite{2020arXiv200913530L} that 
\begin{equation}
    \delta\rho^{\text{b}}_{\vec{q}} \ket{\Psi_0^{\text{b}}(\nu_b)} =  \sum_{\vec{G}\in\text{RL}} \delta_{\vec{q},\vec{G}} R_{\vec{G}} \ket{\Psi_0^{\text{b}}(\nu_b)}, \label{EigenValueEquationMainText}
\end{equation}
with $R_{\vec{G}} = \nu_b \sum_{\vec{k}} \text{tr}[F^{bb}(\vec{k},\vec{G})]/8$ when $w_0=0$, i.e., in the chiral limit. As such, all of these states are exact eigenstates of $H^{\text{b}}$ in \equref{FlatBandTBGHamiltonian}. \refcite{2020arXiv200913530L} further showed that these states will always be ground states of $H^{\text{b}}$ for $\nu_b=0$; the same holds for all other integer $\nu_b$ as long as the flat-metric condition,
\begin{equation}
        F^{bb}_{\vec{k},\vec{G}} =  \mathbbm{1} f(\vec{G}),\qquad \forall\,\vec{k},\vec{G}, \label{FlatMetricConditionMainText}
\end{equation}
is not violated by a significantly large amount. Furthermore, when turning on $w_0\neq 0$, the subset of states in \equref{ExactGroundStatesMainText} with $\eta_j^+=\eta_j^-$ and $s_j^+=s_j^-$, which are necessarily at even integer $\nu_b$, still obey \equref{EigenValueEquationMainText} and remain ground states of $H_b$ [unless $\nu_b=\pm 2$ and \equref{FlatMetricConditionMainText} is sufficiently violated].

Having established the spectrum of $H_{\text{FD}}$ in \equref{ThreeTermsInHFD} for $\lambda=0$, let us next discuss what happens once $\lambda$ is turned on. Using the fact that the graphene form factors obey \equref{SimplePropertyOfGrapheneFormFactorsMainText}, we show in \appref{TurningOnLambda} that $H^{\text{gb}}$ can be rewritten in the low-energy spectrum of MSTG as
\begin{equation}
    H^{\text{gb}} =\frac{1}{N}\sum_{\vec{q}} V(\vec{q}) \delta\rho^{\text{g}}_{-\vec{q}}   \left(\delta\rho^{\text{b}}_{\vec{q}} - \sum_{\vec{G}\in\text{RL}} \delta_{\vec{q},\vec{G}} R_{\vec{G}} \right) + E_0(\nu_b),
\end{equation}
where $E_0(\nu_b)$ is just a constant energy.
So we immediately see that the property (\ref{EigenValueEquationMainText}) of all of the exact TBG states $\ket{\Psi_0^{\text{b}}(\nu_b)}$ defined above, implies
\begin{equation}
    H^{\text{gb}} \ket{\Psi^{\text{g}}_{j}(\nu_b)}\ket{\Psi_0^{\text{b}}(\nu_b)} = E_0(\nu_b) \ket{\Psi^{\text{g}}_{j}(\nu_b)}\ket{\Psi_0^{\text{b}}(\nu_b)}.
\end{equation}
Consequently, all of the states $\ket{\Psi^{\text{g}}_{j}(\nu_b)}\ket{\Psi_0^{\text{b}}(\nu_b)}$ remain \textit{exact} eigenstates of the full MSTG Hamiltonian $H_{\text{FD}}$ in the flat-decoupled limit, at arbitrary $\lambda$.

Whether the states $\ket{\Psi^{\text{g}}_{0}(\nu_b)}\ket{\Psi_0^{\text{b}}(\nu_b)}$ will also remain the exact ground states is a more subtle question: since the states $\ket{\Psi_0^{\text{b}}(\nu_b)}$ break symmetries, $H_{\text{FD}}|_{\lambda=0}$ will have a gapless Goldstone spectrum. In principle, an arbitrarily small $\lambda$ could lower the energy of some of those states below that of $\ket{\Psi^{\text{g}}_{0}(\nu_b)}\ket{\Psi_0^{\text{b}}(\nu_b)}$. However, we show in \appref{GoldstoneModes} that is not the case if $\nu_b=0$ or $w_0=0$ or \equref{FlatMetricConditionMainText} holds. Therefore, a finite $\lambda > 0$ is required before $\ket{\Psi^{\text{g}}_{0}(\nu_b)}\ket{\Psi_0^{\text{b}}(\nu_b)}$ cease to be the exact ground states. 

\subsection{Discussion of ground states}
Taken together, we have shown that the states $\ket{\Psi^{\text{g}}_{j}(\nu_b)}\ket{\Psi_0^{\text{b}}(\nu_b)}$ where $\ket{\Psi^{\text{g}}_{j}(\nu_b)}$ is just the spectrum of single-layer graphene at filling $\nu_g=\nu_g(\nu_{\text{b}})$ and $\ket{\Psi_0^{\text{b}}(\nu_b)}$ is any of the states in \equref{ExactGroundStatesMainText} are exact eigenstates of the MSTG Hamiltonian in the chiral-flat-decoupled limit ($w_0=W_{\text{TBG}}=D_0=0$), $H_{\text{FD}}|_{w_0=0}$, for any integer $\nu_b$. Furthermore, there is a finite region of $\lambda$ for which $\ket{\Psi^{\text{g}}_{0}(\nu_b)}\ket{\Psi_0^{\text{b}}(\nu_b)}$ will remain a groundstate of MSTG if $\ket{\Psi_0^{\text{b}}(\nu_b)}$ is a groundstate of TBG (recall that $\ket{\Psi_0^{\text{b}}(\nu_b)}$ is guaranteed to be a groundstate of TBG for $\nu_b=0$ without further assumptions while it requires that the flat-metric condition is not too strongly violated for $\nu_b\neq 0$). Finally, away from the chiral limit $w_0\neq 0$, the subset of states in \equref{ExactGroundStatesMainText} with $\eta_j^+=\eta_j^-$ and $s_j^+=s_j^-$ are known to be ground states of TBG in the flat limit for integer $\nu_b$ and if the flat metric condition holds \cite{2020arXiv200913530L}. Our analysis shows that these states remain exact eigenstates for $\lambda \neq 0$ and also ground states for $|\lambda| < \lambda_c > 0$ in the non-chiral-flat-decoupled limit ($W_{\text{TBG}}=D_0=0$, $w_0\neq 0$).

In all of these limits, we see that the graphene subsystem retains its (correlated but symmetry-unbroken and, depending on $\nu_b$, doped) semimetallic properties for all integer filling fractions $\nu_b$. This is consistent with experiment, where quantum oscillations indicate a dispersive Dirac cone at $D_0=0$ \cite{Park_2021,Hao_2021}. Furthermore, the exact eigenstates established above will be used as our starting point for further analytical considerations in \secref{AnalyticalPerturbationTheory} and their product-state nature motivates our HF numerical study of the problem in \secref{HFNumericsSection}. Both numerics and analytics will complement the discussion presented above by (i) validating the stability of the Dirac cones at $D_0=0$ in schemes that do not rely on $\lambda$ being small and (ii) by tuning away from the exactly solvable limits ($W_{\text{TBG}}=D_0\neq 0$) and (iii), for the numerics, including additional remote bands.

\begin{table*}[tb]
\begin{center}
\caption{We list the different candidate phases in the TBG-like subspace, constructed as the discrete set of states that are part of the large manifold of exact ground states in the chiral-flat-decoupled limit, see \secref{ExactGroundStates}, but transform under the irreducible representations of the symmetries of the real system. Here $\vec{1}$ ($\vec{3}$) is the singlet (triplet) representation of $\text{SU}(2)_s$ and $0$ ($1$) the one(two)-dimensional representation of $\text{U}(1)_v$. For future reference in \secref{Superconductivity}, we list the behavior ($\pm$ denoting even/odd, and \xmark~indicating absence) under both spinful ($\Theta_s$), and valley ($\widetilde{\Theta}$) time-reversal symmetry, see \tableref{ActionOfSymmetriesMainText}. The last two columns indicate which states are Hund's partners \cite{PhysRevResearch.2.033062}, i.e., transform into each other when reversing the sign of the Hund’s  coupling $J_H$ while being exactly degenerate in the $\text{SU}(2)_+ \times \text{SU}(2)_-$-symmetric limit, and which sign of $J_H$ favors the respective state.}
\label{ListOfOrderParameters}
\begin{ruledtabular}
 \begin{tabular} {cccccccccc} 
Type & Short form & $Q^{b}$ & $\widetilde{Q}^{b}$ & SU(2)$_s$ & U(1)$_v$ & $C_{2z}$ & $\Theta_s/\widetilde{\Theta}$ & Hund's part. & $J_H$ \\ \hline
spin polarized & SP & $\sigma_0\eta_0\vec{s}$ & $\widetilde{\sigma}_0\eta_0\vec{s}$ & $\vec{3}$ & $0$ & \cmark & $-$/\xmark & SVP & $<0$   \\
valley polarized & VP & $\sigma_0\eta_z s_0$ & $\widetilde{\sigma}_0\eta_z s_0$ & $\vec{1}$ & $0$ & \xmark & $-$/$-$ & --- & $0$ \\
$\Theta$-even IVC & IVC$_{+}$ & $\sigma_0\eta_{x,y}s_0$ & $\widetilde{\sigma}_0\eta_{x,y}s_0$ & $\vec{1}$ & $1$ & \cmark & $+$/$-$ & SIVC$_{+}$ & $>0$ \\
$\Theta$-odd IVC & IVC$_{-}$ & $\sigma_y\eta_{x,y}s_0$ & $\widetilde{\sigma}_z\eta_{x,y}s_0$ & $\vec{1}$ & $1$ & \cmark & $-$/$+$ & SIVC$_{-}$ & $>0$  \\
$\Theta$-odd, sublattice pol./Hall & SLP$_{-}$ & $\sigma_y \eta_0 s_0$ & $\widetilde{\sigma}_z \eta_0 s_0$ & $\vec{1}$ & $0$ & \cmark & $-$/$-$ &  --- & $0$  \\
$\Theta$-even, sublattice pol./valley Hall & SLP$_{+}$ & $\sigma_y \eta_z  s_0 $ & $\widetilde{\sigma}_z \eta_z  s_0 $ & $\vec{1}$ & $0$ & \xmark & $+$/$+$ &  --- & $0$  \\ 
$\Theta$-odd, spin-sublattice-pol./spin Hall & SSLP$_{-}$ &  $\sigma_y \eta_0 \vec{s}$ & $\widetilde{\sigma}_z \eta_0 \vec{s}$ & $\vec{3}$ & $0$ & \cmark & $+$/\xmark  & SSLP$_{+}$ & $<0$   \\ \hline
spin-valley polarized & SVP & $\sigma_0 \eta_z\vec{s}$ & $\widetilde{\sigma}_0 \eta_z \vec{s}$ & $\vec{3}$ & $0$ & \xmark & $+$/\xmark &  SP & $>0$  \\
$\Theta$-even, spin-pol. IVC & SIVC$_{+}$ & $\sigma_0\eta_{x,y}\vec{s}$ & $\widetilde{\sigma}_0\eta_{x,y}\vec{s}$ & $\vec{3}$ & $1$ & \cmark & $-$/\xmark & IVC$_{+}$ & $<0$  \\
$\Theta$-odd, spin-pol. IVC & SIVC$_{-}$ & $\sigma_y\eta_{x,y}\vec{s}$ & $\widetilde{\sigma}_z\eta_{x,y}\vec{s}$ & $\vec{3}$ & $1$ & \cmark & $+$/\xmark & IVC$_{-}$ & $<0$ \\
$\Theta$-even, spin-subl. pol./spin-valley Hall & SSLP$_{+}$ &  $\sigma_y \eta_z \vec{s}$ & $\widetilde{\sigma}_z \eta_z \vec{s}$ & $\vec{3}$ & $0$ & \xmark & $-$/\xmark & SSLP$_{-}$ & $>0$ \\
 \end{tabular}
 \end{ruledtabular}
\end{center}
\end{table*}

\subsection{Resultant candidate states}\label{CandidateStatesConstruction}
To build the foundation for these additional analytical and numerical computations, we will use the exact (and highly degenerate) ground states established above in the chiral-flat-decoupled limit to construct a finite set of candidate phases and their respective order parameters. 

We first define the \textit{correlation matrix} $P_{\vec{k}}$ with elements
\begin{equation}
    \left(P_{\vec{k}}\right)_{(t,p,\eta,s),(t',p',\eta',s')} := \braket{\Psi_0|f^\dagger_{\vec{k};(t,p),\eta,s} f^\pdagger_{\vec{k};(t',p'),\eta',s'}|\Psi_0} \label{DefinitionOfCorrelator}
\end{equation}
to characterize a given ground state $\ket{\Psi_0}$ of MSTG. Hermiticity implies $P^\dagger_{\vec{k}} = P^\pdagger_{\vec{k}}$. As is common, we further write $P_{\vec{k}}=\frac{1}{2}(\mathbbm{1}+Q_{\vec{k}})$ and will use $Q_{\vec{k}}$ as our ``order parameter'' to characterize the (potentially symmetry-broken) structure of $\ket{\Psi_0}$. It must obey
\begin{equation}
    Q_{\vec{k}}^\dagger=Q^\pdagger_{\vec{k}}, \quad \frac{1}{N}\sum_{\vec{k}}\text{tr}\left[Q_{\vec{k}}\right]=\nu_b+\nu_g \equiv \nu.
\end{equation}

As we have seen above, the ground states of MSTG in the flat-decoupled limit ($W_{\text{TBG}}=D_0=0$) obey
\begin{equation}
    \left(Q_{\vec{k}}\right)_{(t,p,\eta,s),(t',p',\eta',s')} = \delta_{t,t'} \left(Q^t_{\vec{k}}\right)_{(p,\eta,s),(p',\eta',s')}, \label{DiagonalFormOfOrderParameter}
\end{equation}
i.e., do not exhibit any ``coherence'' between the graphene and TBG sectors. This is expected as the presence of $\sigma_h$ requires any order parameter to be either even (diagonal in $t$ space) or odd (off-diagonal) under $\sigma_h$; due to the large density of states in the TBG sector, we expect the former to dominate. Once $D_0\neq 0$, mixing is allowed, as we will see in our numerics below and discuss in detail analytically in \secref{MixingBetweenTheBands}.

Furthermore, the analysis above reveals that the ground state in the TBG sector will be of the form of \equref{ExactGroundStatesMainText}. For instance, for $\nu_b=0$ with $\eta_j^{+}=\eta_j^{-}=(-1)^j$ and $s_j^{+}=s_j^{-}=\,\uparrow$, $j=1,2$, it holds $\widetilde{Q}^b_{\vec{k}} = \widetilde{\sigma}_0\eta_0s_z$ where $\widetilde{Q}^b_{\vec{k}} = U^*Q^b_{\vec{k}} U^T$ is the order parameter $Q^b_{\vec{k}}$ in \equref{DiagonalFormOfOrderParameter} in the TBG subspace transformed to the chiral basis of \equref{TransformationInChiralBasis}; in the basis of \equref{DefinitionOfCorrelator}, it holds $Q^b_{\vec{k}} = \sigma_0\eta_0s_z$, which we will refer to as spin polarized (SP) state. Here and in the following we will use $\sigma_j$ ($\widetilde{\sigma}_j$) to denote Pauli matrices in the band-space with index $p$ (in the chiral basis with index $c$).
Besides the SP state, \equref{ExactGroundStatesMainText} describes many other possible ground states, that are exactly degenerate in the chiral-flat-decoupled limit. A systematic way of seeing this proceeds by noting that the $\text{U}(4)\times \text{U}(4)$ identified \cite{bultinck2019ground,PhysRevLett.122.246401,bernevig2020tbg} for TBG also persists as a symmetry of MSTG in chiral-flat-decoupled limit \cite{2021PhRvB.103s5411C}; this immediately follows from the structure of $H_{\text{FD}}$ in \equref{ThreeTermsInHFD}. We will here refer to this symmetry group as $(\text{U}(4)\times \text{U}(4))_{\text{b,cf}}$ and its action is particularly simple in the chiral basis \cite{bernevig2020tbg},
\begin{equation}
    \widetilde{b}_{\vec{k}}\,\rightarrow \, \mathcal{U} \widetilde{b}_{\vec{k}}, \quad \mathcal{U} = e^{i \sum_{j=0,3}\sum_{\mu,\mu'=0}^3 \varphi_{j,\mu,\mu'} \widetilde{\sigma}_j \eta_\mu s_{\mu'}}. \label{FulUFourUFourTrafo}
\end{equation}
The form of these transformations is readily inferred from \equref{DensityOperatorChiralLimit} which indicates that all $\mathcal{U}$ with $[\mathcal{U},\widetilde{\sigma}_3]=0$ will leave $H_{\text{FD}}$ invariant. Under \equref{FulUFourUFourTrafo}, the order parameter defined above transforms as $\widetilde{Q}^b_{\vec{k}} \rightarrow \mathcal{U}^*\widetilde{Q}^b_{\vec{k}} \mathcal{U}^T$, which allows us to generate the entire (continuous) set of exactly degenerate ground states from one ``seed'' state, such as the SP state, $\widetilde{Q}^b_{\vec{k}} = \widetilde{\sigma}_0\eta_0s_z$. Since this seed state and $\mathcal{U}$ commute with $\widetilde{\sigma}_3$, we know that $[\widetilde{Q}^b_{\vec{k}},\widetilde{\sigma}_3]=0$ for all ground states. Further noting that $(\widetilde{Q}^b_{\vec{k}})^2 =\mathbbm{1}$ (physically related to the Slater-determinant nature), we can, thus, summarize the properties in the original basis as
\begin{equation}
    Q^b_{\vec{k}}=Q^b, \quad [Q^b,\sigma_2]=0, \quad (Q^b)^2 = \mathbbm{1}, \quad \text{tr}\left[Q^b\right]=0, \label{ConstraintsAtCNP}
\end{equation}
at charge neutrality. 

The actual Hamiltonian of MSTG is not in the chiral-flat-decoupled limit and does not exhibit an exact $(\text{U}(4)\times \text{U}(4))_{\text{b,cf}}$ symmetry. 
Intuitively, this can be thought of as generating an easy axis in this multi-dimensional space of degenerate states, favoring a specific (subspace of) state(s) in \equref{ConstraintsAtCNP}. 
While energetics is required to decide which phase is ultimately preferred by the system---the aim of the subsequent sections---we can use symmetries to derive the discrete and finite set of possible ``candidate states'': to this end, we impose only $\text{U}(1)_v$ and global spin rotations, $\text{SU}(2)_s$, as exact continuous symmetries. We then know that the candidate order parameters must transform under the irreducible representations of these symmetry groups (and be even or odd under the exact discrete symmetries $C_{2z}$ and $\Theta$), leading to the $11$ options listed in \tableref{ListOfOrderParameters}. In order to connect smoothly to the limit $D_0\rightarrow 0$, we take here $Q^g_{\vec{k}}=-\sigma_z$ in the graphene subspace but emphasize that these order parameters are only used to define the different states and characterize their symmetries; for our numerical and analytical discussion below, they are only taken to be the starting point and we will allow for (and also find) mixing between the TBG-like and graphene-like sectors when $D_0\neq 0$ as well as momentum dependence in $Q_{\vec{k}}$.

Finally, we point out that the model introduced in \secref{ModelAndSymmetries}, and which we study energetics in below, has an exact $\text{SU}(2)_+ \times \text{SU}(2)_-$ symmetry. Therefore, certain pairs of states, which we call \textit{Hund's partners} following \cite{PhysRevResearch.2.033062}, have to be exactly degenerate, see \tableref{ListOfOrderParameters}. As we noted earlier, in the realistic system, there is a non-zero \textit{intervalley Hund's coupling}, $J_H\neq0$, which will break $\text{SU}(2)_+ \times \text{SU}(2)_-$ down to $\text{SU}(2)_s$, albeit weakly, and favor one member of each of the pairs over the other. For most of the following study we will focus on the $\text{SU}(2)_+ \times \text{SU}(2)_-$ limit and, hence, can, without loss of generality, restrict the discussion to the first $7$ states above the line in \tableref{ListOfOrderParameters}. However, one has to keep in mind that the real system will realize only one state of each Hund's pair, which will depend on the (unknown) sign of $J_H$ and precise form of the Hund's coupling [see \appref{HundsCouplingAppendix} for more details]. Taking, for concreteness, the intervalley Hund's coupling to be of the form
\begin{equation}
    H_{2} = \frac{J_{H}}{N} \sum_{\vec{q}} \vec{S}_{\vec{q}}^+ \cdot \vec{S}_{\vec{q}}^-, \quad  \vec{S}_{\vec{q}}^{\pm} = \frac{1}{2}c^\dagger_{\vec{k}+\vec{q}} (\mathbbm{1}\pm\eta_z) c_{\vec{k}}, \label{HundsCouplingInteraction}
\end{equation}
we can, in the chiral-flat-decoupled limit, uniquely associate a single state of each Hund's pair with a given sign of $J_H$; this is indicated in the last column in \tableref{ListOfOrderParameters}.

\section{Hartree-Fock numerics}\label{HFNumericsSection}
As it facilitates the presentation of the results, we will begin the discussion of correlated phases away from the chiral-flat-decoupled limit with the HF numerics and postpone the complementary analytics to \secref{AnalyticalPerturbationTheory}. Furthermore, we will first focus on the charge-neutrality point, $\nu=0$.

\subsection{Hartree-Fock Method}
In the HF approximation, one focuses on Slater-determinant states $\ket{\Psi[P_{\vec{k}}]}$ characterized by the correlation matrix $P_{\vec{k}}$ as defined in \equref{DefinitionOfCorrelator}. Consequently, it holds $P^2_{\vec{k}} = P^{\phantom{2}}_{\vec{k}}$ or, equivalently, $Q_{\vec{k}}^2 = \mathbbm{1}$; this is also true for the exact candidate ground states constructed in \secref{CandidateStatesConstruction} for the chiral-flat-decoupled limit and, hence, the HF approximation is expected to provide reliable results.  

The goal of our HF numerics will be to determine the optimal $Q_{\vec{k}}$ that yields the lowest energy expectation value with respect to the interacting Hamiltonian for MSTG introduced in \secref{ModelAndSymmetries}. To be more specific, we start from the full Hamiltonian, $H=H^{\text{Full}}_{0,1}+H^{\text{Full}}_{0,2}+H_1$, consisting of the continuum model in \equref{NoninteractingModelWithoutDisplacement}, the displacement-field term in \equref{DisplacementField}, supplemented by the density-density interaction in \equref{GeneralFormOfDensityDensityInteraction}, and perform a mean-field decoupling. Using the same notation as in \equref{ExpressionJustDeltaRho}, the resulting HF mean-field Hamiltonian reads as   
\begin{equation}
\begin{split}
    H^{\text{MF}}&=\sum_{\vec{k}\in\text{MBZ}}\epsilon_{n,\eta}(\vec{k})f^\dagger_{\vec{k};n,\eta,s}f^\pdagger_{\vec{k};n,\eta,s}\\&+\sum_{\vec{k}\in\text{MBZ}}f^\dagger_{\vec{k}} \left[h_{\text{H}}[P](\vec{k})+h_{\text{F}}[P](\vec{k})\right]f_{\vec{k}} \\&-\frac{1}{2}\sum_{\vec{k}\in\text{MBZ}}\text{Tr}\left[h_{\text{H}}[P](\vec{k})P_{\vec{k}}^T+h_{\text{F}}[P](\vec{k})P_{\vec{k}}^T\right], \label{MeanFieldHartreeFock}\end{split}\end{equation}\label{MeanField}where the Hartree and Fock contributions to the mean-field Hamiltonian can be written in terms of the projector $P_{\vec{k}}$ as 
\begin{equation}
    h_{\text{H}}[P](\vec{k})=\frac{1}{N}\sum_{\vec{G}\in\text{RL}}V(\vec{G})  F_{\vec{k},\vec{G}}\sum_{\vec{k}'\in\text{MBZ}}\text{Tr}\left[ F^*_{\vec{k}',\vec{G}}P_{\vec{k}'}\right]
    \label{MFHartree}
\end{equation}
and
\begin{equation}
    h_{\text{F}}[P](\vec{k})=-\frac{1}{N}\sum_{\vec{q}}V(\vec{q})   F^\dagger_{\vec{k},\vec{q}} P^T_{\vec{k}+\vec{q}} F_{\vec{k},\vec{q}},
    \label{MFFock}
\end{equation}
respectively. This form of the HF mean-field Hamiltonian is valid for an arbitrary number of bands kept. In the numerics presented here, we will focus on the four bands for each spin and valley flavor that are closest to the Fermi level, which contains the graphene-like and TBG-like bands we focus on in the analytics. We verify for representative values of $D_0$ and $w_0$ that the solutions we obtain are stable against doubling the number of remote bands in our self-consistent calculation in \appref{VaryingParameterInNumerics}.

As pointed out in several HF works on TBG \cite{bultinck2019ground,Liu_2021,Xie_2020,Liao_2021}, it is important to note that the continuum model $H_{0,1}^{\text{Full}}$ already references
electron-electron interactions in the experimentally
determined values for microscopic model parameters and therefore
we must define a reference subtraction projector
$P_0$ such that interactions will not be double counted
in our numerics. We here choose $P_0$ such that the projected low-energy Hamiltonian exhibits the manifestly particle-hole symmetric interaction in \equref{FormOfProjectedInteractionUsed}. As shown in \cite{bernevig2020tbg} for TBG, this ansatz has the natural interpretation of effectively taking into account the HF contributions from all remote bands that have been projected out.

\begin{figure}[ht!]
    \centering
    \begin{subfigure}{0.54\columnwidth}
    \includegraphics[width=1\linewidth]{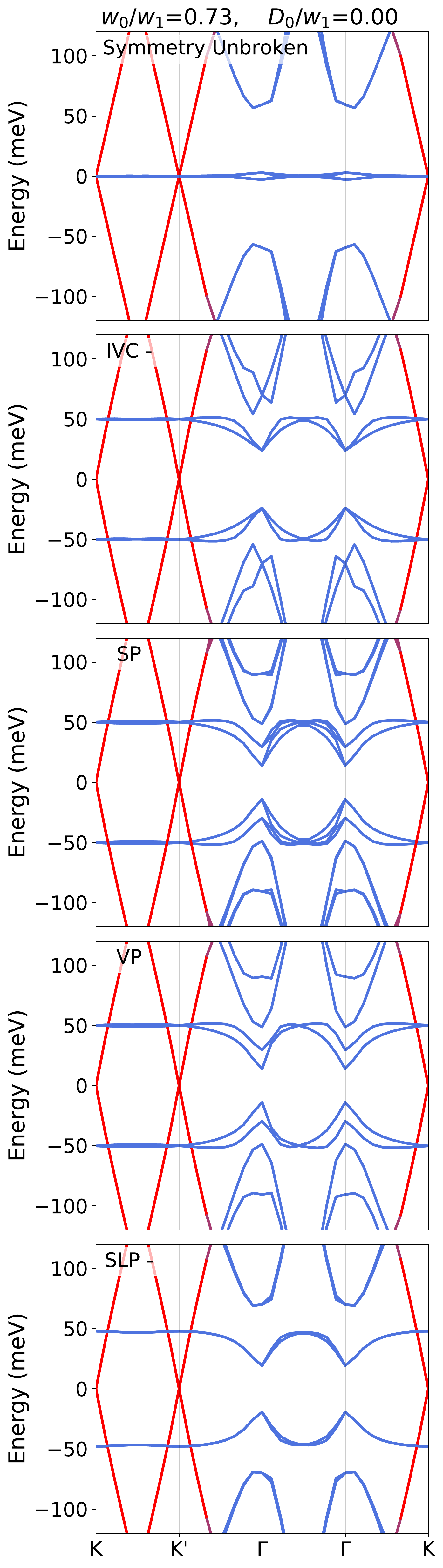}

    \end{subfigure}
    \begin{subfigure}{0.44\columnwidth}
    \includegraphics[width=1\linewidth]{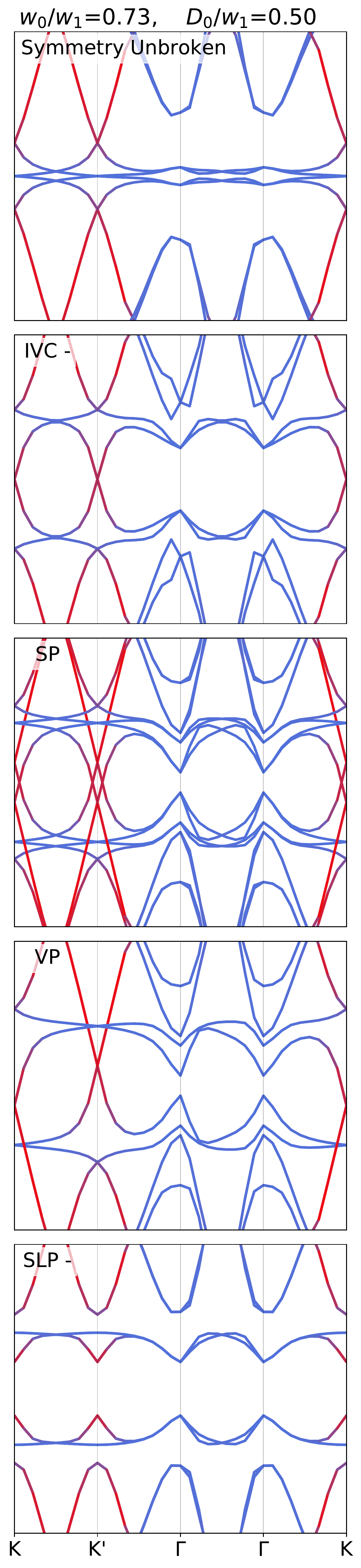}
    \end{subfigure}
    \caption{HF band structures in the limit $D_0=0$ (left) and nonzero $D_0$ (right) as obtained from \equref{MeanFieldHartreeFock} for the optimal $P_{\vec{k}}$ in the respective symmetry-breaking channel defined in \tableref{ListOfOrderParameters}. We show only the states which are either ground states or subleading energy states for some region of our phase diagram, with a full set of band structures available in \appref{VaryingParameterInNumerics}. Each band of the HF Hamiltonian is colored according to the expectation value of the mirror symmetry operator within each band as a function of $\vec{k}$, with the expectation value ranging from -1 (red) to +1 (blue).}
    \label{Bands}
\end{figure}

To determine the optimal $P_{\vec{k}}$, we start with an initial guess for it with the symmetries of a given candidate order in \tableref{ListOfOrderParameters}. We then use the HF Hamiltonian in \equref{MeanFieldHartreeFock} to compute a new projector $P_{\vec{k}}$ and iterate until $P_{\vec{k}}$ converges. More details on our iterative HF procedure and subtraction point are given in 
\appref{Numerics}.

\subsection{Band Structures}
In this section, we will discuss the band structures of the self-consistent solutions we find at $D_0=0$ for each of the states in \tableref{ListOfOrderParameters} and  how these band structures evolve as $D_0$ increases.

\subsubsection{Self-Consistent Band Structures at $D_0=0$}
At $D_0=0$, we can separately describe the behavior of the graphene-like and TBG-like bands for each type of state, since, for all states we consider, the graphene-like bands near the Fermi-level at $D_0=0$ do not mix with any other bands near the K and K$'$ points. They have a bandwidth larger than the scale of Coulomb interactions. For this reason, the Dirac cones of the graphene-like bands prefer a $Q^g_{\vec{k}}$ which equally fills the lower bands of the continuum model and preserves all point group symmetries for every class of solution we study. The Dirac cones thus remain semimetallic and are, in this sense, ``spectators'' at $D_0=0$, in agreement with \secref{ExactGroundStates}. 
On the other hand, the TBG-like bands have a bandwidth (5-10 meV) smaller than the scale of the Coulomb energy at $D_0=0$ and therefore become insulating as they are polarized for a given symmetry-breaking $P_{\vec{k}}$. 

We find converged solutions for each ansatz in \tableref{ListOfOrderParameters} and show representative band structures for those states which have the lowest energy in the leftmost panels of \figref{Bands}. Additional band structures for solutions not shown in \figref{Bands} can be found in \appref{VaryingParameterInNumerics}. We note the similarity of the IVC$_-$ band structure in the TBG-like bands to the band structure of the ground state in \refcite{bultinck2019ground}.

\begin{figure}[tb]
    \centering
    \includegraphics[width=0.44\textwidth]{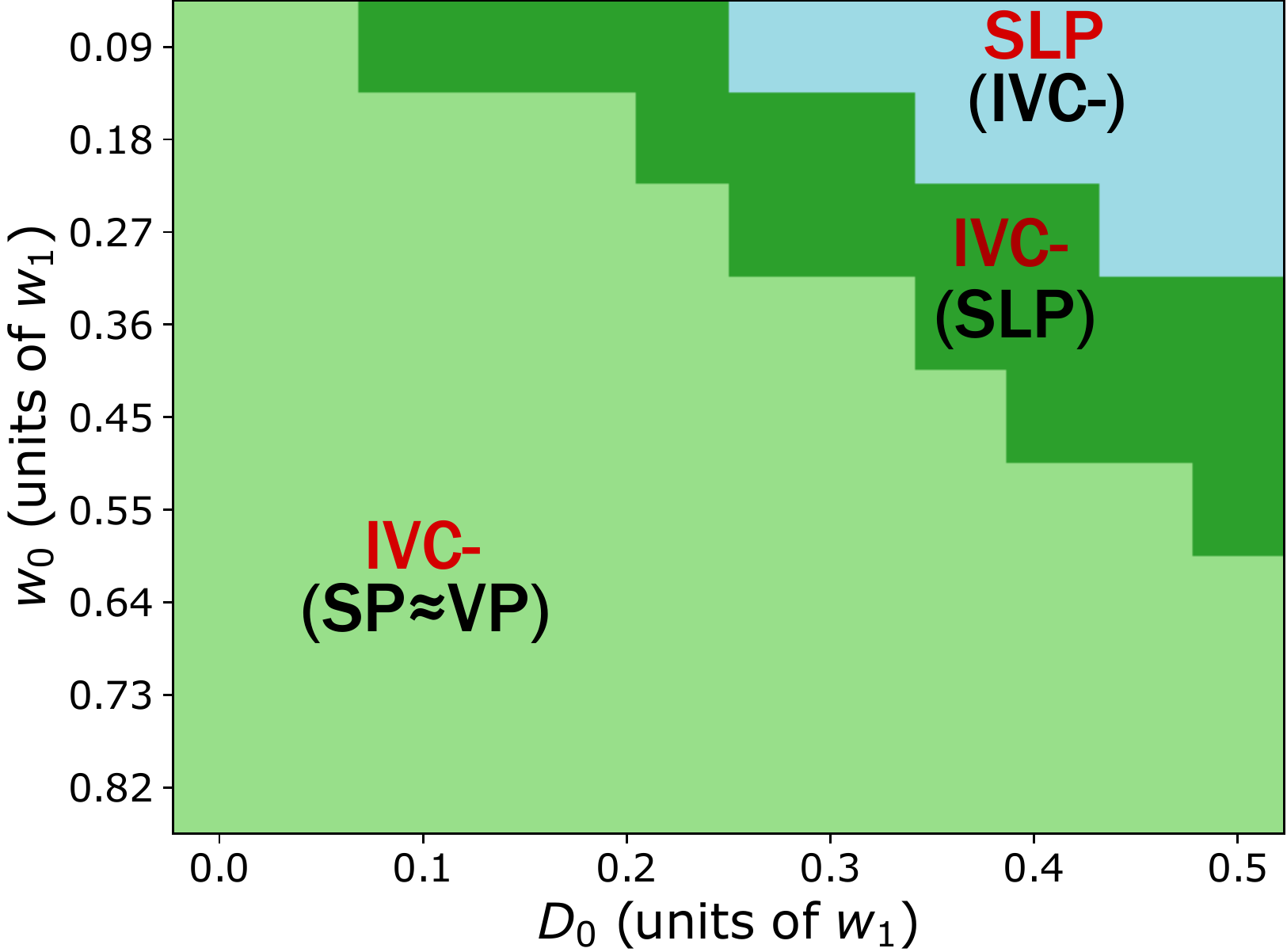}
    \caption{Phases obtained by self-consistent HF method as a function of $w_0$ and $D_0$. Both the leading instability (red) and the subleading ones (black, in paraenthesis) are shown. The phases are identified by the symmetries they break, see \tableref{ListOfOrderParameters}, and representative band structures can be found in \figref{Bands}. We show only one state as a representative of each pair of Hund's partners which are degenerate in our Hartree-Fock procedure. We label the nearly degenerate SLP$_+$, SLP$_-$, SSLP$_+$, and SSLP$_-$ phases with "SLP", though note the preference in our numerics is for the SLP- state, from a Hartree contribution. 
    }
    \label{PhaseDiagram}
\end{figure}

\subsubsection{Self-Consistent Band Structures for $D_0>0$}
All the solutions described for $D_0=0$ are insulating in the TBG-like bands and semimetallic in the graphene-like bands. However, as $D_0$ increases and the TBG-like and graphene-like bands begin to hybridize, the graphene-like bands begin to play a more important role. 

For the spin and valley polarized states (SP and VP) which preserve $C_{2z}\Theta$ and U(1)$_v$, the hybridized Dirac crossings are protected (and pinned to the K/K$'$ points due to $C_{3z}$), meaning if the TBG-like bands acquire a spin or valley polarization, the Dirac crossings of the graphene-like bands must connect to these polarized bands as they are pushed away from the Fermi level. Away from the K points of the MBZ, the graphene-like bands will likely still prefer to fill the lowest bands of the non-interacting model. 
Therefore, the VP and SP are generically expected to be metallic for $D_0>0$. 
This is indeed what we find, as shown in the right panels for the SP and VP states in \figref{Bands}. 

The band structures of the IVC$_\pm$ states also retain Dirac crossings at the K/K$'$ points for nonzero $D_0$, as they exhibit $C_{3z}$ and a $\vec{k}$-local anti-unitary symmetry that commutes with $C_{3z}$ (for the IVC$_+$ and IVC$_-$ these are $C_{2z} \Theta$ and $C_{2z} \widetilde{\Theta}$, respectively). However, unlike the SP and VP states, the IVC$_\pm$ states also preserve SU(2)$_s$ and $C_{2z}$, which pin the Dirac crossings at the Fermi level at $\nu=0$.
We therefore expect the intervalley coherent states will remain semimetallic as $D_0$ increases. We observe this to be true for the self consistent solutions, as can be seen in the right IVC$_-$ panel in \figref{Bands}.

The last class of states are the sublattice-polarized, $C_{2z}\Theta$-symmetry-breaking states (SLP$_{\pm}$, SSLP$_-$) which preserve U(1)$_v$. We expect these states  will generally be insulating for nonzero $D_0$ as there are no protected Dirac crossings and both the TBG-like and graphene-like bands can be gapped out. This is indeed seen in our numerics, with insulating band structures for the SLP$_+$, SLP$_-$, and SSLP$_-$ states. The band structure of the SLP$_-$ state is shown in \figref{Bands}.

\subsection{Energies and phase diagram}
Having established the band structures of the different possible phases, we next turn to their relative energetics and discuss which states are expected to be favored energetically. 

\begin{figure*}[tb]
    \centering
    \includegraphics[scale=.65]{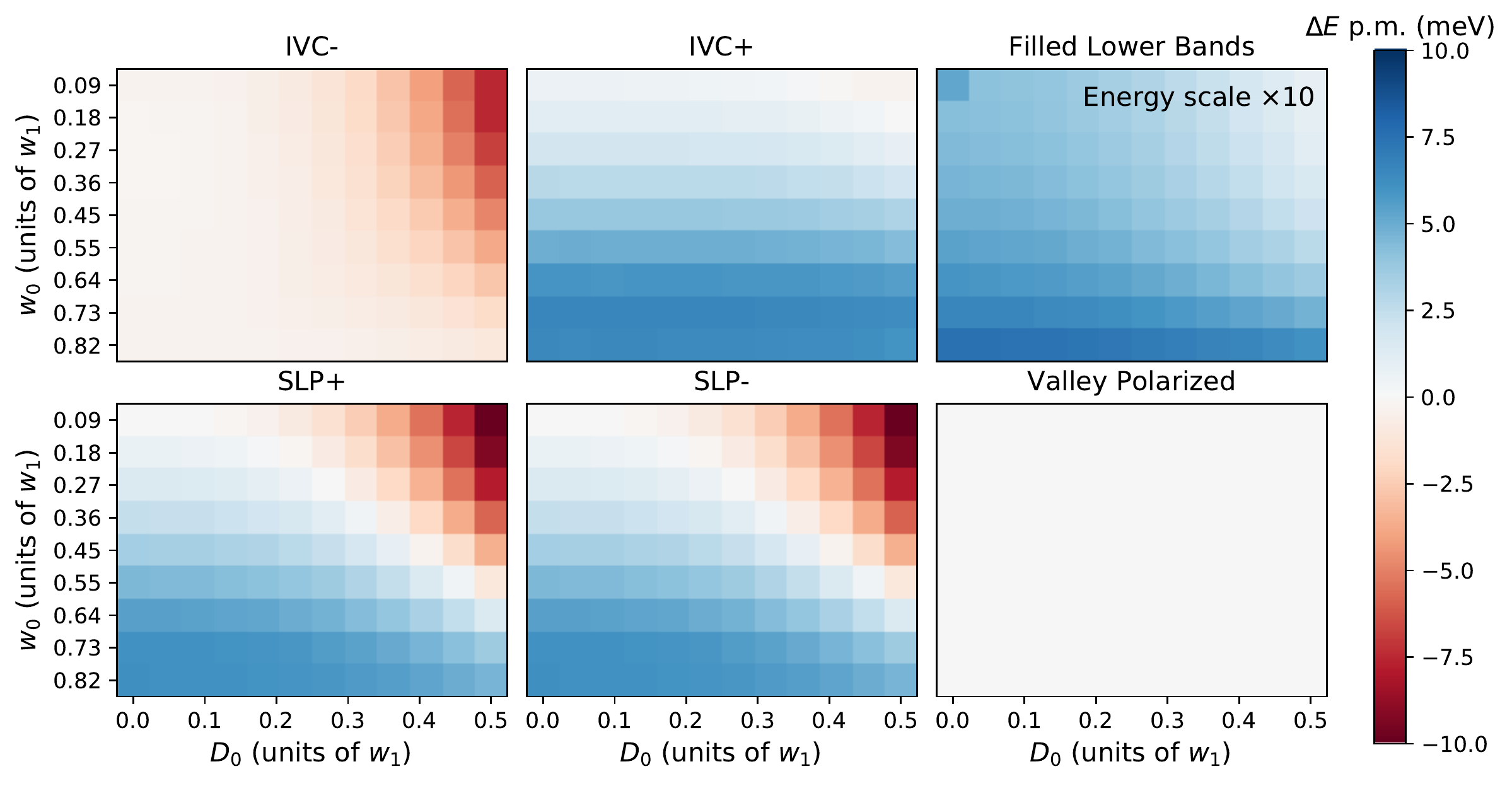}
    \caption{Energies obtained from self-consistent HF calculations for selected candidate states of \tableref{ListOfOrderParameters}. All energies shown are relative to the energy of the self-consistent SP solution and measured in meV per moir\'e unit cell. We take $\epsilon=7$, $d_s=40$ nm, and $w_1=124$ meV and 4 bands per spin and valley. We find the IVC$_-$ is lower in energy than the SP state for all values of $w_0$ and $D_0$. The IVC$_+$ state is always higher in energy than the SP in agreement with \tableref{BehaviorOfEnergetics}. The VP state is nearly exactly degenerate with the SP state, even as $D_0$ increases. Both the SLP$_+$ and SLP$_-$ states have a region of energy lower than both the SP and IVC$_-$ states at small $w_0$ and large $D_0$. While they appear nearly degenerate, a small Hartree contribution favors the time-reversal odd SLP$_-$ over the SLP$_+$ state. However, this difference between the two sublattice polarized states (< 1 meV) is not visible in the energies we show.}
    \label{Energies}
\end{figure*}

\begin{figure*}[t]
    \centering
    \includegraphics[width=\textwidth]{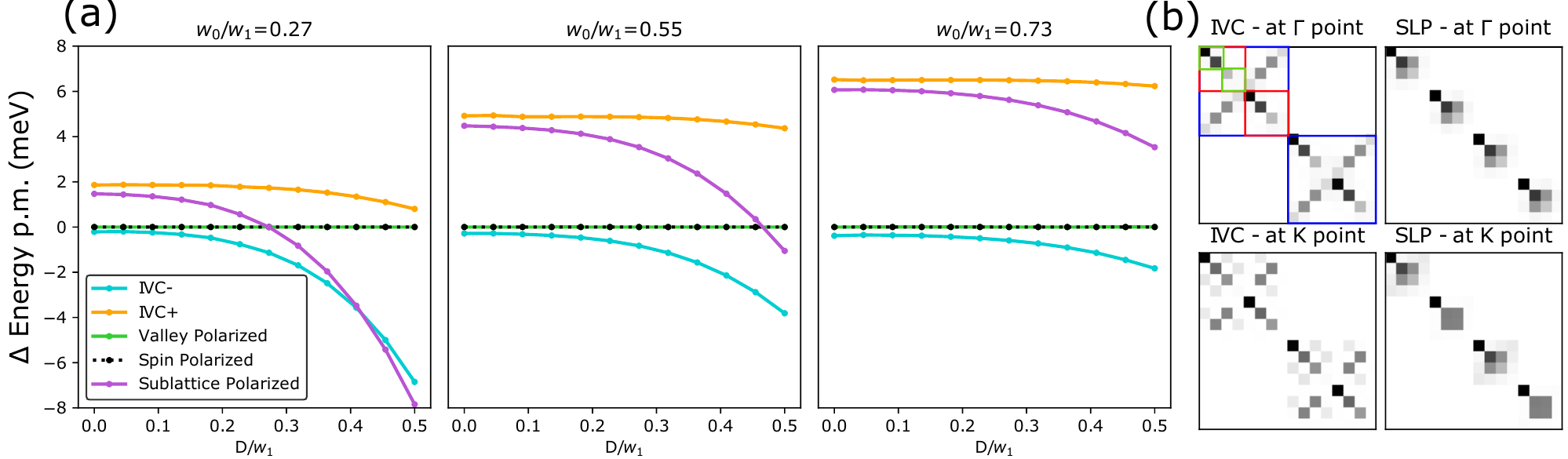}
    \caption{(a) Linecuts for fixed $w_0$ showing the energy of our self consistent solutions relative to a SP state as a function of $D_0$. We use the same parameters as in \figref{Energies}. Note that the SLP$_\pm$ and SSLP$_{-}$ are not exactly degenerate, as discussed in the main text, but are plotted as a single line here since the splitting is too small to be visible on the scale of the plot. (b) Matrix form of our converged IVC$_-$ and SLP$_-$ orders $P_{\vec{k}}$ for $\vec{k}$ near the $\Gamma$ point and near the K point. The matrix structure of $P_{\vec{k}}$ is organized such that the largest block outlined in blue denotes spin flavor, the next largest block outlined in red denotes valley flavor, the green block denotes upper and lower bands in the continuum model and the final two boxes denote the graphene-like and TBG-like band in the upper-band box and the TBG-like then graphene-like bands in the lower-band box.}
    \label{Energycut}
\end{figure*}

The evolution of the energies of each of our self-consistent solutions as a function of $w_0$ and $D_0$ is shown in Figs.~\ref{Energies} and \ref{Energycut}. 
As mentioned before, at $D_0=0$, the Hamiltonian of the system is given by the sum of the Hamiltonian of TBG and that of graphene, both with Coulomb interactions, which are further coupled to each other by a density-density interaction. While we only constructed exact eigenstates in \secref{ExactGroundStates} for the chiral-flat-decoupled limit, we expect a similar picture when $w_0,W_{\text{TBG}} \neq 0$: given the bandwidth of the graphene-like bands is large compared to the scale of the Coulomb interactions, we expect the graphene bands will prefer to fill the lower bands of the continuum model. As the graphene density of states is small compared to that of the flat bands of TBG, it should not crucially alter the ground state in the TBG sector---at least close to the magic angle. Based on previous work \cite{bultinck2019ground,Liao_2021}, we thus expect that the IVC$_-$ state has the lowest energy for $D_0=0$ (though with a smaller energy difference than in previous works between our IVC$_-$ and spin polarized phase due to our choice of subtraction point). Both expectations for the graphene-like and TBG-like bands are confirmed by our numerics which finds the IVC$_-$ state has the lowest energy of all our candidates for all values of $w_0$ studied in the decoupled limit, $D_0=0$. We will also recover these observations analytically in \secref{AnalyticalPerturbationTheory}. 

The lowest energy state for $D_0\neq 0$ cannot be directly inferred from knowledge of the physics of TBG as a finite $D_0$ induces hybridization between the TBG-like and graphene-like bands near the K/K$'$ points of the MBZ; it further breaks symmetries in the TBG sector and, hence, changes the basic form of its dispersion and interaction matrix elements (form factors). Consequently, it is not clear whether the ground state in the TBG-like and/or graphene-like sector changes with increasing $D_0$. As can be seen in Figs.~\ref{Energies} and \ref{Energycut} as well as in the corresponding phase diagram in \figref{PhaseDiagram}, we find within HF that the IVC$_-$ remains the ground state for an extended range of $D_0$, which increases with $w_0$. For reference, the range of $D_0$ in Figs.~\ref{PhaseDiagram}, \ref{Energies}, and \ref{Energycut} when combined with additional studies at larger $D_0$ in \appref{LargerValuesOfD} corresponds roughly to the range of displacement fields studied experimentally in \refcite{Park_2021}.

Beyond the critical value of $D_0$ for the IVC$_-$, the sublattice-polarized (SLP) group of states (SLP$_{\pm}$, SSLP$_-$) dominates. While these latter three states are almost degenerate for all parameters $D_0$, $w_0$ studied, there is a slight preference towards the time-reversal-odd SLP$_-$ (quantum Hall), predominantly associated with the Hartree energy. 
While a SP or VP phase does not appear as a ground state in \figref{PhaseDiagram}, we find the energetically close SP or VP states are either the second or third lowest energy state to the IVC$_-$ and SLP group across the phase diagram. 
In the next subsection, we will recover many of these features analytically by investigating the aforementioned energetic contributions perturbatively.

\section{Analytical perturbation theory}\label{AnalyticalPerturbationTheory}
Finally, we complement the HF numerics with an analytical study of the behavior of the energies and the order parameters of the degenerate ground states of the chiral-flat-decoupled limit ($w_0=W_{\text{TBG}}=D_0=0$) of \secref{ExactGroundStates} and \tableref{ListOfOrderParameters} when turning on $W_{\text{TBG}}$, $D_0$, and $w_0$. 
We first investigate the ordering tendencies of the graphene-like bands (\secref{OrderingGrapheneLikeBands}) and their mixing with the TBG-like bands (\secref{MixingBetweenTheBands}), before addressing the energetic contributions coming from the $D_0$,$w_0$-induced distortion of the TBG-like form factors (\secref{DeformedFormFactors}) and band structure (\secref{PerturbationTheoryInBandwidth}).

\subsection{Ordering in the graphene-like bands}\label{OrderingGrapheneLikeBands}
Since the graphene and graphene-like bands in \figref{TrilayerBands}(b) are highly dispersive, starting from a flat-band limit, as is natural for the TBG-like sector, is not possible for the graphene(-like) bands. 
Since their bands are coupled, without any band gap, to the TBG-like bands with a high density of states, treating the interactions between the two subsystems as a perturbation is also not necessarily controlled for realistic parameters. Instead, we here use a different control parameter: based on the band structure, we expect the effect of the TBG(-like) bands on the graphene(-like) bands to be the strongest around the K/K$'$ points and very weak away from it. To formalize this, let us assume that the graphene-like bands remain in their filled-lower-bands state away from the K/K$'$ points, but allow them to be ``deformed'' in the region $\mathcal{A} = \mathcal{A}_+ \cup \mathcal{A}_-$ of the MBZ, where $\mathcal{A}_+$ ($\mathcal{A}_-$) are simply connected and centered around the K (K$'$) point. As explained in detail in \appref{PolarizationOfGrapheneLikeBands}, we study the energy of symmetry-allowed ordering tendencies of the graphene-like bands for the different candidate phases in \tableref{ListOfOrderParameters} in the limit where the area of $\mathcal{A}$ is small (compared to that of the MBZ).  

To illustrate this procedure, let us consider the SP state. Since the TBG-like bands break spin-rotation symmetry, it is natural to assume that the same happens to the graphene-like bands in $\mathcal{A}$; postponing the discussion of mixing between the bands to \secref{MixingBetweenTheBands}, this means that $(Q^g_{\vec{k}})_{\eta,\eta'} = \delta_{\eta,\eta'} \sigma_0 \vec{s}$, $\vec{k}\in\mathcal{A}_\eta$, in the notation introduced in \secref{CandidateStatesConstruction}, while $(Q^g_{\vec{k}})_{\eta,\eta'} = - \delta_{\eta,\eta'} \sigma_3$ for all other $\vec{k}$. Here we already anticipated (as is also readily checked within this formalism) that it is energetically more favorable if only the graphene valley $\eta=+$ ($\eta=-$) that is at low energies in the region $\mathcal{A}_+$ ($\mathcal{A}_-$) exhibits spin polarization. 
Denoting the linear size of $\mathcal{A}_{\pm}$ by $\Delta k>0$, the structure of the energetic change associated with the deformation of the graphene order is asymptotically given by
\begin{align}\begin{split}
    \Delta E & \sim d_1|D_0| (\Delta k)^2 + d_2 (\Delta k)^3  
    \\ &\qquad + I_{gg} (\Delta k)^3 - I_{gb} D_0^2 (\Delta k)^2 \label{ExplicitForm}
\end{split}\end{align}
for small $\Delta k$ and $D_0$; here, $d_{1,2}>0$ are positive constants (independent of $\Delta k$ and $D_0$, but dependent on $w_0$) associated with the graphene-like dispersion, while $I_{gg}$ depends on the graphene-graphene form factors $F^{gg}$ and $I_{gb}>0$ on $F^{gb}$ and $F^{gg}$. The explicit form and derivation can be found in \appref{PolarizationOfGrapheneLikeBands}.

From \equref{ExplicitForm}, we can read off the behavior of the graphene-like bands of the SP state. For $D_0=0$, we see that $\Delta E > 0$ (deformation is energetically disfavored) as long as $d_2 > I_{gg}$. While this is what our numerics shows, we point out that $d_2<I_{gg}$ would imply that single-layer graphene spontaneously magnetizes, which is known to be not the case. This agrees with our result in \secref{ExactGroundStates} based on adibatically turning on the coupling between the TBG and graphene system and our numerics which displays unpolarized graphene Dirac cones in \figref{Bands} at $D_0=0$.
When $D_0\neq 0$, we see that the energetic cost coming from the dispersion in the first line of \equref{ExplicitForm} will always overcompensate the energetic gain coming from the scattering between the graphene- and TBG-like bands (at least in a finite range of $D_0 \neq 0$). This is also consistent with our HF numerics: as can be seen in the band structure in \figref{Bands}, the graphene-like bands are not spin-polarized for non-zero $D_0$ [see also \figref{MatrixOrders} where $Q_{\vec{k}}$ as obtained in HF are shown]. 

The ordering tendencies in the graphene-like bands for all other states in \tableref{ListOfOrderParameters} can be analyzed in a similar way, see \appref{PolarizationOfGrapheneLikeBands}. For instance, for the VP we find that the graphene-like bands are found to not develop valley polarization either---both for $D_0=0$ and $D_0\neq 0$ and again consistent with the HF numerics. The same holds for the IVC$_\pm$ states, where breaking of $\text{U}(1)_v$ in the graphene-like bands is strongly suppressed due to the fact that, at the K (K$'$) point, the $\eta=-$ ($\eta=+$) graphene-like band energies are large (already at $D_0=0$), see \figref{TrilayerBands}.

The situation is different for the SLP group of states (SLP$_\pm$ and SSLP$_{\pm}$). To begin with the SLP$_\pm$ states, we have $(Q^g_{\vec{k}})_{\eta,\eta'} = \delta_{\eta,\eta'} (-\sigma_3 \cos \theta_{\vec{k}} +\sigma_2 \sin\theta_{\vec{k}})$, $\vec{k}\in\mathcal{A}_\eta$, and we need to determine $\theta_{\vec{k}}$, with $\theta_{\vec{k}}= \mp \theta_{-\vec{k}}$ for SLP$_\pm$ due to $\Theta$ and $C_{2z}$, by minimizing the energy. The change of the energy as a consequence of this ``deformation'' is of the form
\begin{equation}
    \Delta E_{gg}[\theta_{\vec{k}}] = \sum_\eta \sum_{\vec{k}\in\mathcal{A}_\eta} \left[ A_{\vec{k},\eta} (1 - \cos \theta_{\vec{k}}) + B_{\vec{k},\eta} \sin \theta_{\vec{k}} \right] \label{EnergyGainTheta}
\end{equation}
for both SLP$_+$ and SLP$_-$. The explicit expressions for $A_{\vec{k},\eta}$ and $B_{\vec{k},\eta}$, which are functions of $w_0$ and $D_0$, are given in \appref{PolarizationOfGrapheneLikeBands}. From these expressions, it follows that $B_{\vec{k},\eta}=0$ and $A_{\vec{k},\eta}>0$ for $D_0=0$ such that $\Delta E[\theta_{\vec{k}}]$ is minimized when $\sin \theta_{\vec{k}}=0$ (with $\Delta E=0$) and there is no order in the graphene bands---again in agreement with \secref{ExactGroundStates} and the HF numerics. Once $D_0\neq 0$, we get $B_{\vec{k},\eta}\neq 0$ and $\Delta E<0$ by choosing a profile with $\sin \theta_{\vec{k}}\neq 0$. In other words, the graphene-like bands will develop SLP$_{\pm}$ order for any non-zero $D_0$, as is visible in the $Q_{\vec{k}}$ for the SLP$_-$ state close to the K point shown in \figref{Energycut}(b). This gaps out the graphene cone, as can also be seen in our HF band structure in \figref{Bands}. The energetic gain scales as $g_{\pm} (\Delta k)^2 |D_0|^3$, $g_{\pm} >0$, for small $D_0$ and $\Delta k$ for the SLP$_\pm$ state; the prefactors  differ, $g_+ \neq g_-$, due to the symmetry-imposed constraint $\theta_{\vec{k}}= \mp \theta_{-\vec{k}}$. For small $w_0$, we can show that $B_{\vec{k},\eta} > 0$ such that $g_- > g_+$, i.e., the SLP$_-$ state can gain more energy than the SLP$_+$.

For the SSLP$_-$ state and its Hund's partner SSLP$_+$, which we discuss here explicitly for reasons that will become clear shortly, we have $(Q^g_{\vec{k}})_{\eta,\eta'} = \delta_{\eta,\eta'} (-\sigma_3 \cos \theta_{\vec{k}} +\sigma_2 \vec{s} \sin\theta_{\vec{k}})$, $\vec{k}\in\mathcal{A}_\eta$. Symmetry imposes $\theta_{\vec{k}} = \mp \theta_{-\vec{k}}$ for SSLP$_{\pm}$.
The deformation-related energy change $\Delta E$ is found to be again of the form of \equref{EnergyGainTheta}; however, we here obtain $B_{\vec{k},\eta}=0$ for any $D_0$ or $w_0$. This is consistent with $\text{SU}(2)_+ \times \text{SU}(2)_-$ which requires that SSLP$_{\pm}$ have the same energy. We, hence, have shown that the graphene-like bands do not develop any direct SSLP$_{\pm}$ order and gain energy in the process. They will, however, hybridize with the TBG-like bands, as we discuss in the next subsection, which also gaps out the Dirac cones.

In \tableref{BehaviorOfEnergetics}, we summarize our findings that only the SLP$_{\pm}$ states can benefit from ordering the graphene-like bands and how the respective energy gain scales.

\subsection{Mixing between the bands}\label{MixingBetweenTheBands}
As anticipated above, we next look into the mixing between the TBG-like and graphene-like bands. While it is clear by symmetry that the mixing has to vanish for $D_0=0$ and be generically present for $D_0\neq 0$, we here investigate the associated energetic gain and precise from of the band mixing for our candidate states in \tableref{ListOfOrderParameters}. As before, we here outline the basic strategy and discuss the major results of this calculation and defer the details to \appref{MixingBetweenBands}.

Our starting point are product states characterized by a correlator with $Q_{\vec{k}}$ as given in \equref{DiagonalFormOfOrderParameter}, i.e., without any coherence between the TBG-like and graphene-like bands. In accordance with our analysis of \secref{ExactGroundStates}, we will take $Q^g_{\vec{k}}=-\sigma_z$ and let $Q^b_{\vec{k}}$ be any of the candidate orders. To introduce momentum-dependent coherence between these sets of bands, we ``deform'' $Q_{\vec{k}}$ by a unitary transformation, $U_{\vec{k}}$, and take
\begin{equation}
    Q'_{\vec{k}} = U^\pdagger_{\vec{k}} Q_{\vec{k}} U^\dagger_{\vec{k}}, \quad U_{\vec{k}} = e^{i \Lambda_{\vec{k}}}, \quad \Lambda_{\vec{k}}^\dagger = \Lambda_{\vec{k}}^\pdagger \in \mathbb{C}^{16\times 16}, \label{MixingDeformation}
\end{equation}
as ansatz for the correlator. Our goal will be to find the optimal momentum-dependent $\Lambda_{\vec{k}}$ to minimize the energy. 
Since we are interested in band mixing, we will restrict $\Lambda_{\vec{k}}$ to act as a superposition of $\zeta_1$ and $\zeta_2$, $\Lambda_{\vec{k}} = \sum_{j=1,2} \mathcal{M}_{\vec{k},j} \zeta_j$, $\mathcal{M}_{\vec{k},j} \in \mathbb{C}^{8\times 8}$, with $\zeta_j$ denoting Pauli matrices acting in the space of TBG-like and graphene-like bands (with index $t=b,g$). Furthermore, $\Lambda_{\vec{k}}$ will be constrained by the symmetries of the state under consideration.

To illustrate the procedure, let us focus on the IVC$_-$ since this state was found to be dominant in the HF numerics. Choosing $Q^b_{\vec{k}} = \sigma_y\eta_y$ for concreteness, this state preserves the $C_{2z}\Theta$ symmetry of \tableref{ActionOfSymmetriesMainText}, which forces $\Lambda_{\vec{k}}$ to obey $(\Lambda_{\vec{k}})^* = -\Lambda_{\vec{k}}$. Furthermore noting that the IVC$_-$ state does not break the SU(2)$_s$ symmetry (while postponing the consequences of $C_{2z}$ which will relate $\Lambda_{\vec{k}}$ and $\Lambda_{-\vec{k}}$), it follows that $\Lambda_{\vec{k}}$ has to be a momentum-dependent superposition of the $16$ generators
\begin{equation}
    \zeta_x \sigma_y \eta_{0,x,z}, \quad \zeta_x \sigma_{0,x,z} \eta_y, \quad \zeta_y\sigma_{0,x,z} \eta_{0,x,z}, \quad \zeta_y\sigma_y\eta_y. \label{SetOfGenerators}
\end{equation}
To simplify further, we can focus on those $8$ linear combinations [see \equref{GeneratorsIVCM} for their explicit form] of the terms in \equref{SetOfGenerators} that anti-commute rather than commute with $Q_{\vec{k}}$ of the IVC$_-$ state. 

\begin{table*}[tb]
\begin{center}
\caption{Summary of the different energetic contributions, as discussed in \secref{AnalyticalPerturbationTheory}, for the candidate orders in \tableref{ListOfOrderParameters} when tuning away from the chiral-flat-decoupled limit ($w_0=W_{\text{TBG}}=D_0=0$). By construction, the energies are identical for the respective Hund's partners and are, hence, omitted. Here $\Delta E_{bb}(W_{\text{TBG}}=0)$ is the change of energy, relative to SP phase, coming from the modifications of the form factors in the TBG-like bands when turning on $w_0$ and $D_0$ in the flat limit ($W_{\text{TBG}}=0$). As indicated in the column labelled $\Delta E_{gg}$, only the SLP$_\pm$ states gain energy by ordering in the graphene-like bands immediately when $D_0 \neq 0$. In all cases, $D_0\neq 0$ leads to mixing between the bands, with energetic gain as listed in the chiral limit, $w_0=0$, in the column $\Delta E_{gb}$; here $\Delta k$ is the linear size of the fraction $\mathcal{A}_\pm$ of the MBS where ordering in the graphene-like bands and mixing take place. The four columns with $\bar{E}_j^b$ indicate which of the four contributions to the TBG dispersion in \equref{TBGLikeDispersion} can lower the energies to second order in $W_{\text{TBG}}$. Finally, in the last columns, we list the energy change associated with a finite value of the Hund's coupling in \equref{HundsCouplingInteraction} for the respective state ($\Delta E_J$) and its Hund's partner ($\Delta E^H_J$), if it exists, in the chiral-flat-decoupled limit. All coefficients obey $c_j, g_{\pm}, g_j, \beta_j > 0$ and explicit expressions can be found in \appref{app:Energetics}.}
\label{BehaviorOfEnergetics}
\begin{ruledtabular}
 \begin{tabular} {cccccccccccc} 
Type & $Q^b$ & $\widetilde{Q}^b$ & $\Delta E_{bb}(W_{\text{TBG}}=0)$ & $\Delta E_{gg}$ & $\Delta E_{gb}$ & $\,\bar{E}^b_0\,$ & $\bar{E}^b_1 w_0$ & $\bar{E}^b_2D_0$ & $\bar{E}^b_3 w_0D_0$ & $\Delta E^{\phantom{H}}_J$ & $\Delta E^H_J$ \\ \hline
SP & $\sigma_0\eta_0\vec{s}$ & $\widetilde{\sigma}_0\eta_0\vec{s}$ & $0$ (by definition) & $0$ &  $-g_1 (\Delta k)^2 D_0^2$ & \xmark & \xmark & \xmark & \xmark & $2J_H\beta_1$ & $-2J_H\beta_1$   \\

VP & $\sigma_0\eta_z s_0$ & $\widetilde{\sigma}_0\eta_z s_0$ & $0$ & $0$ & $-g_1 (\Delta k)^2 D_0^2$ & \xmark & \xmark & \xmark & \xmark & $0$ & --- \\
IVC$_+$ & $\sigma_0\eta_{x,y}$ & $\widetilde{\sigma}_0\eta_{x,y}$ & $c_1 D_0^4 + c_3 w_0^2$ &  $0$ & $-g_2 (\Delta k)^2 D_0^2$ & \xmark & \cmark & \cmark & \xmark & $-3J_H \beta_2$ & $J_H \beta_2$ \\
IVC$_-$ & $\sigma_y\eta_{x,y}$ & $\widetilde{\sigma}_z\eta_{x,y}$ & $c_1 D_0^4 + c_4 w_0^2 D_0^4$ & $0$ &  $-g_2 (\Delta k)^2 D_0^2$ & \cmark & \cmark & \xmark & \xmark & $-3J_H \beta_2$ & $J_H \beta_2$ \\
SLP$_-$ & $\sigma_y$ & $\widetilde{\sigma}_z$ & $c_3 w_0^2 + c_4 w_0^2 D_0^4$ & $-g_- (\Delta k)^2 |D_0|^3$ &  $-g_1 (\Delta k)^2 D_0^2$ & \cmark & \xmark & \cmark & \xmark & $0$ & ---  \\
SLP$_+$ & $\sigma_y \eta_z$ & $\widetilde{\sigma}_z \eta_z$ & $c_2 D_0^4 + c_3 w_0^2 + c_4 w_0^2 D_0^4$ & $-g_+ (\Delta k)^2 |D_0|^3$ &   $-g_3 (\Delta k)^2 D_0^2$ & \cmark & \xmark & \cmark & \xmark & $0$ & --- \\ 
SSLP$_-$ & $\sigma_y s_z$ & $\widetilde{\sigma}_z s_z$ & $c_3 w_0^2 + c_4 w_0^2 D_0^4$ & $0$ & $-g_1 (\Delta k)^2 D_0^2$ & \cmark & \xmark & \cmark & \xmark & $2J_H \beta_3$ & $-2J_H \beta_3$ \\
 \end{tabular}
 \end{ruledtabular}
\end{center}
\end{table*}

The first energetic constraint we will take into account is related to the fact that the graphene-like bands of valley $\eta=-$ ($\eta=+$) are far away from the Fermi level at the K (K$'$) point, see \figref{TrilayerBands}. This means that all mixing processes in $\Lambda_{\vec{k}}$ that induce a finite occupation of the upper or unoccupied states in the lower graphene-like band of valley $\eta=-$ ($\eta=+$) at the K (K$'$) point are suppressed. In \appref{MixingBetweenBands}, we show that this is equivalent to demanding that $(\mathcal{M}_{\vec{k}})_{(p',\eta',s'),(p,\mp,s)} = 0$ for $\vec{k} \in \mathcal{A}_{\pm}$. This reduces the number of generators further from $8$ to only $4$ [given in \equref{FourGeneratorsForIVCm}]. For each of these $4$ generators, $\Lambda_j$, $j=1,2,3,4$, we compute the change of energy associated with the deformation in \equref{MixingDeformation} where $U_{\vec{k}} = e^{i \varphi_{\vec{k}} \Lambda_j}$. It is found to be of the form
\begin{equation}
    \Delta E_{gb}[\varphi_{\vec{k}}] = \sum_{\vec{k}\in\mathcal{A}_+} \alpha_{\vec{k}} \sin^2 \varphi_{\vec{k}} + \beta_{\vec{k}} \sin \varphi_{\vec{k}} \cos \varphi_{\vec{k}}, \label{MixingEnergyChange}
\end{equation}
where $\alpha_{\vec{k}}$ and $\beta_{\vec{k}}$ are expressions involving the form factors, the interaction $V(\vec{q})$, and the band structure [see \appref{MixingBetweenBands} for explicit form], and thus depend on $D_0$ and $w_0$. In \equref{MixingEnergyChange}, we have already taken into account the $C_{2z}$ constraint that allowed us to write it as a sum over $\vec{k}\in\mathcal{A}_+$ only.

First, we find $\beta_{\vec{k}} = 0$ for $D_0=0$ and, hence, no mixing between the bands as expected by symmetry. However, even when $D_0\neq 0$, we still obtain $\beta_{\vec{k}} = 0$ for two of the four generators, while $\beta_{\vec{k}} \neq 0$ for the other two (say $\Lambda_{1,2}$) in the chiral limit ($w_0=0$). Consequently, at least when $w_0$ is not too large, the generators $\Lambda_{1,2}$ have to dominate. Which of those remaining two is dominant, cannot be determined purely analytically as it will depend on non-universal values of the form factors, so let us compare with numerics. The \textit{mixing matrix}, defined as 
\begin{equation}
    \left(M^Q_{\vec{k}}\right)_{(p,\eta,s),(p',\eta',s')} := \left(Q_{\vec{k}}\right)_{(b,p,\eta,s),(g,p',\eta',s')} \label{DefinitionOfMixingMatrix}
\end{equation}
and can be computed straightforwardly for the two candidate generators $\Lambda_{1,2}$. We find
\begin{subequations}\begin{equation}
    M^Q_{\vec{k}} = 2\sqrt{2} i \left( \sigma_x P_{\pm} \pm \sigma_0 \eta^{\mp} \right) \varphi_{\vec{k}}, \quad \vec{k}\in\mathcal{A}_{\pm}, \label{FirstMixingMatrixMQ}
\end{equation}
and
\begin{equation}
    M^Q_{\vec{k}} = 2\sqrt{2} i \left( \sigma_z P_{\pm} \pm i \sigma_y \eta^{\mp} \right) \varphi_{\vec{k}}, \quad \vec{k}\in\mathcal{A}_{\pm}, \label{SecondMixingMatrixMQ}
\end{equation}\end{subequations}
for $\Lambda_1$ and $\Lambda_2$, respectively. Here, we defined $P_{\pm}=(\eta_0\pm \eta_z)/2$ and $\eta^{\pm} = (\eta_x \pm i \eta_y)/2$. Intuitively, the first mixing matrix means that the lower (upper) graphene-like bands of the valley that is at low energies at K or K$'$ mixes with the upper (lower) and lower (upper) TBG-like bands of the same and opposite valley, respectively. The second option in \equref{SecondMixingMatrixMQ} describes the ``twisted'' situation where the lower (upper) graphene-like band mixes with the lower (upper) TBG-like bands of the same valley and the upper (lower) band in the opposite valley. The HF result for $Q_{\vec{k}}$ close to the K point, shown in \figref{Energycut}(b), is consistent with \equref{FirstMixingMatrixMQ}.   

Minimizing \equref{MixingEnergyChange} and noting that $\beta_{\vec{k}}$ scales linearly with $D_0$ for small $D_0$ while $\alpha_{\vec{k}} > 0$ at $D_0=0$, we find the scaling of the energetic gain to be $\Delta E_{gb} \sim -g_3 (\Delta k)^2 D_0^2$, where $g_3>0$; this is indicated in \tableref{BehaviorOfEnergetics}.  

In the same way, all other candidate states can be studied, see \appref{MixingBetweenBands}. In accordance with our expectation based on symmetry, we find no mixing and vanishing energetic gain, $\Delta E_{gb} = 0$, when $D_0=0$, while mixing immediately sets in and $\Delta E_{gb} < 0$ once $D_0\neq 0$. In all cases, $\Delta E_{gb}$ scales as $\Delta E_{gb} \sim -g_j (\Delta k)^2 D_0^2$, $g_j > 0$, as $D_0\rightarrow 0$. While $g_j = g_j(w_0)$ depend on $w_0$, we list $\Delta E_{gb}$ in the chiral limit, $w_0=0$, in \tableref{BehaviorOfEnergetics} as it reveals some structure: we see that the energetic gain coming from hybridization is identical for the two IVCs and it is the same for the four states SP, VP, SLP$_-$, and SSLP$_-$ in the chiral limit.

The mixing matrices $M^Q$ in \equref{DefinitionOfMixingMatrix} for these states provide some additional consistency checks between analytics and numerics. For instance, the IVC$_+$'s analysis closely parallels the one outlined above for the IVC$_-$: out of $16$ symmetry-allowed generators, only two candidate combinations remain. Interestingly, their associated $\alpha_{\vec{k}}$ and $\beta_{\vec{k}}$ in \equref{MixingEnergyChange} are the same (to leading order in $D_0$) as those of the two IVC$_{-}$ candidates. Having identified $M^Q_{\vec{k}}$ in \equref{FirstMixingMatrixMQ} as being dominant by comparison with numerics, we can read off which of the two analytical candidates for the IVC$_+$ must be realized; indeed, we find the same one in numerics [compare \equref{SecondMixingMatrix} and \figref{MatrixOrders}]. To provide a second example, we find a mixing matrix for the SLP$_-$ given by
\begin{equation}
    M^Q_{\vec{k}} = 2i(\sigma_y + \sigma_z) P_{\pm}\varphi_{\vec{k}}, \quad \vec{k}\in\mathcal{A}_{\pm}.
\end{equation}
This means that there is only mixing in the valley for which the graphene-like bands are at low energies and that both upper and lower graphene-like bands mix with both TBG-like bands. This is consistent with the $Q_{\vec{k}}$ in \figref{Energycut}(b) we found for the SLP$_-$ in HF close to the K point.

\subsection{Deforming the form factors}\label{DeformedFormFactors}
Apart from the mixing with and the ``proximity-induced'' order in the graphene-like bands, there are also important energetic contributions directly within the TBG-like bands, when tuning away from the chiral-flat-decoupled limit. While some of these contributions are analogous to TBG \cite{2020arXiv200913530L,bultinck2019ground}, some others are not: non-zero $D_0$ strongly breaks $P$ (see \tableref{ActionOfSymmetriesMainText}) and, hence, induces terms in the TBG-like band structure, see second line in \equref{TBGLikeDispersion}, without any analogue in TBG; it also leads to terms in the form factors [see \equref{bbMatrixElements} for details] that cannot be present in TBG and have not been studied in the literature.

We here begin with the impact of the form factors and restrict ourselves, for now, to the flat limit, $W_{\text{TBG}}=0$. In the column labeled $\Delta E_{bb}$ in \tableref{BehaviorOfEnergetics} we show the energy of the candidate states for non-zero $w_0$ and $D_0$ relative to the SP [see \appref{DeformationOfFormFactors} for explicit expressions for the prefactors $c_j > 0$]. First, for $D_0=0$, we recover the previous result \cite{bultinck2019ground} that the IVC$_-$ and VP are the only states besides the SP that are not penalized when turning on $w_0$. Second, we see that this changes once $D_0\neq 0$: if $w_0=0$, it is the SLP group of states that is not suppressed by breaking $P$ with $D_0$ \footnote{Note that the term $c_2$, which comes from the Hartree term, turns out to be numerically very small.}, while both IVCs have increasing energy. When both $w_0$ and $D_0$ are simultaneously non-zero, also the SLP group is suppressed (by the exact same amount as the IVC$_-$) compared to the SP and VP. Algebraically, this is related to the fact that the SP and VP order parameters are the only ones that commute with all form factors once $w_0,D_0\neq 0$ [cf.~\equref{bbMatrixElements}].

\subsection{Finite TBG bandwidth}\label{PerturbationTheoryInBandwidth}
Finally, we take into account the finite bandwidth of the TBG-like bands by doing perturbation theory in $W_{\text{TBG}}$ in \equref{FormOfNoninteractingHam}, starting from the product states associated with the candidate orders of \tableref{ListOfOrderParameters}. We will only outline the results here and refer the interested reader to \appref{Superexchange}.

Since the correction to the relative energies of the different candidate orders vanishes to first order in $W_{\text{TBG}}$, we focus on second order perturbation theory, which, if non-zero, will always lower the energy of the states and can be thought of as ``superexchange''. We consider the superexchange processes associated with all four terms, $E^b_{j}$, $j=0,1,2,3$, in \equref{TBGLikeDispersion} of the TBG-like dispersion; the corresponding energetic gain will scale as $W_{\text{TBG}}^2/U$ for $j=0$, $W_{\text{TBG}}^2 w_0^2/U$ for $j=1$, $W_{\text{TBG}}^2D_0^2/U$ for $j=2$, and $W_{\text{TBG}}^2w_0^2D_0^2/U$ for $j=3$ to leading order in $D_0$ and $w_0$, where $U$ is the energy scale associated with occupying a $\vec{k}$-state of an unoccupied flavor in a given ground state. Which of these four superexchange processes are ``active'' for the candidate states is listed in \tableref{BehaviorOfEnergetics}. We find that, by virtue of being proportional to the identity in \equref{TBGLikeDispersion}, $E^b_{3}$ does not affect the energy of the states. Furthermore, we see that the displacement-induced superexchange process favors the IVC$_+$ and the SLP group of states.

\subsection{Comparison of energetics}
Taken together, the energetics obtained analytically as summarized in \tableref{BehaviorOfEnergetics} agrees well with the numerics in several aspects: we can see that the energies of the SLP$_+$, SLP$_-$, and SSLP$_-$ are expected to be very close, exactly as seen in HF, see \figref{Energycut}. There is a very small splitting between the three associated with the fact that the SSLP$_-$ cannot benefit from ordering in the graphene-like bands, that that effect is weaker for the SLP$_+$ than for the SLP$_-$ (recall $g_- > g_+$, for small $w_0$), and that the SLP$_+$ is slightly suppressed by the Hartree term. We have seen that there is a small energetic preference towards the SLP$_-$ in the numeris as well.
In addition, we can also read off from \tableref{BehaviorOfEnergetics} that the SLP group of states should be preferred for small $w_0$ and large $D_0$, which is consistently seen in our HF phase diagram in \figref{PhaseDiagram}.

Notwithstanding the good agreement between the HF and the analytics, concerning the graphene-ordering, mixing between the bands, and the energetics, there are also differences. These can be traced back to the additional presence of remote bands, in particular, the bands with $\sigma_h$ eigenvalues $+1$ just above (below) the almost flat TBG-like band $p=+$ ($p=-$) away from the K/K$'$ points. While these are included in our HF numerics, they are not taken into account in the analytics. Most notably, we have seen that mixing between these bands and the almost flat TBG-like bands further lowers the energy of IVC$_-$ state relative to the SP and VP states.

\subsection{Breaking SU(2)$_+ \times$ SU(2)$_-$}
\label{BreakingSU2xSU2}

Finally, we come back to the fact that certain Hund's partners of states are degenerate in the model we have focused on so far and defined in \secref{ModelAndSymmetries}; this can be traced back to the presence of the SU(2)$_+ \times$ SU(2)$_-$ spin symmetry. Turning on a finite Hund's coupling, $J_H\neq 0$, will lift this degeneracy and also slightly affect the relative energetics of the candidate states. For the form of the Hund's coupling defined in \equref{HundsCouplingInteraction}, we have computed the respective change of energy. With more general and explicit expressions available in \appref{HundsCouplingAppendix}, we present their impact on the energies of the states in the last two columns of \tableref{BehaviorOfEnergetics} in the chiral-flat-decoupled limit. Here $\beta_j$ are positive constants obeying $\beta_2>\beta_1,\beta_3$. 
While estimates based on the Coulomb interaction yield an energy change (per unit cell) of $2J_H\beta_1 \approx -0.2\,\text{meV}$ for the SP state in TBG \cite{PhysRevB.101.165141}, its actual effective value might be smaller due to screening processes and additional electron-phonon coupling, which can also change the sign of $J_H$.  

However, irrespective of the sign of $J_H$ and its precise magnitude, we see that it will favor the SSLP$_{\pm}$ states in the otherwise almost degenerate SLP group of states, since the SLP$_{\pm}$ cannot lower their energy with $J_H\neq 0$. This is due to the fact that SLP$_{\pm}$ are their own Hund's partners. Similarly, the SP and SVP states will be favored over the VP phase. Finally, we note that the energetic gain coming from $J_H$ is of the same order of magnitude (even larger) for the SIVC$_-$ (IVC$_-$) compared to the SP (SVP) state. Consequently, we do not expect that $J_H$ will change the fact that the IVC$_-$ dominates over the SP in the phase diagram in \figref{PhaseDiagram}.

\section{Numerics for $\nu=\pm 2$}\label{nu2}
In this section, we will discuss the numerical results for correlated states at half-filling of the lower or upper TBG-like bands, $\nu=\pm 2$. 

\subsection{Procedure and results}

We find self-consistent HF solutions with energies lower than the symmetry-unbroken normal state where, at $D_0=0$ ($D_0\neq 0$), the TBG (TBG-like and graphene-like) bands are spin polarized and, on top of this, exhibit any of the candidate orders defined for $\nu=0$ in \tableref{ListOfOrderParameters}. The obtained spin polarization is our explanation for the observed \cite{Park_2021,Hao_2021} reduced flavor degeneracy setting in around $|\nu|=2$.

To be more explicit, our self-consistent solutions are found by starting, say for $\nu=-2$, from a correlator $P^{b}_{\nu=-2}=\frac{1}{2}(\mathbbm{1}+s_z)\frac{1}{2}(\mathbbm{1}+Q^b)$ in the TBG-like subspace, where $Q^b$ is any of the candidate orders in \tableref{ListOfOrderParameters}; the initial correlator for the graphene-like sector is taken to be filled lower bands of the continuum model at charge neutrality, $P^{g}=\frac{1}{2}(\mathbbm{1}-\sigma_z)$. The corresponding correlator at $\nu=+2$ in the TBG-like sector is simply given by $P^b_{\nu=+2}=\mathbbm{1}_{8\times8}-P^{b}_{\nu=-2}$. We apply the same iterative procedure as for $\nu=0$ to obtain self-consistent solutions. 

We show the band structures resulting from our self-consistent calculation for the spin-polarized IVC$_-$ and spin-polarized SLP states in \figref{Bands2}. The energies we obtain up to $D_0/w_1=0.5$ are shown in \appref{NumericsAtNuIsTwo}, but are qualitatively similar to the numerical energies we computed at $\nu=0$; we note though the spin-valley-polarized and spin-polarized IVC$_-$ are closer in energy at $\nu=2$ than the SVP and IVC$_-$ state at $\nu=0$. As can be seen in \figref{Bands2}, the graphene bands at $D_0=0$ remain unpolarized, in agreement with the analytics in \secref{ExactGroundStates}. Once a finite displacement field is applied, also the graphene-like bands develop some small polarization. However, exactly as for $\nu=0$, the Dirac cones of the IVC$_-$ state are not gapped out. We note the IVC$_-$, which was semimetallic at $\nu=0$, is metallic at $\nu=\pm 2$ (albeit with small Fermi surfaces). This follows from the additional spin polarization at $|\nu|=2$, breaking SU(2)$_s$ which guaranteed its semimetallic character at $\nu=0$. In \figref{Bands2}(a,b), we see that the SLP state, which was insulating at $\nu=0$, becomes semimetallic at $|\nu|=2$, since the spin-polarization of the TBG-like bands only allows for SLP order in one of the spin species of the graphene-like bands.

\subsection{Connection to experiment}\label{Nu2ConnectionToExperiment}

An important feature in the experimental data on MSTG \cite{Park_2021,Hao_2021}, which may be related to our numerics, is the observation of enhanced resistivity at $\nu=0$ for both $D_0=0$ and $D_0>0$ and a state with high resistivity observed only at a finite value of $D_0$ at $\nu=2$. Taking our results for band structures at $\nu=0$ and $|\nu|=2$ together, we note that at $\nu=0$, the leading phase in our numerics is either a semimetallic IVC$_-$ state or an insulating SLP state, both of which may be compatible with the observed high resistivity state at $\nu=0$. At $\nu=2$, we find our leading solution is a metallic IVC$_-$ state for small values of $D_0$ and a semimetallic SLP state for large enough $D_0$---a possible explanation for why high resistivity at $|\nu|=2$ only sets in above a finite value of $D_0$ in experiment.

\begin{figure}[t]
    \centering
    \includegraphics[width=1\linewidth]{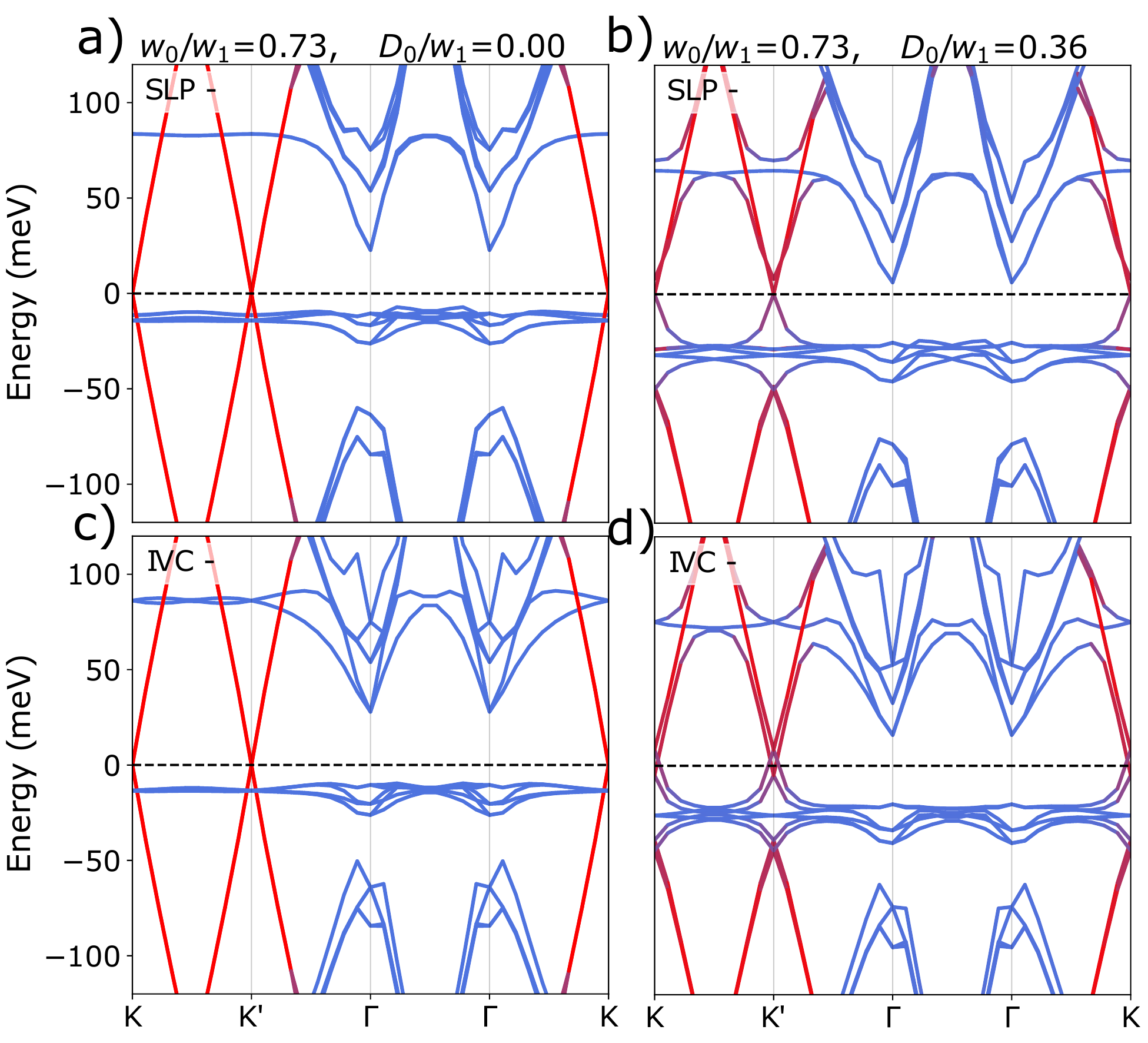}

    \caption{HF band structures at $\nu=2$ for the spin polarized SLP$_-$ state (a) at $D_0=0$ and (b) $D_0>0$ and spin polarized IVC$_-$ state (c) at $D_0=0$ and (d) $D_0>0$.}
    \label{Bands2}
\end{figure}

Another notable experimental observation is that the reset of the band structure at $|\nu|=2$ splits into a Dirac-like feature at $|\nu|=2$ and a van-Hove singulartiy, associated with a rapid increase and sign change in the Hall density, at $|\nu|=2-\delta(D_0)$, $\delta(D_0) \ll 1$ \cite{Park_2021}. If we assume that the spin polarization we found persists for a finite range of $|\nu|$ below $2$, the bandstructures in \figref{Bands2} provide a natural explanation: lowering the chemical potential, e.g., in \figref{Bands2}(d), until it hits the lower, almost flat set of bands will lead to a Lifshitz transition where the hole pockets around the K and K$'$ points merge. This could explain the observed behavior of the Hall density. We note that the tendency, visible in \figref{Bands2}, that increasing $D_0$ pushes these almost flat bands away from the Fermi level at $|\nu|=2$, is consistent with this feature only being visible at non-zero $D_0$ and the associated $\delta(D_0)$ increasing with $|D_0|$ in experiment \cite{Park_2021}.
Numerical results for larger values of $D_0$ are discussed in \appref{LargerValuesOfD}.

\section{Superconductivity}\label{Superconductivity}
Having established the nature of the correlated phases in MSTG at various filling fractions, we next study the consequences for the superconducting states.

\subsection{Pairing in the presence of polarization}\label{SCinPresenceOfPolarization}
Let us begin with the range of electron filling, $2 < |\nu| < 3$, where superconductivity is most prominently observed in experiment \cite{Park_2021,Hao_2021,2021arXiv210312083C}. We have seen in \secref{nu2} that spin polarization supplemented with IVC$_-$ order is favored for realistic parameters at $|\nu|=2$. Because the associated reduction of the number of flavors in the normal state is seen in experiment over the entire or most of the superconducting range of $\nu$, we assume that spin polarization and superconductivity co-exist at least in part of the phase diagram. As follows from the analysis in \refcite{PhysRevResearch.2.033062}, where a classification of pairing in almost SU(2)$_+ \times$ SU(2)$_-$-symmetric graphene moir\'e systems in the presence and absence of flavor polarization can be found, the superconducting state has to be in a non-unitary triplet phase---irrespective of the precise pairing mechanism. We emphasize that this also holds if the additional IVC$_-$ ordering found in the HF at $|\nu|=2$ co-exists with superconductivity in a finite range of $\nu$:  the bands above the Fermi level in the IVC$_-$ band structure in \figref{Bands2} still exhibit Kramers partners at momenta $\vec{k}$ and $-\vec{k}$ with the same spin, due to the preserved spinless $\widetilde{\Theta}$ symmetry; these degenerate states can form Cooper pairs with non-unitary triplet vector. 
Note that this would not be the case, e.g., for the VP state [shown in \figref{fullnu2bands}], which does not exhibit exactly degenerate energy levels at $\vec{k}$ and $-\vec{k}$ above the Fermi level. Given the strong tendency of MSTG towards superconductivity, this VP phase is, thus, a less natural candidate order. Indeed, we find it to be subleading in our HF numerics, see \secref{nu2}. 

It is important to note, however, that the spin polarization (and additional IVC$_-$ state) is only realized for $J_H<0$. While this is expected to be the case if $J_H$ stems entirely from the Coulomb interaction, its sign is unknown. For $J_H>0$, we would, instead, obtain the SVP phase with additional SIVC$_{-}$ order (cf.~\tableref{ListOfOrderParameters}). In that case, the associated superconductor would also be the corresponding Hund's partner, which is an admixture between a singlet and unitary triplet \cite{PhysRevResearch.2.033062}.
In the presence of a magnetic field there is a crucial difference between the two scenarios: while, for $J_H<0$, the SP will just align with the Zeemann field and the superconductor will remain a non-unitary triplet, the two antiparallel spin polarizations in the two valleys of the SVP state will be canted gradually; a coexisting singlet-unitary-triplet superconductor will continuously transition into a non-unitary triplet \cite{PhysRevResearch.2.033062}. For completeness, we demonstrate this explicitly in \appref{SCInPolarizedState}. Based on the estimate $J_H \lesssim 0.2\,\text{meV}$ \cite{PhysRevB.101.165141}, we obtain $3\,\textrm{T}$ as the characteristic magnetic field strength of the transition.

\subsection{Pairing without polarization}\label{NoSpinPolarization}
In experiment \cite{Park_2021,Hao_2021,2021arXiv210312083C}, there are also regions of $\nu$ and $D_0$ with superconductivity but without any signs of flavor polarization in the corresponding normal state (SC II in \figref{SchematicPhaseDiagram}). We here discuss the nature and origin of these superconducting phases.

In light of recent experiments in TBG \cite{StepanovTuning,Liu1261,SaitoTuning}, which indicate that electron-phonon coupling plays an important role for pairing in graphene moir\'e systems, the picture proposed in \refcite{PhysRevB.102.064501} provides a very promising microscopic scenario: while electron-phonon coupling is likely important to stabilize superconductivity and crucially determines the critical temperature, it might predominantly mediate an SU(2)$_+ \times$ SU(2)$_-$ symmetric pairing interaction. This leads to the near degeneracy of singlet and triplet pairing. In the additional presence of flavor polarization, its structure determines whether singlet or triplet is realized, as discussed above.
In the absence of polarization, electron-phonon coupling by itself can never favor triplet over singlet and at most make the two degenerate \cite{PhysRevB.90.184512,PhysRevB.93.174509}; however, additional Cooper-channel interactions coming from the fluctuation of particle-hole orders can determine whether the superconductor will be in a singlet or triplet state \cite{PhysRevB.102.064501}. 
To analyze this, we will consider the action
\begin{equation}
    \mathcal{S} = \mathcal{S}_0 + \mathcal{S}_{\phi} + \mathcal{S}_{\phi f} +  \mathcal{S}_{\text{phonon}}, \label{FullActionForFluctuations}
\end{equation}
which consists of the bare non-interacting action $\mathcal{S}_0$ associated with $H_0$ in \equref{FormOfNoninteractingHam} and $\mathcal{S}_{\phi}$, given by
\begin{equation}
    \mathcal{S}_{\phi} = \frac{1}{2}\int_q \phi_{q}^j \left[\chi^{-1}(i\Omega_n,\vec{q})\right]_{j,j'} \phi_{-q}^{j'}, \label{SphiAction}
\end{equation}
where $\phi_{q}^j$ is the set of real bosonic fields (labelled by $j$) describing the fluctuations of a given candidate particle-hole order in \tableref{ListOfOrderParameters}. We use $q=(i\Omega_n,\vec{q})$ labeling bosonic Matsubara frequencies $\Omega_n$ and momentum $\vec{q}$, introduce the short-cut $\int_q \ldots \equiv T\sum_{\Omega_n} \sum_{\vect{q}} \ldots$, and denote the (fully renormalized, low-energy) susceptibility in the particle-hole channel under consideration by $\chi(i\Omega_n,\vec{q})$. The bosons $\phi_{q}^j$ are coupled to the low-energy electron fields $f_{k;n,\eta,s}$, with $k=(i\omega_n,\vec{k})$, via 
\begin{equation}
    \mathcal{S}_{\phi f} = \int_q\int_k f^\dagger_{k+q} \lambda^j(\vec{k}+\vec{q},\vec{k})  f^\pdagger_{k} \, \phi_{q}^j. \label{phifcoupling}
\end{equation}

Note that the coupling vertex $\lambda^j(\vec{k}+\vec{q},\vec{k})$ is in general a matrix in valley, spin, and band space. For instance, a minimal description of fluctuations of the IVC$_-$ state is given by the two-component boson $\phi^{j}_{q}$, $j=x,y$, with  $[\chi(i\Omega_n,\vec{q})]_{j,j'} \propto \delta_{j,j'}/(\Omega_n^2 + c^2 \vec{q}^2 +\xi^{-2})$, where the coherence length $\xi$ parametrizes the proximity to the critical point. Furthermore, $\lambda^j(\vec{k},\vec{k}') = s_0\eta_j f_{\vec{k},\vec{k}'}$, where $f_{\vec{k},\vec{k}'}$ is a matrix in band space ($p=\pm$, $t=b,g$) only, which obeys  $f^\pdagger_{\vec{k},\vec{k}'} = f^\dagger_{\vec{k}',\vec{k}} = -f^T_{\vec{k}',\vec{k}}$ due to Hermiticity and $\Theta C_{2z}$ symmetry.
Finally, the last part $\mathcal{S}_{\text{phonon}}$ in \equref{FullActionForFluctuations} stands for the action of the phonons and their coupling to the electrons.

We integrate out the bosonic modes $\phi_q^j$ in the action of \equref{FullActionForFluctuations}, leading to an interaction between the fermions $f$. In the saddle-point equations of the Cooper channel, this interaction can be viewed as a SU(2)$_+ \times$ SU(2)$_-$-symmetry breaking correction to the SU(2)$_+ \times$ SU(2)$_-$-symmetric interaction coming from the phonons in $\mathcal{S}_{\text{phonon}}$. 

As shown in \refcite{PhysRevB.102.064501}, whether this tips the balance towards singlet or triplet is determined by the behavior of the fluctuating modes $\phi_q^j$ under \textit{spinful} time-reversal $\Theta_s$: if the bosonic mode is even (odd) under $\Theta_s$, as indicated by $+$ ($-$) in the column $\Theta_s$ in \tableref{ListOfOrderParameters}, it will generically favor singlet (triplet) over triplet (singlet) pairing---in the presence of fine-tuning or additional symmetries, the two might remain degenerate. 
We emphasize that this conclusion does not depend on microscopic details such the precise form of $\chi$ in \equref{SphiAction} or of $\lambda^j$ in \equref{phifcoupling}. 
Due to U(1)$_v$ symmetry, the system also exhibits the spinless time-reversal symmetry $\widetilde{\Theta}$, see \tableref{ActionOfSymmetriesMainText}. We have generalized the analysis of \refcite{PhysRevB.102.064501} in \appref{AppSuperconductivity} to also include this form of time-reversal symmetry and proved that any $\phi_q^j$ that is even (odd) under $\widetilde{\Theta}$ will generically favor triplet (singlet). This also implies that any $\phi_q^j$ which has the same behavior under $\Theta_s$ and $\widetilde{\Theta}$ will keep singlet and triplet degenerate. As expected, this is precisely the case for all order parameters in \tableref{ListOfOrderParameters} that are their own Hund's partner.

First, we focus on the superconducting domes (indicated by SC II in \figref{SchematicPhaseDiagram}) that emerge in the vicinity of but outside the region with flavor polarization and reconstructed band structure \cite{Park_2021}. As mentioned above, with Coulomb interactions only, the IVC$_-$ order with additional spin polarization is favored over the Hund's partner, the SIVC$_-$ with additional SVP order. We can read off from \tableref{ListOfOrderParameters} that triplet will then be favored over singlet pairing. As shown in \refcite{PhysRevResearch.2.033062}, the resulting triplet will be unitary within mean-field theory, but can become non-unitary when spin fluctuations corrections become significant. These two scenarios can be distinguished experimentally since the unitary triplet will exhibit a BKT transition (of a charge-$4e$, spin-rotation-invariant order parameter combination), while the non-unitary state will not. For the other sign of $J_H$, singlet pairing (if mean-field theory applies) or a mixed singlet-triplet phase (if the SVP fluctuations dominate) will be realized \cite{PhysRevResearch.2.033062}.

Second, we also comment on possible pairing phases close to charge neutrality, although these have not been seen experimentally. For the same sign, $J_H < 0$, that favors SP at $\nu=2$, we obtain the SIVC$_{-}$ at $\nu=0$ as the dominant, semimetallic instability at small $D_0$ and large $w_0$, see \figref{PhaseDiagram}. As can be seen from \tableref{ListOfOrderParameters}, it is even under $\Theta_s$ and will, hence, favor singlet pairing.
Among the SLP group of states, realized for larger $D_0$, only fluctuations of the SSLP$_\pm$ states will break the SU(2)$_+ \times$ SU(2)$_-$-symmetry; the SSLP$_-$, realized for $J_H < 0$, will also favor singlet.

Finally, if electron-phonon coupling does not play any role for pairing in MSTG, i.e., the term $\mathcal{S}_{\text{phonon}}$ in \equref{FullActionForFluctuations} can be neglected, we still obtain the same results (see \appref{AppFluctSC} for a derivation): if $\phi_q^j$ are even/odd (odd/even) under $\Theta_s$ ($\widetilde{\Theta}$), singlet/triplet pairing will be favored. Consequently, the above statements about the singlet-triplet competition still apply in this scenario as well.

\subsection{Relevance of Dirac cones and topology}\label{WZWTerms}
Motivated by recent theoretical works \cite{Khalafeabf5299,Christos_2020,2020arXiv201001144C} discussing the potential importance of WZW terms for superconductivity and insulating behavior in TBG, we here analyze under which conditions this can also be relevant for MSTG. 
We note that the WZW term is a purely kinematic term associating a Berry phase with spatio-temporal textures of the orders defined on the TBG-like bands, and is independent of the interactions between the electrons in these bands.
The presence of such a topological term  has crucial consequences for direct transitions between superconductivity and correlated insulators, without closing a single-particle gap at integer fillings. 
We explain in \appref{AppendixOnWZW} how the following results can be formally derived from the exhaustive classification of WZW physics in TBG in \refcite{Christos_2020} and focus here on the main picture and implications.

Although the main superconducting phase is found between $|\nu|=2$ and $|\nu|=3$ in experiment \cite{Park_2021,Hao_2021,2021arXiv210312083C}, we begin our analysis at the charge neutrality point, $\nu=0$, with associated superconducting phase labelled SC III in \figref{SchematicPhaseDiagram}. Since we expect superconductivity to survive finite displacement fields, we focus on Dirac cones that remain close to the Fermi level for $D_0\neq 0$.
Inspection of the band structure in \figref{TrilayerBands}(b) reveals that the effect of $D_0$ is to push the two Dirac cones of the graphene and TBG bands of valley $\eta=+$ ($\eta=-$) at K (K$'$) away from the Fermi level, leaving only a single TBG-like Dirac cone of valley $\eta=-$ ($\eta=+$) at low energies. Denoting the fermionic fields by $\psi_{\vec{q}}$ of these two Dirac cones (per spin), their Hamiltonian reads as 
\begin{equation}
    H^{\text{D}}_0=\sum_{\vec{q}} \psi^\dagger_{\vec{q}}\left[\rho_x\mu_z s_0 q_x+\rho_y s_0 q_y\right] \psi^\pdagger_{\vec{q}}, \label{D0Hamiltonian}
\end{equation} 
where $\vec{q}$ is momentum measured relative to the K (K$'$) point for the Dirac cone at ``mini-valley'' $\mu_z=+$ ($\mu_z=-$). Furthermore, $\rho_j$ are Pauli matrices acting in the Dirac space, which are related to the band-space matrices $\sigma_j$ used above. In this notation, $C_{2z}$ and (anti-unitary) spinfull time-reversal act as $C_{2z}:\,\psi_{\vec{q}}\rightarrow \rho_x \mu_x \psi_{-\vec{q}}$ and $\Theta_s:\,\psi_{\vec{q}}\rightarrow T \psi_{-\vec{q}}$, $T=s_y \mu_x$, respectively.

Similar to \refcite{Christos_2020}, we ask what different types of particle-hole orders, $m_j$, and superconducting order parameters, $\Delta$, coupling to $\psi_{\vec{q}}$ as
\begin{equation}
    H^{\text{D}}_1=\sum_{\vec{q},j} \psi^\dagger_{\vec{q}} m_j \psi^\pdagger_{\vec{q}}+\sum_{\vec{q}} \left(\psi^\dagger_{\vec{q}} \Delta T \psi^\dagger_{-\vec{q}}+\text{H.c.}\right), \label{DeltaAndmj}
\end{equation}
can form WZW terms. 
The Hamiltonian $H^{\text{D}}_0+H^{\text{D}}_1$ is equivalent to the low-energy Dirac Hamiltonian for TBG projected into a valley-minivalley locked subspace, meaning that all pairings and insulating orders with WZW derived in \cite{Christos_2020} which survive projection to the same subspace will also be viable in MSTG. This is also the reason why the number of possible WZW terms we find here is significantly reduced as compared to \refcite{Christos_2020}.

\begin{table}[tb]
\begin{center}
\caption{Possible particle-hole, $m_j$, and superconducting order parameters, $\Delta$, that can exhibit mutual WZW terms at the indicated filling fractions $\nu=0$ and $|\nu|=2$. We here use the Dirac notation of \equsref{D0Hamiltonian}{DeltaAndmj}. As in \refcite{Christos_2020}, moir\'e density wave (MDW) indicates that the state breaks moir\'e translational symmetry; $A$ and $B$ are refer to the irreducible representations of the point group $C_6$ of the superconductor (SC). The MDW with singlet pairing row below corresponds to the first row of Table IV in \refcite{Christos_2020}, while the MDW with triplet SC corresponds to the first row of Table VII (or more explicitly, in lines 5 and 9 of Table XV).}
\label{PossibleWZWTerms}
\begin{ruledtabular}
 \begin{tabular} {cccccc} 
$|\nu|$ & $m_j$ & type & $\Delta$ & SC type \\ \hline
$0,2$ & $\mu_z\rho_z\vec{s}$ & SSLP$_-$ & $\mathbbm{1}$ & $A$ singlet \\
$0,2$ & $\rho_x(\mu_x,\mu_y); \rho_z$ & MDW; SLP$_+$ & $\mathbbm{1}$ & $A$ singlet \\ \hline
$2$ & $\rho_z(s_x,s_y);\rho_z\mu_z s_z$& SSLP$_+$/SSLP$_-$ & $\mu_z s_z$ & $B$ unit. triplet\\ 
$2$ & $\rho_xs_z(\mu_x,\mu_y);\rho_z$& MDW/SLP$_+$ & $\mu_z s_z$ & $B$ unit. triplet

 \end{tabular}
\end{ruledtabular}
\end{center}
\end{table}

Based on our analysis of particle-hole instabilities, the Dirac cones in \equref{D0Hamiltonian} can either be those of the non-interacting bands or those of our leading instability at small $D_0$---the semimetallic IVC$_-$ state with band structure shown in \figref{Bands}---which could persist for an extended range of $|\nu|>0$; the resulting WZW terms we discuss next are identical in both scenarios.

Due to spin-rotation invariance, triplet pairing is not consistent with a WZW term \cite{Christos_2020}. We find that singlet pairing, $\Delta= \rho_0 \mu_0 s_0$, which transforms under the irreducible representation $A$ of the point group, is the only possible superconducting state with WZW term at $\nu=0$. This is consistent with the singlet we established in \secref{NoSpinPolarization} near $\nu=0$ due to electron-phonon coupling and particle-hole fluctuations. 
Considering all possible compatible insulating orders $m_j$, we find there are two types: a spin Hall order of the form $\mu_z\rho_z s_{x,y,z}$, which aligns with our SSLP$_-$ order in \tableref{ListOfOrderParameters}, and a moir\'e density wave (MDW) state, $\rho_x\mu_{x,y}$, which breaks translations on the moir\'e lattice scale together with an SLP$_+$ state ($\rho_z$), see \tableref{PossibleWZWTerms}. We note that since this order is defined in the minivalley-valley locked space and, in the full space of MSTG, thus also breaks U(1)$_v$ symmetry in the same way as our IVC states.
Interestingly, the SSLP$_-$ is precisely our leading instability for larger $D_0$ and, hence, constitutes indeed a natural candidate $m_j$.
The MDW state goes beyond our analysis in this work, as we have not considered states which break moir\'e translational symmetry. We will leave this for future work.

At $|\nu|=2$, the situation is more complicated, because the entire band structure is reconstructed as a consequence of interaction-induced flavor polarization. Nonetheless, we expect the low-energy Dirac theory to be still of the form of \equref{D0Hamiltonian}. While there are other possible microscopic realizations, this is true for our leading instability of the SP (or SVP) with additional IVC$_-$ (or SIVC$_-$) order with band structure in \figref{Bands2}(c,d). Neglecting the small $D_0$-induced SP (or SVP) in the graphene-like bands, \equref{D0Hamiltonian} still applies.

As summarized in \tableref{PossibleWZWTerms}, all options for insulators with singlet pairing then carry over to $|\nu|=2$. A difference between the two cases is the broken SU(2)$_s$ spin symmetry, which also allows for triplet pairing. In this case, there are two additional options compatible with the unitary triplet $\Delta =\mu_z s_z$: one set with insulating orders $\rho_z s_{x,y}$ and $\rho_z\mu_z s_z$ and another set with $\rho_x s_z\mu_{x,y}$ and $\rho_z$. The first two orders correspond to our SSLP$_+$ and SSLP$_-$, respectively. The second two states are a spin polarized MDW state and the SLP$_+$ state. We find no non-unitary state to be consistent with WZW terms. 
Since only either singlet or unitary triplet are possible, the underlying flavor polarization of the TBG-like bands in \figref{Bands}(c,d) must be SVP rather than SP. Consequently, WZW terms at $|\nu|=2$ are more likely relevant if $J_H>0$.

\section{Conclusion and discussion}\label{SummaryOutlook}

To summarize, we have studied particle-hole instabilities and superconductivity in MSTG, for different filling fractions $\nu$ and displacement field values $D_0$, using a combination of analytical arguments and HF numerics. We started, in \secref{ExactGroundStates}, in the limit $D_0=0$, where the non-interacting band structure is just given by the spectrum of TBG and single-layer graphene. In the interacting Hamiltonian (\ref{ThreeTermsInHFD}), these two subsystems are coupled by the density-density interaction in \equref{FormOfCouplingBetweenTheTwoSubSys}. We showed that states of the form $\ket{\Psi^{\text{g}}_{j}(\nu_b)}\ket{\Psi_0^{\text{b}}(\nu_b)}$, where $\ket{\Psi^{\text{g}}_{j}(\nu_b)}$ are the correlated semimetallic eigenstates of single-layer graphene and $\ket{\Psi_0^{\text{b}}(\nu_b)}$ are the eigenstates \cite{2020arXiv200913530L} in \equref{ExactGroundStatesMainText} of the TBG Hamiltonian in the flat limit, are also exact eigenstates of the MSTG in \equref{ThreeTermsInHFD}. Furthermore, if $\ket{\Psi_0^{\text{b}}(\nu_b)}$ are groundstates of the TBG Hamiltonian, $\ket{\Psi^{\text{g}}_{j}(\nu_b)}\ket{\Psi_0^{\text{b}}(\nu_b)}$ are shown to be exact groundstates of MSTG for a finite range of the coupling strength, quantified by $\lambda$ in \equref{FormOfCouplingBetweenTheTwoSubSys}, between the two sectors. In this sense, not only the bare band structure but also the \textit{interacting} physics of MSTG can be separated into that of single-layer graphene and TBG in the flat-decoupled limit.

We used these results to construct the set of candidate particle-hole orders summarized in \tableref{ListOfOrderParameters}, which are exactly degenerate in the chiral-flat-decoupled limit [defined by further setting $w_0=0$ in \equref{InterlayerCouplingMatrixElements}]. We took these states as our starting point of the HF numerics and analytical perturbation theory, which are not based on $\lambda$ being small and allow to tune away from the chiral-flat-decoupled limit. 
The resulting band structures for $\nu=0$ of the most important candidate orders are shown in \figref{Bands} for zero and non-zero $D_0$, with $w_0$ close to what is believed to describe the real system \cite{PhysRevB.96.075311,2019PhRvR...1a3001C} and finite TBG bandwidth.  
We see three distinct types of behavior, with different experimental signatures, when turning on $D_0$: the IVC$_{\pm}$ and their Hund's partners SIVC$_{\pm}$ retain their semimetallic behavior for $D_0\neq 0$, while the graphene Dirac cones of the SLP$_{\pm}$ and SSLP$_-$ (and its Hund's partner SSLP$_+$) are gapped out when turning on $D_0$. Interestingly, for the VP and SP (and its Hund's partner SVP), a finite $D_0$ induces small Fermi surfaces. In combination with further transport and, in particular, scanning tunneling microscopy experiments, which are sensitive to the local spectrum of the system, the computed spectra could help shed light on the correlated physics of MSTG.

Also the relative energetics between these candidate states as a function of $D_0$, $w_0$, and the TBG bandwidth is very rich: as summarized in \tableref{BehaviorOfEnergetics}, there are many contributions without any analogue in TBG---the mixing between the TBG and graphene bands, additional ordering in the graphene bands, and the $D_0$-induced breaking of symmetries in the TBG sector, which changes interaction matrix elements and induces new superexchange processes.
The resultant phase diagram in the $D_0$-$w_0$ plane with leading and subleading phases is presented in \figref{PhaseDiagram}: at small $D_0$, a semimetallic intervalley coherent phase is favored, which transitions into a sublattice polarized phase at larger $D_0$; among the latter set of states, we expect the SSLP$_\pm$ to dominate as they are the only states in this otherwise almost degenerate manifold that can benefit from the intervalley Hund's coupling $J_H$ in \equref{HundsCouplingInteraction}.

At $|\nu|=2$, we found self-consistent HF solutions for all of the candidate states in \tableref{ListOfOrderParameters}, which coexists with additional spin polarization (either of the SP or SVP type, depending on the sign of $J_H$). This can explain the experimentally observed \cite{Park_2021,Hao_2021} band resetting for $2 \lesssim |\nu| \lesssim 3$. For instance, the spectrum for the intervalley coherent state is shown in \figref{Bands2}(c) and (d) for $\nu=2$: increasing the filling fraction $\nu$ slightly will lead to Fermi surfaces of completely spin-polarized TBG-like bands, i.e., with half the number of flavors. 
In magnetic fields, the SP band resetting ($J_H<0$) will not change its form, while the SVP-related resetting ($J_H>0$) will continuously develop a finite canting and transform into a SP configuration; we estimate the associated magnetic-field scale to be of order of $3\,T$.
Since $J_H<0$ follows for pure Coulomb interactions \cite{PhysRevB.101.165141}, we expect the SP to be a more natural candidate; however, also $J_H>0$ is possible, both theoretically and experimentally, and so we have studied both signs of $J_H$ in our analysis.

Building on our results for the correlated normal states of MSTG, we analyzed the superconducting order parameters in \secref{Superconductivity} in the different regimes indicated by SC I--III in \figref{SchematicPhaseDiagram}: when superconductivity co-exists with flavor polarization (SC I), the nature of the pairing state depends crucially on the form of the flavor reduction; for the SP ($J_H<0$) and SVP ($J_H>0$) polarization, that we find around $|\nu|=2$, we obtain, respectively, a non-unitary triplet and its Hund's partner---a singlet-unitary-triplet admixed phase \cite{PhysRevResearch.2.033062}. 
For the superconducting phases (SC II) close to but not in the flavor-polarized region, we find triplet pairing to dominate for $J_H<0$, while the state will either be a singlet or admixed singlet-triplet phases for $J_H>0$.

The behavior of these superconducting phases in in-plane magnetic fields, $B_{\parallel}$, follows from \refcite{PhysRevResearch.2.033062} where their respective phase diagrams in the temperature-$B_{\parallel}$-plane have been worked out: in the presence of SP polarization ($J_H<0$), the critical temperature, $T_c$, of the non-unitary triplet is not affected by the Zeemann coupling and suppressed in quadratic order in the in-plane orbital coupling. This naturally explains the strong violation of the Pauli limit \cite{2021arXiv210312083C}. 
In the other case of SVP polarization ($J_H>0$), the behavior of $T_c$ of the associated singlet-triplet phase is the same, with the only difference that also the singlet-triplet admixture will gradually transform into a non-unitary triplet with increasing magnetic field.
Understanding the reentrant superconducting behavior seen at even larger magnetic fields \cite{2021arXiv210312083C} will require understanding the fate of the correlated parent states, e.g., in \figref{Bands2}, in large magnetic fields, which we leave for future work.
Irrespective of whether the order parameter of SC II is a triplet or a singlet-triplet admixed phase, it will continuously transform into a non-unitary triplet \cite{PhysRevResearch.2.033062} upon applying $B_{\parallel}$, while being eventually suppressed by the orbital coupling.
We emphasize that all of the superconducting states we find, including the triplets, are protected \cite{PhysRevB.102.064501} against non-magnetic impurities on the moir\'e scale, i.e., exhibit an analogue of the ``Anderson theorem'', which is typically only expected for singlet superconductors \cite{RevModPhys.75.657}.

For completeness, we also investigated superconductivity close to the charge neutrality point (SC III), although not prominently seen in current experiments. Here we predict singlet pairing to dominate for the same sign of the Hund's coupling, $J_H>0$, we found triplet pairing near $\nu=2$.

Because the band structure of MSTG also exhibits Dirac cones, a subset of which persist in most of our dominant particle-hole instabilities [see, e.g., IVC$_-$ in \figref{Bands} and \figref{Bands2}(c,d); related to the preserved $C_{2z}\Theta$ symmetry], we have also studied the possible WZW terms between superconducting and insulating orders; as discussed in \cite{Khalafeabf5299,Christos_2020} for TBG, these topological terms are associated with and sensitive to the chirality of the Dirac cones in the normal-state band structure.
We have shown here that the set of possibilities in MSTG is greatly reduced as compared to TBG \cite{Christos_2020}, resulting from the reduced flavor degeneracy at $|\nu| \gtrsim 2$ and the impact of the displacement field at $\nu=0$. As compiled in \tableref{PossibleWZWTerms}, only $s$-wave singlet pairing is consistent with WZW terms close to $\nu=0$ (SC I), while both singlet and triplet pairing can have WZW terms in the flavor-polarized region (SC III). The corresponding particle-hole order parameters that form a mutual WZW term with these superconductors feature the sublattice-polarized states that we find to dominate at finite $D_0$  (in particular, the SSLP$_-$ state) and intervalley-coherent moir\'e density wave (MDW) phases. 

Unlike all the states studied in the HF analyses of this paper, the MDW phases break translational symmetry at the scale of the moir\'e period. The MDW states were
previously studied \cite{Christos_2020} in the context of WZW terms in TBG, break the moir\'e translational symmetry and the valley U(1)$_v$ symmetry. A closely related `incommensurate Kekul\'e spiral' (IKS) appeared in a recent
HF numerics study \cite{2021arXiv210505857K} of TBG. (We also note that Kekul\'e states have been observed in single layer graphene on a Cu substrate \cite{Pasupathy_kekule}, and in the zeroth Landau level \cite{He_kekule}.) Motivated by these results, and our study of WZW terms here, we believe that these types of states are very promising possible additional instabilities in MSTG as well, in models which include the breaking of the $\text{SU}(2)_+ \times \text{SU}(2)_-$ symmetry to the physical SU(2)$_s$ spin rotation symmetry. We leave a detailed energetic study of these states for future work.

\begin{acknowledgments}
M.S.S. thanks R.~Samajdar for discussions and previous collaborations \cite{PhysRevResearch.2.033062,PhysRevB.102.064501} on unconventional superconductivity in moir\'e systems, and acknowledges discussions with T.~Lang, A.~L\"auchli, and R.~Fernandes. This research was supported by the National Science Foundation under Grant No. DMR-2002850. This work was also supported by the Simons Collaboration on Ultra-Quantum Matter, which is a grant from the Simons Foundation (651440, S.S.).

\end{acknowledgments}

\bibliography{draft_Refs}

\begin{thebibliography}{73}%
\makeatletter
\providecommand \@ifxundefined [1]{%
 \@ifx{#1\undefined}
}%
\providecommand \@ifnum [1]{%
 \ifnum #1\expandafter \@firstoftwo
 \else \expandafter \@secondoftwo
 \fi
}%
\providecommand \@ifx [1]{%
 \ifx #1\expandafter \@firstoftwo
 \else \expandafter \@secondoftwo
 \fi
}%
\providecommand \natexlab [1]{#1}%
\providecommand \enquote  [1]{``#1''}%
\providecommand \bibnamefont  [1]{#1}%
\providecommand \bibfnamefont [1]{#1}%
\providecommand \citenamefont [1]{#1}%
\providecommand \href@noop [0]{\@secondoftwo}%
\providecommand \href [0]{\begingroup \@sanitize@url \@href}%
\providecommand \@href[1]{\@@startlink{#1}\@@href}%
\providecommand \@@href[1]{\endgroup#1\@@endlink}%
\providecommand \@sanitize@url [0]{\catcode `\\12\catcode `\$12\catcode
  `\&12\catcode `\#12\catcode `\^12\catcode `\_12\catcode `\%12\relax}%
\providecommand \@@startlink[1]{}%
\providecommand \@@endlink[0]{}%
\providecommand \url  [0]{\begingroup\@sanitize@url \@url }%
\providecommand \@url [1]{\endgroup\@href {#1}{\urlprefix }}%
\providecommand \urlprefix  [0]{URL }%
\providecommand \Eprint [0]{\href }%
\providecommand \doibase [0]{http://dx.doi.org/}%
\providecommand \selectlanguage [0]{\@gobble}%
\providecommand \bibinfo  [0]{\@secondoftwo}%
\providecommand \bibfield  [0]{\@secondoftwo}%
\providecommand \translation [1]{[#1]}%
\providecommand \BibitemOpen [0]{}%
\providecommand \bibitemStop [0]{}%
\providecommand \bibitemNoStop [0]{.\EOS\space}%
\providecommand \EOS [0]{\spacefactor3000\relax}%
\providecommand \BibitemShut  [1]{\csname bibitem#1\endcsname}%
\let\auto@bib@innerbib\@empty
\bibitem [{\citenamefont {MacDonald}(2019)}]{macdonald2019bilayer}%
  \BibitemOpen
  \bibfield  {author} {\bibinfo {author} {\bibfnamefont {A.~H.}\ \bibnamefont
  {MacDonald}},\ }\bibfield  {title} {\enquote {\bibinfo {title} {{Bilayer
  Graphene's Wicked, Twisted Road}},}\ }\href {\doibase 10.1103/physics.12.12}
  {\bibfield  {journal} {\bibinfo  {journal} {Physics}\ }\textbf {\bibinfo
  {volume} {12}},\ \bibinfo {pages} {12} (\bibinfo {year} {2019})}\BibitemShut
  {NoStop}%
\bibitem [{\citenamefont {Andrei}\ and\ \citenamefont
  {MacDonald}(2020)}]{andrei2020graphene}%
  \BibitemOpen
  \bibfield  {author} {\bibinfo {author} {\bibfnamefont {E.~Y.}\ \bibnamefont
  {Andrei}}\ and\ \bibinfo {author} {\bibfnamefont {A.~H.}\ \bibnamefont
  {MacDonald}},\ }\bibfield  {title} {\enquote {\bibinfo {title} {Graphene
  bilayers with a twist},}\ }\href {\doibase 10.1038/s41563-020-00840-0}
  {\bibfield  {journal} {\bibinfo  {journal} {Nat. Mater.}\ }\textbf {\bibinfo
  {volume} {19}},\ \bibinfo {pages} {1265} (\bibinfo {year}
  {2020})}\BibitemShut {NoStop}%
\bibitem [{\citenamefont {Kennes}\ \emph {et~al.}(2021)\citenamefont {Kennes},
  \citenamefont {Claassen}, \citenamefont {Xian}, \citenamefont {Georges},
  \citenamefont {Millis}, \citenamefont {Hone}, \citenamefont {Dean},
  \citenamefont {Basov}, \citenamefont {Pasupathy},\ and\ \citenamefont
  {Rubio}}]{kennes2020moir}%
  \BibitemOpen
  \bibfield  {author} {\bibinfo {author} {\bibfnamefont {D.~M.}\ \bibnamefont
  {Kennes}}, \bibinfo {author} {\bibfnamefont {M.}~\bibnamefont {Claassen}},
  \bibinfo {author} {\bibfnamefont {L.}~\bibnamefont {Xian}}, \bibinfo {author}
  {\bibfnamefont {A.}~\bibnamefont {Georges}}, \bibinfo {author} {\bibfnamefont
  {A.~J.}\ \bibnamefont {Millis}}, \bibinfo {author} {\bibfnamefont
  {J.}~\bibnamefont {Hone}}, \bibinfo {author} {\bibfnamefont {C.~R.}\
  \bibnamefont {Dean}}, \bibinfo {author} {\bibfnamefont {D.~N.}\ \bibnamefont
  {Basov}}, \bibinfo {author} {\bibfnamefont {A.~N.}\ \bibnamefont
  {Pasupathy}}, \ and\ \bibinfo {author} {\bibfnamefont {A.}~\bibnamefont
  {Rubio}},\ }\bibfield  {title} {\enquote {\bibinfo {title} {Moiré
  heterostructures as a condensed-matter quantum simulator},}\ }\href {\doibase
  10.1038/s41567-020-01154-3} {\bibfield  {journal} {\bibinfo  {journal} {Nat.
  Phys.}\ }\textbf {\bibinfo {volume} {17}},\ \bibinfo {pages} {155} (\bibinfo
  {year} {2021})}\BibitemShut {NoStop}%
\bibitem [{\citenamefont {Balents}\ \emph {et~al.}(2020)\citenamefont
  {Balents}, \citenamefont {Dean}, \citenamefont {Efetov},\ and\ \citenamefont
  {Young}}]{balents2020superconductivity}%
  \BibitemOpen
  \bibfield  {author} {\bibinfo {author} {\bibfnamefont {L.}~\bibnamefont
  {Balents}}, \bibinfo {author} {\bibfnamefont {C.~R.}\ \bibnamefont {Dean}},
  \bibinfo {author} {\bibfnamefont {D.~K.}\ \bibnamefont {Efetov}}, \ and\
  \bibinfo {author} {\bibfnamefont {A.~F.}\ \bibnamefont {Young}},\ }\bibfield
  {title} {\enquote {\bibinfo {title} {Superconductivity and strong
  correlations in moir{\'e} flat bands},}\ }\href {\doibase
  10.1038/s41567-020-0906-9} {\bibfield  {journal} {\bibinfo  {journal} {Nat.
  Phys.}\ }\textbf {\bibinfo {volume} {16}},\ \bibinfo {pages} {725} (\bibinfo
  {year} {2020})}\BibitemShut {NoStop}%
\bibitem [{\citenamefont {Scheurer}(2019)}]{STMReview}%
  \BibitemOpen
  \bibfield  {author} {\bibinfo {author} {\bibfnamefont {M.~S.}\ \bibnamefont
  {Scheurer}},\ }\bibfield  {title} {\enquote {\bibinfo {title} {Spectroscopy
  of graphene with a magic twist},}\ }\href {\doibase
  10.1038/d41586-019-02285-1} {\bibfield  {journal} {\bibinfo  {journal}
  {Nature}\ }\textbf {\bibinfo {volume} {572}},\ \bibinfo {pages} {40}
  (\bibinfo {year} {2019})}\BibitemShut {NoStop}%
\bibitem [{\citenamefont {{Zaletel}}(2021)}]{ZaletelJournalClub}%
  \BibitemOpen
  \bibfield  {author} {\bibinfo {author} {\bibfnamefont {T.}~\bibnamefont
  {{Zaletel}}},\ }\bibfield  {title} {\enquote {\bibinfo {title}
  {{Stronger-correlated superconductivity in magic-angle twisted trilayer
  graphene}},}\ }\href@noop {} {\bibfield  {journal} {\bibinfo  {journal}
  {Journal Club in Condensed Matter Physics}\ } (\bibinfo {year}
  {2021})}\BibitemShut {NoStop}%
\bibitem [{\citenamefont {{Cao}}\ \emph
  {et~al.}(2018{\natexlab{a}})\citenamefont {{Cao}}, \citenamefont {{Fatemi}},
  \citenamefont {{Demir}}, \citenamefont {{Fang}}, \citenamefont {{Tomarken}},
  \citenamefont {{Luo}}, \citenamefont {{Sanchez-Yamagishi}}, \citenamefont
  {{Watanabe}}, \citenamefont {{Taniguchi}}, \citenamefont {{Kaxiras}},
  \citenamefont {{Ashoori}},\ and\ \citenamefont
  {{Jarillo-Herrero}}}]{2018Natur.556...80C}%
  \BibitemOpen
  \bibfield  {author} {\bibinfo {author} {\bibfnamefont {Y.}~\bibnamefont
  {{Cao}}}, \bibinfo {author} {\bibfnamefont {V.}~\bibnamefont {{Fatemi}}},
  \bibinfo {author} {\bibfnamefont {A.}~\bibnamefont {{Demir}}}, \bibinfo
  {author} {\bibfnamefont {S.}~\bibnamefont {{Fang}}}, \bibinfo {author}
  {\bibfnamefont {S.~L.}\ \bibnamefont {{Tomarken}}}, \bibinfo {author}
  {\bibfnamefont {J.~Y.}\ \bibnamefont {{Luo}}}, \bibinfo {author}
  {\bibfnamefont {J.~D.}\ \bibnamefont {{Sanchez-Yamagishi}}}, \bibinfo
  {author} {\bibfnamefont {K.}~\bibnamefont {{Watanabe}}}, \bibinfo {author}
  {\bibfnamefont {T.}~\bibnamefont {{Taniguchi}}}, \bibinfo {author}
  {\bibfnamefont {E.}~\bibnamefont {{Kaxiras}}}, \bibinfo {author}
  {\bibfnamefont {R.~C.}\ \bibnamefont {{Ashoori}}}, \ and\ \bibinfo {author}
  {\bibfnamefont {P.}~\bibnamefont {{Jarillo-Herrero}}},\ }\bibfield  {title}
  {\enquote {\bibinfo {title} {{Correlated insulator behaviour at half-filling
  in magic-angle graphene superlattices}},}\ }\href {\doibase
  10.1038/nature26154} {\bibfield  {journal} {\bibinfo  {journal} {Nature}\
  }\textbf {\bibinfo {volume} {556}},\ \bibinfo {pages} {80} (\bibinfo {year}
  {2018}{\natexlab{a}})}\BibitemShut {NoStop}%
\bibitem [{\citenamefont {{Cao}}\ \emph
  {et~al.}(2018{\natexlab{b}})\citenamefont {{Cao}}, \citenamefont {{Fatemi}},
  \citenamefont {{Fang}}, \citenamefont {{Watanabe}}, \citenamefont
  {{Taniguchi}}, \citenamefont {{Kaxiras}},\ and\ \citenamefont
  {{Jarillo-Herrero}}}]{SuperconductivityTBG}%
  \BibitemOpen
  \bibfield  {author} {\bibinfo {author} {\bibfnamefont {Y.}~\bibnamefont
  {{Cao}}}, \bibinfo {author} {\bibfnamefont {V.}~\bibnamefont {{Fatemi}}},
  \bibinfo {author} {\bibfnamefont {S.}~\bibnamefont {{Fang}}}, \bibinfo
  {author} {\bibfnamefont {K.}~\bibnamefont {{Watanabe}}}, \bibinfo {author}
  {\bibfnamefont {T.}~\bibnamefont {{Taniguchi}}}, \bibinfo {author}
  {\bibfnamefont {E.}~\bibnamefont {{Kaxiras}}}, \ and\ \bibinfo {author}
  {\bibfnamefont {P.}~\bibnamefont {{Jarillo-Herrero}}},\ }\bibfield  {title}
  {\enquote {\bibinfo {title} {{Unconventional superconductivity in magic-angle
  graphene superlattices}},}\ }\href {\doibase 10.1038/nature26160} {\bibfield
  {journal} {\bibinfo  {journal} {Nature}\ }\textbf {\bibinfo {volume} {556}},\
  \bibinfo {pages} {43} (\bibinfo {year} {2018}{\natexlab{b}})}\BibitemShut
  {NoStop}%
\bibitem [{\citenamefont {{Shen}}\ \emph {et~al.}(2020)\citenamefont {{Shen}},
  \citenamefont {{Li}}, \citenamefont {{Wang}}, \citenamefont {{Zhao}},
  \citenamefont {{Tang}}, \citenamefont {{Liu}}, \citenamefont {{Tian}},
  \citenamefont {{Chu}}, \citenamefont {{Watanabe}}, \citenamefont
  {{Taniguchi}}, \citenamefont {{Yang}}, \citenamefont {{Meng}}, \citenamefont
  {{Shi}},\ and\ \citenamefont {{Zhang}}}]{2019arXiv190306952S}%
  \BibitemOpen
  \bibfield  {author} {\bibinfo {author} {\bibfnamefont {C.}~\bibnamefont
  {{Shen}}}, \bibinfo {author} {\bibfnamefont {N.}~\bibnamefont {{Li}}},
  \bibinfo {author} {\bibfnamefont {S.}~\bibnamefont {{Wang}}}, \bibinfo
  {author} {\bibfnamefont {Y.}~\bibnamefont {{Zhao}}}, \bibinfo {author}
  {\bibfnamefont {J.}~\bibnamefont {{Tang}}}, \bibinfo {author} {\bibfnamefont
  {J.}~\bibnamefont {{Liu}}}, \bibinfo {author} {\bibfnamefont
  {J.}~\bibnamefont {{Tian}}}, \bibinfo {author} {\bibfnamefont
  {Y.}~\bibnamefont {{Chu}}}, \bibinfo {author} {\bibfnamefont
  {K.}~\bibnamefont {{Watanabe}}}, \bibinfo {author} {\bibfnamefont
  {T.}~\bibnamefont {{Taniguchi}}}, \bibinfo {author} {\bibfnamefont
  {R.}~\bibnamefont {{Yang}}}, \bibinfo {author} {\bibfnamefont {Z.~Y.}\
  \bibnamefont {{Meng}}}, \bibinfo {author} {\bibfnamefont {D.}~\bibnamefont
  {{Shi}}}, \ and\ \bibinfo {author} {\bibfnamefont {G.}~\bibnamefont
  {{Zhang}}},\ }\bibfield  {title} {\enquote {\bibinfo {title} {{Correlated
  states in twisted double bilayer graphene}},}\ }\href {\doibase
  10.1038/s41567-020-0825-9} {\bibfield  {journal} {\bibinfo  {journal} {Nat.
  Phys.}\ }\textbf {\bibinfo {volume} {16}},\ \bibinfo {pages} {520} (\bibinfo
  {year} {2020})}\BibitemShut {NoStop}%
\bibitem [{\citenamefont {Liu}\ \emph {et~al.}(2020)\citenamefont {Liu},
  \citenamefont {Hao}, \citenamefont {Khalaf}, \citenamefont {Lee},
  \citenamefont {Ronen}, \citenamefont {Yoo}, \citenamefont {Najafabadi},
  \citenamefont {Watanabe}, \citenamefont {Taniguchi}, \citenamefont
  {Vishwanath},\ and\ \citenamefont {Kim}}]{ExperimentKim}%
  \BibitemOpen
  \bibfield  {author} {\bibinfo {author} {\bibfnamefont {X.}~\bibnamefont
  {Liu}}, \bibinfo {author} {\bibfnamefont {Z.}~\bibnamefont {Hao}}, \bibinfo
  {author} {\bibfnamefont {E.}~\bibnamefont {Khalaf}}, \bibinfo {author}
  {\bibfnamefont {J.~Y.}\ \bibnamefont {Lee}}, \bibinfo {author} {\bibfnamefont
  {Y.}~\bibnamefont {Ronen}}, \bibinfo {author} {\bibfnamefont
  {H.}~\bibnamefont {Yoo}}, \bibinfo {author} {\bibfnamefont {D.~H.}\
  \bibnamefont {Najafabadi}}, \bibinfo {author} {\bibfnamefont
  {K.}~\bibnamefont {Watanabe}}, \bibinfo {author} {\bibfnamefont
  {T.}~\bibnamefont {Taniguchi}}, \bibinfo {author} {\bibfnamefont
  {A.}~\bibnamefont {Vishwanath}}, \ and\ \bibinfo {author} {\bibfnamefont
  {P.}~\bibnamefont {Kim}},\ }\bibfield  {title} {\enquote {\bibinfo {title}
  {{Tunable spin-polarized correlated states in twisted double bilayer
  graphene}},}\ }\href {\doibase 10.1038/s41586-020-2458-7} {\bibfield
  {journal} {\bibinfo  {journal} {Nature}\ }\textbf {\bibinfo {volume} {583}},\
  \bibinfo {pages} {221} (\bibinfo {year} {2020})}\BibitemShut {NoStop}%
\bibitem [{\citenamefont {Cao}\ \emph {et~al.}(2020)\citenamefont {Cao},
  \citenamefont {Rodan-Legrain}, \citenamefont {Rubies-Bigorda}, \citenamefont
  {Park}, \citenamefont {Watanabe}, \citenamefont {Taniguchi},\ and\
  \citenamefont {Jarillo-Herrero}}]{PabllosExperiment}%
  \BibitemOpen
  \bibfield  {author} {\bibinfo {author} {\bibfnamefont {Y.}~\bibnamefont
  {Cao}}, \bibinfo {author} {\bibfnamefont {D.}~\bibnamefont {Rodan-Legrain}},
  \bibinfo {author} {\bibfnamefont {O.}~\bibnamefont {Rubies-Bigorda}},
  \bibinfo {author} {\bibfnamefont {J.~M.}\ \bibnamefont {Park}}, \bibinfo
  {author} {\bibfnamefont {K.}~\bibnamefont {Watanabe}}, \bibinfo {author}
  {\bibfnamefont {T.}~\bibnamefont {Taniguchi}}, \ and\ \bibinfo {author}
  {\bibfnamefont {P.}~\bibnamefont {Jarillo-Herrero}},\ }\bibfield  {title}
  {\enquote {\bibinfo {title} {Tunable correlated states and spin-polarized
  phases in twisted bilayer--bilayer graphene},}\ }\href {\doibase
  10.1038/s41586-020-2260-6} {\bibfield  {journal} {\bibinfo  {journal}
  {Nature}\ }\textbf {\bibinfo {volume} {583}},\ \bibinfo {pages} {215}
  (\bibinfo {year} {2020})}\BibitemShut {NoStop}%
\bibitem [{\citenamefont {Burg}\ \emph {et~al.}(2019)\citenamefont {Burg},
  \citenamefont {Zhu}, \citenamefont {Taniguchi}, \citenamefont {Watanabe},
  \citenamefont {MacDonald},\ and\ \citenamefont {Tutuc}}]{burg2019correlated}%
  \BibitemOpen
  \bibfield  {author} {\bibinfo {author} {\bibfnamefont {G.~W.}\ \bibnamefont
  {Burg}}, \bibinfo {author} {\bibfnamefont {J.}~\bibnamefont {Zhu}}, \bibinfo
  {author} {\bibfnamefont {T.}~\bibnamefont {Taniguchi}}, \bibinfo {author}
  {\bibfnamefont {K.}~\bibnamefont {Watanabe}}, \bibinfo {author}
  {\bibfnamefont {A.~H.}\ \bibnamefont {MacDonald}}, \ and\ \bibinfo {author}
  {\bibfnamefont {E.}~\bibnamefont {Tutuc}},\ }\bibfield  {title} {\enquote
  {\bibinfo {title} {{Correlated Insulating States in Twisted Double Bilayer
  Graphene}},}\ }\href {\doibase 10.1103/PhysRevLett.123.197702} {\bibfield
  {journal} {\bibinfo  {journal} {Phys. Rev. Lett.}\ }\textbf {\bibinfo
  {volume} {123}},\ \bibinfo {pages} {197702} (\bibinfo {year}
  {2019})}\BibitemShut {NoStop}%
\bibitem [{\citenamefont {{Chen}}\ \emph {et~al.}(2019)\citenamefont {{Chen}},
  \citenamefont {{Jiang}}, \citenamefont {{Wu}}, \citenamefont {{Lyu}},
  \citenamefont {{Li}}, \citenamefont {{Chittari}}, \citenamefont {{Watanabe}},
  \citenamefont {{Taniguchi}}, \citenamefont {{Shi}}, \citenamefont {{Jung}},
  \citenamefont {{Zhang}},\ and\ \citenamefont {{Wang}}}]{2019NatPh..15..237C}%
  \BibitemOpen
  \bibfield  {author} {\bibinfo {author} {\bibfnamefont {G.}~\bibnamefont
  {{Chen}}}, \bibinfo {author} {\bibfnamefont {L.}~\bibnamefont {{Jiang}}},
  \bibinfo {author} {\bibfnamefont {S.}~\bibnamefont {{Wu}}}, \bibinfo {author}
  {\bibfnamefont {B.}~\bibnamefont {{Lyu}}}, \bibinfo {author} {\bibfnamefont
  {H.}~\bibnamefont {{Li}}}, \bibinfo {author} {\bibfnamefont {B.~L.}\
  \bibnamefont {{Chittari}}}, \bibinfo {author} {\bibfnamefont
  {K.}~\bibnamefont {{Watanabe}}}, \bibinfo {author} {\bibfnamefont
  {T.}~\bibnamefont {{Taniguchi}}}, \bibinfo {author} {\bibfnamefont
  {Z.}~\bibnamefont {{Shi}}}, \bibinfo {author} {\bibfnamefont
  {J.}~\bibnamefont {{Jung}}}, \bibinfo {author} {\bibfnamefont
  {Y.}~\bibnamefont {{Zhang}}}, \ and\ \bibinfo {author} {\bibfnamefont
  {F.}~\bibnamefont {{Wang}}},\ }\bibfield  {title} {\enquote {\bibinfo {title}
  {{Evidence of a gate-tunable Mott insulator in a trilayer graphene moir{\'e}
  superlattice}},}\ }\href {\doibase 10.1038/s41567-018-0387-2} {\bibfield
  {journal} {\bibinfo  {journal} {Nat. Phys.}\ }\textbf {\bibinfo {volume}
  {15}},\ \bibinfo {pages} {237} (\bibinfo {year} {2019})},\ \Eprint
  {http://arxiv.org/abs/1803.01985} {arXiv:1803.01985 [cond-mat.mes-hall]}
  \BibitemShut {NoStop}%
\bibitem [{\citenamefont {Chen}\ \emph {et~al.}(2019)\citenamefont {Chen},
  \citenamefont {Sharpe}, \citenamefont {Gallagher}, \citenamefont {Rosen},
  \citenamefont {Fox}, \citenamefont {Jiang}, \citenamefont {Lyu},
  \citenamefont {Li}, \citenamefont {Watanabe}, \citenamefont {Taniguchi},
  \citenamefont {Jung}, \citenamefont {Shi}, \citenamefont {Goldhaber-Gordon},
  \citenamefont {Zhang},\ and\ \citenamefont
  {Wang}}]{SuperconductivityInTrilayer}%
  \BibitemOpen
  \bibfield  {author} {\bibinfo {author} {\bibfnamefont {G.}~\bibnamefont
  {Chen}}, \bibinfo {author} {\bibfnamefont {A.~L.}\ \bibnamefont {Sharpe}},
  \bibinfo {author} {\bibfnamefont {P.}~\bibnamefont {Gallagher}}, \bibinfo
  {author} {\bibfnamefont {I.~T.}\ \bibnamefont {Rosen}}, \bibinfo {author}
  {\bibfnamefont {E.~J.}\ \bibnamefont {Fox}}, \bibinfo {author} {\bibfnamefont
  {L.}~\bibnamefont {Jiang}}, \bibinfo {author} {\bibfnamefont
  {B.}~\bibnamefont {Lyu}}, \bibinfo {author} {\bibfnamefont {H.}~\bibnamefont
  {Li}}, \bibinfo {author} {\bibfnamefont {K.}~\bibnamefont {Watanabe}},
  \bibinfo {author} {\bibfnamefont {T.}~\bibnamefont {Taniguchi}}, \bibinfo
  {author} {\bibfnamefont {J.}~\bibnamefont {Jung}}, \bibinfo {author}
  {\bibfnamefont {Z.}~\bibnamefont {Shi}}, \bibinfo {author} {\bibfnamefont
  {D.}~\bibnamefont {Goldhaber-Gordon}}, \bibinfo {author} {\bibfnamefont
  {Y.}~\bibnamefont {Zhang}}, \ and\ \bibinfo {author} {\bibfnamefont
  {F.}~\bibnamefont {Wang}},\ }\bibfield  {title} {\enquote {\bibinfo {title}
  {Signatures of tunable superconductivity in a trilayer graphene moir{\'e}
  superlattice},}\ }\href {\doibase 10.1038/s41586-019-1393-y} {\bibfield
  {journal} {\bibinfo  {journal} {Nature}\ }\textbf {\bibinfo {volume} {572}},\
  \bibinfo {pages} {215} (\bibinfo {year} {2019})}\BibitemShut {NoStop}%
\bibitem [{\citenamefont {Chen}\ \emph {et~al.}(2020)\citenamefont {Chen},
  \citenamefont {Sharpe}, \citenamefont {Fox}, \citenamefont {Zhang},
  \citenamefont {Wang}, \citenamefont {Jiang}, \citenamefont {Lyu},
  \citenamefont {Li}, \citenamefont {Watanabe}, \citenamefont {Taniguchi},
  \citenamefont {Shi}, \citenamefont {Senthil}, \citenamefont
  {Goldhaber-Gordon}, \citenamefont {Zhang},\ and\ \citenamefont
  {Wang}}]{chen2020tunable}%
  \BibitemOpen
  \bibfield  {author} {\bibinfo {author} {\bibfnamefont {G.}~\bibnamefont
  {Chen}}, \bibinfo {author} {\bibfnamefont {A.~L.}\ \bibnamefont {Sharpe}},
  \bibinfo {author} {\bibfnamefont {E.~J.}\ \bibnamefont {Fox}}, \bibinfo
  {author} {\bibfnamefont {Y.-H.}\ \bibnamefont {Zhang}}, \bibinfo {author}
  {\bibfnamefont {S.}~\bibnamefont {Wang}}, \bibinfo {author} {\bibfnamefont
  {L.}~\bibnamefont {Jiang}}, \bibinfo {author} {\bibfnamefont
  {B.}~\bibnamefont {Lyu}}, \bibinfo {author} {\bibfnamefont {H.}~\bibnamefont
  {Li}}, \bibinfo {author} {\bibfnamefont {K.}~\bibnamefont {Watanabe}},
  \bibinfo {author} {\bibfnamefont {T.}~\bibnamefont {Taniguchi}}, \bibinfo
  {author} {\bibfnamefont {Z.}~\bibnamefont {Shi}}, \bibinfo {author}
  {\bibfnamefont {T.}~\bibnamefont {Senthil}}, \bibinfo {author} {\bibfnamefont
  {D.}~\bibnamefont {Goldhaber-Gordon}}, \bibinfo {author} {\bibfnamefont
  {Y.}~\bibnamefont {Zhang}}, \ and\ \bibinfo {author} {\bibfnamefont
  {F.}~\bibnamefont {Wang}},\ }\bibfield  {title} {\enquote {\bibinfo {title}
  {{Tunable correlated Chern insulator and ferromagnetism in a moir{\'e}
  superlattice}},}\ }\href {\doibase 10.1038/s41586-020-2049-7} {\bibfield
  {journal} {\bibinfo  {journal} {Nature}\ }\textbf {\bibinfo {volume} {579}},\
  \bibinfo {pages} {56} (\bibinfo {year} {2020})}\BibitemShut {NoStop}%
\bibitem [{\citenamefont {Dos~Santos}\ \emph {et~al.}(2007)\citenamefont
  {Dos~Santos}, \citenamefont {Peres},\ and\ \citenamefont
  {Neto}}]{dos2007graphene}%
  \BibitemOpen
  \bibfield  {author} {\bibinfo {author} {\bibfnamefont {J.~M. B.~L.}\
  \bibnamefont {Dos~Santos}}, \bibinfo {author} {\bibfnamefont {N.~M.~R.}\
  \bibnamefont {Peres}}, \ and\ \bibinfo {author} {\bibfnamefont {A.~H.~C.}\
  \bibnamefont {Neto}},\ }\bibfield  {title} {\enquote {\bibinfo {title}
  {Graphene bilayer with a twist: electronic structure},}\ }\href {\doibase
  10.1103/PhysRevLett.99.256802} {\bibfield  {journal} {\bibinfo  {journal}
  {Phys. Rev. Lett.}\ }\textbf {\bibinfo {volume} {99}},\ \bibinfo {pages}
  {256802} (\bibinfo {year} {2007})}\BibitemShut {NoStop}%
\bibitem [{\citenamefont {Bistritzer}\ and\ \citenamefont
  {MacDonald}(2011)}]{bistritzer2011moire}%
  \BibitemOpen
  \bibfield  {author} {\bibinfo {author} {\bibfnamefont {R.}~\bibnamefont
  {Bistritzer}}\ and\ \bibinfo {author} {\bibfnamefont {A.~H.}\ \bibnamefont
  {MacDonald}},\ }\bibfield  {title} {\enquote {\bibinfo {title} {Moir{\'e}
  bands in twisted double-layer graphene},}\ }\href {\doibase
  10.1073/pnas.1108174108} {\bibfield  {journal} {\bibinfo  {journal} {Proc.
  Natl. Acad. Sci. U.S.A.}\ }\textbf {\bibinfo {volume} {108}},\ \bibinfo
  {pages} {12233} (\bibinfo {year} {2011})}\BibitemShut {NoStop}%
\bibitem [{\citenamefont {Dos~Santos}\ \emph {et~al.}(2012)\citenamefont
  {Dos~Santos}, \citenamefont {Peres},\ and\ \citenamefont
  {Neto}}]{dos2012continuum}%
  \BibitemOpen
  \bibfield  {author} {\bibinfo {author} {\bibfnamefont {J.~M. B.~L.}\
  \bibnamefont {Dos~Santos}}, \bibinfo {author} {\bibfnamefont {N.~M.~R.}\
  \bibnamefont {Peres}}, \ and\ \bibinfo {author} {\bibfnamefont {A.~H.~C.}\
  \bibnamefont {Neto}},\ }\bibfield  {title} {\enquote {\bibinfo {title}
  {Continuum model of the twisted graphene bilayer},}\ }\href {\doibase
  10.1103/PhysRevB.86.155449} {\bibfield  {journal} {\bibinfo  {journal} {Phys.
  Rev. B}\ }\textbf {\bibinfo {volume} {86}},\ \bibinfo {pages} {155449}
  (\bibinfo {year} {2012})}\BibitemShut {NoStop}%
\bibitem [{\citenamefont {Kerelsky}\ \emph {et~al.}(2019)\citenamefont
  {Kerelsky}, \citenamefont {McGilly}, \citenamefont {Kennes}, \citenamefont
  {Xian}, \citenamefont {Yankowitz}, \citenamefont {Chen}, \citenamefont
  {Watanabe}, \citenamefont {Taniguchi}, \citenamefont {Hone}, \citenamefont
  {Dean}, \citenamefont {Rubio},\ and\ \citenamefont
  {Pasupathy}}]{PasupathySTM}%
  \BibitemOpen
  \bibfield  {author} {\bibinfo {author} {\bibfnamefont {A.}~\bibnamefont
  {Kerelsky}}, \bibinfo {author} {\bibfnamefont {L.~J.}\ \bibnamefont
  {McGilly}}, \bibinfo {author} {\bibfnamefont {D.~M.}\ \bibnamefont {Kennes}},
  \bibinfo {author} {\bibfnamefont {L.}~\bibnamefont {Xian}}, \bibinfo {author}
  {\bibfnamefont {M.}~\bibnamefont {Yankowitz}}, \bibinfo {author}
  {\bibfnamefont {S.}~\bibnamefont {Chen}}, \bibinfo {author} {\bibfnamefont
  {K.}~\bibnamefont {Watanabe}}, \bibinfo {author} {\bibfnamefont
  {T.}~\bibnamefont {Taniguchi}}, \bibinfo {author} {\bibfnamefont
  {J.}~\bibnamefont {Hone}}, \bibinfo {author} {\bibfnamefont {C.}~\bibnamefont
  {Dean}}, \bibinfo {author} {\bibfnamefont {A.}~\bibnamefont {Rubio}}, \ and\
  \bibinfo {author} {\bibfnamefont {A.~N.}\ \bibnamefont {Pasupathy}},\
  }\bibfield  {title} {\enquote {\bibinfo {title} {Maximized electron
  interactions at the magic angle in twisted bilayer graphene},}\ }\href
  {\doibase 10.1038/s41586-019-1431-9} {\bibfield  {journal} {\bibinfo
  {journal} {Nature}\ }\textbf {\bibinfo {volume} {572}},\ \bibinfo {pages}
  {95} (\bibinfo {year} {2019})}\BibitemShut {NoStop}%
\bibitem [{\citenamefont {{Cao}}\ \emph {et~al.}(2020)\citenamefont {{Cao}},
  \citenamefont {{Rodan-Legrain}}, \citenamefont {{Park}}, \citenamefont {{Noah
  Yuan}}, \citenamefont {{Watanabe}}, \citenamefont {{Taniguchi}},
  \citenamefont {{Fernandes}}, \citenamefont {{Fu}},\ and\ \citenamefont
  {{Jarillo-Herrero}}}]{Cao2020_nematics}%
  \BibitemOpen
  \bibfield  {author} {\bibinfo {author} {\bibfnamefont {Y.}~\bibnamefont
  {{Cao}}}, \bibinfo {author} {\bibfnamefont {D.}~\bibnamefont
  {{Rodan-Legrain}}}, \bibinfo {author} {\bibfnamefont {J.~M.}\ \bibnamefont
  {{Park}}}, \bibinfo {author} {\bibfnamefont {F.}~\bibnamefont {{Noah Yuan}}},
  \bibinfo {author} {\bibfnamefont {K.}~\bibnamefont {{Watanabe}}}, \bibinfo
  {author} {\bibfnamefont {T.}~\bibnamefont {{Taniguchi}}}, \bibinfo {author}
  {\bibfnamefont {R.~M.}\ \bibnamefont {{Fernandes}}}, \bibinfo {author}
  {\bibfnamefont {L.}~\bibnamefont {{Fu}}}, \ and\ \bibinfo {author}
  {\bibfnamefont {P.}~\bibnamefont {{Jarillo-Herrero}}},\ }\bibfield  {title}
  {\enquote {\bibinfo {title} {{Nematicity and Competing Orders in
  Superconducting Magic-Angle Graphene}},}\ }\href@noop {} {\bibfield
  {journal} {\bibinfo  {journal} {arXiv e-prints}\ } (\bibinfo {year}
  {2020})},\ \Eprint {http://arxiv.org/abs/2004.04148} {arXiv:2004.04148
  [cond-mat.mes-hall]} \BibitemShut {NoStop}%
\bibitem [{\citenamefont {Rubio-Verd\'{u}}\ \emph {et~al.}(2020)\citenamefont
  {Rubio-Verd\'{u}}, \citenamefont {Turkel}, \citenamefont {Song},
  \citenamefont {Klebl}, \citenamefont {Samajdar}, \citenamefont {Scheurer},
  \citenamefont {Venderbos}, \citenamefont {Watanabe}, \citenamefont
  {Taniguchi}, \citenamefont {Ochoa}, \citenamefont {Xian}, \citenamefont
  {Kennes}, \citenamefont {Fernandes}, \citenamefont {\'{A}ngel Rubio},\ and\
  \citenamefont {Pasupathy}}]{rubioverdu2020universal}%
  \BibitemOpen
  \bibfield  {author} {\bibinfo {author} {\bibfnamefont {C.}~\bibnamefont
  {Rubio-Verd\'{u}}}, \bibinfo {author} {\bibfnamefont {S.}~\bibnamefont
  {Turkel}}, \bibinfo {author} {\bibfnamefont {L.}~\bibnamefont {Song}},
  \bibinfo {author} {\bibfnamefont {L.}~\bibnamefont {Klebl}}, \bibinfo
  {author} {\bibfnamefont {R.}~\bibnamefont {Samajdar}}, \bibinfo {author}
  {\bibfnamefont {M.~S.}\ \bibnamefont {Scheurer}}, \bibinfo {author}
  {\bibfnamefont {J.~W.~F.}\ \bibnamefont {Venderbos}}, \bibinfo {author}
  {\bibfnamefont {K.}~\bibnamefont {Watanabe}}, \bibinfo {author}
  {\bibfnamefont {T.}~\bibnamefont {Taniguchi}}, \bibinfo {author}
  {\bibfnamefont {H.}~\bibnamefont {Ochoa}}, \bibinfo {author} {\bibfnamefont
  {L.}~\bibnamefont {Xian}}, \bibinfo {author} {\bibfnamefont {D.}~\bibnamefont
  {Kennes}}, \bibinfo {author} {\bibfnamefont {R.~M.}\ \bibnamefont
  {Fernandes}}, \bibinfo {author} {\bibnamefont {\'{A}ngel Rubio}}, \ and\
  \bibinfo {author} {\bibfnamefont {A.~N.}\ \bibnamefont {Pasupathy}},\
  }\href@noop {} {\enquote {\bibinfo {title} {Universal moir\'e nematic phase
  in twisted graphitic systems},}\ } (\bibinfo {year} {2020}),\ \Eprint
  {http://arxiv.org/abs/2009.11645} {arXiv:2009.11645 [cond-mat.str-el]}
  \BibitemShut {NoStop}%
\bibitem [{\citenamefont {Samajdar}\ \emph {et~al.}(2021)\citenamefont
  {Samajdar}, \citenamefont {Scheurer}, \citenamefont {Turkel}, \citenamefont
  {Rubio-Verd{\'{u}}}, \citenamefont {Pasupathy}, \citenamefont {Venderbos},\
  and\ \citenamefont {Fernandes}}]{10.1088/2053-1583/abfcd6}%
  \BibitemOpen
  \bibfield  {author} {\bibinfo {author} {\bibfnamefont {R.}~\bibnamefont
  {Samajdar}}, \bibinfo {author} {\bibfnamefont {M.}~\bibnamefont {Scheurer}},
  \bibinfo {author} {\bibfnamefont {S.}~\bibnamefont {Turkel}}, \bibinfo
  {author} {\bibfnamefont {C.}~\bibnamefont {Rubio-Verd{\'{u}}}}, \bibinfo
  {author} {\bibfnamefont {A.}~\bibnamefont {Pasupathy}}, \bibinfo {author}
  {\bibfnamefont {J.}~\bibnamefont {Venderbos}}, \ and\ \bibinfo {author}
  {\bibfnamefont {R.~M.}\ \bibnamefont {Fernandes}},\ }\bibfield  {title}
  {\enquote {\bibinfo {title} {Electric-field-tunable electronic nematic order
  in twisted double-bilayer graphene},}\ }\href
  {https://doi.org/10.1088/2053-1583/abfcd6} {\bibfield  {journal} {\bibinfo
  {journal} {2D Materials}\ }\textbf {\bibinfo {volume} {8}} (\bibinfo {year}
  {2021})}\BibitemShut {NoStop}%
\bibitem [{\citenamefont {Zondiner}\ \emph {et~al.}(2020)\citenamefont
  {Zondiner}, \citenamefont {Rozen}, \citenamefont {Rodan-Legrain},
  \citenamefont {Cao}, \citenamefont {Queiroz}, \citenamefont {Taniguchi},
  \citenamefont {Watanabe}, \citenamefont {Oreg}, \citenamefont {von Oppen},
  \citenamefont {Stern}, \citenamefont {Berg}, \citenamefont
  {Jarillo-Herrero},\ and\ \citenamefont {Ilani}}]{2019arXiv191206150Z}%
  \BibitemOpen
  \bibfield  {author} {\bibinfo {author} {\bibfnamefont {U.}~\bibnamefont
  {Zondiner}}, \bibinfo {author} {\bibfnamefont {A.}~\bibnamefont {Rozen}},
  \bibinfo {author} {\bibfnamefont {D.}~\bibnamefont {Rodan-Legrain}}, \bibinfo
  {author} {\bibfnamefont {Y.}~\bibnamefont {Cao}}, \bibinfo {author}
  {\bibfnamefont {R.}~\bibnamefont {Queiroz}}, \bibinfo {author} {\bibfnamefont
  {T.}~\bibnamefont {Taniguchi}}, \bibinfo {author} {\bibfnamefont
  {K.}~\bibnamefont {Watanabe}}, \bibinfo {author} {\bibfnamefont
  {Y.}~\bibnamefont {Oreg}}, \bibinfo {author} {\bibfnamefont {F.}~\bibnamefont
  {von Oppen}}, \bibinfo {author} {\bibfnamefont {A.}~\bibnamefont {Stern}},
  \bibinfo {author} {\bibfnamefont {E.}~\bibnamefont {Berg}}, \bibinfo {author}
  {\bibfnamefont {P.}~\bibnamefont {Jarillo-Herrero}}, \ and\ \bibinfo {author}
  {\bibfnamefont {S.}~\bibnamefont {Ilani}},\ }\bibfield  {title} {\enquote
  {\bibinfo {title} {Cascade of phase transitions and dirac revivals in
  magic-angle graphene},}\ }\href {\doibase 10.1038/s41586-020-2373-y}
  {\bibfield  {journal} {\bibinfo  {journal} {Nature}\ }\textbf {\bibinfo
  {volume} {582}},\ \bibinfo {pages} {203} (\bibinfo {year}
  {2020})}\BibitemShut {NoStop}%
\bibitem [{\citenamefont {Wong}\ \emph {et~al.}(2020)\citenamefont {Wong},
  \citenamefont {Nuckolls}, \citenamefont {Oh}, \citenamefont {Lian},
  \citenamefont {Xie}, \citenamefont {Jeon}, \citenamefont {Watanabe},
  \citenamefont {Taniguchi}, \citenamefont {Bernevig},\ and\ \citenamefont
  {Yazdani}}]{2019arXiv191206145W}%
  \BibitemOpen
  \bibfield  {author} {\bibinfo {author} {\bibfnamefont {D.}~\bibnamefont
  {Wong}}, \bibinfo {author} {\bibfnamefont {K.~P.}\ \bibnamefont {Nuckolls}},
  \bibinfo {author} {\bibfnamefont {M.}~\bibnamefont {Oh}}, \bibinfo {author}
  {\bibfnamefont {B.}~\bibnamefont {Lian}}, \bibinfo {author} {\bibfnamefont
  {Y.}~\bibnamefont {Xie}}, \bibinfo {author} {\bibfnamefont {S.}~\bibnamefont
  {Jeon}}, \bibinfo {author} {\bibfnamefont {K.}~\bibnamefont {Watanabe}},
  \bibinfo {author} {\bibfnamefont {T.}~\bibnamefont {Taniguchi}}, \bibinfo
  {author} {\bibfnamefont {B.~A.}\ \bibnamefont {Bernevig}}, \ and\ \bibinfo
  {author} {\bibfnamefont {A.}~\bibnamefont {Yazdani}},\ }\bibfield  {title}
  {\enquote {\bibinfo {title} {Cascade of electronic transitions in magic-angle
  twisted bilayer graphene},}\ }\href {\doibase 10.1038/s41586-020-2339-0}
  {\bibfield  {journal} {\bibinfo  {journal} {Nature}\ }\textbf {\bibinfo
  {volume} {582}},\ \bibinfo {pages} {198} (\bibinfo {year}
  {2020})}\BibitemShut {NoStop}%
\bibitem [{\citenamefont {Christos}\ \emph {et~al.}(2020)\citenamefont
  {Christos}, \citenamefont {Sachdev},\ and\ \citenamefont
  {Scheurer}}]{Christos_2020}%
  \BibitemOpen
  \bibfield  {author} {\bibinfo {author} {\bibfnamefont {M.}~\bibnamefont
  {Christos}}, \bibinfo {author} {\bibfnamefont {S.}~\bibnamefont {Sachdev}}, \
  and\ \bibinfo {author} {\bibfnamefont {M.~S.}\ \bibnamefont {Scheurer}},\
  }\bibfield  {title} {\enquote {\bibinfo {title} {Superconductivity,
  correlated insulators, and wess–zumino–witten terms in twisted bilayer
  graphene},}\ }\href {\doibase 10.1073/pnas.2014691117} {\bibfield  {journal}
  {\bibinfo  {journal} {Proceedings of the National Academy of Sciences}\
  }\textbf {\bibinfo {volume} {117}},\ \bibinfo {pages} {29543–29554}
  (\bibinfo {year} {2020})}\BibitemShut {NoStop}%
\bibitem [{\citenamefont {{Kang}}\ \emph {et~al.}(2021)\citenamefont {{Kang}},
  \citenamefont {{Bernevig}},\ and\ \citenamefont
  {{Vafek}}}]{2021arXiv210401145K}%
  \BibitemOpen
  \bibfield  {author} {\bibinfo {author} {\bibfnamefont {J.}~\bibnamefont
  {{Kang}}}, \bibinfo {author} {\bibfnamefont {B.~A.}\ \bibnamefont
  {{Bernevig}}}, \ and\ \bibinfo {author} {\bibfnamefont {O.}~\bibnamefont
  {{Vafek}}},\ }\bibfield  {title} {\enquote {\bibinfo {title} {{Cascades
  between light and heavy fermions in the normal state of magic angle twisted
  bilayer graphene}},}\ }\href@noop {} {\bibfield  {journal} {\bibinfo
  {journal} {arXiv e-prints}\ ,\ \bibinfo {eid} {arXiv:2104.01145}} (\bibinfo
  {year} {2021})},\ \Eprint {http://arxiv.org/abs/2104.01145} {arXiv:2104.01145
  [cond-mat.str-el]} \BibitemShut {NoStop}%
\bibitem [{\citenamefont {Scheurer}\ and\ \citenamefont
  {Samajdar}(2020)}]{PhysRevResearch.2.033062}%
  \BibitemOpen
  \bibfield  {author} {\bibinfo {author} {\bibfnamefont {M.~S.}\ \bibnamefont
  {Scheurer}}\ and\ \bibinfo {author} {\bibfnamefont {R.}~\bibnamefont
  {Samajdar}},\ }\bibfield  {title} {\enquote {\bibinfo {title} {Pairing in
  graphene-based moir\'e superlattices},}\ }\href {\doibase
  10.1103/PhysRevResearch.2.033062} {\bibfield  {journal} {\bibinfo  {journal}
  {Phys. Rev. Research}\ }\textbf {\bibinfo {volume} {2}},\ \bibinfo {pages}
  {033062} (\bibinfo {year} {2020})}\BibitemShut {NoStop}%
\bibitem [{\citenamefont {Zhang}\ \emph {et~al.}(2021)\citenamefont {Zhang},
  \citenamefont {Zhu}, \citenamefont {Kahn}, \citenamefont {Li}, \citenamefont
  {Yang}, \citenamefont {Herbig}, \citenamefont {Wu}, \citenamefont {Li},
  \citenamefont {Watanabe}, \citenamefont {Taniguchi}, \citenamefont {Cabrini},
  \citenamefont {Zettl}, \citenamefont {Zaletel}, \citenamefont {Wang},\ and\
  \citenamefont {Crommie}}]{zhang2020visualizing}%
  \BibitemOpen
  \bibfield  {author} {\bibinfo {author} {\bibfnamefont {C.}~\bibnamefont
  {Zhang}}, \bibinfo {author} {\bibfnamefont {T.}~\bibnamefont {Zhu}}, \bibinfo
  {author} {\bibfnamefont {S.}~\bibnamefont {Kahn}}, \bibinfo {author}
  {\bibfnamefont {S.}~\bibnamefont {Li}}, \bibinfo {author} {\bibfnamefont
  {B.}~\bibnamefont {Yang}}, \bibinfo {author} {\bibfnamefont {C.}~\bibnamefont
  {Herbig}}, \bibinfo {author} {\bibfnamefont {X.}~\bibnamefont {Wu}}, \bibinfo
  {author} {\bibfnamefont {H.}~\bibnamefont {Li}}, \bibinfo {author}
  {\bibfnamefont {K.}~\bibnamefont {Watanabe}}, \bibinfo {author}
  {\bibfnamefont {T.}~\bibnamefont {Taniguchi}}, \bibinfo {author}
  {\bibfnamefont {S.}~\bibnamefont {Cabrini}}, \bibinfo {author} {\bibfnamefont
  {A.}~\bibnamefont {Zettl}}, \bibinfo {author} {\bibfnamefont {M.~P.}\
  \bibnamefont {Zaletel}}, \bibinfo {author} {\bibfnamefont {F.}~\bibnamefont
  {Wang}}, \ and\ \bibinfo {author} {\bibfnamefont {M.~F.}\ \bibnamefont
  {Crommie}},\ }\bibfield  {title} {\enquote {\bibinfo {title} {Visualizing
  delocalized correlated electronic states in twisted double bilayer
  graphene},}\ }\href {\doibase 10.1038/s41467-021-22711-1} {\bibfield
  {journal} {\bibinfo  {journal} {Nature Communications}\ }\textbf {\bibinfo
  {volume} {12}},\ \bibinfo {pages} {2516} (\bibinfo {year}
  {2021})}\BibitemShut {NoStop}%
\bibitem [{\citenamefont {Liu}\ \emph {et~al.}(2021{\natexlab{a}})\citenamefont
  {Liu}, \citenamefont {Chiu}, \citenamefont {Lee}, \citenamefont {Farahi},
  \citenamefont {Watanabe}, \citenamefont {Taniguchi}, \citenamefont
  {Vishwanath},\ and\ \citenamefont {Yazdani}}]{STMTDBGYazdani}%
  \BibitemOpen
  \bibfield  {author} {\bibinfo {author} {\bibfnamefont {X.}~\bibnamefont
  {Liu}}, \bibinfo {author} {\bibfnamefont {C.-L.}\ \bibnamefont {Chiu}},
  \bibinfo {author} {\bibfnamefont {J.~Y.}\ \bibnamefont {Lee}}, \bibinfo
  {author} {\bibfnamefont {G.}~\bibnamefont {Farahi}}, \bibinfo {author}
  {\bibfnamefont {K.}~\bibnamefont {Watanabe}}, \bibinfo {author}
  {\bibfnamefont {T.}~\bibnamefont {Taniguchi}}, \bibinfo {author}
  {\bibfnamefont {A.}~\bibnamefont {Vishwanath}}, \ and\ \bibinfo {author}
  {\bibfnamefont {A.}~\bibnamefont {Yazdani}},\ }\bibfield  {title} {\enquote
  {\bibinfo {title} {Spectroscopy of a tunable moir{\'e}system with a
  correlated and topological flat band},}\ }\href {\doibase
  10.1038/s41467-021-23031-0} {\bibfield  {journal} {\bibinfo  {journal}
  {Nature Communications}\ }\textbf {\bibinfo {volume} {12}},\ \bibinfo {pages}
  {2732} (\bibinfo {year} {2021}{\natexlab{a}})}\BibitemShut {NoStop}%
\bibitem [{\citenamefont {Park}\ \emph {et~al.}(2021)\citenamefont {Park},
  \citenamefont {Cao}, \citenamefont {Watanabe}, \citenamefont {Taniguchi},\
  and\ \citenamefont {Jarillo-Herrero}}]{Park_2021}%
  \BibitemOpen
  \bibfield  {author} {\bibinfo {author} {\bibfnamefont {J.~M.}\ \bibnamefont
  {Park}}, \bibinfo {author} {\bibfnamefont {Y.}~\bibnamefont {Cao}}, \bibinfo
  {author} {\bibfnamefont {K.}~\bibnamefont {Watanabe}}, \bibinfo {author}
  {\bibfnamefont {T.}~\bibnamefont {Taniguchi}}, \ and\ \bibinfo {author}
  {\bibfnamefont {P.}~\bibnamefont {Jarillo-Herrero}},\ }\bibfield  {title}
  {\enquote {\bibinfo {title} {Tunable strongly coupled superconductivity in
  magic-angle twisted trilayer graphene},}\ }\href {\doibase
  10.1038/s41586-021-03192-0} {\bibfield  {journal} {\bibinfo  {journal}
  {Nature}\ }\textbf {\bibinfo {volume} {590}},\ \bibinfo {pages} {249–255}
  (\bibinfo {year} {2021})}\BibitemShut {NoStop}%
\bibitem [{\citenamefont {Hao}\ \emph {et~al.}(2021)\citenamefont {Hao},
  \citenamefont {Zimmerman}, \citenamefont {Ledwith}, \citenamefont {Khalaf},
  \citenamefont {Najafabadi}, \citenamefont {Watanabe}, \citenamefont
  {Taniguchi}, \citenamefont {Vishwanath},\ and\ \citenamefont
  {Kim}}]{Hao_2021}%
  \BibitemOpen
  \bibfield  {author} {\bibinfo {author} {\bibfnamefont {Z.}~\bibnamefont
  {Hao}}, \bibinfo {author} {\bibfnamefont {A.~M.}\ \bibnamefont {Zimmerman}},
  \bibinfo {author} {\bibfnamefont {P.}~\bibnamefont {Ledwith}}, \bibinfo
  {author} {\bibfnamefont {E.}~\bibnamefont {Khalaf}}, \bibinfo {author}
  {\bibfnamefont {D.~H.}\ \bibnamefont {Najafabadi}}, \bibinfo {author}
  {\bibfnamefont {K.}~\bibnamefont {Watanabe}}, \bibinfo {author}
  {\bibfnamefont {T.}~\bibnamefont {Taniguchi}}, \bibinfo {author}
  {\bibfnamefont {A.}~\bibnamefont {Vishwanath}}, \ and\ \bibinfo {author}
  {\bibfnamefont {P.}~\bibnamefont {Kim}},\ }\bibfield  {title} {\enquote
  {\bibinfo {title} {Electric field–tunable superconductivity in
  alternating-twist magic-angle trilayer graphene},}\ }\href {\doibase
  10.1126/science.abg0399} {\bibfield  {journal} {\bibinfo  {journal}
  {Science}\ }\textbf {\bibinfo {volume} {371}},\ \bibinfo {pages}
  {1133–1138} (\bibinfo {year} {2021})}\BibitemShut {NoStop}%
\bibitem [{\citenamefont {{Cao}}\ \emph {et~al.}(2021)\citenamefont {{Cao}},
  \citenamefont {{Park}}, \citenamefont {{Watanabe}}, \citenamefont
  {{Taniguchi}},\ and\ \citenamefont
  {{Jarillo-Herrero}}}]{2021arXiv210312083C}%
  \BibitemOpen
  \bibfield  {author} {\bibinfo {author} {\bibfnamefont {Y.}~\bibnamefont
  {{Cao}}}, \bibinfo {author} {\bibfnamefont {J.~M.}\ \bibnamefont {{Park}}},
  \bibinfo {author} {\bibfnamefont {K.}~\bibnamefont {{Watanabe}}}, \bibinfo
  {author} {\bibfnamefont {T.}~\bibnamefont {{Taniguchi}}}, \ and\ \bibinfo
  {author} {\bibfnamefont {P.}~\bibnamefont {{Jarillo-Herrero}}},\ }\bibfield
  {title} {\enquote {\bibinfo {title} {{Large Pauli Limit Violation and
  Reentrant Superconductivity in Magic-Angle Twisted Trilayer Graphene}},}\
  }\href@noop {} {\bibfield  {journal} {\bibinfo  {journal} {arXiv e-prints}\
  ,\ \bibinfo {eid} {arXiv:2103.12083}} (\bibinfo {year} {2021})},\ \Eprint
  {http://arxiv.org/abs/2103.12083} {arXiv:2103.12083 [cond-mat.mes-hall]}
  \BibitemShut {NoStop}%
\bibitem [{\citenamefont {Khalaf}\ \emph {et~al.}(2019)\citenamefont {Khalaf},
  \citenamefont {Kruchkov}, \citenamefont {Tarnopolsky},\ and\ \citenamefont
  {Vishwanath}}]{Khalaf_2019}%
  \BibitemOpen
  \bibfield  {author} {\bibinfo {author} {\bibfnamefont {E.}~\bibnamefont
  {Khalaf}}, \bibinfo {author} {\bibfnamefont {A.~J.}\ \bibnamefont
  {Kruchkov}}, \bibinfo {author} {\bibfnamefont {G.}~\bibnamefont
  {Tarnopolsky}}, \ and\ \bibinfo {author} {\bibfnamefont {A.}~\bibnamefont
  {Vishwanath}},\ }\bibfield  {title} {\enquote {\bibinfo {title} {Magic angle
  hierarchy in twisted graphene multilayers},}\ }\href {\doibase
  10.1103/physrevb.100.085109} {\bibfield  {journal} {\bibinfo  {journal}
  {Physical Review B}\ }\textbf {\bibinfo {volume} {100}} (\bibinfo {year}
  {2019}),\ 10.1103/physrevb.100.085109}\BibitemShut {NoStop}%
\bibitem [{\citenamefont {Carr}\ \emph {et~al.}(2020)\citenamefont {Carr},
  \citenamefont {Li}, \citenamefont {Zhu}, \citenamefont {Kaxiras},
  \citenamefont {Sachdev},\ and\ \citenamefont {Kruchkov}}]{Carr_2020}%
  \BibitemOpen
  \bibfield  {author} {\bibinfo {author} {\bibfnamefont {S.}~\bibnamefont
  {Carr}}, \bibinfo {author} {\bibfnamefont {C.}~\bibnamefont {Li}}, \bibinfo
  {author} {\bibfnamefont {Z.}~\bibnamefont {Zhu}}, \bibinfo {author}
  {\bibfnamefont {E.}~\bibnamefont {Kaxiras}}, \bibinfo {author} {\bibfnamefont
  {S.}~\bibnamefont {Sachdev}}, \ and\ \bibinfo {author} {\bibfnamefont
  {A.}~\bibnamefont {Kruchkov}},\ }\bibfield  {title} {\enquote {\bibinfo
  {title} {Ultraheavy and ultrarelativistic dirac quasiparticles in sandwiched
  graphenes},}\ }\href {\doibase 10.1021/acs.nanolett.9b04979} {\bibfield
  {journal} {\bibinfo  {journal} {Nano Letters}\ }\textbf {\bibinfo {volume}
  {20}},\ \bibinfo {pages} {3030–3038} (\bibinfo {year} {2020})}\BibitemShut
  {NoStop}%
\bibitem [{\citenamefont {Mora}\ \emph {et~al.}(2019)\citenamefont {Mora},
  \citenamefont {Regnault},\ and\ \citenamefont
  {Bernevig}}]{PhysRevLett.123.026402}%
  \BibitemOpen
  \bibfield  {author} {\bibinfo {author} {\bibfnamefont {C.}~\bibnamefont
  {Mora}}, \bibinfo {author} {\bibfnamefont {N.}~\bibnamefont {Regnault}}, \
  and\ \bibinfo {author} {\bibfnamefont {B.~A.}\ \bibnamefont {Bernevig}},\
  }\bibfield  {title} {\enquote {\bibinfo {title} {Flatbands and perfect metal
  in trilayer moir\'e graphene},}\ }\href {\doibase
  10.1103/PhysRevLett.123.026402} {\bibfield  {journal} {\bibinfo  {journal}
  {Phys. Rev. Lett.}\ }\textbf {\bibinfo {volume} {123}},\ \bibinfo {pages}
  {026402} (\bibinfo {year} {2019})}\BibitemShut {NoStop}%
\bibitem [{\citenamefont {{C{\v{a}}lug{\v{a}}ru}}\ \emph
  {et~al.}(2021)\citenamefont {{C{\v{a}}lug{\v{a}}ru}}, \citenamefont {{Xie}},
  \citenamefont {{Song}}, \citenamefont {{Lian}}, \citenamefont {{Regnault}},\
  and\ \citenamefont {{Bernevig}}}]{2021PhRvB.103s5411C}%
  \BibitemOpen
  \bibfield  {author} {\bibinfo {author} {\bibfnamefont {D.}~\bibnamefont
  {{C{\v{a}}lug{\v{a}}ru}}}, \bibinfo {author} {\bibfnamefont {F.}~\bibnamefont
  {{Xie}}}, \bibinfo {author} {\bibfnamefont {Z.-D.}\ \bibnamefont {{Song}}},
  \bibinfo {author} {\bibfnamefont {B.}~\bibnamefont {{Lian}}}, \bibinfo
  {author} {\bibfnamefont {N.}~\bibnamefont {{Regnault}}}, \ and\ \bibinfo
  {author} {\bibfnamefont {B.~A.}\ \bibnamefont {{Bernevig}}},\ }\bibfield
  {title} {\enquote {\bibinfo {title} {{Twisted symmetric trilayer graphene:
  Single-particle and many-body Hamiltonians and hidden nonlocal symmetries of
  trilayer moir{\'e} systems with and without displacement field}},}\ }\href
  {\doibase 10.1103/PhysRevB.103.195411} {\bibfield  {journal} {\bibinfo
  {journal} {\prb}\ }\textbf {\bibinfo {volume} {103}},\ \bibinfo {eid}
  {195411} (\bibinfo {year} {2021})},\ \Eprint
  {http://arxiv.org/abs/2102.06201} {arXiv:2102.06201 [cond-mat.str-el]}
  \BibitemShut {NoStop}%
\bibitem [{\citenamefont {Shin}\ \emph {et~al.}(2021)\citenamefont {Shin},
  \citenamefont {Chittari},\ and\ \citenamefont {Jung}}]{shin2021stacking}%
  \BibitemOpen
  \bibfield  {author} {\bibinfo {author} {\bibfnamefont {J.}~\bibnamefont
  {Shin}}, \bibinfo {author} {\bibfnamefont {B.~L.}\ \bibnamefont {Chittari}},
  \ and\ \bibinfo {author} {\bibfnamefont {J.}~\bibnamefont {Jung}},\
  }\href@noop {} {\enquote {\bibinfo {title} {Stacking and gate tunable
  topological flat bands, gaps and anisotropic strip patterns in twisted
  trilayer graphene},}\ } (\bibinfo {year} {2021}),\ \Eprint
  {http://arxiv.org/abs/2104.01570} {arXiv:2104.01570 [cond-mat.mes-hall]}
  \BibitemShut {NoStop}%
\bibitem [{\citenamefont {Lei}\ \emph {et~al.}(2021)\citenamefont {Lei},
  \citenamefont {Linhart}, \citenamefont {Qin}, \citenamefont {Libisch},\ and\
  \citenamefont {MacDonald}}]{lei2021mirror}%
  \BibitemOpen
  \bibfield  {author} {\bibinfo {author} {\bibfnamefont {C.}~\bibnamefont
  {Lei}}, \bibinfo {author} {\bibfnamefont {L.}~\bibnamefont {Linhart}},
  \bibinfo {author} {\bibfnamefont {W.}~\bibnamefont {Qin}}, \bibinfo {author}
  {\bibfnamefont {F.}~\bibnamefont {Libisch}}, \ and\ \bibinfo {author}
  {\bibfnamefont {A.~H.}\ \bibnamefont {MacDonald}},\ }\href@noop {} {\enquote
  {\bibinfo {title} {Mirror symmetry breaking and lateral stacking shifts in
  twisted trilayer graphene},}\ } (\bibinfo {year} {2021}),\ \Eprint
  {http://arxiv.org/abs/2010.05787} {arXiv:2010.05787 [cond-mat.mes-hall]}
  \BibitemShut {NoStop}%
\bibitem [{\citenamefont {{Ramires}}\ and\ \citenamefont
  {{Lado}}(2021)}]{2021arXiv210203312R}%
  \BibitemOpen
  \bibfield  {author} {\bibinfo {author} {\bibfnamefont {A.}~\bibnamefont
  {{Ramires}}}\ and\ \bibinfo {author} {\bibfnamefont {J.~L.}\ \bibnamefont
  {{Lado}}},\ }\bibfield  {title} {\enquote {\bibinfo {title} {{Emulating heavy
  fermions in twisted trilayer graphene}},}\ }\href@noop {} {\bibfield
  {journal} {\bibinfo  {journal} {arXiv e-prints}\ ,\ \bibinfo {eid}
  {arXiv:2102.03312}} (\bibinfo {year} {2021})},\ \Eprint
  {http://arxiv.org/abs/2102.03312} {arXiv:2102.03312 [cond-mat.mes-hall]}
  \BibitemShut {NoStop}%
\bibitem [{\citenamefont {{Choi}}\ and\ \citenamefont
  {{Choi}}(2021)}]{2021arXiv210316132C}%
  \BibitemOpen
  \bibfield  {author} {\bibinfo {author} {\bibfnamefont {Y.~W.}\ \bibnamefont
  {{Choi}}}\ and\ \bibinfo {author} {\bibfnamefont {H.~J.}\ \bibnamefont
  {{Choi}}},\ }\bibfield  {title} {\enquote {\bibinfo {title} {{Dichotomy of
  Electron-Phonon Coupling in Graphene Moire Flat Bands}},}\ }\href@noop {}
  {\bibfield  {journal} {\bibinfo  {journal} {arXiv e-prints}\ ,\ \bibinfo
  {eid} {arXiv:2103.16132}} (\bibinfo {year} {2021})},\ \Eprint
  {http://arxiv.org/abs/2103.16132} {arXiv:2103.16132 [cond-mat.mes-hall]}
  \BibitemShut {NoStop}%
\bibitem [{\citenamefont {{Lake}}\ and\ \citenamefont
  {{Senthil}}(2021)}]{2021arXiv210413920L}%
  \BibitemOpen
  \bibfield  {author} {\bibinfo {author} {\bibfnamefont {E.}~\bibnamefont
  {{Lake}}}\ and\ \bibinfo {author} {\bibfnamefont {T.}~\bibnamefont
  {{Senthil}}},\ }\bibfield  {title} {\enquote {\bibinfo {title} {{Re-entrant
  Superconductivity Through a Quantum Lifshitz Transition in Twisted Trilayer
  Graphene}},}\ }\href@noop {} {\bibfield  {journal} {\bibinfo  {journal}
  {arXiv e-prints}\ ,\ \bibinfo {eid} {arXiv:2104.13920}} (\bibinfo {year}
  {2021})},\ \Eprint {http://arxiv.org/abs/2104.13920} {arXiv:2104.13920
  [cond-mat.supr-con]} \BibitemShut {NoStop}%
\bibitem [{\citenamefont {{Qin}}\ and\ \citenamefont
  {{MacDonald}}(2021)}]{2021arXiv210414026Q}%
  \BibitemOpen
  \bibfield  {author} {\bibinfo {author} {\bibfnamefont {W.}~\bibnamefont
  {{Qin}}}\ and\ \bibinfo {author} {\bibfnamefont {A.~H.}\ \bibnamefont
  {{MacDonald}}},\ }\bibfield  {title} {\enquote {\bibinfo {title} {{In-plane
  critical magnetic fields in magic-angle twisted trilayer graphene}},}\
  }\href@noop {} {\bibfield  {journal} {\bibinfo  {journal} {arXiv e-prints}\
  ,\ \bibinfo {eid} {arXiv:2104.14026}} (\bibinfo {year} {2021})},\ \Eprint
  {http://arxiv.org/abs/2104.14026} {arXiv:2104.14026 [cond-mat.mes-hall]}
  \BibitemShut {NoStop}%
\bibitem [{\citenamefont {{Fischer}}\ \emph {et~al.}(2021)\citenamefont
  {{Fischer}}, \citenamefont {{Goodwin}}, \citenamefont {{Mostofi}},
  \citenamefont {{Lischner}}, \citenamefont {{Kennes}},\ and\ \citenamefont
  {{Klebl}}}]{2021arXiv210410176F}%
  \BibitemOpen
  \bibfield  {author} {\bibinfo {author} {\bibfnamefont {A.}~\bibnamefont
  {{Fischer}}}, \bibinfo {author} {\bibfnamefont {Z.~A.~H.}\ \bibnamefont
  {{Goodwin}}}, \bibinfo {author} {\bibfnamefont {A.~A.}\ \bibnamefont
  {{Mostofi}}}, \bibinfo {author} {\bibfnamefont {J.}~\bibnamefont
  {{Lischner}}}, \bibinfo {author} {\bibfnamefont {D.~M.}\ \bibnamefont
  {{Kennes}}}, \ and\ \bibinfo {author} {\bibfnamefont {L.}~\bibnamefont
  {{Klebl}}},\ }\bibfield  {title} {\enquote {\bibinfo {title} {{Unconventional
  Superconductivity in Magic-Angle Twisted Trilayer Graphene}},}\ }\href@noop
  {} {\bibfield  {journal} {\bibinfo  {journal} {arXiv e-prints}\ ,\ \bibinfo
  {eid} {arXiv:2104.10176}} (\bibinfo {year} {2021})},\ \Eprint
  {http://arxiv.org/abs/2104.10176} {arXiv:2104.10176 [cond-mat.supr-con]}
  \BibitemShut {NoStop}%
\bibitem [{\citenamefont {Chou}\ \emph {et~al.}(2021)\citenamefont {Chou},
  \citenamefont {Wu}, \citenamefont {Sau},\ and\ \citenamefont
  {Sarma}}]{chou2021correlationinduced}%
  \BibitemOpen
  \bibfield  {author} {\bibinfo {author} {\bibfnamefont {Y.-Z.}\ \bibnamefont
  {Chou}}, \bibinfo {author} {\bibfnamefont {F.}~\bibnamefont {Wu}}, \bibinfo
  {author} {\bibfnamefont {J.~D.}\ \bibnamefont {Sau}}, \ and\ \bibinfo
  {author} {\bibfnamefont {S.~D.}\ \bibnamefont {Sarma}},\ }\href@noop {}
  {\enquote {\bibinfo {title} {Correlation-induced triplet pairing
  superconductivity in graphene-based moir\'e systems},}\ } (\bibinfo {year}
  {2021}),\ \Eprint {http://arxiv.org/abs/2105.00561} {arXiv:2105.00561
  [cond-mat.supr-con]} \BibitemShut {NoStop}%
\bibitem [{\citenamefont {Xie}\ and\ \citenamefont
  {MacDonald}(2020)}]{Xie_2020}%
  \BibitemOpen
  \bibfield  {author} {\bibinfo {author} {\bibfnamefont {M.}~\bibnamefont
  {Xie}}\ and\ \bibinfo {author} {\bibfnamefont {A.}~\bibnamefont
  {MacDonald}},\ }\bibfield  {title} {\enquote {\bibinfo {title} {Nature of the
  correlated insulator states in twisted bilayer graphene},}\ }\href {\doibase
  10.1103/physrevlett.124.097601} {\bibfield  {journal} {\bibinfo  {journal}
  {Physical Review Letters}\ }\textbf {\bibinfo {volume} {124}} (\bibinfo
  {year} {2020}),\ 10.1103/physrevlett.124.097601}\BibitemShut {NoStop}%
\bibitem [{\citenamefont {Bultinck}\ \emph {et~al.}(2019)\citenamefont
  {Bultinck}, \citenamefont {Khalaf}, \citenamefont {Liu}, \citenamefont
  {Chatterjee}, \citenamefont {Vishwanath},\ and\ \citenamefont
  {Zaletel}}]{bultinck2019ground}%
  \BibitemOpen
  \bibfield  {author} {\bibinfo {author} {\bibfnamefont {N.}~\bibnamefont
  {Bultinck}}, \bibinfo {author} {\bibfnamefont {E.}~\bibnamefont {Khalaf}},
  \bibinfo {author} {\bibfnamefont {S.}~\bibnamefont {Liu}}, \bibinfo {author}
  {\bibfnamefont {S.}~\bibnamefont {Chatterjee}}, \bibinfo {author}
  {\bibfnamefont {A.}~\bibnamefont {Vishwanath}}, \ and\ \bibinfo {author}
  {\bibfnamefont {M.~P.}\ \bibnamefont {Zaletel}},\ }\href@noop {} {\enquote
  {\bibinfo {title} {Ground state and hidden symmetry of magic angle graphene
  at even integer filling},}\ } (\bibinfo {year} {2019}),\ \Eprint
  {http://arxiv.org/abs/1911.02045} {arXiv:1911.02045 [cond-mat.str-el]}
  \BibitemShut {NoStop}%
\bibitem [{\citenamefont {Liu}\ \emph {et~al.}(2021{\natexlab{b}})\citenamefont
  {Liu}, \citenamefont {Khalaf}, \citenamefont {Lee},\ and\ \citenamefont
  {Vishwanath}}]{Liu_2021}%
  \BibitemOpen
  \bibfield  {author} {\bibinfo {author} {\bibfnamefont {S.}~\bibnamefont
  {Liu}}, \bibinfo {author} {\bibfnamefont {E.}~\bibnamefont {Khalaf}},
  \bibinfo {author} {\bibfnamefont {J.~Y.}\ \bibnamefont {Lee}}, \ and\
  \bibinfo {author} {\bibfnamefont {A.}~\bibnamefont {Vishwanath}},\ }\bibfield
   {title} {\enquote {\bibinfo {title} {Nematic topological semimetal and
  insulator in magic-angle bilayer graphene at charge neutrality},}\ }\href
  {\doibase 10.1103/physrevresearch.3.013033} {\bibfield  {journal} {\bibinfo
  {journal} {Physical Review Research}\ }\textbf {\bibinfo {volume} {3}}
  (\bibinfo {year} {2021}{\natexlab{b}}),\
  10.1103/physrevresearch.3.013033}\BibitemShut {NoStop}%
\bibitem [{\citenamefont {Liao}\ \emph {et~al.}(2021)\citenamefont {Liao},
  \citenamefont {Xu}, \citenamefont {Meng},\ and\ \citenamefont
  {Kang}}]{Liao_2021}%
  \BibitemOpen
  \bibfield  {author} {\bibinfo {author} {\bibfnamefont {Y.-D.}\ \bibnamefont
  {Liao}}, \bibinfo {author} {\bibfnamefont {X.-Y.}\ \bibnamefont {Xu}},
  \bibinfo {author} {\bibfnamefont {Z.-Y.}\ \bibnamefont {Meng}}, \ and\
  \bibinfo {author} {\bibfnamefont {J.}~\bibnamefont {Kang}},\ }\bibfield
  {title} {\enquote {\bibinfo {title} {Correlated insulating phases in the
  twisted bilayer graphene*},}\ }\href {\doibase 10.1088/1674-1056/abcfa3}
  {\bibfield  {journal} {\bibinfo  {journal} {Chinese Physics B}\ }\textbf
  {\bibinfo {volume} {30}},\ \bibinfo {pages} {017305} (\bibinfo {year}
  {2021})}\BibitemShut {NoStop}%
\bibitem [{\citenamefont {Bultinck}\ \emph {et~al.}(2020)\citenamefont
  {Bultinck}, \citenamefont {Chatterjee},\ and\ \citenamefont
  {Zaletel}}]{Bultinck_2020}%
  \BibitemOpen
  \bibfield  {author} {\bibinfo {author} {\bibfnamefont {N.}~\bibnamefont
  {Bultinck}}, \bibinfo {author} {\bibfnamefont {S.}~\bibnamefont
  {Chatterjee}}, \ and\ \bibinfo {author} {\bibfnamefont {M.~P.}\ \bibnamefont
  {Zaletel}},\ }\bibfield  {title} {\enquote {\bibinfo {title} {Mechanism for
  anomalous hall ferromagnetism in twisted bilayer graphene},}\ }\href
  {\doibase 10.1103/physrevlett.124.166601} {\bibfield  {journal} {\bibinfo
  {journal} {Physical Review Letters}\ }\textbf {\bibinfo {volume} {124}}
  (\bibinfo {year} {2020}),\ 10.1103/physrevlett.124.166601}\BibitemShut
  {NoStop}%
\bibitem [{\citenamefont {Zhang}\ \emph {et~al.}(2020)\citenamefont {Zhang},
  \citenamefont {Jiang}, \citenamefont {Wang},\ and\ \citenamefont
  {Zhang}}]{Zhang_2020}%
  \BibitemOpen
  \bibfield  {author} {\bibinfo {author} {\bibfnamefont {Y.}~\bibnamefont
  {Zhang}}, \bibinfo {author} {\bibfnamefont {K.}~\bibnamefont {Jiang}},
  \bibinfo {author} {\bibfnamefont {Z.}~\bibnamefont {Wang}}, \ and\ \bibinfo
  {author} {\bibfnamefont {F.}~\bibnamefont {Zhang}},\ }\bibfield  {title}
  {\enquote {\bibinfo {title} {Correlated insulating phases of twisted bilayer
  graphene at commensurate filling fractions: A hartree-fock study},}\ }\href
  {\doibase 10.1103/physrevb.102.035136} {\bibfield  {journal} {\bibinfo
  {journal} {Physical Review B}\ }\textbf {\bibinfo {volume} {102}} (\bibinfo
  {year} {2020}),\ 10.1103/physrevb.102.035136}\BibitemShut {NoStop}%
\bibitem [{\citenamefont {Stepanov}\ \emph {et~al.}(2020)\citenamefont
  {Stepanov}, \citenamefont {Das}, \citenamefont {Lu}, \citenamefont
  {Fahimniya}, \citenamefont {Watanabe}, \citenamefont {Taniguchi},
  \citenamefont {Koppens}, \citenamefont {Lischner}, \citenamefont {Levitov},\
  and\ \citenamefont {Efetov}}]{StepanovTuning}%
  \BibitemOpen
  \bibfield  {author} {\bibinfo {author} {\bibfnamefont {P.}~\bibnamefont
  {Stepanov}}, \bibinfo {author} {\bibfnamefont {I.}~\bibnamefont {Das}},
  \bibinfo {author} {\bibfnamefont {X.}~\bibnamefont {Lu}}, \bibinfo {author}
  {\bibfnamefont {A.}~\bibnamefont {Fahimniya}}, \bibinfo {author}
  {\bibfnamefont {K.}~\bibnamefont {Watanabe}}, \bibinfo {author}
  {\bibfnamefont {T.}~\bibnamefont {Taniguchi}}, \bibinfo {author}
  {\bibfnamefont {F.~H.~L.}\ \bibnamefont {Koppens}}, \bibinfo {author}
  {\bibfnamefont {J.}~\bibnamefont {Lischner}}, \bibinfo {author}
  {\bibfnamefont {L.}~\bibnamefont {Levitov}}, \ and\ \bibinfo {author}
  {\bibfnamefont {D.~K.}\ \bibnamefont {Efetov}},\ }\bibfield  {title}
  {\enquote {\bibinfo {title} {Untying the insulating and superconducting
  orders in magic-angle graphene},}\ }\href {\doibase
  10.1038/s41586-020-2459-6} {\bibfield  {journal} {\bibinfo  {journal}
  {Nature}\ }\textbf {\bibinfo {volume} {583}},\ \bibinfo {pages} {375}
  (\bibinfo {year} {2020})}\BibitemShut {NoStop}%
\bibitem [{\citenamefont {Liu}\ \emph {et~al.}(2021{\natexlab{c}})\citenamefont
  {Liu}, \citenamefont {Wang}, \citenamefont {Watanabe}, \citenamefont
  {Taniguchi}, \citenamefont {Vafek},\ and\ \citenamefont {Li}}]{Liu1261}%
  \BibitemOpen
  \bibfield  {author} {\bibinfo {author} {\bibfnamefont {X.}~\bibnamefont
  {Liu}}, \bibinfo {author} {\bibfnamefont {Z.}~\bibnamefont {Wang}}, \bibinfo
  {author} {\bibfnamefont {K.}~\bibnamefont {Watanabe}}, \bibinfo {author}
  {\bibfnamefont {T.}~\bibnamefont {Taniguchi}}, \bibinfo {author}
  {\bibfnamefont {O.}~\bibnamefont {Vafek}}, \ and\ \bibinfo {author}
  {\bibfnamefont {J.~I.~A.}\ \bibnamefont {Li}},\ }\bibfield  {title} {\enquote
  {\bibinfo {title} {Tuning electron correlation in magic-angle twisted bilayer
  graphene using coulomb screening},}\ }\href {\doibase
  10.1126/science.abb8754} {\bibfield  {journal} {\bibinfo  {journal}
  {Science}\ }\textbf {\bibinfo {volume} {371}},\ \bibinfo {pages} {1261}
  (\bibinfo {year} {2021}{\natexlab{c}})}\BibitemShut {NoStop}%
\bibitem [{\citenamefont {Saito}\ \emph {et~al.}(2020)\citenamefont {Saito},
  \citenamefont {Ge}, \citenamefont {Watanabe}, \citenamefont {Taniguchi},\
  and\ \citenamefont {Young}}]{SaitoTuning}%
  \BibitemOpen
  \bibfield  {author} {\bibinfo {author} {\bibfnamefont {Y.}~\bibnamefont
  {Saito}}, \bibinfo {author} {\bibfnamefont {J.}~\bibnamefont {Ge}}, \bibinfo
  {author} {\bibfnamefont {K.}~\bibnamefont {Watanabe}}, \bibinfo {author}
  {\bibfnamefont {T.}~\bibnamefont {Taniguchi}}, \ and\ \bibinfo {author}
  {\bibfnamefont {A.~F.}\ \bibnamefont {Young}},\ }\bibfield  {title} {\enquote
  {\bibinfo {title} {Independent superconductors and correlated insulators in
  twisted bilayer graphene},}\ }\href {\doibase 10.1038/s41567-020-0928-3}
  {\bibfield  {journal} {\bibinfo  {journal} {Nature Physics}\ }\textbf
  {\bibinfo {volume} {16}},\ \bibinfo {pages} {926} (\bibinfo {year}
  {2020})}\BibitemShut {NoStop}%
\bibitem [{\citenamefont {Samajdar}\ and\ \citenamefont
  {Scheurer}(2020)}]{PhysRevB.102.064501}%
  \BibitemOpen
  \bibfield  {author} {\bibinfo {author} {\bibfnamefont {R.}~\bibnamefont
  {Samajdar}}\ and\ \bibinfo {author} {\bibfnamefont {M.~S.}\ \bibnamefont
  {Scheurer}},\ }\bibfield  {title} {\enquote {\bibinfo {title} {Microscopic
  pairing mechanism, order parameter, and disorder sensitivity in moir\'e
  superlattices: Applications to twisted double-bilayer graphene},}\ }\href
  {\doibase 10.1103/PhysRevB.102.064501} {\bibfield  {journal} {\bibinfo
  {journal} {Phys. Rev. B}\ }\textbf {\bibinfo {volume} {102}},\ \bibinfo
  {pages} {064501} (\bibinfo {year} {2020})}\BibitemShut {NoStop}%
\bibitem [{\citenamefont {Khalaf}\ \emph {et~al.}(2021)\citenamefont {Khalaf},
  \citenamefont {Chatterjee}, \citenamefont {Bultinck}, \citenamefont
  {Zaletel},\ and\ \citenamefont {Vishwanath}}]{Khalafeabf5299}%
  \BibitemOpen
  \bibfield  {author} {\bibinfo {author} {\bibfnamefont {E.}~\bibnamefont
  {Khalaf}}, \bibinfo {author} {\bibfnamefont {S.}~\bibnamefont {Chatterjee}},
  \bibinfo {author} {\bibfnamefont {N.}~\bibnamefont {Bultinck}}, \bibinfo
  {author} {\bibfnamefont {M.~P.}\ \bibnamefont {Zaletel}}, \ and\ \bibinfo
  {author} {\bibfnamefont {A.}~\bibnamefont {Vishwanath}},\ }\bibfield  {title}
  {\enquote {\bibinfo {title} {Charged skyrmions and topological origin of
  superconductivity in magic-angle graphene},}\ }\href {\doibase
  10.1126/sciadv.abf5299} {\bibfield  {journal} {\bibinfo  {journal} {Science
  Advances}\ }\textbf {\bibinfo {volume} {7}} (\bibinfo {year} {2021}),\
  10.1126/sciadv.abf5299}\BibitemShut {NoStop}%
\bibitem [{\citenamefont {Nam}\ and\ \citenamefont
  {Koshino}(2017)}]{PhysRevB.96.075311}%
  \BibitemOpen
  \bibfield  {author} {\bibinfo {author} {\bibfnamefont {N.~N.~T.}\
  \bibnamefont {Nam}}\ and\ \bibinfo {author} {\bibfnamefont {M.}~\bibnamefont
  {Koshino}},\ }\bibfield  {title} {\enquote {\bibinfo {title} {Lattice
  relaxation and energy band modulation in twisted bilayer graphene},}\ }\href
  {\doibase 10.1103/PhysRevB.96.075311} {\bibfield  {journal} {\bibinfo
  {journal} {Phys. Rev. B}\ }\textbf {\bibinfo {volume} {96}},\ \bibinfo
  {pages} {075311} (\bibinfo {year} {2017})}\BibitemShut {NoStop}%
\bibitem [{\citenamefont {{Carr}}\ \emph {et~al.}(2019)\citenamefont {{Carr}},
  \citenamefont {{Fang}}, \citenamefont {{Zhu}},\ and\ \citenamefont
  {{Kaxiras}}}]{2019PhRvR...1a3001C}%
  \BibitemOpen
  \bibfield  {author} {\bibinfo {author} {\bibfnamefont {S.}~\bibnamefont
  {{Carr}}}, \bibinfo {author} {\bibfnamefont {S.}~\bibnamefont {{Fang}}},
  \bibinfo {author} {\bibfnamefont {Z.}~\bibnamefont {{Zhu}}}, \ and\ \bibinfo
  {author} {\bibfnamefont {E.}~\bibnamefont {{Kaxiras}}},\ }\bibfield  {title}
  {\enquote {\bibinfo {title} {{Exact continuum model for low-energy electronic
  states of twisted bilayer graphene}},}\ }\href {\doibase
  10.1103/PhysRevResearch.1.013001} {\bibfield  {journal} {\bibinfo  {journal}
  {Physical Review Research}\ }\textbf {\bibinfo {volume} {1}},\ \bibinfo {eid}
  {013001} (\bibinfo {year} {2019})},\ \Eprint
  {http://arxiv.org/abs/1901.03420} {arXiv:1901.03420 [cond-mat.mes-hall]}
  \BibitemShut {NoStop}%
\bibitem [{\citenamefont {Tarnopolsky}\ \emph {et~al.}(2019)\citenamefont
  {Tarnopolsky}, \citenamefont {Kruchkov},\ and\ \citenamefont
  {Vishwanath}}]{Tarnopolsky_2019}%
  \BibitemOpen
  \bibfield  {author} {\bibinfo {author} {\bibfnamefont {G.}~\bibnamefont
  {Tarnopolsky}}, \bibinfo {author} {\bibfnamefont {A.~J.}\ \bibnamefont
  {Kruchkov}}, \ and\ \bibinfo {author} {\bibfnamefont {A.}~\bibnamefont
  {Vishwanath}},\ }\bibfield  {title} {\enquote {\bibinfo {title} {Origin of
  magic angles in twisted bilayer graphene},}\ }\href {\doibase
  10.1103/physrevlett.122.106405} {\bibfield  {journal} {\bibinfo  {journal}
  {Physical Review Letters}\ }\textbf {\bibinfo {volume} {122}} (\bibinfo
  {year} {2019}),\ 10.1103/physrevlett.122.106405}\BibitemShut {NoStop}%
\bibitem [{\citenamefont {Bernevig}\ \emph {et~al.}(2020)\citenamefont
  {Bernevig}, \citenamefont {Song}, \citenamefont {Regnault},\ and\
  \citenamefont {Lian}}]{bernevig2020tbg}%
  \BibitemOpen
  \bibfield  {author} {\bibinfo {author} {\bibfnamefont {B.~A.}\ \bibnamefont
  {Bernevig}}, \bibinfo {author} {\bibfnamefont {Z.-D.}\ \bibnamefont {Song}},
  \bibinfo {author} {\bibfnamefont {N.}~\bibnamefont {Regnault}}, \ and\
  \bibinfo {author} {\bibfnamefont {B.}~\bibnamefont {Lian}},\ }\href@noop {}
  {\enquote {\bibinfo {title} {Tbg iii: Interacting hamiltonian and exact
  symmetries of twisted bilayer graphene},}\ } (\bibinfo {year} {2020}),\
  \Eprint {http://arxiv.org/abs/2009.12376} {arXiv:2009.12376
  [cond-mat.str-el]} \BibitemShut {NoStop}%
\bibitem [{\citenamefont {{Lian}}\ \emph {et~al.}(2020)\citenamefont {{Lian}},
  \citenamefont {{Song}}, \citenamefont {{Regnault}}, \citenamefont {{Efetov}},
  \citenamefont {{Yazdani}},\ and\ \citenamefont
  {{Bernevig}}}]{2020arXiv200913530L}%
  \BibitemOpen
  \bibfield  {author} {\bibinfo {author} {\bibfnamefont {B.}~\bibnamefont
  {{Lian}}}, \bibinfo {author} {\bibfnamefont {Z.-D.}\ \bibnamefont {{Song}}},
  \bibinfo {author} {\bibfnamefont {N.}~\bibnamefont {{Regnault}}}, \bibinfo
  {author} {\bibfnamefont {D.~K.}\ \bibnamefont {{Efetov}}}, \bibinfo {author}
  {\bibfnamefont {A.}~\bibnamefont {{Yazdani}}}, \ and\ \bibinfo {author}
  {\bibfnamefont {B.~A.}\ \bibnamefont {{Bernevig}}},\ }\bibfield  {title}
  {\enquote {\bibinfo {title} {{TBG IV: Exact Insulator Ground States and Phase
  Diagram of Twisted Bilayer Graphene}},}\ }\href@noop {} {\bibfield  {journal}
  {\bibinfo  {journal} {arXiv e-prints}\ ,\ \bibinfo {eid} {arXiv:2009.13530}}
  (\bibinfo {year} {2020})},\ \Eprint {http://arxiv.org/abs/2009.13530}
  {arXiv:2009.13530 [cond-mat.str-el]} \BibitemShut {NoStop}%
\bibitem [{\citenamefont {Kang}\ and\ \citenamefont
  {Vafek}(2019)}]{PhysRevLett.122.246401}%
  \BibitemOpen
  \bibfield  {author} {\bibinfo {author} {\bibfnamefont {J.}~\bibnamefont
  {Kang}}\ and\ \bibinfo {author} {\bibfnamefont {O.}~\bibnamefont {Vafek}},\
  }\bibfield  {title} {\enquote {\bibinfo {title} {Strong coupling phases of
  partially filled twisted bilayer graphene narrow bands},}\ }\href {\doibase
  10.1103/PhysRevLett.122.246401} {\bibfield  {journal} {\bibinfo  {journal}
  {Phys. Rev. Lett.}\ }\textbf {\bibinfo {volume} {122}},\ \bibinfo {pages}
  {246401} (\bibinfo {year} {2019})}\BibitemShut {NoStop}%
\bibitem [{\citenamefont {Castro~Neto}\ \emph {et~al.}(2009)\citenamefont
  {Castro~Neto}, \citenamefont {Guinea}, \citenamefont {Peres}, \citenamefont
  {Novoselov},\ and\ \citenamefont {Geim}}]{RevModPhys.81.109}%
  \BibitemOpen
  \bibfield  {author} {\bibinfo {author} {\bibfnamefont {A.~H.}\ \bibnamefont
  {Castro~Neto}}, \bibinfo {author} {\bibfnamefont {F.}~\bibnamefont {Guinea}},
  \bibinfo {author} {\bibfnamefont {N.~M.~R.}\ \bibnamefont {Peres}}, \bibinfo
  {author} {\bibfnamefont {K.~S.}\ \bibnamefont {Novoselov}}, \ and\ \bibinfo
  {author} {\bibfnamefont {A.~K.}\ \bibnamefont {Geim}},\ }\bibfield  {title}
  {\enquote {\bibinfo {title} {The electronic properties of graphene},}\ }\href
  {\doibase 10.1103/RevModPhys.81.109} {\bibfield  {journal} {\bibinfo
  {journal} {Rev. Mod. Phys.}\ }\textbf {\bibinfo {volume} {81}},\ \bibinfo
  {pages} {109} (\bibinfo {year} {2009})}\BibitemShut {NoStop}%
\bibitem [{Note1()}]{Note1}%
  \BibitemOpen
  \bibinfo {note} {Note that the term $c_2$, which comes from the Hartree term,
  turns out to be numerically very small.}\BibitemShut {Stop}%
\bibitem [{\citenamefont {Chatterjee}\ \emph {et~al.}(2020)\citenamefont
  {Chatterjee}, \citenamefont {Bultinck},\ and\ \citenamefont
  {Zaletel}}]{PhysRevB.101.165141}%
  \BibitemOpen
  \bibfield  {author} {\bibinfo {author} {\bibfnamefont {S.}~\bibnamefont
  {Chatterjee}}, \bibinfo {author} {\bibfnamefont {N.}~\bibnamefont
  {Bultinck}}, \ and\ \bibinfo {author} {\bibfnamefont {M.~P.}\ \bibnamefont
  {Zaletel}},\ }\bibfield  {title} {\enquote {\bibinfo {title} {Symmetry
  breaking and skyrmionic transport in twisted bilayer graphene},}\ }\href
  {\doibase 10.1103/PhysRevB.101.165141} {\bibfield  {journal} {\bibinfo
  {journal} {Phys. Rev. B}\ }\textbf {\bibinfo {volume} {101}},\ \bibinfo
  {pages} {165141} (\bibinfo {year} {2020})}\BibitemShut {NoStop}%
\bibitem [{\citenamefont {Brydon}\ \emph {et~al.}(2014)\citenamefont {Brydon},
  \citenamefont {Das~Sarma}, \citenamefont {Hui},\ and\ \citenamefont
  {Sau}}]{PhysRevB.90.184512}%
  \BibitemOpen
  \bibfield  {author} {\bibinfo {author} {\bibfnamefont {P.~M.~R.}\
  \bibnamefont {Brydon}}, \bibinfo {author} {\bibfnamefont {S.}~\bibnamefont
  {Das~Sarma}}, \bibinfo {author} {\bibfnamefont {H.-Y.}\ \bibnamefont {Hui}},
  \ and\ \bibinfo {author} {\bibfnamefont {J.~D.}\ \bibnamefont {Sau}},\
  }\bibfield  {title} {\enquote {\bibinfo {title} {Odd-parity superconductivity
  from phonon-mediated pairing: Application to
  ${\mathrm{cu}}_{x}{\mathrm{bi}}_{2}{\mathrm{se}}_{3}$},}\ }\href {\doibase
  10.1103/PhysRevB.90.184512} {\bibfield  {journal} {\bibinfo  {journal} {Phys.
  Rev. B}\ }\textbf {\bibinfo {volume} {90}},\ \bibinfo {pages} {184512}
  (\bibinfo {year} {2014})}\BibitemShut {NoStop}%
\bibitem [{\citenamefont {Scheurer}(2016)}]{PhysRevB.93.174509}%
  \BibitemOpen
  \bibfield  {author} {\bibinfo {author} {\bibfnamefont {M.~S.}\ \bibnamefont
  {Scheurer}},\ }\bibfield  {title} {\enquote {\bibinfo {title} {Mechanism,
  time-reversal symmetry, and topology of superconductivity in
  noncentrosymmetric systems},}\ }\href {\doibase 10.1103/PhysRevB.93.174509}
  {\bibfield  {journal} {\bibinfo  {journal} {Phys. Rev. B}\ }\textbf {\bibinfo
  {volume} {93}},\ \bibinfo {pages} {174509} (\bibinfo {year}
  {2016})}\BibitemShut {NoStop}%
\bibitem [{\citenamefont {{Chatterjee}}\ \emph {et~al.}(2020)\citenamefont
  {{Chatterjee}}, \citenamefont {{Ippoliti}},\ and\ \citenamefont
  {{Zaletel}}}]{2020arXiv201001144C}%
  \BibitemOpen
  \bibfield  {author} {\bibinfo {author} {\bibfnamefont {S.}~\bibnamefont
  {{Chatterjee}}}, \bibinfo {author} {\bibfnamefont {M.}~\bibnamefont
  {{Ippoliti}}}, \ and\ \bibinfo {author} {\bibfnamefont {M.~P.}\ \bibnamefont
  {{Zaletel}}},\ }\bibfield  {title} {\enquote {\bibinfo {title} {{Skyrmion
  Superconductivity: DMRG evidence for a topological route to
  superconductivity}},}\ }\href@noop {} {\bibfield  {journal} {\bibinfo
  {journal} {arXiv e-prints}\ ,\ \bibinfo {eid} {arXiv:2010.01144}} (\bibinfo
  {year} {2020})},\ \Eprint {http://arxiv.org/abs/2010.01144} {arXiv:2010.01144
  [cond-mat.str-el]} \BibitemShut {NoStop}%
\bibitem [{\citenamefont {Mackenzie}\ and\ \citenamefont
  {Maeno}(2003)}]{RevModPhys.75.657}%
  \BibitemOpen
  \bibfield  {author} {\bibinfo {author} {\bibfnamefont {A.~P.}\ \bibnamefont
  {Mackenzie}}\ and\ \bibinfo {author} {\bibfnamefont {Y.}~\bibnamefont
  {Maeno}},\ }\bibfield  {title} {\enquote {\bibinfo {title} {The
  superconductivity of ${\mathrm{sr}}_{2}{\mathrm{ruo}}_{4}$ and the physics of
  spin-triplet pairing},}\ }\href {\doibase 10.1103/RevModPhys.75.657}
  {\bibfield  {journal} {\bibinfo  {journal} {Rev. Mod. Phys.}\ }\textbf
  {\bibinfo {volume} {75}},\ \bibinfo {pages} {657} (\bibinfo {year}
  {2003})}\BibitemShut {NoStop}%
\bibitem [{\citenamefont {{Kwan}}\ \emph {et~al.}(2021)\citenamefont {{Kwan}},
  \citenamefont {{Wagner}}, \citenamefont {{Soejima}}, \citenamefont
  {{Zaletel}}, \citenamefont {{Simon}}, \citenamefont {{Parameswaran}},\ and\
  \citenamefont {{Bultinck}}}]{2021arXiv210505857K}%
  \BibitemOpen
  \bibfield  {author} {\bibinfo {author} {\bibfnamefont {Y.~H.}\ \bibnamefont
  {{Kwan}}}, \bibinfo {author} {\bibfnamefont {G.}~\bibnamefont {{Wagner}}},
  \bibinfo {author} {\bibfnamefont {T.}~\bibnamefont {{Soejima}}}, \bibinfo
  {author} {\bibfnamefont {M.~P.}\ \bibnamefont {{Zaletel}}}, \bibinfo {author}
  {\bibfnamefont {S.~H.}\ \bibnamefont {{Simon}}}, \bibinfo {author}
  {\bibfnamefont {S.~A.}\ \bibnamefont {{Parameswaran}}}, \ and\ \bibinfo
  {author} {\bibfnamefont {N.}~\bibnamefont {{Bultinck}}},\ }\bibfield  {title}
  {\enquote {\bibinfo {title} {{Kekul{\'e} spiral order at all nonzero integer
  fillings in twisted bilayer graphene}},}\ }\href@noop {} {\bibfield
  {journal} {\bibinfo  {journal} {arXiv e-prints}\ ,\ \bibinfo {eid}
  {arXiv:2105.05857}} (\bibinfo {year} {2021})},\ \Eprint
  {http://arxiv.org/abs/2105.05857} {arXiv:2105.05857 [cond-mat.str-el]}
  \BibitemShut {NoStop}%
\bibitem [{\citenamefont {Guti{\'e}rrez}\ \emph {et~al.}(2016)\citenamefont
  {Guti{\'e}rrez}, \citenamefont {Kim}, \citenamefont {Brown}, \citenamefont
  {Schiros}, \citenamefont {Nordlund}, \citenamefont {Lochocki}, \citenamefont
  {Shen}, \citenamefont {Park},\ and\ \citenamefont
  {Pasupathy}}]{Pasupathy_kekule}%
  \BibitemOpen
  \bibfield  {author} {\bibinfo {author} {\bibfnamefont {C.}~\bibnamefont
  {Guti{\'e}rrez}}, \bibinfo {author} {\bibfnamefont {C.-J.}\ \bibnamefont
  {Kim}}, \bibinfo {author} {\bibfnamefont {L.}~\bibnamefont {Brown}}, \bibinfo
  {author} {\bibfnamefont {T.}~\bibnamefont {Schiros}}, \bibinfo {author}
  {\bibfnamefont {D.}~\bibnamefont {Nordlund}}, \bibinfo {author}
  {\bibfnamefont {E.~B.}\ \bibnamefont {Lochocki}}, \bibinfo {author}
  {\bibfnamefont {K.~M.}\ \bibnamefont {Shen}}, \bibinfo {author}
  {\bibfnamefont {J.}~\bibnamefont {Park}}, \ and\ \bibinfo {author}
  {\bibfnamefont {A.~N.}\ \bibnamefont {Pasupathy}},\ }\bibfield  {title}
  {\enquote {\bibinfo {title} {{Imaging chiral symmetry breaking from
  Kekul{\'e} bond order in graphene}},}\ }\href {\doibase 10.1038/nphys3776}
  {\bibfield  {journal} {\bibinfo  {journal} {Nature Physics}\ }\textbf
  {\bibinfo {volume} {12}},\ \bibinfo {pages} {950} (\bibinfo {year}
  {2016})}\BibitemShut {NoStop}%
\bibitem [{\citenamefont {{Li}}\ \emph {et~al.}(2019)\citenamefont {{Li}},
  \citenamefont {{Zhang}}, \citenamefont {{Yin}},\ and\ \citenamefont
  {{He}}}]{He_kekule}%
  \BibitemOpen
  \bibfield  {author} {\bibinfo {author} {\bibfnamefont {S.-Y.}\ \bibnamefont
  {{Li}}}, \bibinfo {author} {\bibfnamefont {Y.}~\bibnamefont {{Zhang}}},
  \bibinfo {author} {\bibfnamefont {L.-J.}\ \bibnamefont {{Yin}}}, \ and\
  \bibinfo {author} {\bibfnamefont {L.}~\bibnamefont {{He}}},\ }\bibfield
  {title} {\enquote {\bibinfo {title} {{Scanning tunneling microscope study of
  quantum Hall isospin ferromagnetic states in the zero Landau level in a
  graphene monolayer}},}\ }\href {\doibase 10.1103/PhysRevB.100.085437}
  {\bibfield  {journal} {\bibinfo  {journal} {Phys. Rev. B}\ }\textbf {\bibinfo
  {volume} {100}},\ \bibinfo {eid} {085437} (\bibinfo {year} {2019})},\ \Eprint
  {http://arxiv.org/abs/1904.06902} {arXiv:1904.06902 [cond-mat.mes-hall]}
  \BibitemShut {NoStop}%
\bibitem [{\citenamefont {{Bernevig}}\ \emph {et~al.}(2020)\citenamefont
  {{Bernevig}}, \citenamefont {{Lian}}, \citenamefont {{Cowsik}}, \citenamefont
  {{Xie}}, \citenamefont {{Regnault}},\ and\ \citenamefont
  {{Song}}}]{2020arXiv200914200B}%
  \BibitemOpen
  \bibfield  {author} {\bibinfo {author} {\bibfnamefont {B.~A.}\ \bibnamefont
  {{Bernevig}}}, \bibinfo {author} {\bibfnamefont {B.}~\bibnamefont {{Lian}}},
  \bibinfo {author} {\bibfnamefont {A.}~\bibnamefont {{Cowsik}}}, \bibinfo
  {author} {\bibfnamefont {F.}~\bibnamefont {{Xie}}}, \bibinfo {author}
  {\bibfnamefont {N.}~\bibnamefont {{Regnault}}}, \ and\ \bibinfo {author}
  {\bibfnamefont {Z.-D.}\ \bibnamefont {{Song}}},\ }\bibfield  {title}
  {\enquote {\bibinfo {title} {{TBG V: Exact Analytic Many-Body Excitations In
  Twisted Bilayer Graphene Coulomb Hamiltonians: Charge Gap, Goldstone Modes
  and Absence of Cooper Pairing}},}\ }\href@noop {} {\bibfield  {journal}
  {\bibinfo  {journal} {arXiv e-prints}\ ,\ \bibinfo {eid} {arXiv:2009.14200}}
  (\bibinfo {year} {2020})},\ \Eprint {http://arxiv.org/abs/2009.14200}
  {arXiv:2009.14200 [cond-mat.str-el]} \BibitemShut {NoStop}%
\bibitem [{Note2()}]{Note2}%
  \BibitemOpen
  \bibinfo {note} {This means that its adjacency graph is strongly connected.
  In our case, it corresponds to the situation that one can scatter between any
  pair of momenta via some number of virtual states; at least for $D_0\not =0$,
  this is generically expected to be the case.}\BibitemShut {Stop}%
\end{thebibliography}%

\onecolumngrid

\begin{appendix}

\section{Hamiltonian for the system}
\label{app:model}
Here we will state the detailed form of the interacting Hamiltonian of MSTG used in our work, discuss the relevant symmetries and their representation in the basis employed in this paper, and their consequences for the form factors.

\subsection{Continuum model and low-energy degrees of freedom}\label{ContinuumModel}
To describe the non-interacting bands, we use a continuum model, which is just the three-layer version of the continuum model of TBG \cite{dos2007graphene,bistritzer2011moire,dos2012continuum}. 
Let $c_{\vec{k};\rho,l,\eta,s,\vec{G}}$ denote the operator annihilating an electron at crystalline momentum $\vec{k}$ in the moir\'e Brillouin zone (MBZ), in sublattice $\rho=A,B$ and valley $\eta = \pm$ of the microscopic graphene sheets, of spin $s=\uparrow,\downarrow$, and with reciprocal lattice (RL) vector $\vec{G}$ of the moir\'e lattice, thus, forming a triangular lattice; $\vec{G} = \sum_{j} n_j \vec{G}^{\text{M}}_j$, $n_j \in \mathbbm{Z}$, with $\vec{G}_1^{\text{M}} = -\sqrt{3}k_\theta (1,\sqrt{3})^T/2$ and $\vec{G}_2^{\text{M}} = \sqrt{3}k_\theta(1,0)^T$, where $k_\theta$ represents the magnitude of the vector connecting the K and K$'$ points in the MBZ.

Before writing down the Hamiltonian, we follow \cite{Khalaf_2019} and perform a unitary transformation in layer space,
\begin{equation}
    c_{\vec{k};\rho,l,\eta,s,\vec{G}} = V_{l,\ell} \psi_{\vec{k};\rho,\ell,\eta,s,\vec{G}}, \qquad V=\frac{1}{\sqrt{2}} \begin{pmatrix} 1 & 0 & -1 \\ 0 & \sqrt{2} & 0 \\ 1 & 0 & 1 \end{pmatrix}
\end{equation}
that decomposes the system into mirror-even ($\ell = 1,2$) and mirror-odd ($\ell = 3$) subspaces (for $D_0=0$), as will become apparent below. The full continuum model reads as
\begin{equation}
    H_0^{\text{Full}} = \sum_{\vec{k} \in \text{MBZ}} \sum_{\rho,\rho'=A,B} \sum_{\ell,\ell'=1,2,3} \sum_{\eta=\pm} \sum_{s=\uparrow,\downarrow}  \sum_{\vec{G},\vec{G}' \in \text{RL}} \psi^\dagger_{\vec{k};\rho,\ell,\eta,s,\vec{G}} \left(h_{\vec{k},\eta}\right)_{\rho,\ell,\vec{G};\rho',\ell',\vec{G}'} \psi^\pdagger_{\vec{k};\rho',\ell',\eta,s,\vec{G}'}, \label{FullContinuumModel}
\end{equation}
where $h_{\vec{k},\eta} = h^{(g)}_{\vec{k},\eta} + h^{(t)}_{\vec{k},\eta} + h^{(D)}_{\vec{k}}$ with the decoupled-layer graphene Hamiltonian
\begin{align}
    \left(h^{(g)}_{\vec{k},+}\right)_{\rho,\ell,\vec{G};\rho',\ell',\vec{G}'} &= \delta_{\ell,\ell'} \delta_{\vec{G},\vec{G}'} v_F (\vec{\rho}_{\theta_\ell})_{\rho,\rho'} \left(\vec{k} + \vec{G} - (-1)^\ell \vec{q}_{1}/2 \right), \label{DiracCones} \\ \left(h^{(g)}_{\vec{k},-}\right)_{\rho,\ell,\vec{G};\rho',\ell',\vec{G}'} &= \left(h^{(g)}_{-\vec{k},+}\right)^*_{\rho,\ell,-\vec{G};\rho',\ell',-\vec{G}'},
\end{align}
$\vec{\rho}_{\theta} = e^{i \theta \rho_3/2} \vec{\rho} e^{-i \theta \rho_3/2}$, $\vec{q}_1$ connecting the K and K' points in the MBZ. Throughout the appendix, $\vec{k}\in$ MBZ is measured relative to the M point of the moir\'e Brillouin zone.
The contribution
\begin{equation}
    \left(h^{(t)}_{\vec{k},+}\right)_{\rho,\ell,\vec{G};\rho',\ell',\vec{G}'} = \sqrt{2} \begin{pmatrix} 0 & (T_{\vec{G}-\vec{G}'})_{\rho,\rho'} & 0 \\  (T_{\vec{G}'-\vec{G}}^*)_{\rho',\rho} & 0 & 0 \\ 0 & 0 & 0 \end{pmatrix}, \qquad \left(h^{(t)}_{\vec{k},-}\right)_{\rho,\ell,\vec{G};\rho',\ell',\vec{G}'} = \left(h^{(t)}_{-\vec{k},+}\right)^*_{\rho,\ell,-\vec{G};\rho',\ell',-\vec{G}'}
\end{equation}
to the Hamiltonian describes the tunnelling that is modulated on the moir\'e lattice. We use the usual BM form \cite{bistritzer2011moire},
\begin{align}
    T_{\delta\vec{G}} = \sum_{j=0,1,2}\delta_{\delta\vec{G}+\vec{A}_j,0} \left[w_0 \rho_0 + w_1 \begin{pmatrix} 0 & \omega^j \label{FormOfT} \\  \omega^{-j} & 0 \end{pmatrix} \right], \\ \omega = e^{-i \frac{2\pi}{3}}, \quad \vec{A}_0 =0, \quad \vec{A}_1 = \vec{G}_1, \quad \vec{A}_2 = \vec{G}_1 + \vec{G}_2. 
\end{align}
Note that it holds $T_{\delta\vec{G}}^\dagger = T_{\delta\vec{G}}^\pdagger$ and $\rho_x T_{\delta\vec{G}} \rho_x = T^*_{\delta\vec{G}}$. So we see very clearly in this basis that the Hamiltonian is given by the sum of the TBG continuum model (with rescaled interlayer couplings) and a single graphene cone. This is different once a displacement field is applied, $D_0\neq0$; its contribution to the Hamiltonian reads as
\begin{equation}
    \left(h^{(D)}_{\vec{k}}\right)_{\rho,\ell,\vec{G};\rho',\ell',\vec{G}'} = -D_0 \delta_{\rho,\rho'}\delta_{\vec{G},\vec{G}'} \begin{pmatrix} 0 & 0 & 1 \\ 0 & 0 & 0 \\ 1 & 0 & 0 \end{pmatrix}_{\ell,\ell'},
\end{equation}
which is seen to couple the different mirror-eigenvalue sectors, as it breaks the mirror symmetry.

Here and in the following we use the same symbol with subscript $j=0,1,2,3$ for Pauli matrices and the associated quantum numbers: $\rho_j$ are Pauli matrices in sublattice, $s_j$ in spin, and $\eta_j$ in valley space.

The Hamiltonian $H_0^{\text{Full}}$ can be diagonalized by solving the eigenvalue problem
\begin{equation}
    h_{\vec{k},\eta} u_{n,\eta}(\vec{k}) = \epsilon_{n}(\vec{k}) u_{n,\eta}(\vec{k}) \label{EigenvaluesEquation}
\end{equation}
at every $\vec{k}\in\text{MBZ}$ and defining new fermionic operators $f_{\vec{k};n,\eta,s} = \sum_{\rho,\ell,\vec{G}} [u_{n,\eta}(\vec{k})]_{\rho,\ell,\vec{G}} \psi^\pdagger_{\vec{k};\rho,\ell,\eta,s,\vec{G}}$. 

In this work, we will mostly focus on the four bands (for each spin and valley flavor) that are closest to the charge-neutrality point. Let us label them by the multi-index $n=(t,p)$ with $p=\pm$, $t=g,b$ indicating whether these bands are \textit{graphene-like} or \textit{TBG-like}. To make the association unique, we use the conventions
\begin{equation}
    p \, \epsilon_{(t,p),\eta}(\vec{k}) \geq 0, \qquad |\epsilon_{(b,p),\eta}(\vec{k})| \leq |\epsilon_{(g,p),\eta}(\vec{k})|. \label{OrderingOfBands}
\end{equation}
In this way, the graphene-like (TBG-like) bands transition to the single-layer graphene (TBG) band structure for $D_0 \rightarrow 0$. But \equref{OrderingOfBands} allows us to generalize this notion to non-zero $D_0$.
We will further introduce the notation
\begin{equation}
    b_{\vec{k};p,\eta,s} := f_{\vec{k};(b,p),\eta,s}, \qquad g_{\vec{k};p,\eta,s} := f_{\vec{k};(g,p),\eta,s}
\end{equation}
on the level of electronic field operators. Our non-interacting, low-energy Hamiltonian then reads as
\begin{equation}
    H_0 = \sum_{\vec{k}\in\text{MBZ}} \sum_{\eta=\pm} \sum_{s=\uparrow,\downarrow} \sum_{p=\pm} \left[ W_{\text{TBG}}\epsilon_{(b,p),\eta}(\vec{k}) b^\dagger_{\vec{k};p,\eta,s}b^\pdagger_{\vec{k};p,\eta,s} + \epsilon_{(g,p),\eta}(\vec{k}) g^\dagger_{\vec{k};p,\eta,s}g^\pdagger_{\vec{k};p,\eta,s}  \right]  \label{NonInteractingHam}
\end{equation}
and in total contains $2^4 = 16$ bands, associated with valley, spin, Dirac, and even-odd mirror space. In \equref{NonInteractingHam}, we introduced the dimensionless parameter $W_{\text{TBG}} \in [0,1]$. In the real system, we have $W_{\text{TBG}}=1$ and we define $W_{\text{TBG}}=0$ as the \textit{flat-band limit}. In \appref{Superexchange} below, we will compute the effect of the dispersion to second order in $W_{\text{TBG}}$.

\subsection{Single-particle symmetries}\label{DiscussionOfSymmetries}
As it will play an important role for our analysis of the correlated phases in the system, we will here list the relevant symmetries of the continuum-model Hamiltonian (\ref{FullContinuumModel}) and their representation on our low-energy degrees of freedom $b^\pdagger_{\vec{k};p,\eta,s}$ and $g^\pdagger_{\vec{k};p,\eta,s}$ in \equref{NonInteractingHam}. 

\vspace{1em}

We distinguish four different classes of symmetries:
\begin{enumerate}
    \item moir\'e translations, which we will assume to be preserved throughout and, hence, will not have to be considered explicitly here.
    \item spin and charge conservation symmetries: since we do not take into account spin-orbit coupling nor the for small twist angles suppressed intervalley tunneling processes, the Hamiltonian is invariant under spin and U(1) phase rotations in each valley $\eta=\pm$ separately. 
    \item point symmetries, forming the point group $C_6$ (for $D_0\neq 0$) or $C_{6h}$ (when $D_0=0$).
    \item additional internal symmetries: the anti-unitary time-reversal symmetry $\Theta$, the chiral symmetry $C$, and the unitary particle-hole symmetry $P$.
\end{enumerate}
While all other symmetries leave $H_0$ invariant, the ``anti-commuting symmetries'' $C$ and $P$ send $H_0 \rightarrow -H_0$. Note that $C$ is only an anti-commuting symmetry if $w_0=0$ in \equref{FormOfT} (but persists for $D_0 \neq 0$), while $P$ is only present if $D_0=0$ (but persists for $w_0 \neq 0$). So the removal of the unitary particle hole symmetry in the TBG-like subspace upon turning on $D_0$ is a novel aspect of the trilayer system as it is always a symmetry in TBG (at least within the analogous continuum-model description). 
Among the point symmetries, the mirror symmetry $\sigma_h$ is only present if $D_0=0$, similar to $C_{2x}$ in TBG.

In \tableref{ActionOfSymmetries}, we list the action of point and internal symmetries on the microscopic operators $\psi_{\vec{k}}$ in \equref{FullContinuumModel} and on the low-energy operators $b^\pdagger_{\vec{k};p,\eta,s}$ and $g^\pdagger_{\vec{k};p,\eta,s}$. The latter representations have to be understood as a gauge-fixing condition; the gauge-fixing conventions we use here, closely follow \cite{bernevig2020tbg}. For instance, we choose the phase of the wavefunctions (modulo remaining $\pm$ sign) in \equref{EigenvaluesEquation} such that $C_{2z}\Theta u_{n,\eta}(\vec{k}) =  \rho_1 [u_{n,\eta}(\vec{k})]^* = u_{n,\eta}(\vec{k})$.

To illustrate the gauge fixing and for later reference, we discuss it explicitly for $D_0=0$ in the mirror-odd sector bands ($\ell = 3$). As detailed in \appref{ContinuumModel}, the Hamiltonian in this sector is simply that of single-layer graphene and, in our notation of \equref{DiracCones}, given by
\begin{equation}
    \left(h^{(e)}_{\vec{k},+}\right)_{\rho,\ell,\vec{G};\rho',\ell',\vec{G}'} = \delta_{\ell,3}\delta_{\ell',3} \delta_{\vec{G},\vec{G}'} v_F (\vec{\rho}_{\theta_\ell})_{\rho,\rho'} \left(\vec{k} + \vec{G} + \vec{q}_{1}/2 \right) \label{SpecificallyGrapheneHam}
\end{equation}
in valley $\eta = +$ (the wavefunctions in the other valley will simply follow from $C_{2z}$ and $\Theta$). Since \equref{SpecificallyGrapheneHam} is already diagonal in the $\vec{G}$ indices, the wavefunctions simply follow from diagonalization of the structure in sublattice space. Focusing, as above, on the lowest bands, we have
\begin{equation}
    \left(u_{(g,p),+}(\vec{k})\right)_{\rho,\ell,\vec{G}} = \delta_{\vec{G},-\vec{G}_{\vec{k}+\vec{q}_{1}/2}} \delta_{\ell,3} \frac{i^{(1-p)/2}}{\sqrt{2}}\begin{pmatrix} p e^{-i (\xi_{\delta\vec{k}}-\theta)/2} \\ e^{i( \xi_{\delta\vec{k}}-\theta)/2} \end{pmatrix} , \quad p=\pm 1, \quad e^{i \xi_{\vec{k}}} = \frac{k_x +i k_y}{|\vec{k}|}, \quad \delta\vec{k} = \text{MBZ}(\vec{k} + \vec{q}_1/2), \label{WavefunctionsOfGraphene}
\end{equation}
where we defined the notation $G_{\vec{k}}$ to indicate the reciprocal lattice vector that folds $\vec{k}$ back to the MBZ, i.e., $\vec{k}-\vec{G}_{\vec{k}} \in \text{MBZ}$. Furthermore, we write  $\text{MBZ}(\vec{k}):=\vec{k}-\vec{G}_{\vec{k}}$ for the momentum folded back into the MBZ.

Modulo a (in general $\vec{k}$ and $p$ dependent) minus sign to be discussed shortly, the phase of the wavefunctions has been chosen such that (``real gauge'')
\begin{equation}
    C_{2z}\Theta u_{(g,p),+}(\vec{k}) \equiv \rho_1 [u_{(g,p),+}(\vec{k})]^* = u_{n,+}(\vec{k}).
\end{equation}
The relative minus sign between $p=+$ and $p=-$ has been fixed in \equref{WavefunctionsOfGraphene} to ensure the behavior,
\begin{equation}
    C u_{(g,p),+}(\vec{k}) \equiv -\rho_3 C u_{(g,p),+}(\vec{k}) = - p \, i \, u_{(g,-p),+}(\vec{k}),
\end{equation}
under $C$ given in \tableref{ActionOfSymmetries}. Note that the relative minus sign at different $\vec{k}$ is simply determined by continuity of $u_{(g,p),+}(\vec{k})$ as a function of $\vec{k}$ (away from K/K$'$).

Finally, we furthermore fixed the relative signs between wavefunctions at $\vec{k}$ and $\vec{k}+\vec{G}_1$, $\vec{G}_1 \in \text{RL}$, to be $+1$ (``periodic gauge''),
\begin{equation}
    \left(u_{(g,p),+}(\vec{k}+\vec{G}_1)\right)_{\rho,\ell,\vec{G}} = \left(u_{(g,p),+}(\vec{k})\right)_{\rho,\ell,\vec{G}+\vec{G}_1},
\end{equation}
which follows readily from \equref{WavefunctionsOfGraphene} by noting that $\vec{G}_{\vec{k}+\vec{G}_1} = \vec{G}_{\vec{k}}+\vec{G}_1$, for all $\vec{G}_1 \in \text{RL}$.

\begin{table*}[tb]
\begin{center}
\caption{Action of the symmetries of the continuum theory on the microscopic field operators ($\psi_{\vec{k}}$) and in the band basis ($b_{\vec{k}}$ and $g_{\vec{k}}$). Although redundant, we have added the relevant $\vec{k}$-local combinations $C_{2z}\Theta$ and $C_{2z}P$ of the symmetries for convenience of the reader. Note that our choice of action of $P$ in the graphene-like subspace is not independent from the other symmetries (it is related to $C$, $C_{2z}$, combined with a valley rotation symmetry). As indicated, $P$ (and, hence, $C_{2z}P$) only anti-commutes with the Hamiltonian if we neglect the rotation of the Pauli matrices $\vec{\rho}_{\theta}$ in \equref{DiracCones}, which becomes asymptotically valid for small twist angles. As in the main text, $\eta_j$ act in valley, $\sigma_j$ in band, $s_j$ in spin, and $\rho_j$ in sublattice space.}
\label{ActionOfSymmetries}
\begin{ruledtabular}
 \begin{tabular} {ccccccc} 
Symmetry $S$ & unitary & $H_0\rightarrow \pm H_0$ & $S\psi_{\vec{k};\ell,\vec{G}}S^\dagger$ & $S b_{\vec{k}}S^\dagger$ & $S g_{\vec{k}}S^\dagger$ & condition \\ \hline
$C_{2z}$ & \cmark  & $+$  & $\eta_1 \rho_1 \psi_{-\vec{k};\ell,-\vec{G}}$  & $\eta_1\sigma_0b_{-\vec{k}}$ &  $\eta_1\sigma_0g_{-\vec{k}}$ & ---  \\
$C_{3z}$ & \cmark  & $+$  & $e^{i\frac{2\pi}{3} \rho_3\eta_3} \psi_{C_{3z}\vec{k};\ell,C_{3z}\vec{G}}$  & $\eta_0\sigma_0b_{C_{3z}\vec{k}}$ &  $\eta_0\sigma_0g_{C_{3z}\vec{k}}$ & ---  \\
$\sigma_{h}$ & \cmark  & $+$  & $ (1,1,-1)_\ell \psi_{\vec{k};\ell,\vec{G}}$  & $\eta_0\sigma_0b_{\vec{k}}$ &  $-\eta_0\sigma_0g_{\vec{k}}$ & $D_0=0$  \\ \hline

$\Theta$ & \xmark  & $+$  & $\eta_1 \psi_{-\vec{k};\ell,-\vec{G}}$  & $\eta_1\sigma_0b_{-\vec{k}}$ &  $\eta_1\sigma_0g_{-\vec{k}}$ & ---  \\
$C$ & \cmark  & $-$  & $ (1,1,-1)_\ell \rho_3 \psi_{\vec{k};\ell,\vec{G}}$  & $\eta_3\sigma_2 b_{\vec{k}}$ &  $\eta_3\sigma_2g_{\vec{k}}$ & $w_0=0$  \\
$P$ & \cmark  & $-$  & $ \begin{pmatrix} 0 & \eta_3 & 0 \\ -\eta_3 & 0 & 0 \\ 0 & 0 & i \rho_2 \eta_2 \end{pmatrix}_{\ell,\ell'} \hspace{-0.6em} \psi_{-\vec{k};\ell',-\vec{G}}$  & $-i \eta_3\sigma_2 b_{-\vec{k}}$ &  $ i\eta_1\sigma_2 g_{-\vec{k}}$ & $D_0=0$, $\theta\rightarrow 0$  \\ \hline
$C_{2z}\Theta$ & \xmark  & $+$  & $\rho_1 \psi_{\vec{k};\ell,\vec{G}}$  & $\eta_0\sigma_0b_{\vec{k}}$ &  $\eta_0\sigma_0g_{\vec{k}}$ & ---  \\
$C_{2z}P$ & \cmark  & $-$  & $ \begin{pmatrix} 0 & -i\eta_2\rho_1 & 0 \\ i\eta_2\rho_1 & 0 & 0 \\ 0 & 0 & -i \rho_3 \eta_3 \end{pmatrix}_{\ell,\ell'} \hspace{-0.6em} \psi_{\vec{k};\ell',\vec{G}}$  & $\eta_2\sigma_2 b_{\vec{k}}$ &  $ i\sigma_2 g_{\vec{k}}$ & $D_0=0$, $\theta\rightarrow 0$  \\
 \end{tabular}
\end{ruledtabular}
\end{center}
\end{table*}

With these symmetry representations at hand, it is very simple to see that the band structure of the TBG-like band in \equref{NonInteractingHam} has the form,
\begin{equation}
    \epsilon_{(b,p),\eta}(\vec{k}) = p\, E^b_{0}(\vec{k}) + \eta \, E^b_{1}(\vec{k}) + p\,\eta\, E^b_{2}(\vec{k}) + E^b_{3}(\vec{k}), \label{ConstraintsOnEnergy}
\end{equation}
where $E^b_{1}$ ($E^b_{2}$) can only be non-zero if the same holds for $w_0$ ($D_0$) such that $C$ ($P$) is broken; $E^b_{3}\neq 0$ requires $w_0,D_0\neq 0$ simultaneously. Furthermore, it holds $E^b_{j}(\vec{k}) = E^b_{j}(-\vec{k})$, for $j=0,3$, and $E^b_{j}(\vec{k}) = -E^b_{j}(-\vec{k})$, for $j=1,2$, due to $\Theta$. Clearly, the band structure for finite $D_0$ shown in \figref{TrilayerBands}(b) can only be consistent with \equref{ConstraintsOnEnergy} if $E^b_{2}(\vec{k})\neq 0$, which is consistent with our observation that $P$ must be broken when $D_0\neq 0$.

For later reference we introduce the following notation: for any quantity $q(\{x_k\})$ (such as $E^b_j$ here) that depends on a set of parameters $x_k$ (here $D_0$ or $w_0$), we introduce $\bar{q}$ via
\begin{equation}
    q(\{x_k\}) =  \bar{q}(\{x_k\})\, \prod_k (x_k)^{n_k}, \quad \text{with maximal $n_k$ such that we still have}\quad \lim_{x_k\rightarrow 0} |\bar{q}(\{x_k\})| < \infty. \label{ScalingNotation}
\end{equation}
This allows us to organize our perturbation theory. For instance, \equref{ConstraintsOnEnergy} can then be written in the more informative way
\begin{equation}
    \epsilon_{(b,p),\eta}(\vec{k}) = p\, \bar{E}^b_{0}(\vec{k}) + \eta \, \bar{E}^b_{1}(\vec{k})\,w_0 + p\,\eta\, \bar{E}^b_{2}(\vec{k})\,D_0 + \bar{E}^b_{3}(\vec{k})\,w_0 D_0.  \label{TBGLikeEnergy}
\end{equation}

Similarly, we can analyze the properties of the graphene-like bands. As anticipated above, $P$, when present, does not lead to additional constraints and one readily concludes that
\begin{equation}
    \epsilon_{(g,p),\eta}(\vec{k}) = p\, \bar{E}^g_{0}(\vec{k};\eta) +  \bar{E}^g_{1}(\vec{k};\eta) \, w_0 D_0, \label{GrapheneLikeEnergy}
\end{equation}
where we noticed that the graphene-like bands still have a chiral symmetry for $D_0=0$ even if $w_0\neq 0$. $\Theta$ dictates that $\bar{E}^g_{j}(\vec{k};\eta) = \bar{E}^g_{j}(-\vec{k};-\eta)$. Note that, throughout this work (cf.~\figref{TrilayerBands}), we will label the valleys such that $|\epsilon_{(g,p),+}(\vec{k})| < |\epsilon_{(g,p),-}(\vec{k})|$ for $\vec{k}$ near the K point.

\subsection{Interaction and form factors}\label{FormFactorsAndInteraction}

We follow a variety of previous works on moir\'e graphene systems \cite{Xie_2020,bultinck2019ground,Liu_2021,Liao_2021,Bultinck_2020,Zhang_2020,2020arXiv200913530L} and study a density-density interaction
\begin{equation}
    H_1^{\text{Full}} = \frac{1}{2N}\sum_{\vec{q}} V(\vec{q}) : \rho^{\text{Full}}_{\vec{q}}   \rho^{\text{Full}}_{-\vec{q}} :, \label{GeneralFormOfInteraction}
\end{equation}
with $:...:$ indicating normal ordering and 
\begin{equation}
    \rho^{\text{Full}}_{\vec{q}} = \sum_{\vec{k} \in \text{MBZ}} \sum_{\rho,\ell,\eta,s,\vec{G}} \psi^\dagger_{\text{MBZ}(\vec{k}+\vec{q});\rho,\ell,\eta,s,\vec{G}+\vec{G}_{\vec{k}+\vec{q}}} \psi^\pdagger_{\vec{k};\rho,\ell,\eta,s,\vec{G}} 
\end{equation}
denoting the Fourier transform of the electron density, where, as above, $\vec{q}=\text{MBZ}(\vec{q})+\vec{G}_{\vec{q}}$ with $\text{MBZ}(\vec{q})\in \text{MBZ}$. Furthermore, $V(\vec{q})$ in \equref{GeneralFormOfInteraction} is the momentum-space interaction potential. For our analytical analysis, we will not have to specify the precise form of $V(\vec{q})$ and will only use that $V(\vec{q})=V(-\vec{q})>0$. For our numerical computations we use $V(\vec{q})$ as given in \equref{ExplicitFormOfVOfq}.

Similar to our discussion of the non-interacting theory, we transform to the eigenbasis of $H_0$ and only keep the lowest four bands per spin and valley; the resulting projected interaction then reads as
\begin{equation}
        H_1 = \frac{1}{2N}\sum_{\vec{q}} V(\vec{q}) : \rho_{\vec{q}}   \rho_{-\vec{q}} :, \qquad \rho_{\vec{q}} = \sum_{\vec{k}\in\text{MBZ}} \sum_{\eta=\pm}\sum_{s=\uparrow,\downarrow}\sum_{t,t'=b,g}\sum_{p,p'=\pm}  f^\dagger_{\text{MBZ}(\vec{k}+\vec{q});(t,p),\eta,s} \left(F^\eta_{\vec{k},\vec{q}}\right)_{(t,p),(t',p')} f^\pdagger_{\vec{k};(t',p'),\eta,s}, \label{ProjectedInteraction}
\end{equation}
where all the microscopic details are encoded in the \textit{form factors}
\begin{equation}
    \left(F^\eta_{\vec{k},\vec{q}}\right)_{n,n'} : = \sum_{\rho,\ell,\vec{G}} \left[u_{n,\eta}(\text{MBZ}(\vec{k}+\vec{q}))\right]^*_{\rho,\ell,\vec{G}+\vec{G}_{\vec{k}+\vec{q}}} \left[u_{n',\eta}(\vec{k})\right]_{\rho,\ell,\vec{G}}. \label{DefinitionOfFormFactors}
\end{equation}

In order to make sure that the projected interaction respects (many-body) particle-hole symmetry, we follow \cite{bernevig2020tbg} and first rewrite the unprojected interaction (\ref{GeneralFormOfInteraction}) as 
\begin{equation}
    \widetilde{H}_1^{\text{Full}} = \frac{1}{2N}\sum_{\vec{q}} V(\vec{q}) \delta \rho^{\text{Full}}_{\vec{q}}   \delta \rho^{\text{Full}}_{-\vec{q}} + \dots,  \quad \delta \rho^{\text{Full}}_{\vec{q}} =  \sum_{\vec{k} \in \text{MBZ}} \sum_{\rho,\ell,\eta,s,\vec{G}} \left( \psi^\dagger_{\text{MBZ}(\vec{k}+\vec{q});\rho,\ell,\eta,s,\vec{G}+\vec{G}_{\vec{k}+\vec{q}}} \psi^\pdagger_{\vec{k};\rho,\ell,\eta,s,\vec{G}} - \frac{1}{2}\delta_{\vec{q},0} \right), \label{TheFullInteractionPHSym}
\end{equation}
where $\dots$ just represent irrelevant constants and chemical potential terms.
The wavefunctions in \equref{EigenvaluesEquation} obey the completeness relation
\begin{equation}
     \sum_{n}[u_{n,\eta}(\vec{k})]_{\rho_1,\ell_1,\vec{G}_1} [u_{n,\eta}(\vec{k})]^*_{\rho_2,\ell_2,\vec{G}_2} = \delta_{\rho_1,\rho_2}\delta_{\ell_1,\ell_2}\delta_{\vec{G}_1,\vec{G}_2}. \label{CompletenessRelation}
\end{equation}
Setting $\rho_1=\rho_2=\rho$, $\ell_1=\ell_2=\ell$, $\vec{G}_1=\vec{G}$, and $\vec{G}_2=\vec{G}+\vec{G}'$ and summing both sides of \equref{CompletenessRelation} over $\rho$, $\ell$, and $\vec{G}$, we get 
\begin{equation}
    \sum_n\left(F^\eta_{\vec{k},\vec{G}'}\right)_{n,n} =  \sum_{\rho,\ell,\vec{G}} \delta_{\vec{G}',0},
\end{equation}
where the sum over $n$ is over \textit{all} bands. Using this property we rewrite the full density in \equref{TheFullInteractionPHSym} as 
\begin{equation}
    \delta\rho^{\text{Full}}_{\vec{q}} =  \sum_{\vec{k} \in \text{MBZ}} \sum_{n,n',\eta,s} \left( f^\dagger_{\text{MBZ}(\vec{k}+\vec{q});n,\eta,s} \left(F^\eta_{\vec{k},\vec{q}}\right)_{n,n'} f^\pdagger_{\vec{k};n',\eta,s} - \frac{1}{2} \delta_{n,n'} \sum_{\vec{G}\in\text{RL}}\left(F^\eta_{\vec{k},\vec{G}}\right)_{n,n'} \delta_{\vec{q},\vec{G}} \right). \label{MoreNaturalFormOfUnProjectedInteraction}
\end{equation}
So far, this is still an exact rewriting of the full interaction. However, in this form, the projection of the interaction to the low-energy bands (graphene-like and TBG-like) in a way that respects particle-hole symmetry is particularly natural: one simply has to restrict the sum over $n,n'$ in \equref{MoreNaturalFormOfUnProjectedInteraction} to those bands. Defining for notational convenience
\begin{equation}
    \left(F_{\vec{k},\vec{q}}\right)_{((t,p),\eta,s),((t',p'),\eta',s')} = \delta_{s,s'}\delta_{\eta,\eta'} \left(F^\eta_{\vec{k},\vec{q}}\right)_{(t,p),(t',p')},
\end{equation}
we then arrive at the final form of the projected interaction
\begin{equation}
    \widetilde{H}_1 = \frac{1}{2N}\sum_{\vec{q}} V(\vec{q})  \delta\rho_{\vec{q}}   \delta\rho_{-\vec{q}}, \qquad \delta \rho_{\vec{q}} = \sum_{\vec{k}\in\text{MBZ}} \left( f^\dagger_{\text{MBZ}(\vec{k}+\vec{q})} F_{\vec{k},\vec{q}} f^\pdagger_{\vec{k}} - \frac{1}{2} \sum_{\vec{G}\in\text{RL}} \delta_{\vec{q},\vec{G}}\text{tr}(F_{\vec{k},\vec{G}})\right). \label{TheInteractionDeltaRho}
\end{equation}

Let us next discuss properties of these form factors.
While we have computed these form factors numerically using the continuum model, we can make analytical progress by exploiting constraints on them. First, Hermiticity, $\rho_{\vec{q}}^\dagger = \rho_{-\vec{q}}$, implies
\begin{equation}
    \left(F^\eta_{\vec{k},\vec{q}}\right)^*_{n',n} =  \left(F^\eta_{\text{MBZ}(\vec{k}+\vec{q}),-\vec{q}}\right)_{n,n'}\quad \text{and, specifically,}\quad \left(F^\eta_{\vec{k},\vec{G}}\right)^*_{n',n} = \left(F^\eta_{\vec{k},-\vec{G}}\right)_{n,n'}. \label{Hermiticity}
\end{equation}
Furthermore, since the density operator $\rho^{\text{Full}}_{\vec{q}}$ and, by design, its projection $\rho_{\vec{q}}$ transforms as a scalar function of the vector $\vec{q}$ under all regular and anti-commuting symmetries of $H_0$ in \tableref{ActionOfSymmetries}, we obtain further constraints on the form factors. Using all $\vec{k}$-local operators listed in \tableref{ActionOfSymmetries} one finds after straightforward algebra:
\begin{subequations}\begin{align}
    \begin{split}\left(F^\eta_{\vec{k},\vec{q}}\right)_{(b,p),(b,p')} &= \bar{F}^{bb}_1(\vec{k},\vec{q}) (\sigma_0)_{p,p'} + i \bar{F}^{bb}_2(\vec{k},\vec{q}) (\sigma_2)_{p,p'} \\ &\quad + \left[\bar{F}^{bb}_3(\vec{k},\vec{q}) \eta (\sigma_0)_{p,p'}  + i \bar{F}^{bb}_4(\vec{k},\vec{q}) \eta (\sigma_2)_{p,p'} \right] D_0^2 \\ & \quad + \left[\bar{F}^{bb}_5(\vec{k},\vec{q}) \eta (\sigma_1)_{p,p'}   + \bar{F}^{bb}_6(\vec{k},\vec{q}) \eta (\sigma_3)_{p,p'} \right] w_0 \\ & \quad  + \left[\bar{F}^{bb}_7(\vec{k},\vec{q})  (\sigma_1)_{p,p'} +   \bar{F}^{bb}_8(\vec{k},\vec{q})  (\sigma_3)_{p,p'} \right] w_0 D_0^2, \label{bbMatrixElements}\end{split}   \\
    \begin{split}\left(F^\eta_{\vec{k},\vec{q}}\right)_{(g,p),(g,p')} &= \bar{F}^{gg}_1(\vec{k},\vec{q}) (\sigma_0)_{p,p'} + i \bar{F}^{gg}_2(\vec{k},\vec{q}) (\sigma_2)_{p,p'} +\bar{F}^{gg}_3(\vec{k},\vec{q}) \eta (\sigma_0)_{p,p'}  + i \bar{F}^{gg}_4(\vec{k},\vec{q}) \eta (\sigma_2)_{p,p'} \\
    &\quad + \left[ \bar{F}^{gg}_5(\vec{k},\vec{q}) \eta (\sigma_1)_{p,p'} + \bar{F}^{gg}_6(\vec{k},\vec{q}) \eta (\sigma_3)_{p,p'}  + \bar{F}^{gg}_7(\vec{k},\vec{q})  (\sigma_1)_{p,p'} +   \bar{F}^{gg}_8(\vec{k},\vec{q})  (\sigma_3)_{p,p'} \right] w_0 D_0^2, \label{ggMatrixElements} \end{split} \\
    \begin{split}\left(F^\eta_{\vec{k},\vec{q}}\right)_{(g,p),(b,p')} &= \left[\bar{F}^{gb}_1(\vec{k},\vec{q}) (\sigma_0)_{p,p'} + i \bar{F}^{gb}_2(\vec{k},\vec{q}) (\sigma_2)_{p,p'} +\bar{F}^{gb}_3(\vec{k},\vec{q}) \eta (\sigma_0)_{p,p'}  + i \bar{F}^{gb}_4(\vec{k},\vec{q}) \eta (\sigma_2)_{p,p'} \right]D_0 \\ &\quad + \left[\bar{F}^{gb}_5(\vec{k},\vec{q}) \eta (\sigma_1)_{p,p'} + \bar{F}^{gb}_6(\vec{k},\vec{q}) \eta (\sigma_3)_{p,p'}  + \bar{F}^{gb}_7(\vec{k},\vec{q})  (\sigma_1)_{p,p'} +   \bar{F}^{gb}_8(\vec{k},\vec{q})  (\sigma_3)_{p,p'}\right] w_0D_0, \label{MatrixElementsgb}\end{split}
\end{align}\label{FormFactors}\end{subequations}
while $(F^\eta_{\vec{k},\vec{q}})_{(b,p),(g,p')}$ follows from \equref{Hermiticity}, i.e.,
\begin{equation}
    \bar{F}^{gb}_{j,\eta}(\vec{k},\vec{q}) = \bar{F}^{bg}_{j,\eta}(\text{MBS}(\vec{k}+\vec{q}),-\vec{q}).
\end{equation}
In \equref{FormFactors}, all expansion co-efficients are real, $F^{tt'}_j(\vec{k},\vec{q})\in\mathbb{R}$, and we again employed the convenient notation defined in \equref{ScalingNotation} to make their leading parameter dependence visible. While, by construction, all of the terms in $F^\eta_{(g,p),(g,p')}$ and $F^\eta_{(g,p),(b,p')}$ are absent in the previously studied TBG system, also half of the terms ($F^{bb}_j$, $j=3,4,7,8$) entering $F^\eta_{(g,p),(g,p')}$ in \equref{bbMatrixElements} are unique to MSTG, as they can only be non-zero if $D_0\neq 0$.

Note that $F^{bb}_j$, $j=3,4,7,8$, have to be even functions of $D_0$ at generic momenta and, hence, start at order $D_0^2$, which can be seen by noting that $\sigma_h$ transforms the continuum model Hamiltonian $h_{\vec{k},\eta}(D_0)$ to $h_{\vec{k},\eta}(-D_0)$; the eigenstates, $u_{n,\eta}(\vec{k};D_0)$, in \equref{EigenvaluesEquation} thus obey $\sigma_h u_{n,\eta}(\vec{k};D_0) = \pm u_{n,\eta}(\vec{k};-D_0)$. Alternatively, it is also clear from the structure of (non-degenerate) perturbation theory in $D_0$: corrections to matrix elements of states that have the same eigenvalue under $\sigma_h$ can only appear in even orders of perturbation theory. The same line of reasoning shows that all $F_j^{gb}$ have to be odd functions of $D_0$, as indicated in \equref{MatrixElementsgb}.

In addition, $C_{2z}$ implies that
\begin{align}\begin{split}
    F^{tt'}_j(\vec{k},\vec{q}) &= F^{tt'}_j(-\vec{k},-\vec{q}), \qquad j=1,2,7,8, \\ 
    F^{tt'}_j(\vec{k},\vec{q}) &= -F^{tt'}_j(-\vec{k},-\vec{q}), \qquad j=3,4,5,6.\label{C2zConstraint}\end{split}
\end{align}
The Hermiticity constraint in \equref{Hermiticity} further implies for the form factors with $t=t'$ that
\begin{align}\begin{split}
    F^{tt}_j(\vec{k},\vec{G}) &= F^{tt}_j(\vec{k},-\vec{G}), \qquad j=1,3,5,6,7,8, \label{GToMinusG} \\ 
    F^{tt}_j(\vec{k},\vec{G}) &= -F^{tt}_j(\vec{k},-\vec{G}), \qquad j=2,4.
\end{split}\end{align}
For later reference, we combine this with \equref{C2zConstraint} above to obtain
\begin{align}\begin{split}
    F^{tt}_j(\vec{k},\vec{G}) &= F^{tt}_j(-\vec{k},\vec{G}), \qquad j=1,4,7,8 \\ 
    F^{tt}_j(\vec{k},\vec{G}) &= -F^{tt}_j(-\vec{k},\vec{G}), \qquad j=2,3,5,6. \label{GConstraint}
\end{split}\end{align}

Although we will not use $C_{3z}$ constraints on the form factors explicitly in our analytical considerations, we note that it further forces the momentum dependence of all pre-factors to obey $F^{tt'}_j(\vec{k},\vec{q})=F^{tt'}_j(C_{3z}\vec{k},C_{3z}\vec{q})$. 

Finally, we point out that $F^\eta_{\vec{k},\vec{q}=0}=\mathbbm{1}$, for all $\vec{k}\in\text{MBZ}$, as readily follows from the definition (\ref{DefinitionOfFormFactors}). As such, we have
\begin{equation}
    F_j^{tt'}(\vec{k},\vec{q}=0) = \delta_{j,1}\delta_{t,t'}. \label{ZeroqLimit}
\end{equation}

The form factors of the mirror-odd (graphene) subsystem for $D_0=0$ can be readily computed from the wavefunctions defined in \equref{WavefunctionsOfGraphene}. One finds
\begin{equation}
    \left(F^+_{\vec{k},\vec{q}}\right)_{(g,p),(g,p')} = \delta_{\vec{G}_{\vec{k}+\vec{q}+\frac{\vec{q}_1}{2}},\vec{G}_{\vec{k}+\frac{\vec{q}_1}{2}}} \left[ \delta_{p,p'} \cos\left(\frac{\xi_{\delta(\vec{k}+\vec{q})}-\xi_{\delta\vec{k}}}{2}\right) + i (\sigma_y)_{p,p'} \sin\left(\frac{\xi_{\delta(\vec{k}+\vec{q})}-\xi_{\delta\vec{k}}}{2}\right) \right]. \label{FormOfGrapheneMatrixElements}
\end{equation}
Note that this agrees with the symmetry-based structure given in \equref{ggMatrixElements}. On top of this, we observe that the matrix elements obey
\begin{equation}
    \left(F^\eta_{\vec{k},\vec{G}}\right)_{(g,p),(g,p')} = \delta_{\vec{G},0} \delta_{p,p'}. \label{SpecialPropoertyGrapheneMatrixElements}
\end{equation}
We will use both the explicit form in \equref{FormOfGrapheneMatrixElements} as well as the special property (\ref{SpecialPropoertyGrapheneMatrixElements}) of the graphene matrix elements below.

\section{Exact statements about the interacting groundstates}\label{ExactStatements}
In this appendix, we will discuss limits where exact many-body groundstates of MSTG can be derived. To this end, we will focus on $D_0=0$, where the \textit{non-interacting} part of the Hamiltonian, $H_0$, is just given by the sum of the TBG continuum model, $H_0^{\text{b}}$, and that of single-layer graphene, $H_0^{\text{g}}$. Importantly, the presence of mirror symmetry $\sigma_h$ also leads to $F^{gb}_j(\vec{k},\vec{q})=F^{bg}_j(\vec{k},\vec{q})=0$, see \equref{MatrixElementsgb}. Consequently, the density operator $\rho_{\vec{q}}$ in \equref{ProjectedInteraction} and its symmetrized cousin $\delta \rho_{\vec{q}}$ in \equref{TheInteractionDeltaRho} are just given by the sum of the contributions from the graphene (g) and TBG (b) subsystem. Focusing on $\delta \rho$, we have $\delta \rho_{\vec{q}} = \delta \rho^{\text{g}}_{\vec{q}} + \delta \rho^{\text{b}}_{\vec{q}}$

with
\begin{align}
    \delta\rho^{\text{g}}_{\vec{q}} = \sum_{\vec{k}\in\text{MBZ}}\left(\sum_{\mu,\mu'}g^\dagger_{\text{MBZ}(\vec{k}+\vec{q});\mu} \left(F^{gg}_{\vec{k},\vec{q}}\right)_{\mu,\mu'} g^\pdagger_{\vec{k};\mu'} - \frac{1}{2} \sum_{\vec{G}\in\text{RL}} \delta_{\vec{q},\vec{G}}\text{tr}(F^{gg}_{\vec{k},\vec{G}}) \right), \\  
    \delta\rho^{\text{b}}_{\vec{q}} = \sum_{\vec{k}\in\text{MBZ}}\left(\sum_{\mu,\mu'} b^\dagger_{\text{MBZ}(\vec{k}+\vec{q});\mu} \left(F^{bb}_{\vec{k},\vec{q}}\right)_{\mu,\mu'} b^\pdagger_{\vec{k};\mu'} - \frac{1}{2} \sum_{\vec{G}\in\text{RL}} \delta_{\vec{q},\vec{G}}\text{tr}(F^{bb}_{\vec{k},\vec{G}}) \right),
\end{align}
where we introduced the multi-indices comprising band, valley, and spin, $\mu=(p,\eta,s)$, 
and the submatrix form factors in the respective subsystem, $(F^{tt}_{\vec{k},\vec{q}})_{(p,\eta,s),(p',\eta',s')}:= \delta_{s,s'}\delta_{\eta,\eta'}(F^\eta_{\vec{k},\vec{q}})_{(t,p),(t,p')}$. Notwithstanding this simplification, the interaction in \equref{TheInteractionDeltaRho} still couples the graphene and TBG subsystem and it is not clear how this will affect the groundstates. 
In particular, it is unclear whether one can still write down exact eigenstates (or even groundstates) of the combined system, as has been done in TBG \cite{2020arXiv200913530L}, and whether the system will also exhibit order in the graphene subspace.

\subsection{Turning on the coupling between the subsystems}\label{TurningOnLambda}
To address these questions, we split the projected Hamiltonian of the system into three parts, 
\begin{equation}
    H_0 + \widetilde{H}_1 = H^{\text{g}} + H^{\text{b}} + \lambda H^{\text{gb}},
\end{equation}
where
\begin{equation}
    H^{\text{g}} = \sum_{\vec{k}\in\text{MBZ}} \sum_{\eta,p=\pm} \sum_{s=\uparrow,\downarrow} p E_0^g(\vec{k};\eta) g^\dagger_{\vec{k};p,\eta,s}g^\pdagger_{\vec{k};p,\eta,s} + \frac{1}{2N}\sum_{\vec{q}} V(\vec{q})  \delta\rho^{\text{g}}_{\vec{q}}   \delta\rho^{\text{g}}_{-\vec{q}} \label{HGraphene}
\end{equation}
is just the usual Hamiltonian of graphene with Coulomb interaction, while 
\begin{equation}
    H^{\text{b}} = W_{\text{TBG}} \sum_{\vec{k}\in\text{MBZ}} \sum_{\eta,p=\pm} \sum_{s=\uparrow,\downarrow}  (p\, \bar{E}^b_{0}(\vec{k}) + \eta \, \bar{E}^b_{1}(\vec{k})\,w_0) b^\dagger_{\vec{k};p,\eta,s}b^\pdagger_{\vec{k};p,\eta,s}  + \frac{1}{2N}\sum_{\vec{q}} V(\vec{q})  \delta\rho^{\text{b}}_{\vec{q}}   \delta\rho^{\text{b}}_{-\vec{q}} \label{HBilayer}
\end{equation}
is the TBG Hamiltonian, consisting of the quasi-flat bands interacting via a projected density-density interaction. We further introduced the dimensionless parameter $\lambda$ that allows us to adiabatically turn on the density-density interaction,
\begin{equation}
    \lambda H^{\text{gb}} = \frac{\lambda}{N}\sum_{\vec{q}} V(\vec{q})  \delta\rho^{\text{g}}_{-\vec{q}}   \delta\rho^{\text{b}}_{\vec{q}} \label{CouplingBetweenTheSystems}
\end{equation}
between the two subsystems, with $\lambda=1$ corresponding to the physical system.

Let us start with $\lambda=0$, where the Hamiltonian is given by $H^{\text{g}} + H^{\text{b}}$. Being the sum of the two commuting subsystem Hamiltonians, its eigenstates are just given by all combinations of the individual eigenstates of graphene and TBG. As discussed in more detail in the main text, the ground state of the graphene subspace is a symmetry-unbroken correlated semimetal. We denote its ground state by $\ket{\Psi^{\text{g}}_{0}(\nu_b)}$ and its (gapless) excited states by $\ket{\Psi^{\text{g}}_{j}(\nu_b)}$, $j>0$, where $\nu_b$ indicates that the filling fraction of the graphene sector is a function of the filling $\nu_b$ in the TBG sector. We will here focus on $\nu_{\text{b}}=0,\pm 1, \pm 2, \pm 3, \pm 4$.

The exact ground states of certain limits of $H^{\text{b}}$ in \equref{HBilayer} at integer filling fractions $\nu_b$ have been discussed in detail in \cite{2020arXiv200913530L} and we next use these insights, within our notation, and transfer them to MSTG. First, one can see from \equref{bbMatrixElements} that the TBG form factors become
\begin{equation}
   \bar{F}^{bb}_{\vec{k},\vec{q}} = \bar{F}^{bb}_1(\vec{k},\vec{q}) \sigma_0 + i \bar{F}^{bb}_2(\vec{k},\vec{q}) \sigma_2 \label{FormFactorsToDiagonalize}
\end{equation}
in the chiral-decoupled limit ($w_0=D_0=0$). We can, hence, bring the form factors and density operators to a diagonal form by introducing another set of field-operators,
\begin{equation}
    \delta\rho^{\text{b}}_{\vec{q}} = \sum_{\vec{k}\in\text{MBZ}}\sum_{\eta,c=\pm,s}\bar{F}^{bb}_c(\vec{k},\vec{q})\left(\tilde{b}^\dagger_{\text{MBZ}(\vec{k}+\vec{q});c,\eta,s} \tilde{b}^\pdagger_{\vec{k};c,\eta,s} - \frac{1}{2} \sum_{\vec{G}\in\text{RL}} \delta_{\vec{q},\vec{G}} \right), \, \tilde{b}^\pdagger_{\vec{k};c,\eta,s} = U_{c,p} b^\pdagger_{\vec{k};p,\eta,s}, \quad U = \frac{1}{\sqrt{2}} \begin{pmatrix} 1 & -i \\ 1 & i \end{pmatrix}, \label{DensityOperatorChiralLimit}
\end{equation}
where $\bar{F}^{bb}_{\pm}(\vec{k},\vec{q}) = \bar{F}^{bb}_1(\vec{k},\vec{q}) \pm i  \bar{F}^{bb}_2(\vec{k},\vec{q})$. 
As is readily verified, the following set of states \cite{2020arXiv200913530L}
\begin{equation}
    \ket{\Psi_0^{\text{b}}(\nu_b)} = \prod_{\vec{k}\in\text{MBZ}} \prod_{c=\pm} \prod_{j_c=1}^{\nu_c} \tilde{b}^\dagger_{\vec{k};c,\eta^c_{j_c},s^c_{j_c}}  \ket{0^{\text{b}}} \label{GeneralFormOfGroundState}
\end{equation}
with $\nu_++\nu_- = \nu_{\text{b}}$ obey
\begin{equation}
    \delta\rho^{\text{b}}_{\vec{q}} \ket{\Psi_0^{\text{b}}(\nu_b)} =  \sum_{\vec{G}\in\text{RL}} \delta_{\vec{q},\vec{G}} R_{\vec{G}} \ket{\Psi_0^{\text{b}}(\nu_b)}, \qquad R_{\vec{G}} = \nu_b \sum_{\vec{k}} F_1^{bb}(\vec{k},\vec{G}), \label{PropertyOfEigenstates}
\end{equation}
in the chiral limit, $w_0=0$, for arbitrary combinations of occupied spin/valley-flavors $\{\eta^{\pm}_j,s^{\pm}_j\}$. Consequently, all of these states are eigenstates of $H^{\text{b}}$ in the flat-limit, i.e., when $W_{\text{TBG}}=0$ (or right at the magic angle for $w_0=0$ where the bands become perfectly flat). It can further be shown \cite{2020arXiv200913530L} that all $\ket{\Psi_0^{\text{b}}(\nu_b)}$ in \equref{GeneralFormOfGroundState} are groundstates of $H^{\text{b}}$ for $\nu_b=0$ and that the same holds for all integer $\nu_b\neq 0$, if the flat-metric condition,
\begin{equation}
    F^{bb}_{\vec{k},\vec{G}} =  \mathbbm{1} f(\vec{G}),\qquad \forall\,\vec{k},\vec{G} \label{FlatMetricCondition}
\end{equation}
applies. As argued in \refcite{2020arXiv200913530L}, \equref{FlatMetricCondition} holds approximately for realistic parameters of TBG and the $\ket{\Psi_0^{\text{b}}(\nu_b)}$ in \equref{GeneralFormOfGroundState} will remain \textit{exact} ground states if it is only weakly violated. The fact that there are several exact ground states is associated with an emergent enhanced continuous symmetry [$(\text{U}(4)\times \text{U}(4))_{\text{b,cf}}$] in the chiral-flat limit \cite{bultinck2019ground,PhysRevLett.122.246401,bernevig2020tbg}.

Once $w_0\neq 0$, i.e., in the nonchiral-flat limit, this symmetry group is reduced [to U(4)] and only the subset of states in \equref{GeneralFormOfGroundState} where $\eta_j^+=\eta_j^-$ and $s_j^+=s_j^-$, which can only be defined for even $\nu_b$, are exact eigenstates of $\delta\rho^{\text{b}}_{\vec{q}}$ and obey \equref{PropertyOfEigenstates}. They are always the exact ground states at $\nu_b=0$, while this only holds if \equref{FlatMetricCondition} is not sufficiently violated for $\nu_b=\pm 2$. We refer to \refcite{2020arXiv200913530L} for the derivations of these statements.

Having established the properties of the spectrum of the graphene subsystem and the exact groundstates of the TBG Hamiltonian, $H^{\text{b}}$, we can next address what happens when the coupling $\lambda H^{\text{gb}}$ is turned on. As a first step, we rewrite the coupling Hamiltonian in \equref{CouplingBetweenTheSystems} as
\begin{equation}
    H^{\text{gb}} = \frac{1}{N}\sum_{\vec{q}} V(\vec{q}) \left[  \delta\rho^{\text{g}}_{-\vec{q}}   \left(\delta\rho^{\text{b}}_{\vec{q}} - \sum_{\vec{G}\in\text{RL}}\delta_{\vec{q},\vec{G}} R_{\vec{G}} \right) + \sum_{\vec{G}\in\text{RL}} \delta_{\vec{q},\vec{G}} R_{\vec{G}} \delta\rho^{\text{g}}_{-\vec{G}} \right].
\end{equation}
Recall from \equref{SpecialPropoertyGrapheneMatrixElements} that $F^{gg}_{\vec{k},\vec{G}}=\delta_{\vec{G},0}\mathbbm{1}$, i.e., the graphene form factors obey the flat-metric condition (\ref{FlatMetricCondition}) exactly (with $f(\vec{G})=\delta_{\vec{G},0}$) in the limit $D_0=0$, we have
\begin{equation}
    \delta\rho^{\text{g}}_{\vec{G}} = N\delta_{\vec{G},0} \hat{\nu}_g, \qquad \hat{\nu}_g = \frac{1}{N}\sum_{\vec{k}}\sum_\mu \left( g^\dagger_{\vec{k};\mu}g^\pdagger_{\vec{k};\mu} - \frac{1}{2}\right)
\end{equation}
and
\begin{equation}
    H^{\text{gb}} = H^{\text{gb},0} + V(0) \hat{\nu}_g, \qquad H^{\text{gb},0} =\frac{1}{N}\sum_{\vec{q}} V(\vec{q}) \delta\rho^{\text{g}}_{-\vec{q}}   \left(\delta\rho^{\text{b}}_{\vec{q}} - \sum_{\vec{G}\in\text{RL}} \delta_{\vec{q},\vec{G}} R_{\vec{G}} \right). \label{RewrittenFormOfHgb}
\end{equation}
Note that the last term proportional to the filling fraction $\hat{\nu}_g$ of the graphene subspace is only a constant in the low-energy subspace of the system since the TBG ground states in \equref{GeneralFormOfGroundState} are known to be insulators and, hence, exhibit a charge gap (in fact, also the two-particle spectrum was shown to be gapped \cite{2020arXiv200914200B}; more generally, this is also expected based on the experimental absence of superconductivity at integer filling in TBG in the vicinity of the magic angle, which implies that potential additional two-particle bound states cannot occur below the gapped particle-hole continuum).

Using the form of the coupling in \equref{RewrittenFormOfHgb} and recalling that the exact groundstates of the TBG sector obey \equref{PropertyOfEigenstates}, we immediately see that
\begin{equation}
    H^{\text{gb},0} \ket{\Psi^{\text{g}}_{j}}\ket{\Psi_0^{\text{b}}(\nu_b)} = 0
\end{equation}
for both the graphene ground state, $j=0$, and its excited states, $j>0$, and all exact TBG groundstates discussed above. As such, all of these states remain \textit{exact} eigenstates for arbitrary $\lambda$. 

Next, we address the question whether they remain the ground states for a finite range of $\lambda >0$. First, as we have already discussed, the TBG sector has a finite charge gap and, hence, states which involve wavefunctions where one (or more) electrons of the TBG sector are transferred to the graphene sector can only become ground states at sufficiently large $\lambda$.
Second, since the TBG ground states break symmetries, there are gapless Goldstone modes which require a more careful analysis. This is presented in the next subsection.

\subsection{Behavior of Goldstone modes}\label{GoldstoneModes}
As follows from $C_{3z}$ rotational symmetry (and has been demonstrated in \refcite{2020arXiv200914200B}), the Goldstone spectrum of TBG must be of the form $E_{\text{G}}(\vec{p}) = \frac{1}{2}\rho\, \vec{p}^2 + \mathcal{O}(\vec{p}^4)$, with (isotropic) stiffness $\rho>0$.
Due to the arbitrarily small energies of the TBG Goldstone modes for $\vec{p}\rightarrow 0$ at $\lambda=0$, we have to make sure that non-zero $\lambda$ will not immediately lower the energy of these states below that of $\ket{\Psi^{\text{g}}_{0}}\ket{\Psi_0^{\text{b}}(\nu_b)}$ in MSTG. For that reason, we have to compute the evolution of the $\vec{p}\rightarrow 0$ Goldstone modes of the TBG sector upon turning on $\lambda$. Using momentum $\vec{p}$ to label those states, we have to compute the behavior of their energy $\Delta E(\lambda,\vec{p})$ (relative to $\ket{\Psi^{\text{g}}_{0}}\ket{\Psi_0^{\text{b}}(\nu_b)}$). Assuming that $\Delta E(\lambda,\vec{p})$ is analytic in $\vec{p}$, $C_{3z}$ again implies that
\begin{equation}
   \Delta E(\lambda,\vec{p}) = \frac{1}{2}\rho\, \vec{p}^2 + m(\lambda) + \delta\rho(\lambda)\vec{p}^2 + \mathcal{O}(\vec{p}^4) \label{ExpansionOfGoldstoneSpec}
\end{equation}
and our goal will be to determine $m(\lambda)$ and $\delta\rho(\lambda)$.

To get started, let us investigate the wavefunctions of the Goldstone modes of TBG for $\lambda=0$. Fortunately, these can be computed exactly in the chiral-flat limit \cite{2020arXiv200914200B}, as the property in \equref{PropertyOfEigenstates} reduces their computation to a one-particle problem. The mathematical reason for this is that $[H^{\text{b}},\tilde{b}^\dagger_{\vec{k}+\vec{p};c_2,\eta_2,s_2}\tilde{b}^\pdagger_{\vec{k};c_1,\eta_1,s_1}]\ket{\Psi_0^{\text{b}}(\nu_b)}$ is a superposition of states of the form $\tilde{b}^\dagger_{\vec{k}'+\vec{p};c_a,\eta_a,s_a}\tilde{b}^\pdagger_{\vec{k}';c_b,\eta_b,s_b}\ket{\Psi_0^{\text{b}}(\nu_b)}$. Further noting that the density operator in \equref{DensityOperatorChiralLimit} does not scatter states between different quantum numbers $c$, $\eta$, $s$ and is independent of $\eta$, $s$, we see that the Goldstone modes with momentum $\vec{p}$ must be of the form
\begin{equation}
    \ket{\Psi_{\vec{p}}^{\text{b}}(\nu_b;c_1,\eta_1,s_1;c_2,\eta_2,s_2)} = \sum_{\vec{k}} \alpha_{\vec{p},c_1,c_2}(\vec{k}) \tilde{b}^\dagger_{\vec{k}+\vec{p};c_2,\eta_2,s_2}\tilde{b}^\pdagger_{\vec{k};c_1,\eta_1,s_1} \ket{\Psi_0^{\text{b}}(\nu_b)}, \label{GeneralFormOfGoldstoneModes}
\end{equation}
where the $(c_1,\eta_1,s_1)$ flavor is occupied and $(c_2,\eta_2,s_2)$ is unoccupied in the specific groundstate $\ket{\Psi_0^{\text{b}}(\nu_b)}$ we start from.
In the limit $\vec{p}\rightarrow 0$, \refcite{2020arXiv200914200B} has shown that one Goldstone zero mode has $c_1=c_2$ and $\alpha_{\vec{p}=0,c,c}(\vec{k})=\alpha$, i.e.,
\begin{equation}
    \ket{\Psi_{\vec{p}\rightarrow 0}^{\text{b}}(\nu_b;c,\eta_1,s_1;c,\eta_2,s_2)} = \frac{1}{\sqrt{N}}\sum_{\vec{k}} \tilde{b}^\dagger_{\vec{k};c,\eta_2,s_2}\tilde{b}^\pdagger_{\vec{k};c,\eta_1,s_1} \ket{\Psi_0^{\text{b}}(\nu_b)}.
\end{equation}
All other zero-energy Goldstone modes follow from $(\text{U}(4)\times \text{U}(4))_{\text{b,cf}}$ rotations.

It is a matter of straightforward algebra to show that
\begin{equation}
    \delta\rho^{\text{b}}_{\vec{q}}\ket{\Psi_{\vec{p}\rightarrow 0}^{\text{b}}(\nu_b;c,\eta_1,s_1;c,\eta_2,s_2)}  =  \sum_{\vec{G}\in\text{RL}} \delta_{\vec{q},\vec{G}} R_{\vec{G}} \ket{\Psi_{\vec{p}\rightarrow 0}^{\text{b}}(\nu_b;c,\eta_1,s_1;c,\eta_2,s_2)}, \quad R_{\vec{G}} = \nu_b \sum_{\vec{k}} F_1^{bb}(\vec{k},\vec{G}) \in \mathbbm{R}. \label{PropertyOfGoldstonemodes}
\end{equation}
Since $\delta\rho^{\text{b}}_{\vec{q}}$ is invariant under $(\text{U}(4)\times \text{U}(4))_{\text{b,cf}}$ rotations, this also holds for all other zero-energy Goldstone modes. In fact, this behavior can be understood more generally and intuitively by rewriting the $H_b$ in \equref{HBilayer} (recall we always focus on the flat-band limit, $W_{\text{TBG}}=0$) as
\begin{align}\begin{split}
    H^{\text{b}} &=  \frac{1}{2N}\sum_{\vec{q}} V(\vec{q})  \left(\delta\rho^{\text{b}}_{\vec{q}} - \sum_{\vec{G}\in\text{RL}}\delta_{\vec{q},\vec{G}} R_{\vec{G}}\right)  \left(\delta\rho^{\text{b}}_{-\vec{q}} - \sum_{\vec{G}\in\text{RL}}\delta_{-\vec{q},\vec{G}} R_{\vec{G}} \right) \\ &\qquad + \frac{1}{N}\sum_{\vec{G}\in\text{RL}} V(\vec{G}) R_{\vec{G}}\delta\rho^{\text{b}}_{\vec{G}} - \frac{1}{2N} \sum_{\vec{G}\in\text{RL}} V(\vec{G}) R_{\vec{G}}R_{-\vec{G}}. \label{RewritingTheHbTerm} \end{split}
\end{align}
While the last term in \equref{RewritingTheHbTerm} is just a constant, it holds
\begin{equation}
    \delta\rho^{\text{b}}_{\vec{G}} = N f(\vec{G}) \hat{\nu}_b, \qquad \hat{\nu}_b = \frac{1}{N}\sum_{\vec{k}}\sum_\mu \left( b^\dagger_{\vec{k};\mu}b^\pdagger_{\vec{k};\mu} - \frac{1}{2}\right),
\end{equation}
if the flat-metric condition (\ref{FlatMetricCondition}) applies. Since $R_{\vec{G}}=R_{-\vec{G}}$, following from \equref{GToMinusG}, and as $\nu_b$ is constant in the low-energy subspace of TBG, we can write in this subspace \cite{PhysRevLett.122.246401}
\begin{equation}
    H^{\text{b}} =  \frac{1}{2N}\sum_{\vec{q}} \mathcal{O}^\dagger_{\vec{q}} \mathcal{O}^\pdagger_{\vec{q}} + \text{const.}, \qquad \mathcal{O}_{\vec{q}} := \sqrt{V(\vec{q})} \left( \delta\rho^{\text{b}}_{\vec{q}} - \sum_{\vec{G}\in\text{RL}}\delta_{\vec{q},\vec{G}} R_{\vec{G}} \right).  
\end{equation}
Since all exact TBG groundstates discussed above obey \equref{PropertyOfEigenstates}, or $\mathcal{O}_{\vec{q}} \ket{\Psi_0^{\text{b}}(\nu_b)} = 0$, all Goldstone modes with vanishing momentum, $\vec{p}\rightarrow 0$, must be annihilated by all $\mathcal{O}_{\vec{q}}$ as their energy must approach that of the groundstates. This is why \equref{PropertyOfGoldstonemodes} must hold. Note that, for $\nu_b=0$, we did not have to assume the flat-metric condition since $R_{\vec{G}}=0$.

Having established \equref{PropertyOfGoldstonemodes}, we also immediately see that 
\begin{equation}
    H^{\text{gb},0} \ket{\Psi^{\text{g}}_{j}}\ket{\Psi_{\vec{p}\rightarrow 0}^{\text{b}}(\nu_b;c_1,\eta_1,s_1;c_2,\eta_2,s_2)}  = 0.
\end{equation}
This shows that also $\ket{\Psi^{\text{g}}_{j}}\ket{\Psi_{\vec{p}\rightarrow 0}^{\text{b}}(\nu_b;c_1,\eta_1,s_1;c_2,\eta_2,s_2)}$ remain exact eigenstates of the Hamiltonian for any $\lambda\neq 0$ and that $m(\lambda) = 0$ in \equref{ExpansionOfGoldstoneSpec}. Making the natural assumption that $\delta\rho(\lambda)$ is analytic in $\lambda$, we can expand
\begin{equation}
    \delta\rho(\lambda) \sim \sum_{n \geq 0} c^{\rho}_n \lambda^n.
\end{equation}
Therefore, it will require a finite value of $\lambda$ before $\Delta E(\lambda,\vec{p})$ can lose its positive semi-definite nature and the groundstates discussed above have to remain the groundstates in a finite region of $\lambda>0$. 

As we discuss in \secref{SummaryOutlook} of the main text, one possibility is that at sufficiently large $\lambda$ a state with finite momentum $\vec{p}$ resulting from a mixing between the graphene and TBG degrees of freedom becomes the ground state. We leave a quantitative analysis of such a MDW state \cite{Christos_2020} for future work.

\section{Hartree-Fock functional}
\label{app:HartreeFock}

\subsection{General form}
Let us assume that the ground state is the Slater determinant $\ket{\psi[P_{\vec{k}}]}$ characterized by the correlator 
\begin{equation}
    (P_{\vec{k}})_{\alpha,\alpha'} = \braket{\psi[P_{\vec{k}}]|f^\dagger_{\vec{k};\alpha}f^\pdagger_{\vec{k};\alpha'}|\psi[P_{\vec{k}}]}, \qquad P_{\vec{k}}^\dagger = P^\pdagger_{\vec{k}} = (P^\pdagger_{\vec{k}})^2, \label{CorrelationMatrixP}
\end{equation} 
where we use the multi-index $\alpha=((t,p),\eta,s)$ to keep the expressions more compact. The HF mean-field Hamiltonian $H^{\text{MF}}$ associated with $H_0 + H_1$, defined in \equsref{NonInteractingHam}{ProjectedInteraction} contains four terms,
\begin{equation}
    H^{\text{MF}}[P_{\vec{k}}] = H^{\text{MF}}_k + H^{\text{MF}}_{\text{H}}[P_{\vec{k}}] + H^{\text{MF}}_{\text{F}}[P_{\vec{k}}] + E_0[P_{\vec{k}}]. \label{FormOfHFHamiltonian}
\end{equation}
Postponing the definition of the energetic off-set, $E_0$, which is proportional to the identity operator, these terms are the kinetic term
\begin{equation}
H^{\text{MF}}_k = H_0 =   \sum_{\vec{k}\in\text{MBZ}} f^\dagger_{\vec{k};\alpha} h^{k}_{\alpha,\alpha'}(\vec{k})f^\pdagger_{\vec{k};\alpha'} \equiv  \sum_{\vec{k}\in\text{MBZ}} f^\dagger_{\vec{k}} h^{k}(\vec{k})f^\pdagger_{\vec{k}}, \label{KineticPartOfMFHam}
\end{equation}
where $h^{k}$ is the diagonal matrix
\begin{equation}
    h^{k}_{((t,p),\eta,s),((t',p'),\eta',s')}(\vec{k}) = \delta_{t,t'}\delta_{p,p'}\delta_{\eta,\eta'}\delta_{s,s'} \epsilon_{(t,p),\eta}(\vec{k}) 
\end{equation}
with $\epsilon_{(t,p),\eta}$ given in \equsref{TBGLikeEnergy}{GrapheneLikeEnergy}, the Hartree term,
\begin{equation}
    H^{\text{MF}}_{\text{H}}[P_{\vec{k}}] = \frac{1}{N} \sum_{\vec{G}\in\text{RL}} V(\vec{G}) \Gamma_{\vec{G}}  \sum_{\vec{k}\in\text{MBZ}} f_{\vec{k}}^\dagger F_{\vec{k},\vec{G}} f_{\vec{k}}^\pdagger, \qquad \Gamma_{\vec{G}} = \sum_{\vec{k}\in\text{MBZ}} \text{tr} \left[ P_{\vec{k}} F^*_{\vec{k},\vec{G}}\right], 
\end{equation}
and the Fock contribution
\begin{equation}
    H^{\text{MF}}_{\text{F}}[P_{\vec{k}}] = -\frac{1}{N} \sum_{\vec{q}} V(\vec{q})\sum_{\vec{k}\in\text{MBZ}} f_{\vec{k}}^\dagger F^\dagger_{\vec{k},\vec{q}} P^T_{\text{MBZ}(\vec{k}+\vec{q})} F_{\vec{k},\vec{q}} f_{\vec{k}}^\pdagger.
\end{equation}

The associated ground state energy can be decomposed into three contributions,
\begin{subequations}\begin{equation}
    E^{\text{HF}}[P_{\vec{k}}] = \braket{\psi[P_{\vec{k}}]|(H_0 + H_1)|\psi[P_{\vec{k}}]} = E_{\text{k}}[P_{\vec{k}}] + E_{\text{H}}[P_{\vec{k}}] + E_{\text{F}}[P_{\vec{k}}],
\end{equation}
which read as
\begin{align}
    E_{\text{k}}[P_{\vec{k}}] & = \sum_{\vec{k}\in\text{MBZ}} \text{tr}\left[ P^T_{\vec{k}} h^k(\vec{k}) \right],\\
    E_{\text{H}}[P_{\vec{k}}] & = \frac{1}{2N} \sum_{\vec{G}\in\text{RL}} V(\vec{G})  \left| \sum_{\vec{k}\in\text{MBZ}} \text{tr}\left[P_{\vec{k}} F^T_{\vec{k},\vec{G}} \right]  \right|^2, \label{HartreeEnergy} \\
    E_{\text{F}}[P_{\vec{k}}] & = -\frac{1}{2N} \sum_{\vec{q}} V(\vec{q})\sum_{\vec{k}\in\text{MBZ}} \text{tr}\left[ P_{\vec{k}} F^T_{\vec{k},\vec{q}} P_{\text{MBZ}(\vec{k}+\vec{q})} F^*_{\vec{k},\vec{q}} \right].\label{FockEnergyOnly}
\end{align}\label{HFEnergy}\end{subequations}
Having established this notation, the energetic off-set $E_0$ in \equref{FormOfHFHamiltonian} can now be conveniently stated as $E_0 = -E_{\text{H}} - E_{\text{F}}$.

As has been discussed in previous works on TBG \cite{bultinck2019ground,Liu_2021,Xie_2020,Liao_2021}, it is important to note that the continuum model defined in \secref{ContinuumModel} already takes into account some correlation effects of the system. So using the continuum model dispersion in $h^{k}(\vec{k})$ in the HF mean-field Hamiltonian as defined in \equref{FormOfHFHamiltonian} would constitute a double counting of these effects. To avoid this, we replace (ignoring the irrelevant constant $E_0[P_{\vec{k}}^0])$
\begin{equation}
    H^{\text{MF}}_k[P_{\vec{k}}] \quad \longrightarrow \quad H^{\text{MF}}_k[P_{\vec{k}}] - \left(H^{\text{MF}}_{\text{H}}[P^0_{\vec{k}}] + H^{\text{MF}}_{\text{F}}[P^0_{\vec{k}}] \right) \label{ShiftOfHFHamiltonian}
\end{equation}
in \equref{FormOfHFHamiltonian}, where $P_{\vec{k}}^0$ is a reference density matrix for which the continuum model dispersion is expected to be valid. The redefinition in \equref{ShiftOfHFHamiltonian} is equivalent to replacing
\begin{equation}
    h^{k}(\vec{k}) \quad \longrightarrow \quad \widetilde{h}^{k}(\vec{k}) = h^{k}(\vec{k}) - h_{\text{HF}}[P_{\vec{k}}^0](\vec{k}), \label{SubtractionPoint}
\end{equation}
in \equref{KineticPartOfMFHam}, where we defined $h_{\text{HF}}[P_{\vec{k}}]$ such that $H^{\text{MF}}_{\text{H}}[P_{\vec{k}}] + H^{\text{MF}}_{\text{F}}[P_{\vec{k}}] = \sum_{\vec{k}} f_{\vec{k}}^\dagger h_{\text{HF}}[P_{\vec{k}}](\vec{k})f_{\vec{k}}^\pdagger$.

To connect to our analytical discussion and with other work on TBG \cite{bernevig2020tbg,2020arXiv200913530L,2020arXiv200914200B}, we will use a subtraction point such that the starting Hamiltonian of our HF analysis is equal to (modulo a constant) the non-normal ordered but manifestly particle-hole symmetric Hamiltonian $H = H_0 + \widetilde{H}_1$ with $\widetilde{H}_1$ given in \equref{TheInteractionDeltaRho}.
It is a matter of straightforward algebra to show that this corresponds to using $P_{\vec{k}}^0 = \mathbbm{1}/2$ in \equref{SubtractionPoint}. 

\subsection{Numerical procedure}\label{Numerics}

\subsubsection{How we fix the phases in numerics}
In this appendix, we describe how we fix the phases of the wavefunctions in the TBG sector in our numerics, using symmetries. As above, we denote the wavefunction in valley $\eta=\pm$, at momentum $\vec{k}$, and of band $n=\pm$ by $u_{n,\eta,\vec{k}}$; these are vectors in layer, sublattice, and $\vec{G}$ space. 

To fix the phase of the wavefunctions (modulo $\pm$) at every $\vec{k}$ point, $\eta$, and $n$, we use that the Hamiltonian commutes with $C_{2z}\Theta$ and enforce that
\begin{equation}
C_{2z}\Theta u_{n,\eta,\vec{k}} = u_{n,\eta,\vec{k}}. \label{C2Theta}
\end{equation}
All that is left to do is remove the relevant parts of the remaining $\vec{k}$, $\eta$, and $n$ dependent sign ambiguity.
As a first step, we fix the relative sign of $u_{n,\eta,\vec{k}}$ and $u_{n,-\eta,-\vec{k}}$ by choosing
\begin{equation}
\Theta u_{n,\eta,\vec{k}} = u_{n,-\eta,-\vec{k}}.
\end{equation}
Because of \equref{C2Theta}, this is equivalent to $C_{2z} u_{n,\eta,\vec{k}} = u_{n,-\eta,-\vec{k}}$.

Next we fix the relative sign of the wavefunctions in the two bands $n=+$ and $n=-$ at each $\vec{k}$ and $\eta$. If we are in the chiral limit and the system has the chiral symmetry $C$ (which anti-commutes with the Hamiltonian at each $\vec{k}$), we can just enforce
\begin{equation}
 C u_{\pm,\eta,\vec{k}} = \pm i \eta \, u_{\mp,\eta,\vec{k}}. \label{ChiralSym}
\end{equation}
Note it is easy to see that the prefactor has to have the form $\pm i \eta$ since $C^2 = \mathbbm{1}$, $\{C_{2z}\Theta,C\}=0$, and $[\Theta,C]=0$. In our numerics, we can readily implement \equref{ChiralSym} by
\begin{equation}
u_{+,\eta,\vec{k}} \,\rightarrow \, u_{+,\eta,\vec{k}}, \qquad u_{-,\eta,\vec{k}} \,\rightarrow \, -i \,\eta \braket{u_{-,\eta,\vec{k}}|C|u_{+,\eta,\vec{k}}} u_{-,\eta,\vec{k}} \label{ChiralImpl}
\end{equation}
at every $\vec{k}$ point.

When $w_0\neq 0$ but not extremely large, we can expect $C u_{\pm,\eta,\vec{k}} \approx \text{const.}\times u_{\mp,\eta,\vec{k}} $. So we replace \equref{ChiralImpl} by the generalized condition
\begin{equation}
u_{+,\eta,\vec{k}} \,\rightarrow \, u_{+,\eta,\vec{k}}, \qquad u_{-,\eta,\vec{k}} \,\rightarrow \,  \eta \, \text{sign}\left[\text{Im} \braket{u_{-,\eta,\vec{k}}|C|u_{+,\eta,\vec{k}}} \right] u_{-,\eta,\vec{k}}. \label{ChiralImplGen}
\end{equation}
Note that \equref{C2Theta} actually implies $ \braket{u_{-,\eta,\vec{k}}|C|u_{+,\eta,\vec{k}}} \in i \mathbbm{R}$ even away from the chiral limit. 

Finally, the only remaining relevant relative sign is that of $u_{n,\eta,\vec{k}}$ and $u_{n,-\eta,\vec{k}}$. We fix this one by making sure that
\begin{equation}
C_{2z}P u_{n,\eta,\vec{k}} = n \, \eta \, u_{-n,-\eta,\vec{k}}. \label{ActionOfC2zP}
\end{equation}

In practice, we make sure that \equref{ActionOfC2zP} holds by adjusting the sign of the wavefunctions of \textit{both} bands in the $\eta=-$ valley according to
\begin{equation}
u_{+,+,\vec{k}} \, \rightarrow \, u_{+,+,\vec{k}} \bra{u_{++,\vec{k}}}C_{2z}P\ket{u_{--,\vec{k}}} \qquad  u_{+,-,\vec{k}} \, \rightarrow \,  -\braket{u_{-,+,\vec{k}}|C_{2z} P|u_{+,-,\vec{k}}} u_{+,-,\vec{k}}, \qquad n=\pm.
\end{equation}

\subsubsection{Iterative procedure}
We will here give the details of how we numerically solve the HF equations in MSTG and find self-consistent solutions $P_{\vec{k}}$ in \equref{CorrelationMatrixP}. Our iterative procedure has the following steps:
\begin{enumerate}
    \item Guess an initial form of the projector $P_{\vec{k}}$. We choose the initial ansatz for $P_{\vec{k}}$ for a given symmetry breaking state to be those given in the band basis in \tableref{BehaviorOfEnergetics}.
    \item $P_{\vec{k}}$ is then substituted into the HF functional expression \eqref{MeanFieldHartreeFock}. The HF functional is then diagonalized at each point via a unitary transformation $U_{\vec{k}}$.
    \item $P_{\vec{k}}$ is then recomputed from the HF functional as $P_{\vec{k}}=U_{\vec{k}}^*D U^T_{\vec{k}}$ where $D$ is the density matrix in the diagonal basis at zero temperature, with 1's on the diagonal elements corresponding to filled bands of the HF Hamiltonian and zeros elsewhere. 
    \item We then check if $P_{\vec{k}}$ has converged from the previous iteration. If it has, the procedure is finished and we have found a self-consistent solution characterized by $P_{\vec{k}}$. If $P_{\vec{k}}$ has not converged, we return to step 2.
\end{enumerate}
Unless otherwise specified, we include 3 shells of moir\'e Brillouin zones in our construction of the trilayer model (37 unit moir\'e unit cells total) of which we include out to 2 shells of moir\'e Brillouin zones in our numerical calculations and a 10$\times$10 grid of $\vec{k}$-points per moir\'e Brillouin zone (243 $\vec{k}$-points total per moir\'e Brillouin zone). Unless otherwise specified, we take $\epsilon=7$ and screening length $d_{s}=40$ nm.
\subsubsection{Ground State dependence on grid size}
Here, we make note of an effect we observe for smaller grid size for a 6$\times$6 $\vec{k}$ grid (75 $\vec{k}$-points per moir\'e unit cell). We find for the smaller grid, a region of spin or valley polarized ground state emerges for intermediate, nonzero $D_0$ and intermediate to large $w_0$ which became disfavored relative to the IVC$_-$ state as we increased the grid size to better resolve the Dirac cones of the graphene-like bands. We attribute this difference to momentum dependent mixing between remote bands in the Fock term.

\subsection{Varying Parameters}\label{VaryingParameterInNumerics}
\begin{figure}[tb]
    \centering
    \includegraphics[scale=.25]{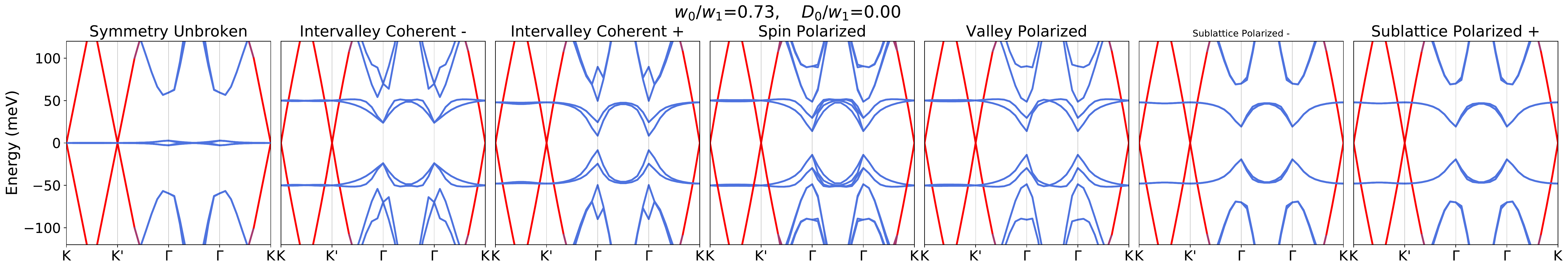}
    \includegraphics[scale=.25]{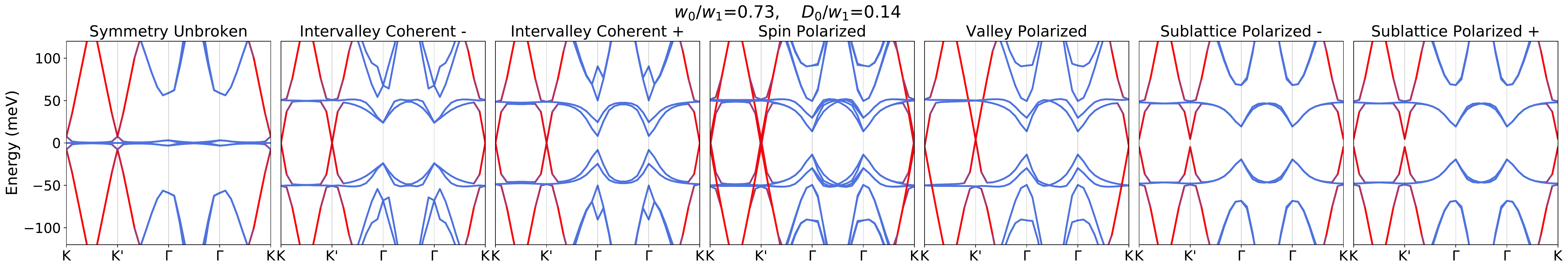}
    \includegraphics[scale=.25]{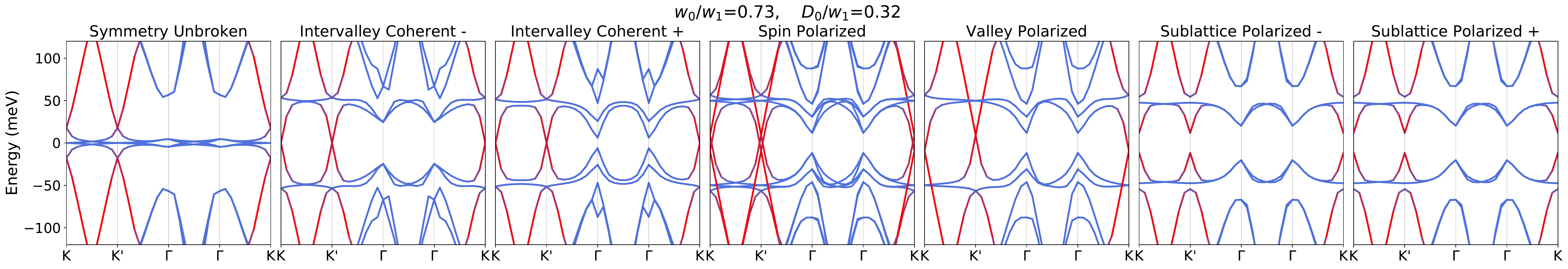}
    \includegraphics[scale=.25]{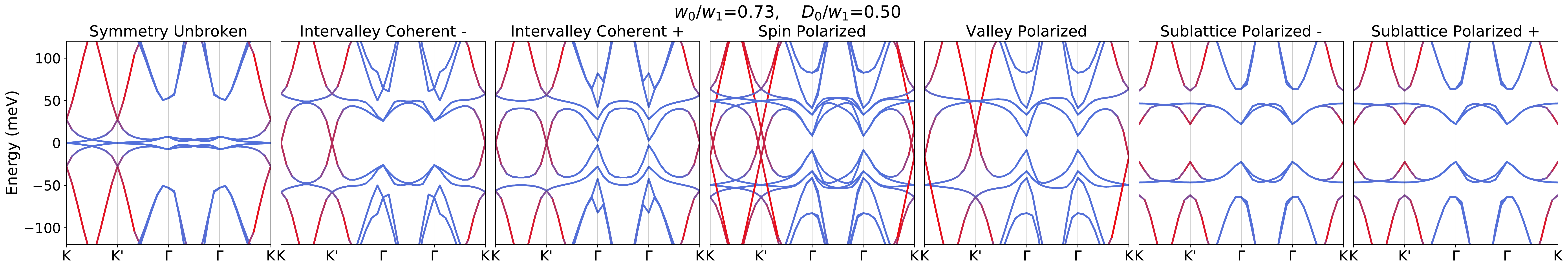}
    \caption{We show band structures of converged solutions at $\nu=0$ corresponding to the ans\"atze in \tableref{BehaviorOfEnergetics}, including those which we do not appear in the phase diagram for dielectric constant $\epsilon=7$ at zero and nonzero $D_0$.}
    \label{AllOrders}
\end{figure}

In \figref{AllOrders}, we show the band structures obtained from our self-consistent calculation for all ans\"atze we attempt, exlcluding the spin Hall state which is quantitatively and qualitatively similar to our SLP$_-$ state which is shown. The contour taken through the moir\'e Brillouin zone to produce the band structures is shown in \figref{TrilayerBands}.

\begin{figure*}[tb]
    \centering
    \includegraphics[scale=.33]{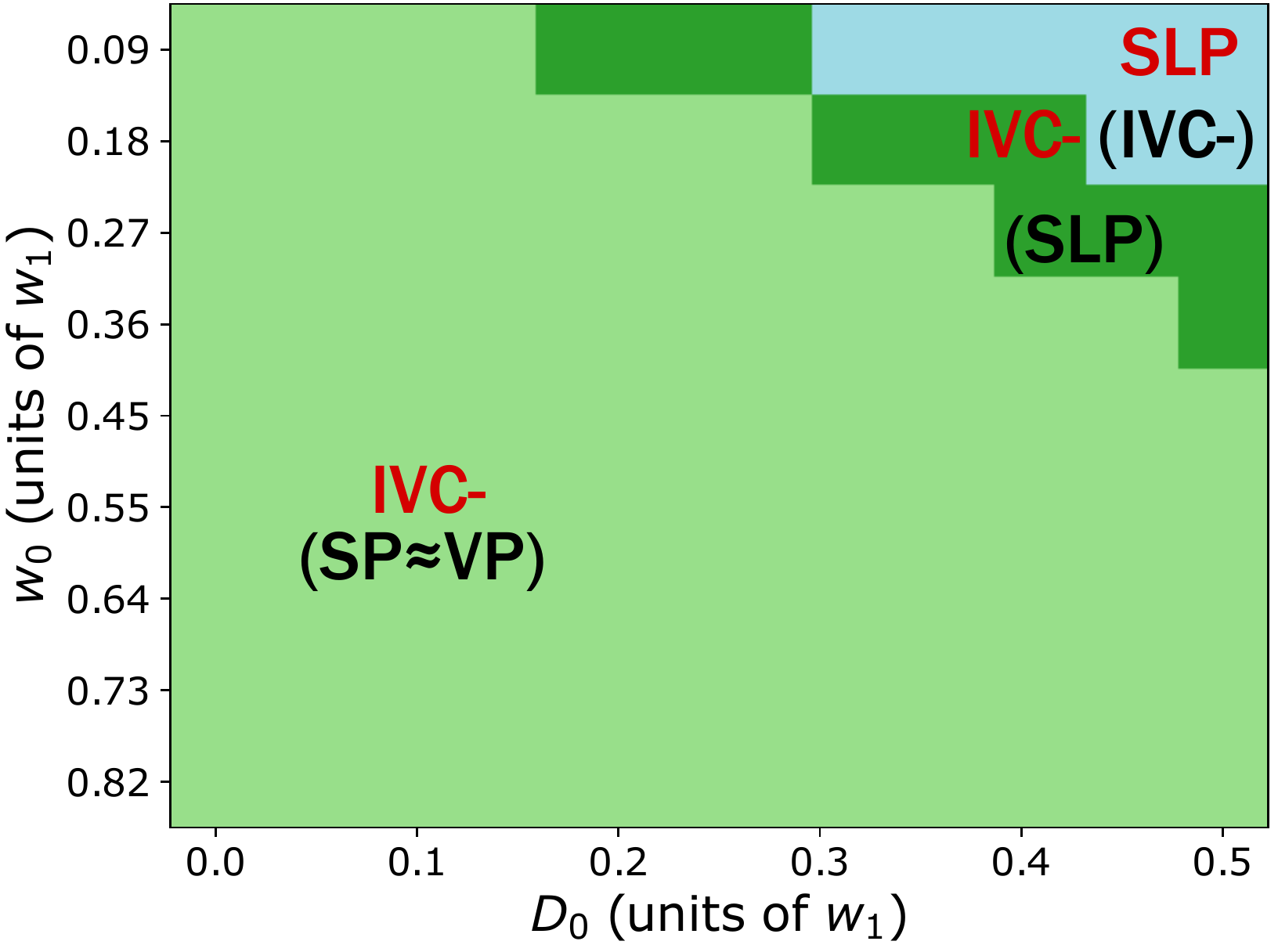}
    \includegraphics[scale=.33]{PhaseDiagram.pdf}
    \includegraphics[scale=.33]{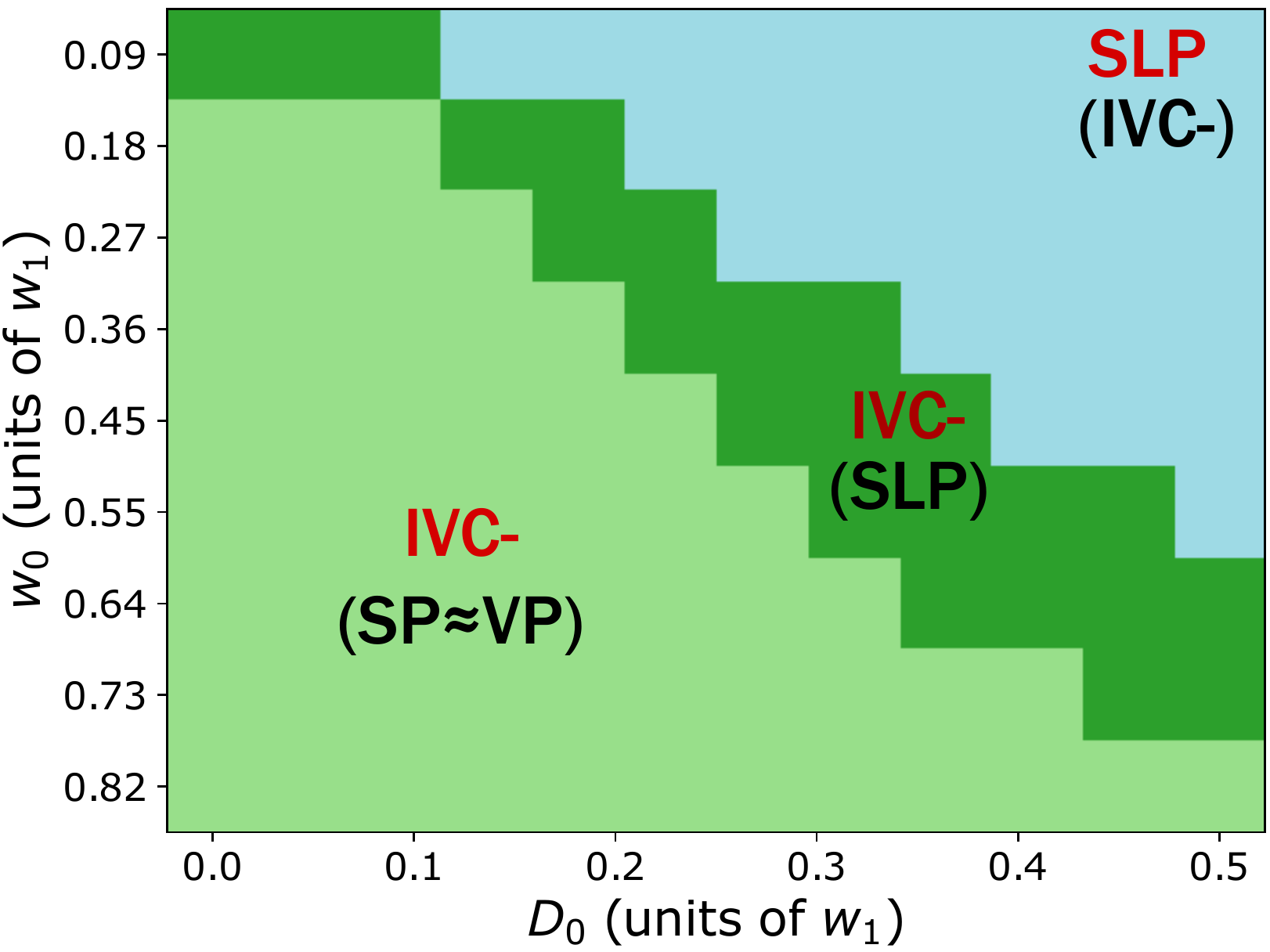}
    \caption{Phases as we vary the dielectric constant in the Coulomb potential in \equref{ExplicitFormOfVOfq} from $\epsilon=4$ (left), $\epsilon=7$ (center), and $\epsilon=15$ (right). As the $\epsilon$ is varied, the phase boundaries shift but the presences of a time-reversal-odd IVC phase and sublattice polarized phase are robust against changes in the potential. As $\epsilon$ increases, the bandwidth of the TBG-like bands relative to the Coulomb potential $V(\vec{q})$ increases; as predicted in \tableref{BehaviorOfEnergetics}, the $D_0$-dependent increases in bandwidth favors a sublattice polarized state over the IVC$_-$.}
    \label{VaryEpsilon}
\end{figure*}

\begin{figure}[tb]
    \centering
    \includegraphics[scale=.28]{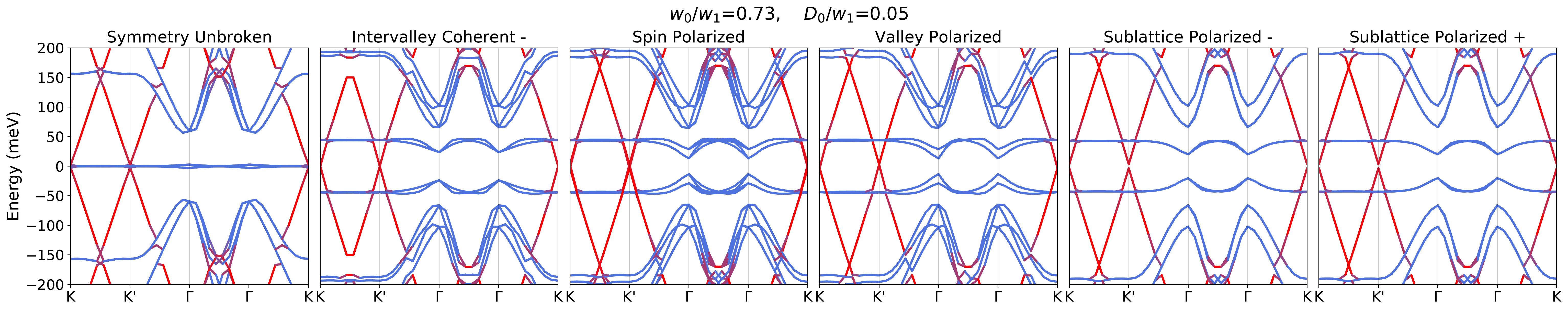}
    \includegraphics[scale=.28]{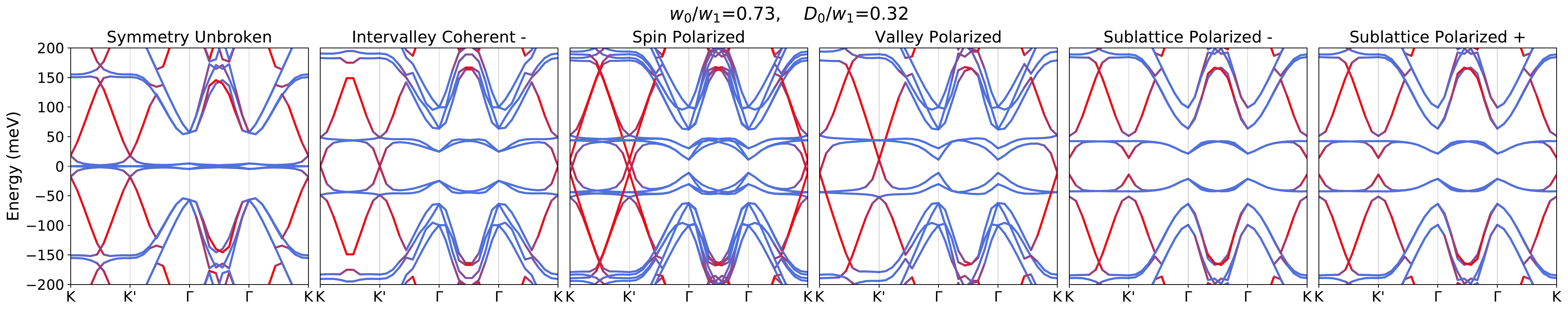}
    \includegraphics[scale=.28]{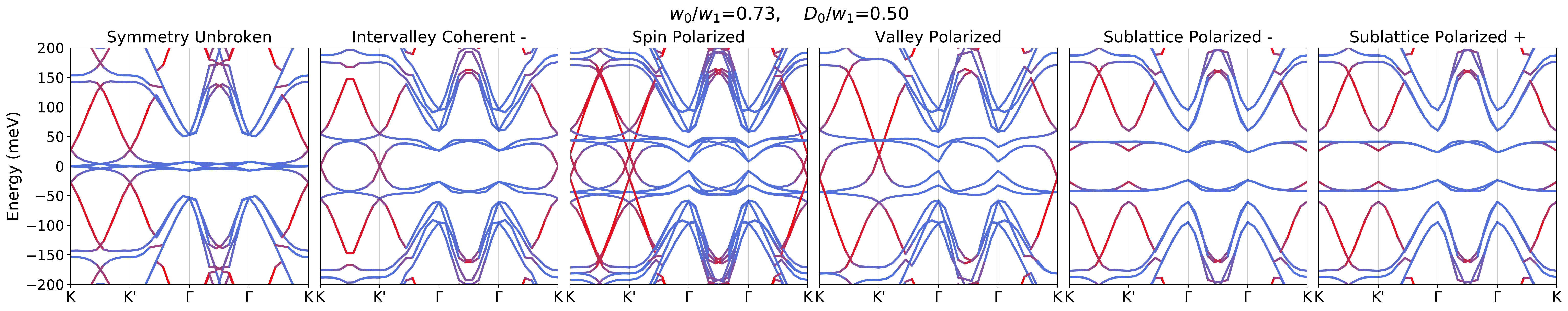}

    \caption{We show band structures for a self-consistent calculation keeping 8 rather than 4 bands per spin-valley flavor, for $\epsilon=7$, $d_s=40$ nm.}
    \label{doubledbandsorders}
\end{figure}

We verify our results are relatively independent  of parameters may vary in experiment. We first consider additional values of the dielectric constant $\epsilon$, which controls the relative contributions to the Hartree-Fock energy of the kinetic and interaction terms. We find the phase boundary between the IVC$_-$ state and sublattice polarized states shifts slightly but the overall features do not depend on the values of $\epsilon$ we test in \figref{VaryEpsilon}.
We also vary the screening length $d_s$ from 10 nm to 80 nm and find while the relative energies shift, the phase diagram does not change when $d_s$ is varied. We also verify our results are stable to increasing the number of bands we keep in our self-consistent calculation, by repeating the same calculation with the closest 8 bands per valley and spin to the Fermi level at charge neutrality. The resulting band structures as $D_0$ is varied are shown in \figref{doubledbandsorders}. We note the IVC$_+$ which we found a converged solution for in the calculation keeping only 4 bands per spin and valley no longer converges when more bands are added and is not shown in \figref{doubledbandsorders}. Given the large energy difference separating the IVC$_+$ from the other lower energy symmetry breaking states, the absence of the IVC$_+$ on doubling the number of bands is not relevant to our main results. We also increase our $\vec{k}$-grid size up to a 14$\times$14 grid to verify convergence.

\begin{figure}[tb]
    \centering
    \includegraphics[scale=.25]{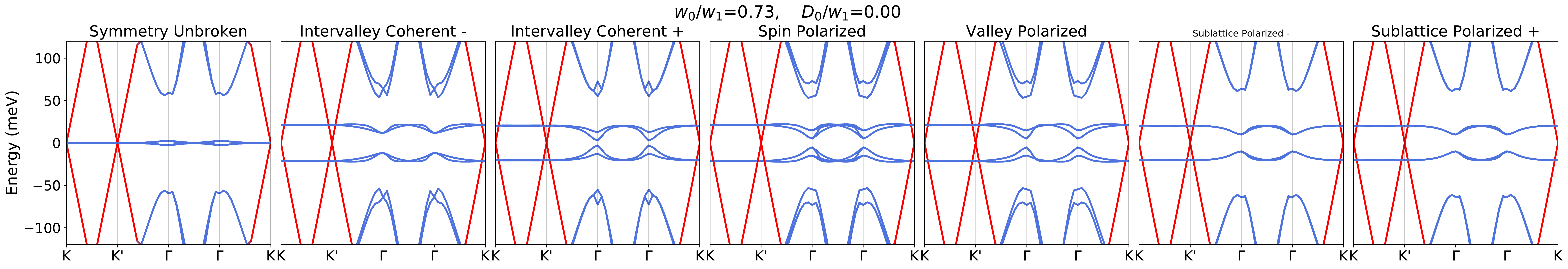}
    \includegraphics[scale=.25]{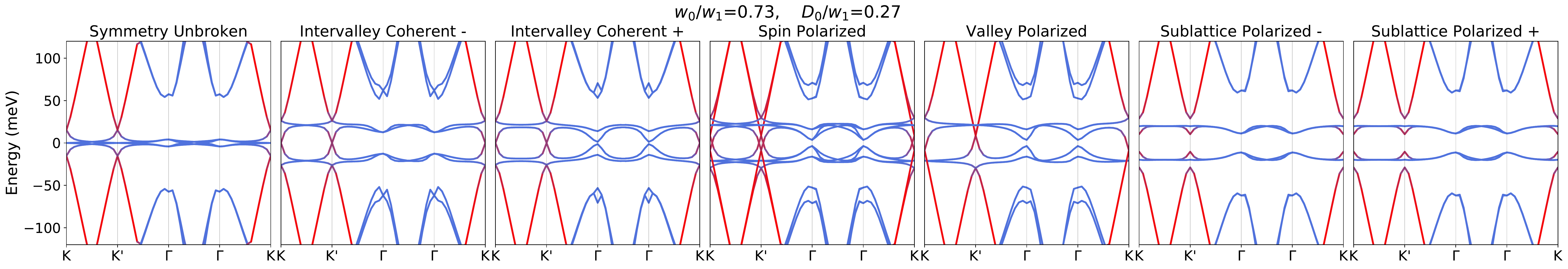}
    \includegraphics[scale=.25]{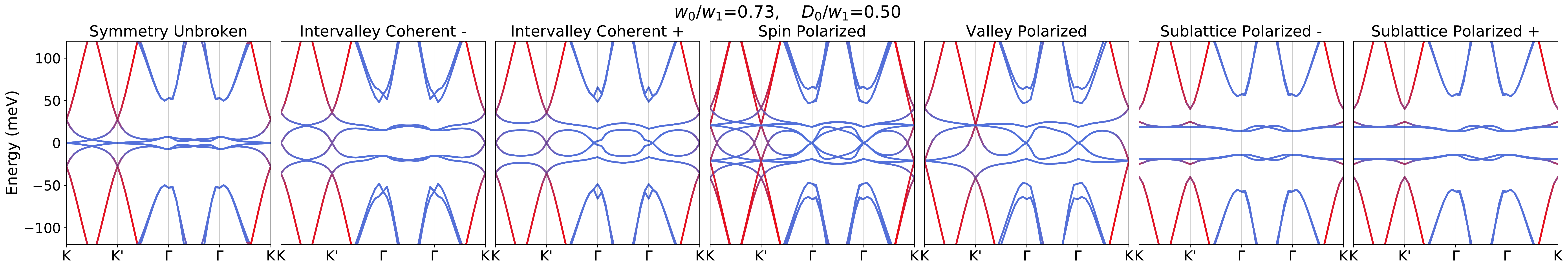}
    \caption{We show band structures for a HF calculation with a 14$\times$14 grid for extreme value $\epsilon=15$, $d_s=40$ nm.}
    \label{epsilon15Bands}
\end{figure}

\section{Numerics at $\nu=2$}\label{NumericsAtNuIsTwo}
\subsection{Stable Phases at $\nu=2$}
\begin{figure}
    \centering
    \includegraphics[scale=.62]{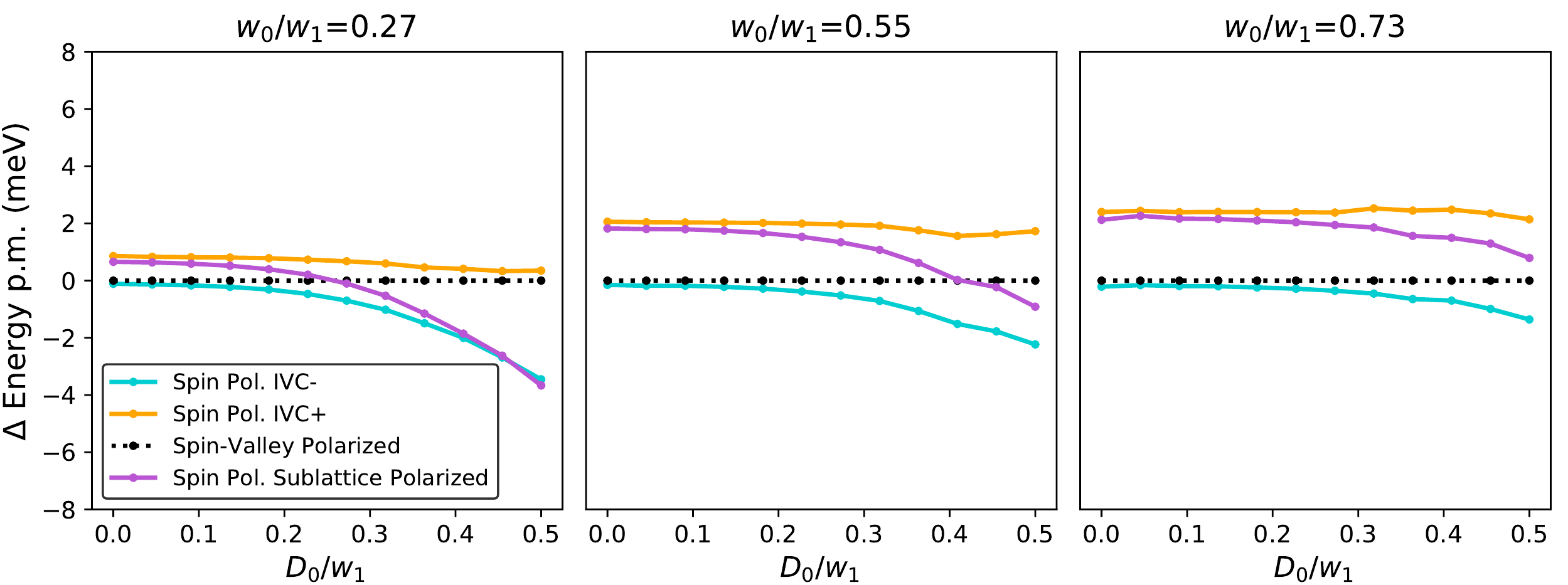}
    \caption{We show the energies resulting from our HF calculations at $\nu=2$. As at $\nu=0$, the spin-polarized IVC$_-$ is dominant relative to a spin-valley polarized state, except for a region of smaller $w_0$ and large $D_0$ where a spin-polarized sublattice polarized state dominates. We note that the spin-valley polarized state is closer in energy to the spin-polarized IVC$_-$ state as we vary $D_0$ in our $\nu=2$ calculation than at $\nu=0$. }
    
\label{Cutsnu2}\end{figure}

\begin{figure}
    \centering
    \includegraphics[scale=.37]{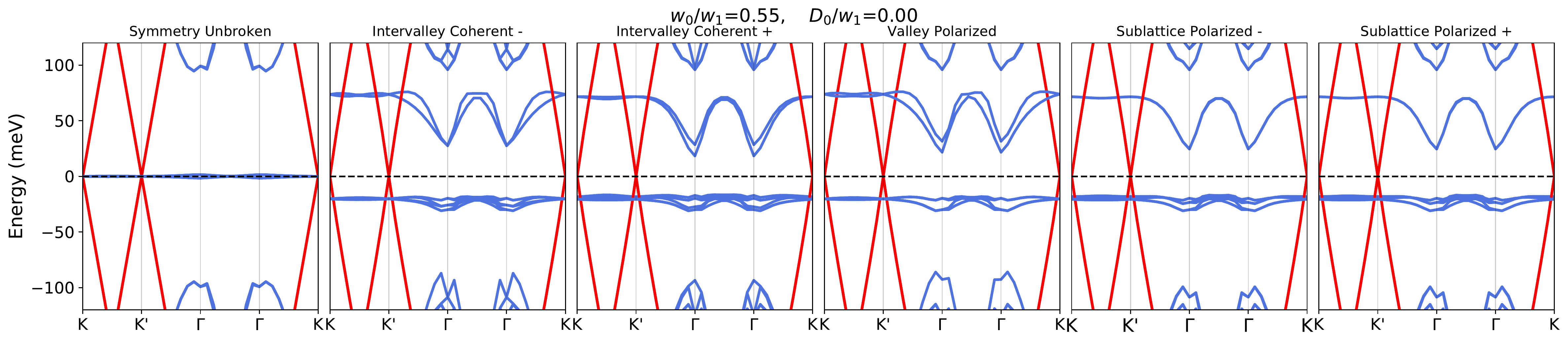}
    \includegraphics[scale=.37]{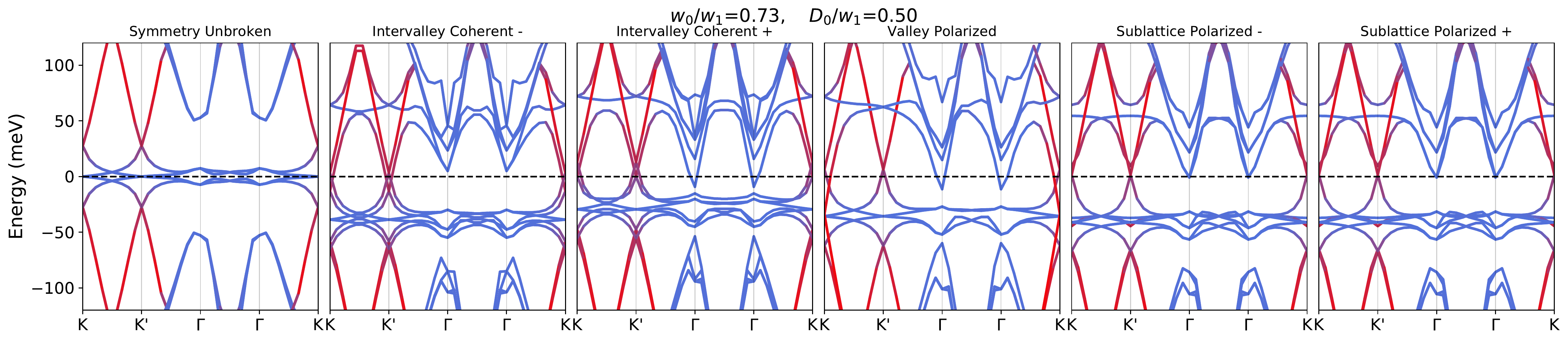}
    \caption{Band structures obtained at $\nu=2$ with $\epsilon=7$, $d_s=40$ nm for the full set of spin-polarized versions of ans\"atze in \tableref{BehaviorOfEnergetics}.}
    
\label{fullnu2bands}\end{figure}

In this appendix we will briefly discuss our numerical results at $\nu=2$. We readjust our numerical procedure such that the chemical potential is recomputed at each iteration to fix the filling at $\nu=+2$ electrons per moir\'e unit cell. In the projected space of the TBG-like and graphene-like bands, this means the trace of $Q_{\vec{k}}$ satisfies $\frac{1}{N}\sum_{\vec{k}}\text{Tr}[Q_{\vec{k}}]=10$. We also adjust the projectors $P$ corresponding to our starting ans\"atze to be compatible with half-filling by setting $P_{b,\nu=2}=\frac{1}{2}(1+s_z)P_{b,\nu=0}$ in the TBG-bands at $\nu=2$. The structure of $P_{g,\nu=2}=\frac{1}{2}(1-\sigma_z)$ is the same as at charge neutrality. The energies resulting from our HF calculations are shown in \figref{Cutsnu2}. We observe no qualitative change from the phases at charge neutrality, but note the spin-polarized IVC$_-$ and spin-valley polarized state are closer in energy at $\nu=2$. We show the band structures obtained for the full set of ans\"atze we check in \figref{fullnu2bands}. We note that where the IVC$_-$ at charge neutrality was a semimetal, at $\nu=2$ the spin polarized IVC$_-$ is a metal with a small Fermi surface. We also note that the spin polarized versions of the sublattice polarized states now are either a metal or semimetal. The change of the SLP states to semimetals at $|\nu|=2$ can be understood as state with an insulating SLP order in the TBG-like bands of one spin flavor and an order which fills both TBG-like bands in the other spin flavor, resulting in a Dirac crossing where the TBG-like and graphene-like bands connect at the Fermi level. 
\subsection{Connection to Experimental Phase Diagram}
We can connect the band structures at $\nu=0$ and $\nu=2$ to the experimental phase diagram; at $\nu=0$, the leading instability is either a semimetallic IVC$_-$ state or an insulating sublattice polarized state. Either of these states can be related to the region of high resisitivity at $\nu=0$ which persists both at $D_0=0$ and $|D_0|>0$. At $\nu=2$, the spin polarized IVC$_-$ is still the leading phase at $D_0=0$ and for a finite range of $D_0$. However, unlike at $\nu=0$, the IVC$_-$ is metallic at $\nu=2$. We note the lack of a strong high-resistivity state at $\nu=2$ when $D_0=0$ in the experimental phase diagram. For large enough $D_0$, the IVC$_-$ state transitions to a semimetallic spin-polarized SLP state. This transition may be related to the appearance of a state with higher resisitivity at $\nu=2$ for a finite value of $D_0$ in the experimental phase diagram.
\section{Band Structure and solution for $D_0\sim w_1$}\label{LargerValuesOfD}
In \secref{HFNumericsSection}, we studied phases varying $D_0$ up to half the value of $w_1$. We will here discuss the behavior of the phases we find varying $D_0$ up to and past $w_1$. We find our numerics begin not to converge for $D_0/w_1>1$ for some values of $w_0$ we study. We
will focus here on the behavior of the valley polarized state at $\nu=0$ and the spin polarized version of this state at $\nu=2$ but note the similarities between the behavior of this state and the sublattice polarized states and IVC states where they converge in the HF numerics.

At $\nu=0$, as stated in the main text, the valley polarized solution becomes metallic as $D_0$ increases and the bands in either minivalley are pushed away from the Fermi level. For $D_0/w_1$ relatively small, the bands in each valley flavor cross the Fermi level only near the K and K' points. However as $D_0/w_1\rightarrow 1$, the bandwidth of the valley-polarized TBG-like bands increases, resulting in additional Fermi surfaces near the $\Gamma$ point. For large $D_0$, the lower TBG-like band which begins above the Fermi level at small $D_0$ is pushed completely below the Fermi level and, for large enough $D_0$, the band structure reverts to a filled lower bands solution. 

\begin{figure}
    \centering
    \includegraphics[scale=.41]{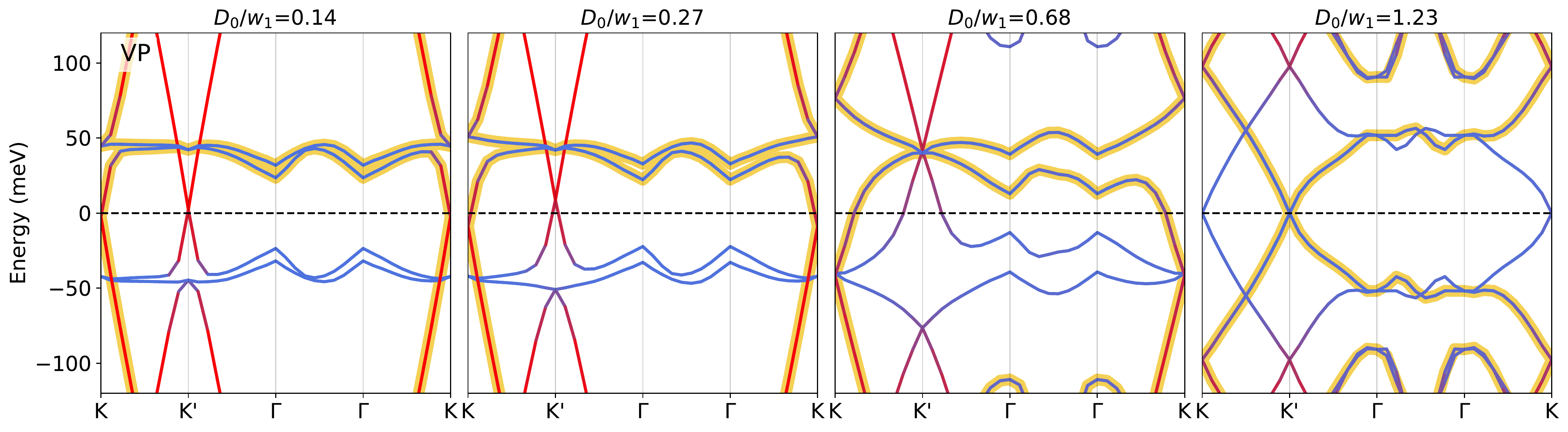}
    \caption{We show the band structure of the VP state at $\nu=0$ as we vary $D_0$ for $w_0/w_1=0.55$. At charge neutrality, the VP state at $D_0$ comparable to $w_1$ acquires additional hole-like Fermi surfaces. We hi-light a single valley-spin flavor band in yellow to show the evolution of a single flavor at large $D_0$.}
    \label{nu0FermiSurfaces}
\end{figure}

\begin{figure}
    \centering
    \includegraphics[scale=.41]{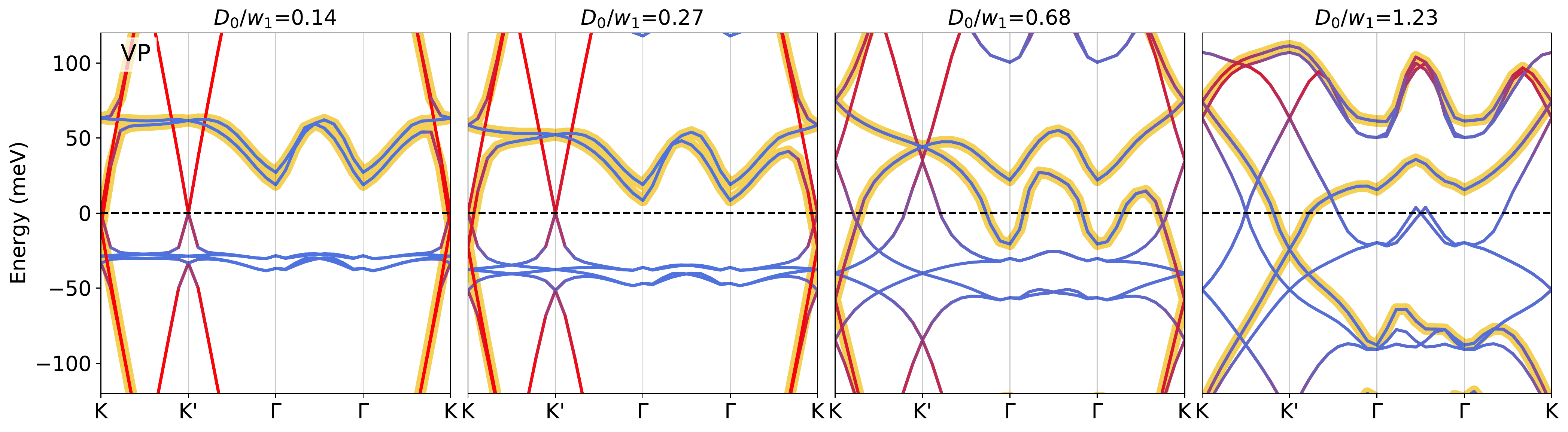}
    \caption{We show the band structure of the spin-VP phase at $\nu=2$ as we vary $D_0$ for $w_0/w_1=0.33$. We note at half-filling both the spin-valley polarized and intervalley coherent states at $D_0$ comparable to $w_1$ acquire additional hole-like Fermi surfaces. We hi-light a single valley-spin flavor band in yellow to show where the additional Fermi surfaces appear as a function of $D_0$.}
    \label{nu2FermiSurfaces}
\end{figure}

At $\nu=2$, the band structure of the solutions we obtain via HF for the spin-valley polarized ansatz also begin with separated Fermi-surfaces around the K and K' points and no Fermi surfaces near $\Gamma$. Similar to the $\nu=0$ case, as $D_0$ increases, the bandwidth of the polarized TBG-like bands increases, leading to changes in the Fermi surface, with additional hole-like Fermi surfaces appearing near the $\Gamma$ point at intermediate values of $D_0$ until the lower polarized band which begins above the Fermi level is pushed below the Fermi level for large $D_0$. Unlike at $\nu=0$, the solution at large $D_0$ retains a finite valley polarization. Experimental Hall densities measured in MSTG show a change in sign of the Hall density at half filling for large $D_0$ at $|\nu|=2$. In the main text we argued this effect could be accounted for by the merging of Fermi surfaces of the IVC$_-$ bands (or a similar transition of the SLP$_-$ phase) near the K and K' point doping slightly away from $\nu=2$ when $D_0>0$. Should the experimental value of the displacement field at which the sign change appears exceed the values of the displacement field discussed in the main text, it is possible that instead the sign change could be related to the dramatic changes in Fermi surfaces we observe and discuss here. We show how the band structure of the valley polarized state evolves at large $D_0$ in \figref{nu0FermiSurfaces} for $\nu=0$ and in \figref{nu2FermiSurfaces} for $|\nu|=2$. The IVC$_-$ and sublattice polarized states display similar behavior at charge-neutrality and $\nu=2$. 

Another general feature we note is a tendency after a critical value of $D_0/w_1>1$ for all of our orders to converge to solutions which are nearly degenerate in energy. While some retain the symmetry breaking in their initial ansatz, the solutions generally have a preference to fill the lower bands up to the chemical potential at $\nu=0$ and $\nu=2$ and exhibit additional Fermi surfaces at large $D_0$. Our numerics do not always converge in this region, so we do not include it in our phase diagrams or energies which are limited to $D_0/w_1<0.5$.

\section{Different contributions to the energy}
\label{app:Energetics}

In this appendix, we detail how the different energetic contributions for particle-hole instabilities, discussed briefly in \secref{AnalyticalPerturbationTheory} of the main text,  are derived and what their analytical form is. We use these expressions to compute their respective scaling with system parameters, summarized in \tableref{BehaviorOfEnergetics}. 

Our starting point are Slater-determinant states with correlation matrix $P^0_{\vec{k}} = (\mathbbm{1} + Q^0_{\vec{k}})/2$, see \equref{DefinitionOfCorrelator} for its definition, where $Q^0_{\vec{k}}$ is block diagonal in the graphene-like-TBG-like space and of the form
\begin{equation}
    Q^0_{\vec{k}} = \begin{pmatrix}  Q^{\text{g},0}_{\vec{k}} & 0 \\ 0 & Q^{\text{b},0}_{\vec{k}}   \end{pmatrix}, \qquad Q^{\text{g},0}_{\vec{k}} = -\sigma_z, \label{DiagonalQMatrix}
\end{equation}
i.e., further assume that the graphene-like bands do not develop any order while we take several different candidate orders, $Q^{\text{b},0}_{\vec{k}}$, in the TBG-like band to be discussed shortly.
Equation (\ref{DiagonalQMatrix}) is motivated by the following observations: first, at $D_0=0$, all order parameters can be classified as mirror-even (diagonal in graphene-TBG space) and mirror-odd (off-diagonal in graphene-TBG space). Based on the band structure, it is natural to expect that the former are energetically favored. In fact, we will see in \appref{MixingBetweenBands} explicitly and in our HF numerics [see second row in \figref{MatrixOrders}] as well as in our analysis of exact eigenstates in \appref{ExactStatements} that off-diagonal components are not favored at $D_0=0$. 
Second, $Q^{\text{g},0}_{\vec{k}} = -\sigma_z$ is motivated by the observations of \appref{ExactStatements} that the manifold of exact ground states of the MSTG Hamiltonian has this form for $D_0=0$ in a finite range of coupling parameters [$\lambda$ in \equref{FormOfCouplingBetweenTheTwoSubSys}] between the graphene cones and the TBG bands. Again, this will be confirmed by our HF numerics, where the lowest-energy solutions are of the form of \equref{DiagonalQMatrix} in the flat-decoupled limit ($D_0=W_{\text{TBG}}=0$), and our discussion of order in the graphene-like bands in \appref{PolarizationOfGrapheneLikeBands} below.

In \secref{CandidateStatesConstruction} of the main text, we have constructed the different possible ans\"atze, $Q^{\text{b},0}_{\vec{k}}$, using the $(\text{U}(4)\times \text{U}(4))_{\text{b,cf}}$ symmetry. We will complement this here by deriving the exact same set of states purely within HF.
As a first step, let us focus on minimizing the Fock energy in \equref{FockEnergyOnly} at $D_0=w_0=0$:
\begin{equation}
    E_{\text{F}}[P^0_{\vec{k}}] =  E_{\text{F,g}} -\frac{1}{2N} \sum_{\vec{q}} V(\vec{q})\sum_{\vec{k}\in\text{MBZ}} \text{tr}\left[ P^{\text{b}}_{\vec{k}} (F^{bb}_{\vec{k},\vec{q}})^T P^{\text{b}}_{\text{MBZ}(\vec{k}+\vec{q})} (F^{bb}_{\vec{k},\vec{q}})^* \right], \quad P^{\text{b}}_{\vec{k}} = \frac{1}{2}(\mathbbm{1} + Q^{\text{b},0}_{\vec{k}}), \label{FockEnergyTrialStates}
\end{equation}
where $E_{\text{F,g}}$ is the Fock energy of the graphene subsystem being in a semimetallic state and $F^{bb}$ the form factors in TBG subspace, which obey [cf.~\equref{bbMatrixElements}]
\begin{equation}
    \bar{F}^{bb}(\vec{k},\vec{q}) = \bar{F}^{bb}_1(\vec{k},\vec{q}) \eta_0\sigma_0s_0 + i \bar{F}^{bb}_2(\vec{k},\vec{q}) \eta_0\sigma_2s_0 + \mathcal{O}(w_0,D_0^2).
\end{equation}
Since $\braket{A,B}= \text{tr}[AB]$ defines an inner product on Hermitian matrices, the Cauchy-Schwarz inequality holds, $\braket{A,B} \leq \sqrt{\braket{A,A}} \sqrt{\braket{B,B}}$ with maximum reached when $A = c B$ with $c>0$. Applying this to the second term in \equref{FockEnergyTrialStates}, we see that this term reaches its minimum when \cite{Liu_2021}
\begin{equation}
    P^{\text{b}}_{\vec{k}} = C_{\vec{k},\vec{q}} (F^{bb}_{\vec{k},\vec{q}})^T P^{\text{b}}_{\text{MBZ}(\vec{k}+\vec{q})} (F^{bb}_{\vec{k},\vec{q}})^*, \quad C_{\vec{k},\vec{q}}>0, \qquad \forall \vec{k},\vec{q}. \label{ConditionOnTBG}
\end{equation}
In general, solving \equref{ConditionOnTBG} can be difficult (or impossible) and the minimum of $E_{\text{F}}[P^0_{\vec{k}}]$ requires a momentum-dependent correlator. However, for $D_0=w_0=0$, it holds $(F^{bb}_{\vec{k},\vec{q}})^T (F^{bb}_{\vec{k},\vec{q}})^* = \gamma_{\vec{k},\vec{q}} \mathbbm{1}$ with $\gamma_{\vec{k},\vec{q}}=\sum_{j=1,2}(\bar{F}^{bb}_j(\vec{k},\vec{q}))^2>0$. Then, \equref{ConditionOnTBG} is obeyed as long as
\begin{equation}
    \left[Q^{\text{b},0}_{\vec{k}}, \eta_0\sigma_2s_0 \right] =0. \label{Commutator}
\end{equation}
This is the same commutator relation as in \equref{ConstraintsAtCNP}. All momentum independent states that obey this property, as well as their symmetries are summarized in \tableref{ListOfOrderParameters}.
Furthermore, we have also checked all of the states in \tableref{ListOfOrderParameters} are also exactly degenerate in the Hartree term (\ref{HartreeEnergy}) and in the additional subtraction point contribution in \equref{SubtractionPoint}, as long as $w_0=D_0=0$.

In the following subsections, we will discuss the various energetic corrections to $Q^0_{\vec{k}}$ in \equref{DiagonalQMatrix} once we turn on $D_0$, $w_0$, and $W_{\text{TBG}}$, and allow the graphene-like bands to develop order and mix with the TBG-like bands in the vicinity of the Dirac cones. To simplify the discussion, we will first focus on the SU(2)$_+ \times$ SU(2)$_-$-symmetric model defined in \secref{ModelAndSymmetries}. In that case, certain pairs of states, referred to Hund's partners in \cite{PhysRevResearch.2.033062}, are guaranteed to be degenerate. Therefore, we will focus, without loss of generality, on only one member of each pair of Hund's partners; for concreteness, we choose those above the vertical line in \tableref{ListOfOrderParameters}. In \secref{HundsCouplingAppendix}, we will study how these degeneracies are lifted for $J_H \neq 0$ in \equref{HundsCouplingInteraction}.

\begin{figure}[tb]
    \centering
    \includegraphics[scale=.63]{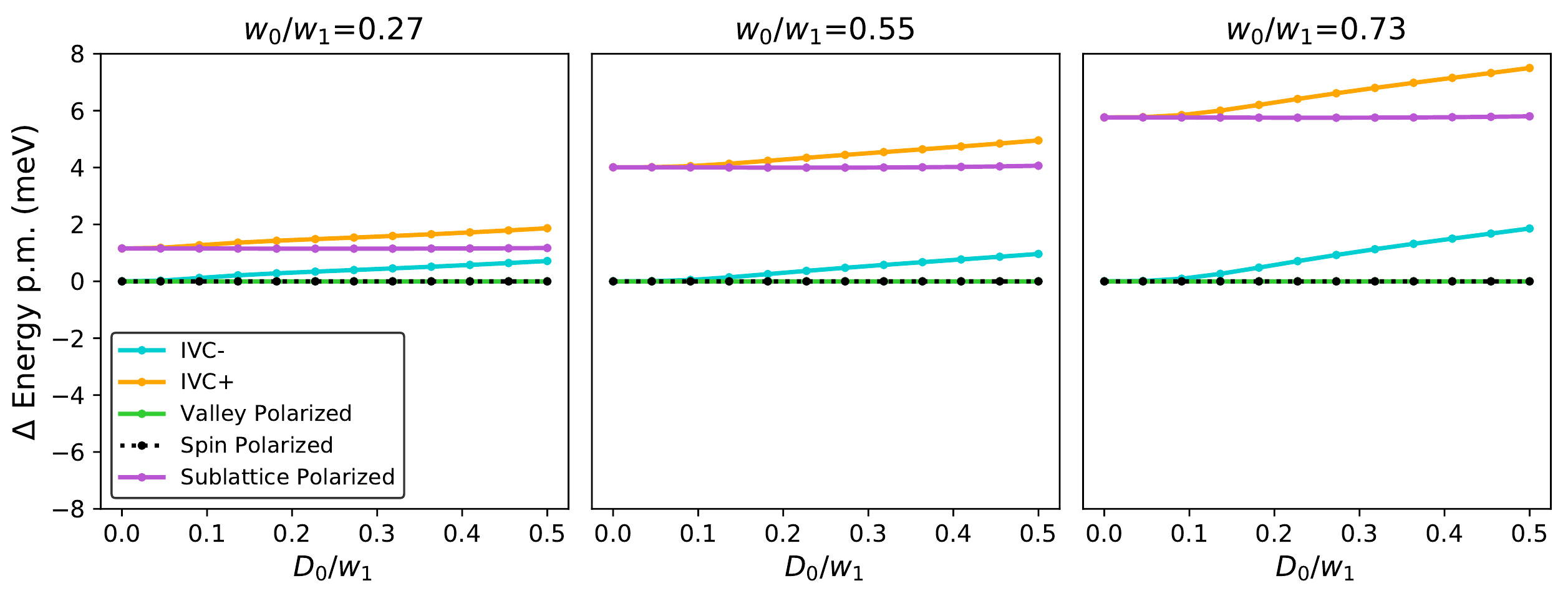}
    \caption{HF energies relative to the SP state obtained by plugging in the ans\"atze in \tableref{ListOfOrderParameters} in the HF energy expression. The behavior agrees with the analytical prediction in the column $\Delta E_{bb}(W_{\text{TBG}}=0)$ in \tableref{BehaviorOfEnergetics}.}
    \label{Energycut_notSC}
\end{figure}

\subsection{Deformation of the form factors}\label{DeformationOfFormFactors}
Let us first analyze the energetic impact for these different phases due to the deformation of the form factors, see \equref{FormFactors}, as a consequence of $D_0$ and $w_0$. To this end, we compute the change of the HF energy $E^{\text{HF}}$ in \equref{HFEnergy} for $Q^0_{\vec{k}}$ in \equref{DiagonalQMatrix} for the different $Q^{\text{b}}_{\vec{k}}$ in \tableref{ListOfOrderParameters}, setting $W_{\text{TBG}}=0$. Using the structure of the form factors in \equref{FormFactors} along with the important constraints in \equsref{C2zConstraint}{GConstraint}, we obtain after straightforward algebra the resultant change of energy $\Delta E_{bb}(W_{\text{TBG}}=0)$ listed in \tableref{BehaviorOfEnergetics}. The prefactors are given by
\begin{align}
    c_1 & =  \frac{2}{N} \sum_{\vec{q},\vec{k}} V(\vec{q}) \left\{ [\bar{F}_3^{bb}(\vec{k},\vec{q})]^2 + [\bar{F}_4^{bb}(\vec{k},\vec{q})]^2 \right\},\\
    c_2 & = \frac{8}{N} \sum_{\vec{G}} V(\vec{G}) \left[\sum_{\vec{k}} \bar{F}_4^{bb}(\vec{k},\vec{G})\right]^2 ,\\
    c_3 & = \frac{2}{N} \sum_{\vec{q},\vec{k}} V(\vec{q}) \left\{ [\bar{F}_5^{bb}(\vec{k},\vec{q})]^2 + [\bar{F}_6^{bb}(\vec{k},\vec{q})]^2 \right\} ,\\
    c_4 & = \frac{2}{N} \sum_{\vec{q},\vec{k}} V(\vec{q}) \left\{ [\bar{F}_7^{bb}(\vec{k},\vec{q})]^2 + [\bar{F}_8^{bb}(\vec{k},\vec{q})]^2 \right\}.
\end{align}
Due to \equref{ZeroqLimit}, we expect the Hartree contribution, $c_2$, to be numerically smaller than $c_1$. We have verified the form of these energy contributions wihtin our HF numerics, see \figref{Energycut_notSC}.

\subsection{Finite bandwidth of the TBG-like bands}
\label{Superexchange}
Next, we allow for $W_{\text{TBG}} \neq 0$ in \equref{NonInteractingHam} by treating it as a perturbation to the different states in \tableref{ListOfOrderParameters}.

\subsubsection{First order perturbation theory}
The first-order contribution is readily found by evaluating the expectation value of $\Delta H_0 := H_0-H_0|_{W_{\text{TBG}}=0}$ with $H_0$ given in \equref{NonInteractingHam} with respect to the states $\ket{\psi[P_{\vec{k}}]}$. Within the matrix notation introduced in \appref{app:HartreeFock}, the correction to the energy can also be written as
\begin{equation}
     \Delta E^{(1)}[P_{\vec{k}}]  = \sum_{\vec{k}\in\text{MBZ}} \text{tr}\left[ P^T_{\vec{k}} ( h^k(\vec{k}) - h^k(\vec{k})|_{W_{\text{TBG}}=0} )\right].
\end{equation}
Using the form of the kinetic energy of the TBG-like bands given in \equref{TBGLikeEnergy} and that $\bar{E}_1^b(\vec{k})=-\bar{E}_1^b(-\vec{k})$, one finds 
\begin{equation}
    \Delta E^{(1)}=\frac{1}{2}\sum_{\vec{k}\in\text{MBZ}} \bar{E}_3^b(\vec{k}) \,w_0D_0
\end{equation}
for all states of \tableref{ListOfOrderParameters}. Consequently, we have to go to second order in $W_{\text{TBG}}$ to find a contribution that favors one of the orders over the others. This is what we will discuss next.

\subsubsection{Second order perturbation theory: superexchange}
For the discussion of second order perturbation theory, it will most convenient to switch back to the second quantization description. The unperturbed Hamiltonian is the full Hamiltonian in the flat-band limit, $H_{\text{I}} = H_0|_{W_{\text{TBG}}=0} + \widetilde{H}_1$, with $\widetilde{H}_1$ given in \equref{TheInteractionDeltaRho}, and the perturbation is the dispersion in the TBG-like bands, i.e.,
\begin{equation}
        H_{\text{II}} = W_{\text{TBG}} \sum_{\vec{k}\in\text{MBZ}} \sum_{\eta=\pm} \sum_{s=\uparrow,\downarrow} \sum_{p=\pm}  \epsilon_{(b,p),\eta}(\vec{k}) b^\dagger_{\vec{k};p,\eta,s}b^\pdagger_{\vec{k};p,\eta,s} = W_{\text{TBG}} \sum_{j=0}^3 H_{\text{II}}^{(j)}, \quad H_{\text{II}}^{(j)}= \sum_{\vec{k}\in\text{MBZ}} E^b_{j}(\vec{k}) \, b^\dagger_{\vec{k}} M_j  b^\pdagger_{\vec{k}} . \label{HIIDefinition}
\end{equation}
Here we used the parameterization in \equref{TBGLikeEnergy} and introduced the matrices
\begin{equation}
    M_0 = \sigma_3\eta_0 s_0, \qquad M_1 = \sigma_0\eta_3 s_0, \qquad M_2 = \sigma_3\eta_3 s_0,\qquad M_3 = \sigma_0\eta_0 s_0 \label{DefinitionOfMj}
\end{equation}
acting in the TBG-like subspace. As will become useful shortly, this decomposition of the kinetic energy has the property that all $M_j$ defined in \equref{DefinitionOfMj} either commute or anticommute with all of the $Q^b$ in \tableref{ListOfOrderParameters} separately.

Let us assume that (one of) the groundstate(s) of $H_{\text{I}}$ with energy $E^0_{Q^b}$ is the product state $\ket{\psi[Q^b]}$, characterized by the correlation matrix $P_{\vec{k}} = (\mathbbm{1} + \text{diag}(-\sigma_z,Q^b))/2$ in \equref{CorrelationMatrixP}. Here, $Q^b$ can be any of the states listed in \tableref{ListOfOrderParameters}.
The energy correction to second order in $W_{\text{TBG}}$ has the form
\begin{equation}
    \Delta E^{(2)} = - \sum_{\ket{n} \neq \ket{\psi[Q^b]}} \frac{|\braket{n|H_{\text{II}}|\psi[Q^b]}|^2}{E^0_{n}- E^0_{Q^b}}, \label{GeneralFormOfSecondOrder}
\end{equation}
where the sum involves all eigenstates $\ket{n}$ of $H_{\text{I}}$ (excluding  $\ket{\psi[Q^b]}$) with energy $E^0_{n}$. Assuming that the system is in an insulating phase in the TBG-like sector, we will have a finite gap, $E^0_{n}>E^0_{Q^b}$, such that perturbation theory is controlled in bandwidth over gap and $\Delta E^{(2)}\leq 0$, i.e., the ``superexchange''-like processes discussed here, if finite, will lower the energy. 

We will next show that for a given $Q^b$, all terms $j$ in \equref{HIIDefinition} with $[M_j,Q^b] = 0$ will not contribute to \equref{GeneralFormOfSecondOrder} and, hence, only those that obey $\{M_j,Q^b\} = 0$ can lower the energy to order $W_{\text{TBG}}^2$. To see this, first note that $[M_{j_0},Q^b] = 0$ implies that there exists a unitary matrix $V$ such that both $V^\dagger Q^b V$ and $V^\dagger M_{j_0} V$ are diagonal. Since $\text{tr} \, Q^b = 0$, we can take $V^\dagger Q^b V = \text{diag}(1,1,1,1,-1,-1,-1,-1)$ without loss of generality at $\nu=0$. Consequently, the associated many-body state assumes the form
\begin{equation}
    \ket{\psi[Q^b]} = \prod_{\vec{k}\in\text{MBZ}}\prod_{v=1}^4 \bar{b}^\dagger_{\vec{k};v}  \ket{\psi_{0,g}}, \qquad \bar{b}_{\vec{k};v}:= \sum_{\alpha }b_{\vec{k};\alpha} V_{\alpha,v}, \label{FormOfGroundstate}
\end{equation}
where $\ket{\psi_{0,g}}$ is the state of empty TBG-like bands and lower-filled graphene-like bands. Upon noting that $M^T_j = M_j$ and $M_j^2=M_j$, we also see
\begin{equation}
    H_{\text{II}}^{(j_0)}= \sum_{\vec{k}\in\text{MBZ}} E^b_{j_0}(\vec{k}) \,\sum_{v=1}^8  r_v \bar{b}^\dagger_{\vec{k};v} \bar{b}^\pdagger_{\vec{k};v}, \qquad r_v = \pm 1.
\end{equation}
Consequently, we get
\begin{equation}
    \braket{n|H_{\text{II}}^{(j_0)}|\psi[Q^b]} = \sum_{\vec{k}\in\text{MBZ}} E^b_{j_0}(\vec{k}) \,\sum_{v=1}^4  r_v \braket{n|\psi[Q^b]} = 0
\end{equation}
since $\ket{\psi[Q^b]}$ and $\ket{n}$ are orthogonal for all terms in the sum of \equref{GeneralFormOfSecondOrder}. 

From this algebraic criterion, we can immediately read off that the SP and VP states have $\Delta E^{(2)}=0$, i.e., cannot gain energy from this ``superexchange'' mechanism. The SLP$_\pm$ and  SSLP$_\pm$ order parameters, however, anticommute with $M_0$ and $M_2$. As such, they can benefit both from the superexchange processes associated with $E_0^b = \bar{E}_0^b$ (already present when $w_0=D_0=0$, but $\theta$ tuned away from the magic angle) and $E_2^b = \bar{E}_2^b D_0$ (unique to the trilayer system and only non-zero for $D_0\neq 0$). Furthermore, while both IVCs can benefit from $E_1^b = \bar{E}_1^b w_0$, the IVC$_-$ (IVC$_+$) also benefits from $E_0^b$ ($E_2^b=\bar{E}_2^b D_0$). This leads to the results shown in \tableref{BehaviorOfEnergetics}.

\subsection{Ordering in the graphene-like bands}\label{PolarizationOfGrapheneLikeBands}
Let us next also take into account that the coupling between the graphene-like and TBG-like bands can modify the ansatz in \equref{DiagonalQMatrix}. Note that a perturbative treatment of bandwidth of the graphene-like bands, similar to our analysis in \secref{Superexchange}, is not controlled as these bands are highly dispersive and their bandwidth and energy, $\epsilon_{(g,p),\eta}(\vec{k})$, at generic momentum points are larger than the interaction energy scale. At the same time, $\epsilon_{(g,p),\eta}(\vec{k})$ is small around the K/K' points such that taking the interaction as a small parameter---as employed in the full quantum mechanical analysis in \appref{ExactStatements}---is not necessarily valid either for realistic system parameters. Therefore, we here use a different approach that takes advantage of these observations: while the diagonal structure of \equref{DiagonalQMatrix} and $Q^{g}_{\vec{k}} = -\sigma_z$ are likely to be a good approximation far from the K/K' points, it is expected to be modified in the vicinity of these points.

To formalize this, we define $\mathcal{A} = \mathcal{A}_+ \cup \mathcal{A}_-$ where $\mathcal{A}_+$ ($\mathcal{A}_-$) is a region of momentum points around the K (K') point. The precise form does not matter, we only need that its area, $V_{\mathcal{A}}$, is small compared to the overall size of the MBZ. We now take
\begin{equation}
    Q_{\vec{k}} = \begin{cases} Q^0_{\vec{k}},\quad \vec{k}\in \bar{\mathcal{A}}, \\
    Q^1_{\vec{k}}, \quad \vec{k}\in \mathcal{A},  \end{cases} \qquad \bar{\mathcal{A}} := \text{MBZ} \setminus \mathcal{A} \label{DeformedQ}
\end{equation}
with $Q^0_{\vec{k}}$ given in \equref{DiagonalQMatrix}, and accordingly for $P_{\vec{k}}$. Expanding the change of the HF energy, $\Delta E^{\text{HF}} = E^{\text{HF}}[P_{\vec{k}}]-E^{\text{HF}}[P_{\vec{k}}^0]$, to leading order in $V_{\mathcal{A}}$, one finds after straightforward algebra

\begin{align}
    \begin{split}\Delta E^{\text{HF}} &= \frac{1}{2}\sum_{\vec{k}\in  \mathcal{A}} \text{tr}[\delta Q^T_{\vec{k}} h^k(\vec{k})] + \frac{1}{N}\sum_{\vec{G}\in\text{RL}} V(\vec{G}) \, \text{Re} \left\{ \left( \sum_{\vec{k}'\in\bar{\mathcal{A}}} \text{tr}[P^0_{\vec{k}'}F^T_{\vec{k}',\vec{G}}] \right) \sum_{\vec{k}\in\mathcal{A}} \text{tr}[\delta Q_{\vec{k}} F^T_{\vec{k},\vec{G}}] \right\} \\
    &\quad - \frac{1}{N} \sum_{\vec{G}\in\text{RL}} \sum_{\vec{k}\in\mathcal{A}}\sum_{\vec{k}'\in\bar{\mathcal{A}}}V(\vec{k}'-\vec{k}+\vec{G}) \text{tr}[\delta Q_{\vec{k}} F^T_{\vec{k},\vec{k}'-\vec{k}+\vec{G}} P^0_{\vec{k}'}F^*_{\vec{k},\vec{k}'-\vec{k}+\vec{G}}] \\
    & \quad  - \frac{1}{2N} \sum_{\vec{G} \in\text{RL}} V(\vec{G}) \Gamma^0_{\vec{G}} \sum_{\vec{k}\in\mathcal{A}} \text{tr}[\delta Q_{\vec{k}} F_{\vec{k},\vec{G}}^T] + \frac{1}{2N} \sum_{\vec{q}} V(\vec{q}) \sum_{\vec{k}\in\mathcal{A}} \text{tr}[\delta Q^T_{\vec{k}} F_{\vec{k},\vec{q}}^\dagger F_{\vec{k},\vec{q}}^\pdagger] +\mathcal{O}(V^2_{\mathcal{A}}), \label{ChangeOfHFEnergy}
\end{split}\end{align}
where we defined the shorthand $\delta Q_{\vec{k}} = Q^1_{\vec{k}}-Q^0_{\vec{k}}$ and $\Gamma^0_{\vec{G}} = 8 \sum_{\vec{k}} \left( F^{bb}_1(\vec{k},\vec{G}) + F^{gg}_1(\vec{k},\vec{G}) \right)$. Here, the first, second, and third terms come from the dispersion, Hartree, and Fock term, while the last line is due to the subtraction point.
To simplify further, note that we can extend the sums over $\vec{k}'$ in \equref{ChangeOfHFEnergy} to the entire MBZ without changing the result at linear order in $V_{\mathcal{A}}$. Focusing on the relevant case of $P_{\vec{k}}^0 = P^0$, we can write
\begin{align}
    \begin{split}\Delta E^{\text{HF}} &= \frac{1}{2}\sum_{\vec{k}\in  \mathcal{A}} \text{tr}[\delta Q^T_{\vec{k}} h^k(\vec{k})] \\ & \quad + \frac{1}{2N}\sum_{\vec{G}\in\text{RL}} V(\vec{G}) \, \text{Re} \left(  \sum_{\vec{k}'} \text{tr}[Q^0 F^T_{\vec{k}',\vec{G}}]  \sum_{\vec{k}\in\mathcal{A}} \text{tr}[\delta Q_{\vec{k}} F^T_{\vec{k},\vec{G}}] \right) \\
    &\quad - \frac{1}{N} \sum_{\vec{q}} \sum_{\vec{k}\in\mathcal{A}}V(\vec{q}) \left( \text{tr}[\delta Q_{\vec{k}} F^T_{\vec{k},\vec{q}} P^0_{\vec{k}'}F^*_{\vec{k},\vec{q}}] - \frac{1}{2} \text{tr}[\delta Q_{\vec{k}} F_{\vec{k},\vec{q}}^T F_{\vec{k},\vec{q}}^*] \right)  +\mathcal{O}(V^2_{\mathcal{A}}). \label{ChangeOfHFEnergy2}
\end{split}\end{align}

Postponing the discussion of $Q_{\vec{k}}$ with non-zero off-diagonal components that mix the graphene-like and TBG-like bands to the next subsection, we here focus on block diagonal
\begin{equation}
    Q^1_{\vec{k}} = \begin{pmatrix}  Q^{\text{g}}_{\vec{k}} & 0 \\ 0 & Q^{b}_{\vec{k}}   \end{pmatrix},  \label{FormOfQ1}
\end{equation}
where $Q^{\text{g}}_{\vec{k}}$ will be chosen to describe the same order as in the TBG-like bands, i.e., to break/keep exactly the same symmetries as the considered order parameter in the TBG-like bands.

Let us begin our discussion with the \textit{spin-polarized} state, i.e., take 
\begin{equation}
    \left(Q^{\text{g}}_{\vec{k}}\right)_{\eta,\eta'} = \delta_{\eta,\eta'}\begin{cases} \sigma_0 s_3, \qquad \vec{k} \in \mathcal{A}_\eta \\ -\sigma_3 s_0, \quad \text{otherwise} \end{cases} \label{SpinPolarizedGrapheneBands}
\end{equation}

in \equref{DeformedQ}. 
Using \equref{ChangeOfHFEnergy2}, it is a matter of straightforward algebra to show that the associated change of the energy is given by
\begin{align}\begin{split}
     \Delta E^{\text{HF}}_{\text{SpinPol}} &\sim 2 \sum_{\eta}\sum_{\vec{k}\in\mathcal{A}_\eta} \bar{E}^{g}_0(\vec{k};\eta) \\ &-\frac{2}{N} \sum_{\vec{q}} \sum_\eta \sum_{\vec{k}\in\mathcal{A}_\eta}V(\vec{q}) \left[ (\bar{F}_{2,\eta}^{gg}(\vec{k},\vec{q}))^2 - (\bar{F}_{1,\eta}^{gg}(\vec{k},\vec{q}))^2 \right]\\
     &-\frac{2}{N} \sum_{\vec{q}} \sum_{\eta} \sum_{\vec{k}\in\mathcal{A}_\eta}V(\vec{q}) \left[ (\bar{F}_{1,\eta}^{bg}(\vec{k},\vec{q}))^2 + (\bar{F}_{2,\eta}^{bg}(\vec{k},\vec{q}))^2 \right] D_0^2 \\
     &-\frac{2}{N} \sum_{\vec{q}} \sum_{\eta} \sum_{\vec{k}\in\mathcal{A}_\eta}V(\vec{q}) \left[ (\bar{F}_{7,\eta}^{bg}(\vec{k},\vec{q}))^2 + (\bar{F}_{8,\eta}^{bg}(\vec{k},\vec{q}))^2 \right] w_0^2D_0^2 \\
     & -\frac{2}{N} \Bigl\{\sum_{\vec{q}} \sum_{\eta} \sum_{\vec{k}\in\mathcal{A}_\eta}V(\vec{q}) \left[ (\bar{F}_{7,\eta}^{gg}(\vec{k},\vec{q}))^2 - (\bar{F}_{8,\eta}^{gg}(\vec{k},\vec{q}))^2 \right] + 2\sum_{\vec{G}} V(\vec{G})  \sum_{\eta',\vec{k}'}\bar{F}^{gg}_{8,\eta'}(\vec{k}',\vec{G}) \sum_{\eta,\vec{k}\in\mathcal{A}_\eta}  \bar{F}^{gg}_{8,\eta}(\vec{k},\vec{G})  \Bigr\} w_0^2 D_0^4 \label{SpinPolarEnergyChange}
\end{split}\end{align}
where $\bar{E}^{g}_0(\vec{k};\eta)$ is (the part of) the dispersion of the graphene-like bands defined in \equref{GrapheneLikeEnergy}. To keep the notation more compact, we further defined the $\eta$-dependent form-factor components as
\begin{align}
    F^{tt'}_{j,\eta}(\vec{k},\vec{q}) &= F^{tt'}_{j}(\vec{k},\vec{q}) + \eta \, F^{tt'}_{j+2}(\vec{k},\vec{q}), \qquad j=1,2, \\
    F^{tt'}_{j,\eta}(\vec{k},\vec{q}) &= F^{tt'}_{j}(\vec{k},\vec{q}) + \eta \, F^{tt'}_{j-2}(\vec{k},\vec{q}), \qquad j=7,8.
\end{align}
To understand the scaling behavior of the first contribution associated with the kinetic term, let us take $\bar{E}^{g}_0(\vec{k};\eta) = \alpha |D_0| + v_D |\delta_\eta \vec{k}|$ where $\delta_\eta \vec{k}$ is the distance from the K (K') point in valley $\eta=+$ ($\eta=-$) and $\alpha > 0$. As such we expect the scaling behavior
\begin{equation}
    \sum_{\eta}\sum_{\vec{k}\in\mathcal{A}_\eta} \bar{E}^{g}_0(\vec{k};\eta) \sim c_1 |D_0| (\Delta k)^2 + c_2 v_D(\Delta k)^3, \qquad c_j > 0 ,
\end{equation}
as a function of $D_0$ and the linear size, $\Delta k$, of $\mathcal{A}$. While we have already indicated the scaling behavior of all remaining terms in \equref{SpinPolarEnergyChange} with $D_0$ and $w_0$, we further note that they scale with the area of $\mathcal{A}$ and, thus, quadratically with $\Delta k$.

To discuss the consequences, let us first focus on $D_0=0$. We see that the energetic penalty due to the kinetic term scales as $\Delta k^3$ while the energetic gain from the second line of \equref{SpinPolarEnergyChange} scales as $\Delta k^2$. One could naively conclude that this implies that the system should lower its energy by polarizing the graphene bands in some vicinity of its Dirac cones. Closer inspection, however, shows the involved energetics is equivalent to that of single-layer graphene: note that, for $D_0=0$, the first two lines of \equref{SpinPolarEnergyChange} involve the dispersion and interaction matrix elements only in the graphene subspace. Since we know that graphene is not a spin-polarized insulator, this spin polarization cannot be preferred by the system. In fact, one can show that the two terms in the second line of \equref{SpinPolarEnergyChange} cancel each other to leading order in $\Delta k$: To see this, let us focus on the contribution of one valley, say $\eta=+$, and consider $\vec{k}\rightarrow -\vec{q}_1/2$ (more precisely $\vec{k}=k (\cos \phi_0,\sin \phi_0)-\vec{q}_1/2$ and $k\rightarrow 0^+$) where we get
\begin{align}
    \sum_{\vec{q}} V(\vec{q}) \left[ (\bar{F}_{2,+}^{gg}(\vec{k},\vec{q}))^2 - (\bar{F}_{1,+}^{gg}(\vec{k},\vec{q}))^2 \right] &\rightarrow \sum_{\vec{q}} V(\vec{q}) \left[ \sin^2\left(\frac{\xi_{\text{MBZ}(\vec{q})}-\phi_0}{2}\right) -\cos^2\left(\frac{\xi_{\text{MBZ}(\vec{q})}-\phi_0}{2}\right) \right] \delta_{\vec{G}_{\vec{q}},0}, \\
    & = \sum_{\vec{q}\in\text{MBZ}} V(\vec{q}) \left[ \sin^2\left(\frac{\xi_{\vec{q}}-\phi_0}{2}\right) -\cos^2\left(\frac{\xi_{\vec{q}}-\phi_0}{2}\right) \right].
\end{align}
Here we inserted the explicit form of the form factors given in \equref{FormOfGrapheneMatrixElements}. This expression has to vanish since $V(\vec{q}) = V(-\vec{q})$, $\xi_{\vec{q}} = \xi_{-\vec{q}} +\pi$, and $\vec{q}\in\text{MBZ}$ implies $-\vec{q}\in\text{MBZ}$. The same analysis can be performed for the other valley $\eta=-$. For this reason, also the interaction correction in \equref{SpinPolarEnergyChange} scales (at least) as $(\Delta k)^3$ for $D_0=0$ and, hence, does not generically dominate the kinetic contribution of the first line.

Turning on $D_0$, we see that while there can be some additional gain starting at order $D_0^2$, this gain is always overcompensated for sufficiently small $D_0$ by the additional cost from the kinetic energy which scales as $D_0$.

These results are completely consistent with our HF numerics: as can be seen in \figref{Bands}, at $D_0=0$, the graphene Dirac cones are not split. Furthermore, while there are spin-split Dirac cones around the Fermi level at K/K' for $D_0\neq 0$, these are not related to the spin-ordering in the graphene-like bands defined in \equref{SpinPolarizedGrapheneBands}. This becomes obvious by noting that both spin flavors of the upper (lower) graphene-like band are unoccopied (occupied), as is clearly visible based on the band connectivity. Furthermore, we can see it in the correlator of the SP around the K/K' points shown in \figref{MatrixOrders}.

All other states in \tableref{ListOfOrderParameters} can be analyzed in a similar way. To begin with the \textit{valley polarized} state, the analogue of \equref{SpinPolarizedGrapheneBands} reads as
\begin{equation}
    Q^{\text{g}}_{\vec{k}} = \begin{cases} \sigma_0 \eta_3 s_0, \qquad \vec{k} \in \mathcal{A}, \\ -\sigma_3 \eta_0 s_0, \quad \,\text{otherwise}, \end{cases} 
\end{equation}
which leads to a kinetic energy contribution [from the first line of \equref{ChangeOfHFEnergy2}]
\begin{equation}
    \Delta E^{\text{kin}}=\frac{1}{2}\sum_{\vec{k}\in  \mathcal{A}} \text{tr}[\delta Q^T_{\vec{k}} h^k(\vec{k})] = 2 \sum_{\vec{k}\in  \mathcal{A}} \sum_\eta \left( \bar{E}_0^g(\vec{k},\eta) + \eta \bar{E}_1^g(\vec{k},\eta) w_0 D_0 \right) = 2 \sum_{\vec{k}\in  \mathcal{A}} \sum_\eta  \bar{E}_0^g(\vec{k},\eta), 
 \end{equation}
where (although not crucial for the following conclusion), we made the natural assumption $\vec{k} \in \mathcal{A} \Leftrightarrow -\vec{k} \in \mathcal{A}$ in the last equality. Most importantly, we see that the kinetic energy contribution now scales as $(\Lambda + D_0)(\Delta k)^2$ to leading order in $\Delta k$, where $\Lambda$ is the large energy of the graphene-like remote band, i.e., in the valley that does not have a Dirac cone at the respective K point $\vec{k}$ is close to. As such, it is energetically not favorable to have valley polarization of the graphene-like bands around the K/K' points.

As is intuitively clear and can be derived in the same way, this suppression due to the kinetic energy also applies to the \textit{IVC states} in \tableref{ListOfOrderParameters}. One also finds exactly the same expression, $\Delta E^{\text{kin}} = 2 \sum_{\vec{k}\in  \mathcal{A}} \sum_\eta  \bar{E}_0^g(\vec{k},\eta) \sim (\Lambda + D_0)(\Delta k)^2$. Consequently, ordering of the graphene-like bands in the vicinity of the K/K' points can also be excluded for the IVC states.

\vspace{1em}

For the \textit{sublattice polarized} states, SLP$_{\pm}$, the situation is slightly more complicated since the structure in the graphene-like bands is in general of the form
\begin{align}
    \left(Q^{\text{g}}_{\vec{k}}\right)_{\eta,\eta'} = \delta_{\eta,\eta'}\begin{cases} (- \sigma_3\cos \theta_{\vec{k}} + \sigma_2 \sin \theta_{\vec{k}}) s_0, \qquad &\vec{k} \in \mathcal{A}_\eta \\ -\sigma_3 s_0, \quad &\text{otherwise}, \end{cases} \label{SLPAnsatzVariation}
\end{align}
and it is left to determine the optimal $\theta_{\vec{k}} \in\mathbb{R}$ subject to the constraint $\theta_{\vec{k}}=\mp \theta_{-\vec{k}}$ for the SLP$_{\pm}$ state resulting from $C_{2z}$ and $\Theta$. Note that $(Q^{\text{g}}_{\vec{k}})^2=\mathbbm{1}$ and $\text{tr}\, Q^{\text{g}}_{\vec{k}} = 0$, hold for any $\theta_{\vec{k}}$ and, hence, does not further constrain it.

Making the natural assumption that $\sin \theta_{\vec{k}}$ is significantly non-zero only in the small region $\mathcal{A}$ around the K/K$'$ points, we can still use \equref{ChangeOfHFEnergy2} which yields 
\begin{equation}
     \Delta E^{\text{HF}}_{\text{SLP}_\pm}[\theta_{\vec{k}}] = \sum_\eta \sum_{\vec{k}\in\mathcal{A}_\eta} \left[ A_{\vec{k},\eta} (1 - \cos \theta_{\vec{k}}) + B_{\vec{k},\eta} \sin \theta_{\vec{k}} \right], \label{EnergyForSLP}
\end{equation}
where 
\begin{subequations}
\begin{align}
    A_{\vec{k},\eta} & = 2 \bar{E}^g_0(\vec{k},\eta) + \frac{2}{N} \sum_{\vec{q}} V(\vec{q})\left[ (\bar{F}_{1,\eta}^{gg}(\vec{k},\vec{q}))^2 - (\bar{F}_{2,\eta}^{gg}(\vec{k},\vec{q}))^2 \right] + \mathcal{O}(D_0^4), \label{SLPsAEn} \\
    B_{\vec{k},\eta} & = - \frac{2}{N} \sum_{\vec{q}} V(\vec{q})\left[ (\bar{F}_{1,\eta}^{gb}(\vec{k},\vec{q}))^2 + (\bar{F}_{2,\eta}^{gb}(\vec{k},\vec{q}))^2 \right] D_0^2 + \frac{2}{N} \sum_{\vec{q}} V(\vec{q})\left[ (\bar{F}_{7,\eta}^{gb}(\vec{k},\vec{q}))^2 + (\bar{F}_{8,\eta}^{gb}(\vec{k},\vec{q}))^2 \right] w_0^2 D_0^2. \label{SLPsBEn}
\end{align}\end{subequations}
First, note that $B_{\vec{k},\eta} =0$ for $D_0=0$. Furthermore, the second term in \equref{SLPsAEn} is suppressed to leading order in $\Delta k$, as already discussed above, such that $A_{\vec{k},\eta} > 0$ and, thus, $\theta_{\vec{k}}=0$ for $D_0 = 0$. This agrees with our result of \appref{ExactStatements} and with the HF band structure in \figref{Bands} which exhibits a gapless graphene Dirac cone at $D_0=0$ for the SLP$_-$ (the same applies for SLP$_+$, not shown). More explicitly, this can also be seen in the upper panel of \figref{MatrixOrders}, where the correlators of the SLP$_\pm$ states close to the K point at $D_0=0$ are shown: the graphene bands are just in the lower-filled-band configuration.

The system behaves differently when $D_0 \neq 0$. We then have $B_{\vec{k},\eta} \neq 0$ and it will become energetically favorable to develop finite SLP$_\pm$ order in the graphene-like bands according to $\tan \theta_{\vec{k}} = \widetilde{B}_{\vec{k}}^\pm / \widetilde{A}_{\vec{k}}$, $\vec{k} \in \mathcal{A}_+$, where we defined
\begin{equation}
    \widetilde{A}_{\vec{k}} = A_{\vec{k},+} + A_{-\vec{k},-}, \qquad \widetilde{B}^\pm_{\vec{k}} = B_{\vec{k},+} \mp B_{-\vec{k},-},
\end{equation}
with $+$ ($-$) for the SLP$_+$ (SLP$_-$) case. The associated minimized energy change is given by
\begin{equation}
    \min_{\theta_{\vec{k}}}\Delta E^{\text{HF}}_{\text{SLP}_\pm}[\theta_{\vec{k}}] = \sum_{\vec{k}\in\mathcal{A}_+} \left[\widetilde{A}_{\vec{k}} - \sqrt{\widetilde{A}^2_{\vec{k}} + (\widetilde{B}_{\vec{k}}^\pm)^2}  \right].
\end{equation}
Taken together, both SLP$_{\pm}$ states will develop order in the graphene-like bands around the K and K' points as long as $D_0$ is non-zero. As $\widetilde{A} \sim D_0$ and $\widetilde{B}^{\pm} \sim D^2_0$, the associated energy gain will scale as $-g_{\pm} V_{\mathcal{A}} |D_0|^3$ for small $D_0$. As follows from \equref{SLPsBEn}, $B_{\vec{k},\eta}$ has a definite sign for sufficiently small $w_0$, such that $|\widetilde{B}_{\vec{k}}^-|>|\widetilde{B}_{\vec{k}}^+|$ and, thus, $g_{-} > g_+$, i.e., the SLP$_-$ state can gain more energy than the SLP$_+$ phase from this process for small $w_0$. Since $\sin \theta_{\vec{k}}\neq 0$ in \equref{SLPAnsatzVariation} corresponds to a mixing of the upper and lower bands of the graphene-like bands, this will gap out the Dirac cones of the SLP$_{\pm}$ states at $D_0\neq 0$ and we obtain an insulator rather than a semimetal; this is also seen in our HF numerics (cf.~\figref{Bands}). Furthermore, it is directly visible in the correlators of the SLP$_{\pm}$ states shown in the lower panel of \figref{MatrixOrders}; the additional mixing of the TBG-like and graphene-like band that can also be seen in the HF data will be analyzed in \appref{MixingBetweenBands}. 

\vspace{1em}

In analogy to these two states, we choose for the \textit{quantum spin Hall state} (SSLP$_{-}$)
\begin{align}
    \left(Q^{\text{g}}_{\vec{k}}\right)_{\eta,\eta'} = \delta_{\eta,\eta'}\begin{cases} - \sigma_3 s_0\cos \theta_{\vec{k}} + \sigma_2 s_3 \sin \theta_{\vec{k}}, \qquad &\vec{k} \in \mathcal{A}_\eta \\ -\sigma_3 s_0, \quad &\text{otherwise}. \end{cases}
\end{align}
The corresponding change of the energy has again the form of \equref{EnergyForSLP}, this time with
\begin{equation}
    A_{\vec{k},\eta} = 2 \bar{E}^g_0(\vec{k},\eta) + \frac{2}{N} \sum_{\vec{q}} V(\vec{q})\left[ (\bar{F}_{1,\eta}^{gg}(\vec{k},\vec{q}))^2 - (\bar{F}_{2,\eta}^{gg}(\vec{k},\vec{q}))^2 \right] + \mathcal{O}(w_0^2D_0^4), \qquad B_{\vec{k},\eta} = 0.
\end{equation}
Note that the vanishing of $B_{\vec{k},\eta}$ is consistent with the fact that the energy should not depend on whether $\theta_{\vec{k}}=\theta_{-\vec{k}}$ (SSLP$_-$) or $\theta_{\vec{k}}=-\theta_{-\vec{k}}$ (SSLP$_{+}$), since these two states are related by a SU(2)$_+ \times$ SU(2)$_-$ transformation. We clearly see that the energy is minimized by $\cos{\theta_{\vec{k}}} = 1$, i.e., no order in the graphene-like bands and no associated energetic gain. We emphasize that the presence of SSLP$_-$ order in the TBG-like bands and, at $D_0\neq 0$, interactions that couple the two subsystems, also induces a gap in the original graphene Dirac cone; we will see this in \appref{MixingBetweenBands} below where we will find a non-zero mixing of the TBG-like and graphene-like bands for the SSLP states.

\subsection{Mixing between the TBG-like and graphene-like bands}\label{MixingBetweenBands}
\begin{figure}[tb]
    \centering
    \includegraphics[scale=.8]{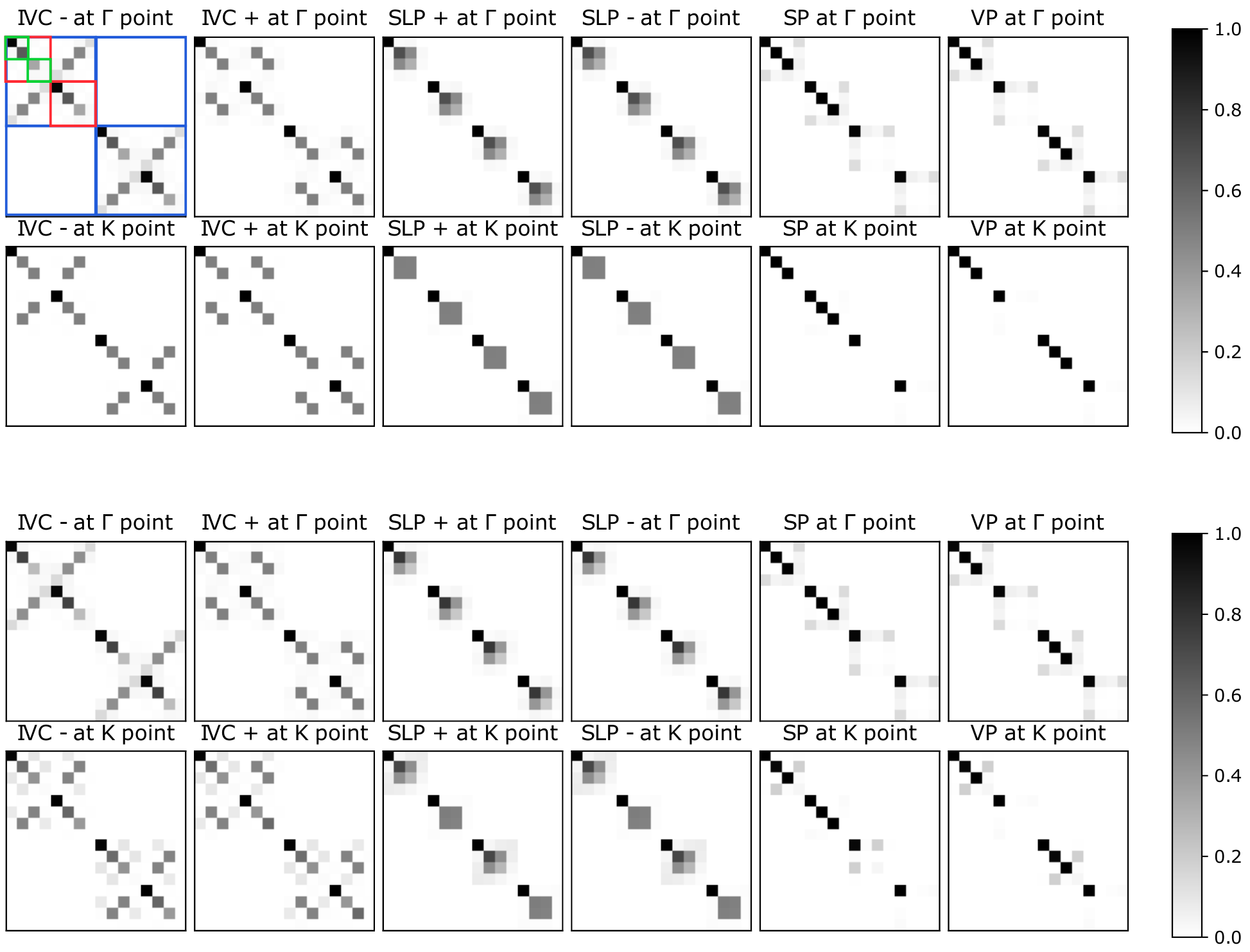}
    \caption{We show the projector $P$ which characterizes each of our symmetry breaking solution for a point near the $\Gamma$ point of the Brillouin zone and $K$ point of the Brillouin zone for $w_0/w_1=0.73$ and $D_0/w_1=0$ (top) and $D_0/w_1=0.5$ (bottom). We plot the matrix form of $P$ such that the largest block shown outlined in blue in the first panel of the top row denotes spin flavor, the next smallest blocks outlined in red denote valley flavor, and the next smallest block outlined in green denotes upper and lower bands  in the continuum model and final two boxes denoting TBG-like or graphene-like band index in the continuum model, where the final two boxes denote the graphene-like and TBG-like band in the upper-band box and the TBG-like then graphene-like bands in the lower-band box.}
    \label{MatrixOrders}
\end{figure}
We have not yet taken into account the possibility that the TBG-like and graphene-like bands can mix or, in other words, develop some coherence in the vicinity of the K/K$'$ points; this corresponds to allowing for off-diagonal components in $Q_{\vec{k}}$. It is clear by symmetry that such a mixing is only possible for $D_0\neq 0$.

To describe the mixing between the bands, we write
\begin{equation}
        Q_{\vec{k}} = U_{\vec{k}}^\pdagger Q^0_{\vec{k}} U_{\vec{k}}^\dagger, \qquad  Q^0_{\vec{k}} = \begin{pmatrix}  -\sigma_z \eta_0 s_0 & 0 \\ 0 & Q^{b}  \end{pmatrix} \qquad   U_{\vec{k}}^\pdagger U_{\vec{k}}^\dagger = \mathbbm{1}, \label{DeformTheMatrix}
\end{equation}
where $Q^{b}$ is any of the order parameters in the TBG-like bands listed in \tableref{ListOfOrderParameters} and the unitary matrix $U_{\vec{k}}$, which mixes the TBG-like and graphene-like bands, is constrained by the symmetries of the state $Q^{b}$. Note that unitarity of $U_{\vec{k}}$ guarantees $Q_{\vec{k}}^2=\mathbbm{1}$, $Q^\dagger_{\vec{k}}=Q^\pdagger_{\vec{k}}$, and $\text{tr} \, Q_{\vec{k}} = 0$. 

Since are interested in mixing of the bands, we will focus on $U_{\vec{k}}$ of the form
\begin{equation}
    U_{\vec{k}} = e^{i \varphi \lambda_{\vec{k}}} = e^{i \varphi \sum_{j=1,2}\sum_{j'} \hat{n}_{\vec{k}}^{(j,j')}  M^{(j,j')} \zeta_{j}}, \qquad \left(M^{(j,j')}\right)^\dagger = M^{(j,j')}, \quad \hat{n}_{\vec{k}}^{(j,j')}\in\mathbbm{R}   \label{VeryGeneralFormOfU}
\end{equation}
where $\zeta_{j}$ are Pauli matrices acting between the TBG-like and graphene-like bands and $M^{(j,j')}$ are matrices in spin, band, and valley space. It, thus, holds $[M^{(j_1,j_2)}, \zeta_{j}]=0$. Before discussing the different candidate states separately, where the different symmetries will constrain the available generators in \equref{VeryGeneralFormOfU}, let us expand \equref{DeformTheMatrix} up to second order in the rotational angle $\varphi$. To this end, define $\mathcal{M}_{j,\vec{k}} = \sum_{j'} \hat{n}_{\vec{k}}^{(j,j')}  M^{(j,j')}$ and $\mathcal{M}_{\vec{k}} = \mathcal{M}_{1,\vec{k}} + i \mathcal{M}_{2,\vec{k}}$. Equation (\ref{DeformTheMatrix}) then becomes
\begin{align}\begin{split}
    Q_{\vec{k}} &\sim \begin{pmatrix}  -\sigma_z  & 0 \\ 0 & Q^{\text{b}}  \end{pmatrix} + \begin{pmatrix} 0 & i\mathcal{M}^\dagger_{\vec{k}} Q^{\text{b}} + i\sigma_z \mathcal{M}^\dagger_{\vec{k}} \\  -iQ^{\text{b}} \mathcal{M}_{\vec{k}} -i \mathcal{M}_{\vec{k}} \sigma_z  & 0  \end{pmatrix} \varphi  \\
    & \quad +\left[  \begin{pmatrix} \mathcal{M}^\dagger_{\vec{k}} Q^{\text{b}}  \mathcal{M}_{\vec{k}} & 0 \\ 0 & - \mathcal{M}^\dagger_{\vec{k}} \sigma_z  \mathcal{M}_{\vec{k}} \end{pmatrix} + \frac{1}{2}\begin{pmatrix}  \{\mathcal{M}^\dagger_{\vec{k}} \mathcal{M}_{\vec{k}},\sigma_z \} & 0 \\ 0 & \{\mathcal{M}_{\vec{k}} \mathcal{M}^\dagger_{\vec{k}} ,Q^{\text{b}}\}  \end{pmatrix} \right]\varphi^2 + \mathcal{O}(\varphi^3). \label{QExpansion} \end{split}
\end{align}
Below, we will use this expression to interpret the mixing matrices we obtain analytically and to constrain the possible generators $M^{(j,j')}$.

Since the IVC$_-$ is favored energetically in most of parameter space, let us start with the \textit{intervalley coherent phases}, which can be discussed simultaneously. Due to the valley U(1)$_v$ symmetry, we can choose their ``undeformed'' correlator in the TBG-like subspace, without loss of generality, to be $Q^{b} = \eta_x$ and  $Q^{b} = \sigma_y\eta_y$ for the IVC$_+$ and IVC$_-$, respectively. As such, the phases still preserve the $C_{2z}\Theta$ symmetry in \tableref{ActionOfSymmetries} (for other choices, it has to be combined with a valley U(1)$_v$ rotation). Consequently, $M^{(j,j')}$ in \equref{VeryGeneralFormOfU} must obey
\begin{equation}
    (M^{(1,j')})^*=-M^{(1,j')}, \quad (M^{(2,j')})^*=M^{(2,j')}, \quad \Leftrightarrow \quad \mathcal{M}^* = -\mathcal{M}. \label{PropertyOfM}
\end{equation}

Furthermore, the preserved spin-rotation symmetry forces $M_j$ to be trivial in spin-space. This leaves the following $3 + 3 = 6$ options for $M_1$ and $3\times 3 + 1 = 10$ possibilities for $M_2$,
\begin{equation}
    M^{(1,j')} \in \{ \sigma_y \eta_{0,x,z}, \sigma_{0,x,z} \eta_{y} \}, \qquad M^{(2,j')} \in \{ \sigma_{0,x,z} \eta_{0,x,z}, \sigma_{y} \eta_{y} \}.
\end{equation}
However, not all (linear combinations) of the $16$ generators are important for the IVC states. First, we can neglect all parts of $\lambda_{\vec{k}}$ in \equref{VeryGeneralFormOfU} that commute with $Q^{0}_{\vec{k}}$. This can be done formally by setting $\lambda_{\vec{k}} \rightarrow (\lambda_{\vec{k}} - Q^{0}_{\vec{k}}\lambda_{\vec{k}}Q^{0}_{\vec{k}})/2$ or, equivalently, by replacing
\begin{equation}
\mathcal{M}_{\vec{k}}\rightarrow \frac{1}{2}(\mathcal{M}_{\vec{k}} + Q^{\text{b}}\mathcal{M}_{\vec{k}}\sigma_z).
\end{equation}
For future reference, this means that it holds
\begin{equation}
    \mathcal{M}_{\vec{k}} \sigma_z = Q^{\text{b}}\mathcal{M}_{\vec{k}}. \label{PropertyDueToSymmetrization}
\end{equation}
This reduces the number of independent generators to $8$. To begin with the IVC$_+$, a complete set of associated generators is given by
\begin{align}\begin{split}
    \frac{1}{2} \zeta_y\left(\sigma_0\eta_0 + \sigma_z \eta_x \right), \quad \frac{1}{2} \zeta_y\left(\sigma_z\eta_0 + \sigma_0 \eta_x \right), \quad \frac{1}{2} \zeta_y\left(\sigma_y\eta_y - \sigma_x \eta_z \right), \quad \frac{1}{2} \zeta_x\left(\sigma_y\eta_z + \sigma_x \eta_y \right), \qquad  \\
    \frac{1}{2} \left(\zeta_y\sigma_x\eta_0 + \zeta_x\sigma_y \eta_x \right), \quad \frac{1}{2} \left(\zeta_y\sigma_0\eta_z + \zeta_x\sigma_z \eta_y \right), \quad \frac{1}{2} \left(\zeta_y\sigma_x\eta_x + \zeta_x\sigma_y \eta_0 \right), \quad \frac{1}{2} \left(\zeta_y\sigma_z\eta_z + \zeta_x\sigma_0 \eta_y \right). \label{GeneratorsIVCP}\end{split}
\end{align}
For the IVC$_-$, they read as 
\begin{align}\begin{split}
    \frac{1}{2} \zeta_y\left(\sigma_z\eta_0 + \sigma_y \eta_y \right), \quad \frac{1}{2} \zeta_y\left(\sigma_0\eta_x + \sigma_x \eta_z \right), \quad \frac{1}{2} \zeta_y\left(\sigma_x\eta_x - \sigma_0 \eta_z \right), \quad \frac{1}{2} \zeta_x\left(\sigma_y\eta_0 + \sigma_z \eta_y \right), \qquad  \\
    \frac{1}{2} \left(\zeta_x\sigma_y\eta_x - \zeta_y\sigma_z \eta_z \right), \quad \frac{1}{2} \left(\zeta_x\sigma_y\eta_z + \zeta_y\sigma_z \eta_x \right), \quad \frac{1}{2} \left(\zeta_x\sigma_0\eta_y + \zeta_y\sigma_x \eta_0 \right), \quad \frac{1}{2} \left(\zeta_x\sigma_x\eta_y - \zeta_y\sigma_0 \eta_0 \right). \label{GeneratorsIVCM}\end{split}
\end{align}

To further constrain the remaining $8$ generators, we note that the graphene-like bands in valley $\eta=-$ (valley $\eta=+$) are at energies far away from the Fermi surface (of scale $\Lambda$) for $\vec{k}\in\mathcal{A}_{+}$ ($\vec{k}\in\mathcal{A}_{-}$). In the following, we will neglect all mixing processes that are suppressed in the limit of large $\Lambda$. Inspection of the first term in \equref{ChangeOfHFEnergy2} shows that this is equivalent to demanding that $U_{\vec{k}}$ obey
\begin{equation}
    \Delta_\Lambda(\eta) = \sum_{\vec{k}\in  \mathcal{A}_\eta} \text{tr}_{\text{g}}[(U_{\vec{k}}Q^0_{\vec{k}}U_{\vec{k}}^\dagger-Q^0_{\vec{k}}) \hat{\Lambda}_\eta] = 0, \qquad \hat{\Lambda}_\pm = \Lambda \sigma_z (\eta_0 \mp \eta_3)/2,  
\end{equation}
where the trace is only over the graphene-like subspace. Using \equref{QExpansion}, we can express this quantity as
\begin{equation}
    \Delta_\Lambda(\eta) = \sum_{\vec{k}\in  \mathcal{A}_\eta}\left( \text{tr}_{\text{g}}[\mathcal{M}^\dagger_{\vec{k}} Q^{\text{b}} \mathcal{M}_{\vec{k}} \hat{\Lambda}_\eta ] + \frac{1}{2}\text{tr}_{\text{g}}[\{\mathcal{M}^\dagger_{\vec{k}} \mathcal{M}_{\vec{k}},\sigma_z \}\hat{\Lambda}_\eta] \right)\varphi^2  + \mathcal{O}(\varphi^3).
\end{equation}
To simplify further, we use \equref{PropertyDueToSymmetrization} and obtain
\begin{equation}
    \Delta_\Lambda(\eta) = 2 \sum_{\vec{k}\in  \mathcal{A}_\eta}\text{tr}_{\text{g},-\eta}[\mathcal{M}^\dagger_{\vec{k}} \mathcal{M}_{\vec{k}}]\varphi^2  + \mathcal{O}(\varphi^3),
\end{equation}
where the trace is only over the graphene-like subspace in valley $-\eta$. As such, $\Delta_\Lambda(\eta)=0$ requires
\begin{equation}
    \text{tr}_{\text{g},\mp}[\mathcal{M}^\dagger_{\vec{k}} \mathcal{M}_{\vec{k}}] = \sum_{s,s',p,p',\eta'} \left|(\mathcal{M}_{\vec{k}})_{(p',\eta',s'),(p,\mp,s)}\right|^2 = 0 \quad \forall \vec{k}\in\mathcal{A}_{\pm} \quad \Leftrightarrow \quad (\mathcal{M}_{\vec{k}})_{(p',\eta',s'),(p,\mp,s)} = 0 \quad \forall \vec{k} \in \mathcal{A}_{\pm}. 
\end{equation}
It is straightforward to see that there are $4$ linear combinations of the $8$ generators in \equref{GeneratorsIVCP} (in \equref{GeneratorsIVCM}) for the IVC$_+$ (IVC$_-$) that satisfy this requirement. For $\vec{k}\in\mathcal{A}_{\pm}$, these are
\begin{subequations}\begin{align}
    \frac{1}{2\sqrt{2}} \left[\zeta_y (\sigma_0(\eta_0 \pm \eta_z) + \sigma_z \eta_x) \pm \zeta_x \sigma_z\eta_y \right], \qquad \frac{1}{2\sqrt{2}} \left[\zeta_x (\sigma_y(\eta_0 \pm \eta_z) \pm \sigma_x \eta_y) + \zeta_y \sigma_x\eta_x \right], \label{FourGeneratorsForIVCpFirst} \\
    \frac{1}{2\sqrt{2}} \left[\zeta_y (\sigma_z(\eta_0 \pm \eta_z) + \sigma_0 \eta_x) \pm \zeta_x \sigma_0\eta_y \right], \qquad \frac{1}{2\sqrt{2}} \left[\zeta_y (\sigma_x(\eta_0 \pm \eta_z) \mp \sigma_y \eta_y) + \zeta_x \sigma_y\eta_x \right], \label{FourGeneratorsForIVCpSecond}
\end{align}\label{FourGeneratorsForIVCp}\end{subequations}
for the IVC$_+$ and 
\begin{subequations}\begin{align}
    \frac{1}{2\sqrt{2}} \left[\zeta_x (\sigma_y(\eta_0 \pm \eta_z) + \sigma_z \eta_y) \pm \zeta_y \sigma_z\eta_x \right], \qquad \frac{1}{2\sqrt{2}} \left[\zeta_y (\sigma_0(\eta_0 \pm \eta_z) \mp \sigma_x \eta_x) - \zeta_x \sigma_x\eta_y \right], \label{FourGeneratorsForIVCmFirst} \\
    \frac{1}{2\sqrt{2}} \left[\zeta_y (\sigma_z(\eta_0 \pm \eta_z) + \sigma_y \eta_y) \mp \zeta_x \sigma_y\eta_x \right], \qquad \frac{1}{2\sqrt{2}} \left[\zeta_y (\sigma_x(\eta_0 \pm \eta_z) \mp \sigma_0 \eta_x) + \zeta_x \sigma_0\eta_y \right], \label{FourGeneratorsForIVCmSecond}
\end{align}\label{FourGeneratorsForIVCm}\end{subequations}
for the IVC$_-$.

Taken together, we use $U_{\vec{k}} = e^{i\varphi_{\vec{k}}\lambda_{\pm}}$ for $\vec{k}\in\mathcal{A}_\pm$ and $U_{\vec{k}} = \mathbbm{1}$ otherwise, where $\lambda_{\pm}$ is any of the four generators in \equref{FourGeneratorsForIVCp} for the IVC$_+$ and any of those in \equref{FourGeneratorsForIVCm} for the IVC$_-$. From \equref{ChangeOfHFEnergy2}, we can compute the associated change of energy that is found to be of the form
\begin{equation}
    \Delta E^{\text{HF}} = \sum_{\vec{k}\in\mathcal{A}_+} \alpha_{\vec{k}} \sin^2 \varphi_{\vec{k}} + \beta_{\vec{k}} \sin \varphi_{\vec{k}} \cos \varphi_{\vec{k}}; \label{ParameterizationOfDeltaEHF}
\end{equation}
to reduce the summation to momenta in $\mathcal{A}_+$ only, we have used $\varphi_{\vec{k}} = \varphi_{-\vec{k}}$. For the IVC$_+$ with $Q^{b} = \eta_x$, this follows from the fact that $C_{2z}$ as given in \tableref{ActionOfSymmetries} acts as $\eta_x$ and that $\eta_x\lambda_{\pm}\eta_x = \lambda_{\mp}$ for all generators in \equref{FourGeneratorsForIVCp}. For the IVC$_-$ with $Q^{b} =\sigma_y\eta_y$, $C_{2z}$ has to be combined with the U(1)$_v$ valley rotation $i\eta_z$ to be a symmetry (with action on the spinors given by $\eta_y\sigma_0$). However, as follows from \equref{FourGeneratorsForIVCm}, it holds $\eta_y\lambda_{\pm}\eta_y = \lambda_{\mp}$ in this case, leading to the same result, $\varphi_{\vec{k}} = \varphi_{-\vec{k}}$.

For ease of presentation, let us discuss the behavior of $\Delta E^{\text{HF}}$ separately for the two IVCs and begin with the IVC$_+$ state. Focusing on the limit $w_0=0$ for now, we obtain $\beta_{\vec{k}} = 0$ (and $\alpha_{\vec{k}} > 0$ at least in a finite range of $D_0$ around $D_0=0$) for the generators in \equref{FourGeneratorsForIVCpSecond} such that $\Delta E^{\text{HF}}$ reaches its minimum $\Delta E^{\text{HF}}=0$ when $\sin \varphi_{\vec{k}}=0$ (no mixing). For the first and the second generator in \equref{FourGeneratorsForIVCpFirst}, we find 
\begin{subequations}\begin{align}
    \alpha_{\vec{k}} &= 4\left(2\epsilon_{(g,+),+}(\vec{k}) - \sum_{\eta=\pm}\epsilon_{(b,+),\eta}(\vec{k})\right) + \frac{8}{N}\sum_{\vec{q}} V(\vec{q}) \left[ \sum_{j=1}^2 \left(\bar{F}^{bb}_{j}(\vec{k},\vec{q})\right)^2 - \sum_{j=1}^4 (-1)^j \left(\bar{F}^{gg}_{j}(\vec{k},\vec{q})\right)^2\right] + \mathcal{O}(D_0^2), \\
    \beta_{\vec{k}} &= \frac{8\sqrt{2}}{N}\sum_{\vec{q}} V(\vec{q}) \left[ \sum_{j=1}^4\bar{F}^{gg}_{j}(\vec{k},\vec{q})\bar{F}^{gb}_{j}(\vec{k},\vec{q}) - \sum_{j=1}^2\bar{F}^{bb}_{j}(\vec{k},\vec{q})\bar{F}^{bg}_{j}(\vec{k},\vec{q}) \right] D_0 + \mathcal{O}(D_0^3), \label{Alpha1}
\end{align}\label{EnIVCp1}\end{subequations}
and 
\begin{subequations}\begin{align}
    \alpha_{\vec{k}} &= 4\left(2\epsilon_{(g,+),+}(\vec{k}) + \sum_{\eta=\pm}\epsilon_{(b,+),\eta}(\vec{k})\right) + \frac{8}{N}\sum_{\vec{q}} V(\vec{q}) \left[ \sum_{j=1}^2 \left(\bar{F}^{bb}_{j}(\vec{k},\vec{q})\right)^2 - \sum_{j=1}^4 (-1)^j \left(\bar{F}^{gg}_{j}(\vec{k},\vec{q})\right)^2\right] + \mathcal{O}(D_0^2), \\
    \begin{split}\beta_{\vec{k}} &= \frac{8\sqrt{2}}{N}\sum_{\vec{q}} V(\vec{q}) \Bigl[ \bar{F}^{bb}_{1}(\vec{k},\vec{q})\bar{F}^{bg}_{2}(\vec{k},\vec{q}) - \bar{F}^{bb}_{2}(\vec{k},\vec{q})\bar{F}^{bg}_{1}(\vec{k},\vec{q}) +\bar{F}^{gg}_{2}(\vec{k},\vec{q})\bar{F}^{gb}_{1}(\vec{k},\vec{q}) \label{Alpha2} \\ 
    &\qquad\qquad - \bar{F}^{gg}_{1}(\vec{k},\vec{q})\bar{F}^{gb}_{2}(\vec{k},\vec{q}) + \bar{F}^{gg}_{4}(\vec{k},\vec{q})\bar{F}^{gb}_{3}(\vec{k},\vec{q}) - \bar{F}^{gg}_{3}(\vec{k},\vec{q})\bar{F}^{gb}_{4}(\vec{k},\vec{q})    \Bigr] D_0 + \mathcal{O}(D_0^3), \end{split}
\end{align}\label{EnIVCp2}\end{subequations}
respectively. To arrive at these expressions, we have used the $C_{2z}$ constraint in \equref{C2zConstraint}. 

We first observe that, in both cases, $\beta_{\vec{k}}=0$ for $D_0=0$ and, hence, the energy (\ref{ParameterizationOfDeltaEHF}) is simply minimized when $\sin \varphi_{\vec{k}}=0$, i.e., no mixing between the graphene and TBG bands occurs. This is expected since the presence of $\sigma_h$ implies that the order parameters have to be either even (intra-system) or odd (inter-system) under $\sigma_h$, prohibiting mixing between the bands. 

When $D_0 \neq 0$, the system can immediately gain energy by allowing for non-zero $\varphi_{\vec{k}}$: $\Delta E^{\text{HF}}$ in \equref{ParameterizationOfDeltaEHF} is minimized by $\varphi_{\vec{k}}=\varphi_{\vec{k}}^0$ obeying $\tan 2 \varphi^0_{\vec{k}} = - \beta_{\vec{k}}/\alpha_{\vec{k}}$ with minimal energy given by
\begin{equation}
    \Delta E^{\text{HF}}_0=\text{min}_{\varphi_{\vec{k}}}\Delta E^{\text{HF}}[\varphi_{\vec{k}}] = \frac{1}{2}\sum_{\vec{k}\in\mathcal{A}_+} \left(  \alpha_{\vec{k}} - \sqrt{\alpha_{\vec{k}}^2+\beta_{\vec{k}}^2} \right) \leq 0.  \label{GeneralEnergyGain}
\end{equation}
As long as $\beta_{\vec{k}}\neq 0$, this lowers the energy, which shows that the system will exhibit hybridization between the bands for $D_0\neq 0$ as described by the two generators in \equref{FourGeneratorsForIVCpFirst}, while those in \equref{FourGeneratorsForIVCpSecond} are disfavored energetically by the interactions. While this only holds for $w_0=0$, we still expect the generators in \equref{FourGeneratorsForIVCpFirst} to be dominant in an extended region of finite $w_0$. As we will see shortly, this is confirmed by our numerics. 

To obtain the scaling of the energy gain with (small) $D_0$, we expand \equref{GeneralEnergyGain} in $\beta_{\vec{k}} \propto D_0$, which yields 
\begin{equation}
    \Delta E^{\text{HF}}_0 \propto -\sum_{\vec{k}\in\mathcal{A}_+} \frac{\beta_{\vec{k}}^2}{4\alpha_{\vec{k}}} \propto -g_3 V_{\mathcal{A}} D_0^2,
\end{equation}
where we indicated that the result will be proportional to the area $V_{\mathcal{A}} \propto (\Delta k)^2$ (our small parameter) and the displacement field squared; the constant $g_1>0$ contains all the non-universal properties and itself depends on $D_0$ and $w_0$. 

Which of these two remaining generators in \equref{FourGeneratorsForIVCpFirst} (or which linear combination) is preferred by the system is the \textit{only} aspect that depends on the microscopic details of the form factors; it is not possible to say which of the $\beta_{\vec{k}}$ in \equsref{Alpha1}{Alpha2} is dominant at a given $\vec{k}$ without computing the form factors in the microscopic model. The associated mixing of the bands for these two cases are readily evaluated from \equref{QExpansion}. We get
\begin{equation}
    Q^{\text{b}} \mathcal{M}_{\vec{k}} + \mathcal{M}_{\vec{k}} \sigma_z = 2 \mathcal{M}_{\vec{k}} \sigma_z = \sqrt{2} i \sigma_z (\eta_0 \pm \eta_z) + \sqrt{2} \sigma_0 (i\eta_x \pm \eta_y)  ,\quad \vec{k}\in\mathcal{A}_\pm
\end{equation}
for the first term in \equref{FourGeneratorsForIVCpFirst}. This means that the graphene-like band with positive (negative) energy in the valley that is closer to the Fermi level mixes with the TBG-like band of both valleys that are also at positive (negative) energies. 

For the second generator in \equref{FourGeneratorsForIVCpFirst}, we find a mixing matrix of the form
\begin{equation}
    Q^{\text{b}} \mathcal{M}_{\vec{k}} + \mathcal{M}_{\vec{k}} \sigma_z = \sqrt{2} i \sigma_y (\eta_0 \pm \eta_z) - \sqrt{2}i \sigma_x (i\eta_x \pm \eta_y)  ,\quad \vec{k}\in\mathcal{A}_\pm. \label{SecondMixingMatrix}
\end{equation}
Instead, we here get that the graphene-like band with positive (negative) energy in the valley that is closer to the Fermi level mixes with the TBG-like bands of both valleys that are at negative (positive) energies. In our numerical HF computations, we find very good agreement with the mixing matrix in \equref{SecondMixingMatrix}, see \figref{MatrixOrders}, even at the moderately large value of $w_0/w_1=0.73$.

The energetic discussion can be done in a similar way for the IVC$_-$ state and we find quite similar structures: first, we get $\beta_{\vec{k}}=0$ for $w_0=0$ for the two generators in \equref{FourGeneratorsForIVCmSecond} and, thus, no energetic gain via mixing. Second, this is different for those in \equref{FourGeneratorsForIVCmFirst}: the first one leads to
\begin{subequations}\begin{align}
    \alpha_{\vec{k}} &= 4\left(2\epsilon_{(g,+),+}(\vec{k}) + \sum_{\eta=\pm}\eta \epsilon_{(b,+),\eta}(\vec{k})\right) + \frac{8}{N}\sum_{\vec{q}} V(\vec{q}) \left[ \sum_{j=1}^2 \left(\bar{F}^{bb}_{j}(\vec{k},\vec{q})\right)^2 - \sum_{j=1}^4 (-1)^j \left(\bar{F}^{gg}_{j}(\vec{k},\vec{q})\right)^2\right] + \mathcal{O}(D_0^2), \\
    \begin{split}\beta_{\vec{k}} &= \frac{8\sqrt{2}}{N}\sum_{\vec{q}} V(\vec{q}) \Bigl[ \bar{F}^{bb}_{1}(\vec{k},\vec{q})\bar{F}^{bg}_{2}(\vec{k},\vec{q}) - \bar{F}^{bb}_{2}(\vec{k},\vec{q})\bar{F}^{bg}_{1}(\vec{k},\vec{q}) +\bar{F}^{gg}_{2}(\vec{k},\vec{q})\bar{F}^{gb}_{1}(\vec{k},\vec{q})  \\ 
    &\qquad\qquad - \bar{F}^{gg}_{1}(\vec{k},\vec{q})\bar{F}^{gb}_{2}(\vec{k},\vec{q}) + \bar{F}^{gg}_{4}(\vec{k},\vec{q})\bar{F}^{gb}_{3}(\vec{k},\vec{q}) - \bar{F}^{gg}_{3}(\vec{k},\vec{q})\bar{F}^{gb}_{4}(\vec{k},\vec{q})    \Bigr] D_0 + \mathcal{O}(D_0^3), \end{split}
\end{align}\label{EnIVCm1}\end{subequations}
while the second one is associated with the coefficients
\begin{subequations}\begin{align}
    \alpha_{\vec{k}} &= 4\left(2\epsilon_{(g,+),+}(\vec{k}) - \sum_{\eta=\pm}\eta\epsilon_{(b,+),\eta}(\vec{k})\right) + \frac{8}{N}\sum_{\vec{q}} V(\vec{q}) \left[ \sum_{j=1}^2 \left(\bar{F}^{bb}_{j}(\vec{k},\vec{q})\right)^2 - \sum_{j=1}^4 (-1)^j \left(\bar{F}^{gg}_{j}(\vec{k},\vec{q})\right)^2\right] + \mathcal{O}(D_0^2), \\
    \beta_{\vec{k}} &= \frac{8\sqrt{2}}{N}\sum_{\vec{q}} V(\vec{q}) \left[ \sum_{j=1}^4\bar{F}^{gg}_{j}(\vec{k},\vec{q})\bar{F}^{gb}_{j}(\vec{k},\vec{q}) - \sum_{j=1}^2\bar{F}^{bb}_{j}(\vec{k},\vec{q})\bar{F}^{bg}_{j}(\vec{k},\vec{q}) \right] D_0 + \mathcal{O}(D_0^3) \label{Alpha1}
\end{align}\label{EnIVCm2}\end{subequations}
in \equref{ParameterizationOfDeltaEHF}. As before, while $D_0=0$ does not lead to mixing, the bands start to mix at non-zero $D_0$; this mixing is described by the two generators in \equref{FourGeneratorsForIVCmFirst} and which of the two dominates depends on microscopic details. 
The nature of the mixing depends on the generator. For the first one in \equref{FourGeneratorsForIVCmFirst}, the mixing matrix reads as
\begin{equation}
    2 \mathcal{M}_{\vec{k}} \sigma_z = \sqrt{2} i \sigma_x (\eta_0 \pm \eta_z) + \sqrt{2} \sigma_0 (\eta_y \pm i \eta_x)  ,\quad \vec{k}\in\mathcal{A}_\pm. \label{FirstMixingMatrixIVCm}
\end{equation}
This means that the lower/upper graphene-like band in the valley that is at low energies hybridizes with the upper/lower (lower/upper) TBG-like band of the same (opposite) valley. For the second generator in \equref{FourGeneratorsForIVCmFirst}, we get instead
\begin{equation}
    2 \mathcal{M}_{\vec{k}} \sigma_z = \sqrt{2} i \sigma_z (\eta_0 \pm \eta_z) + \sqrt{2} i \sigma_y (\eta_y \pm i \eta_x)  ,\quad \vec{k}\in\mathcal{A}_\pm,
\end{equation}
i.e., the lower/upper graphene-like band in the valley that is at low energies hybridizes with the upper/lower (lower/upper) TBG-like band of the  opposite (same) valley.

Comparison of \equsref{EnIVCp2}{EnIVCm1} [the same holds for \equsref{EnIVCp1}{EnIVCm2}] shows that the corresponding prefactors and, hence, the energies are identical to leading order in $D_0$ (note the subleading corrections due to the kinetic term). Since our numerics has identified the mixing matrix in \equref{SecondMixingMatrix} to be realized in the system for the IVC$_+$ and the parameters in \figref{MatrixOrders}, we thus expect \equref{FirstMixingMatrixIVCm} to dominate for the IVC$_-$; this is indeed confirmed by \figref{MatrixOrders}. Another consequence of this observation is that the leading (in $D_0$ and $w_0$) correction to the energy of the IVC$_{-}$ is identical to that of the IVC$_+$, given by $-g_1 V_{\mathcal{A}} D_0^2$. As we have seen above, further corrections to higher orders in $D_0$ (e.g., due to the kinetic terms) are different for these two IVC states. These are not included in \tableref{BehaviorOfEnergetics}.

\vspace{1em}

We next continue with the \textit{spin polarized} state, $Q^{\text{b}}=\sigma_0\eta_0s_3$. Since the valley U(1)$_v$ symmetry is preserved, $U_{\vec{k}}$ has to be diagonal in valley space. We can write 
\begin{equation}
    \left(U_{\vec{k}}\right)_{\eta,\eta'} = s_0 \sigma_0 \delta_{\eta,\eta'} e^{i \sum_{j=1}^3 \varphi^{(j)}_{\vec{k},\eta} \zeta_j}. \label{GeneralFormOfU}
\end{equation}
While this form of $U_{\vec{k}}$ automatically preserves all unitary symmetries in \tableref{ActionOfSymmetries} (except for $P$, which, however, is broken anyway for $D_0\neq 0$), it only preserves $C_{2z}\Theta$ if $\varphi^{(1)}=\varphi^{(3)}=0$ and we are left with
\begin{equation}
    \left(U_{\vec{k}}\right)_{\eta,\eta'} = s_0 \sigma_0 \delta_{\eta,\eta'} e^{i  \varphi_{\vec{k},\eta} \zeta_2} \label{UForSpinPol}
\end{equation}
with the additional constraint $\varphi_{\vec{k},\eta} = \varphi_{-\vec{k},-\eta}$, which comes from $\Theta$ (or, equivalently, $C_{2z}$).
Our task now it to determine the real-valued function $\varphi_{\vec{k},\eta}$ such that the energy is minimized. We expect the mixing to be the strongest (if at all present) around the K and K$'$ points for valley $\eta=+$ and $\eta=-$, respectively, and, thus, assume that $\varphi_{\vec{k},\eta} \neq 0$ only in the region $\mathcal{A}_{\eta}$ and expand in its area leading to \equref{ChangeOfHFEnergy2}. The energy change due to $U_{\vec{k}} \neq \pm \mathbbm{1}$ in \equref{UForSpinPol} for the spin polarized state is found to be
\begin{align}
    \Delta E[\varphi_{\vec{k},\eta}] &= \sum_\eta\sum_{\vec{k}\in\mathcal{A}_\eta} \left( a_{\vec{k},\eta} \sin^2 \varphi_{\vec{k},\eta}  + b_{\vec{k},\eta} \sin \varphi_{\vec{k},\eta} \cos \varphi_{\vec{k},\eta}\right)  \label{GeneralFormOfTheEnergy} \\
    &= \sum_{\vec{k}\in\mathcal{A}_+} \left( a_{\vec{k}} \sin^2 \varphi_{\vec{k},+}  + b_{\vec{k}} \sin \varphi_{\vec{k},+} \cos \varphi_{\vec{k},+}\right) , \qquad a_{\vec{k}} := \sum_{\eta} a_{\eta \vec{k},\eta}, \quad b_{\vec{k}} := \sum_{\eta} b_{\eta\vec{k},\eta}, \label{ReducedFormOfEnergy}
\end{align}
where we used the constraint $\varphi_{\vec{k},\eta} = \varphi_{-\vec{k},-\eta}$ in the second line. 
Here and in all of the following computations of this appendix, we focus on terms that do not vanish for $w_0=0$. This is motivated by the fact that the final energy gain will already be of order $D_0^2$ and so further corrections of order $w_0$ of its prefactor will be subleading. We find
\begin{subequations}\begin{align}
\begin{split}a_{\vec{k}} &= 2\sum_{p=\pm} p \left(  \epsilon_{(g,p),+}(\vec{k}) -  \epsilon_{(b,p),+}(\vec{k}) \right) \\ &+ \frac{4}{N} \sum_{\vec{q}} V(\vec{q}) \Bigl[ (\bar{F}^{bb}_{1,+}(\vec{k},\vec{q}))^2 + (\bar{F}^{bb}_{2,+}(\vec{k},\vec{q}))^2 + (\bar{F}^{gg}_{1,+}(\vec{k},\vec{q}))^2 - (\bar{F}^{gg}_{2,+}(\vec{k},\vec{q}))^2 \Bigr]  ,\\
&- \frac{4}{N} \sum_{\vec{q}} V(\vec{q}) \Bigl[ (\bar{F}^{bg}_{1,+}(\vec{k},\vec{q}))^2 + (\bar{F}^{gb}_{1,+}(\vec{k},\vec{q}))^2 + (\bar{F}^{bg}_{2,+}(\vec{k},\vec{q}))^2 - (\bar{F}^{gb}_{2,+}(\vec{k},\vec{q}))^2 \Bigr] D_0^2 \end{split} \\ 
    b_{\vec{k}} &=  -\frac{8}{N} \sum_{\vec{q}} V(\vec{q}) \sum_{j=1,2} \Bigl[ \bar{F}^{bb}_{j,+} (\vec{k},\vec{q}) \bar{F}^{bg}_{j,+}(\vec{k},\vec{q}) - \bar{F}^{gg}_{j,+} (\vec{k},\vec{q}) \bar{F}^{gb}_{j,+}(\vec{k},\vec{q}) \Bigr] D_0.
\end{align}\label{abForSpinPol}\end{subequations}

We see that $b_{\vec{k},\eta}=0$ if $D_0=0$ while $a_{\vec{k},\eta}>0$. So $\sin \varphi_{\vec{k},\eta}=0$ minimizes the energy and, hence, there is no mixing between the bands for $D_0 = 0$, as expected by symmetry. However, any arbitrarily small non-zero $D_0$ leads to $b_{\vec{k},\eta}\neq 0$ and the energy is lowered when the bands hybridize: \equref{GeneralFormOfTheEnergy} is minimized by $\varphi_{\vec{k},\eta}^0$ obeying $\tan 2 \varphi^0_{\vec{k},\eta} = - b_{\vec{k},\eta}/a_{\vec{k},\eta}$ and with energy change
\begin{equation}
    \text{min}_{\varphi_{\vec{k},+}}\Delta E[\varphi_{\vec{k},+}] = \frac{1}{2}\sum_{\vec{k}\in\mathcal{A}_+} \left(  a_{\vec{k}} - \sqrt{a_{\vec{k}}^2+b_{\vec{k}}^2} \right) \leq 0.  \label{SpinPolEnGain}
\end{equation}
Since $b_{\vec{k}} \propto D_0$, the energy correction scales as $-g_1 V_{\mathcal{A}} D_0^2$, as indicated schematically in \tableref{BehaviorOfEnergetics}.

\vspace{1em}

\textit{Valley polarized.} This state also preserves the $C_{2z}\Theta$ and valley U(1)$_v$ symmetries, such that $U_{\vec{k}}$ is again of the form of \equref{UForSpinPol}; this time, however, without the constraint $\varphi_{\vec{k},\eta} = \varphi_{-\vec{k},-\eta}$ as $C_{2z}$ and $\Theta$ are both broken individually. The energy change is of the form of \equref{GeneralFormOfTheEnergy} with coefficients given by (again, only keeping terms to zeroth order in $w_0$)
\begin{subequations}
\begin{align}
\begin{split}a_{\vec{k},\eta} &= 2 \left(  \epsilon_{(g,+),\eta}(\vec{k}) -  \epsilon_{(b,+),\eta}(\vec{k}) \right) \\ &+ \frac{2}{N} \sum_{\vec{q}} V(\vec{q}) \Bigl[ (\bar{F}^{bb}_{1,\eta}(\vec{k},\vec{q}))^2 + (\bar{F}^{bb}_{2,\eta}(\vec{k},\vec{q}))^2 + (\bar{F}^{gg}_{1,\eta}(\vec{k},\vec{q}))^2 - (\bar{F}^{gg}_{2,\eta}(\vec{k},\vec{q}))^2 \Bigr]  ,\\
&- \frac{2}{N} \sum_{\vec{q}} V(\vec{q}) \Bigl[ (\bar{F}^{bg}_{1,\eta}(\vec{k},\vec{q}))^2 + (\bar{F}^{gb}_{1,\eta}(\vec{k},\vec{q}))^2 + (\bar{F}^{bg}_{2,\eta}(\vec{k},\vec{q}))^2 - (\bar{F}^{gb}_{2,\eta}(\vec{k},\vec{q}))^2 \Bigr] D_0^2 \label{aketaVP} \end{split} \\ 
    b_{\vec{k},\eta} &=-\frac{4}{N} \sum_{\vec{q}} V(\vec{q}) \sum_{j=1,2} \Bigl[ \bar{F}^{bb}_{j,\eta} (\vec{k},\vec{q}) \bar{F}^{bg}_{j,\eta}(\vec{k},\vec{q}) - \bar{F}^{gg}_{j,\eta} (\vec{k},\vec{q}) \bar{F}^{gb}_{j,\eta}(\vec{k},\vec{q}) \Bigr] D_0.
\end{align}\label{FormOfaandb}\end{subequations}
Minimization yields
\begin{equation}
    \text{min}_{\varphi_{\vec{k},\eta}}\Delta E[\varphi_{\vec{k},\eta}] = \frac{1}{2}\sum_\eta\sum_{\vec{k}\in\mathcal{A}_\eta} \left(  a_{\vec{k},\eta}^2 - \sqrt{a_{\vec{k},\eta}^2+b_{\vec{k},\eta}^2} \right) \leq 0. \label{EnergyReduction}
\end{equation}
It is easy to see that $a_{\vec{k},\eta}=a_{-\vec{k},-\eta}$ and $b_{\vec{k},\eta}=b_{-\vec{k},-\eta}$, following from $C_{2z}$ symmetry of the form factors and dispersion, and that $a_{\vec{k}}$ and $b_{\vec{k}}$ in \equref{abForSpinPol} are given by $a_{\vec{k}} = 2a_{\vec{k},+}$ and $b_{\vec{k}}=2b_{\vec{k},+}$. As such the energy gain is the same for the spin and for the valley polarized states at $w_0=0$. 

\vspace{1em}

Next, we discuss the \textit{quantum Hall state} SLP$_-$. While this state also preserves valley U(1)$_v$ symmetry and spin rotations, it breaks the $\vec{k}$-space-local $C_{2z}\Theta$ symmetry. As such, the transformation has to be of the more general form in \equref{GeneralFormOfU}, involving three generators, $\zeta_{x,y,y}$. Due to the preserved $C_{2z}$ symmetry, we have $\varphi_{\vec{k},+}^{(j)} =  \varphi_{-\vec{k},-}^{(j)}$. Exactly as before, general energetic reasoning leads to $\varphi_{\vec{k},\eta}^{(j)} = 0$ if $\vec{k}\notin \mathcal{A}_\eta$.
Instead of optimizing all three associated angles $\varphi_{\vec{k},+}^{(j)}$ simultaneously, let us first investigate them one at a time: first of all, $j=3$ leads to $\Delta E^{\text{HF}} = 0$ (for $w_0=0$) in \equref{ChangeOfHFEnergy2}. The generator $\zeta_x$ does lead to non-zero $\Delta E^{\text{HF}}$, which is found to be of the form of \equref{GeneralFormOfTheEnergy}; this time with $b_{\vec{k},\eta}=0$ and $a_{\vec{k},\eta}$ as given in \equref{aketaVP}. Since $a_{\vec{k},\eta} >0$, at least for sufficiently small $D_0$, the system does not gain any energy as a consequence of hybridization described by $(U_{\vec{k}})_{\eta,\eta'} = \delta_{\eta,\eta'}e^{i  \varphi_{\vec{k},\eta} \zeta_x}$. The only generator left is $j=2$, i.e., $(U_{\vec{k}})_{\eta,\eta'} = \delta_{\eta,\eta'}e^{i  \varphi_{\vec{k},\eta} \zeta_y}$; in fact, this one allows for a reduction of energy and we obtain, to order $w_0^0$, exactly the same of $a_{\vec{k},\eta}$ and $b_{\vec{k},\eta}$ as defined in \equref{FormOfaandb}. In order to take into account the $C_{2z}$ constraint, $\varphi_{\vec{k},\eta} = \varphi_{-\vec{k},-\eta}$, we again simplify from \equref{GeneralFormOfTheEnergy} to \equref{ReducedFormOfEnergy}. Upon noting, as above, that $a_{\vec{k},\eta}=a_{-\vec{k},-\eta}$ and $b_{\vec{k},\eta}=b_{-\vec{k},-\eta}$, we have $a_{\vec{k}} = 2a_{\vec{k},+}$ and $b_{\vec{k}}=2b_{\vec{k},+}$ and the energy reduction due to hybridization of the SLP$_-$ is the same as that of the VP state (at $w_0=0$).   

Upon noting that $\mathcal{M}_{\vec{k}} = i \sigma_0 s_0 (\eta_0 \pm \eta_z)$, $\vec{k}\in\mathcal{A}_{\pm}$, in \equref{QExpansion} for our $U_{\vec{k}}$ studied here, it is easy to compute the mixing matrix: 
\begin{equation}
    Q^{\text{b}} \mathcal{M}_{\vec{k}} + \mathcal{M}_{\vec{k}} \sigma_z = i (\sigma_y + \sigma_z)(\eta_0 \pm \eta_z), \qquad \vec{k}\in\mathcal{A}_{\pm},
\end{equation}
i.e., both graphene-like bands in the valley that is at low energies mixes with both TBG-like bands in the same valley. This agrees well with the correlator near the K point of the SLP$_-$ state obtained in HF, see \figref{MatrixOrders}.

\vspace{1em}

To continue with the \textit{valley Hall state} SLP$_+$, we note that $C_{2z}\Theta$ is also broken in this state, leading again to the general form of $U_{\vec{k}}$ given in \equref{GeneralFormOfU} with three generators. This time $\Theta$ (rather than $C_{2z}$) is preserved and we have $\varphi_{\vec{k},+}^{(j)} = (-1)^j \varphi_{-\vec{k},-}^{(j)}$. As above, we study first each generator, $j=1,2,3$, separately and note that $j=1$ and $j=3$ do not allow to lower the energy (again, for $w_0=0$ and not too large values of $D_0$). However, $j=2$ does allow to lower the energy via hybridization for any non-zero $D_0$. Writing $\varphi_{\vec{k},+}=\varphi_{\vec{k},+}^{(2)}$, the energy gain is of the form (\ref{ReducedFormOfEnergy}) with prefactors
\begin{subequations}\begin{align}
\begin{split}a_{\vec{k}} &= 2\sum_{p=\pm} p \left(  \epsilon_{(g,p),+}(\vec{k}) -  \epsilon_{(b,p),+}(\vec{k}) \right) \\ &+ \frac{4}{N} \sum_{\vec{q}} V(\vec{q}) \Bigl[ (\bar{F}^{bb}_{1,+}(\vec{k},\vec{q}))^2 + (\bar{F}^{bb}_{2,+}(\vec{k},\vec{q}))^2 + (\bar{F}^{gg}_{1,+}(\vec{k},\vec{q}))^2 - (\bar{F}^{gg}_{2,+}(\vec{k},\vec{q}))^2 \Bigr]  ,\\
&- \frac{4}{N} \sum_{\vec{q}} V(\vec{q}) \Bigl[ (\bar{F}^{bg}_{1,+}(\vec{k},\vec{q}))^2 + (\bar{F}^{gb}_{1,+}(\vec{k},\vec{q}))^2 + (\bar{F}^{bg}_{2,+}(\vec{k},\vec{q}))^2 - (\bar{F}^{gb}_{2,+}(\vec{k},\vec{q}))^2 \Bigr] D_0^2  \\
& +\frac{32}{N} \sum_{\vec{G}} V(\vec{G}) \left(\sum_{\vec{k}'} \bar{F}_4^{bb}(\vec{k}',\vec{G}) \right) \left( \bar{F}^{bb}_{2,+}(\vec{k},\vec{G}) - \bar{F}^{gg}_{2,+}(\vec{k},\vec{G}) \right) D_0^2 \end{split} \\ 
    b_{\vec{k}} &=  -\frac{8}{N} \sum_{\vec{q}} V(\vec{q}) \sum_{j=1,2} \Bigl[ \bar{F}^{bb}_{j,+} (\vec{k},\vec{q}) \bar{F}^{bg}_{j,+}(\vec{k},\vec{q}) - \bar{F}^{gg}_{j,+} (\vec{k},\vec{q}) \bar{F}^{gb}_{j,+}(\vec{k},\vec{q}) \Bigr] D_0.
\end{align}\end{subequations}
Comparison with the expressions for the VP and SLP$_-$ shows that the energy reduction due to hybridization is almost the same for the SLP$_+$. The only difference is the additional Hartree term in $a_{\vec{k}}$, which is expected to be only a very small correction due to \equref{ZeroqLimit} and since it is higher order in $D_0$. As such, we have $g_3 \approx g_1$ in \tableref{BehaviorOfEnergetics}.

The corresponding $\mathcal{M}_{\vec{k}} = i \sigma_0 s_0 (\eta_0 \pm \eta_z)$, $\vec{k}\in\mathcal{A}_{\pm}$, in \equref{QExpansion} leads to the mixing contribution in $Q_{\vec{k}}$
\begin{equation}
    Q^{\text{b}} \mathcal{M}_{\vec{k}} + \mathcal{M}_{\vec{k}} \sigma_z = i (\sigma_y \eta_z + \sigma_z)(\eta_0 \pm \eta_z), \qquad \vec{k}\in\mathcal{A}_{\pm}.
\end{equation}
This means that both the upper and the lower graphene-like bands in the valley that is at low energies mixes with both TBG-like bands in the same valley, in perfect agreement with \figref{MatrixOrders}.

\vspace{1em}

Finally, the situation is slightly more complex for the SSLP$_-$ state, as it also breaks SU(2) spin-rotation symmetry down to residual rotation along one spin axis (say $s_z$). Consequently, we will have to allow for six generators,
\begin{equation}
    \left(U_{\vec{k}}\right)_{\eta,\eta'} = \sigma_0 \delta_{\eta,\eta'} e^{i \sum_{j=1}^3 ( \varphi^{(j,0)}_{\vec{k},\eta} \zeta_js_0 + \varphi^{(j,3)}_{\vec{k},\eta} \zeta_js_3 )}, \qquad \varphi^{(j,a)}_{\vec{k},\eta} = \varphi^{(j,a)}_{-\vec{k},-\eta},
\end{equation}
where the constraint comes from $C_{2z}$ symmetry. Let us investigate every generator separately. First, one finds no change in energy for $\zeta_3 s_0$ and $\zeta_3 s_3$. In all other cases, the energy change is again of the form (\ref{GeneralFormOfTheEnergy}). For $\zeta_{1,2} s_3$ and $\zeta_1s_0$ we get $b_{\vec{k},\eta} = 0$ and $a_{\vec{k},\eta} > 0$ (for sufficiently small $D_0$) and, hence, no associated energetic gain. Hence, we are, again, only left with a single generator and $(U_{\vec{k}})_{\eta,\eta'} = \delta_{\eta,\eta'} e^{i  \varphi_{\vec{k},\eta} \zeta_2 s_0}$ with $ \varphi_{\vec{k},\eta} =  \varphi_{-\vec{k},\eta}$. The prefactors are given by \equref{FormOfaandb} and we obtain the same energetic gain as for the SLP$_-$ (or VP and SP) state. As anticipated in \appref{PolarizationOfGrapheneLikeBands}, this mixing induces a gap for the SSLP$_-$ state as well, as also seen in our HF numerics, despite not explicitly ordering in the graphene-like bands (for small $D_0$).

\subsection{Hund's Coupling}\label{HundsCouplingAppendix}
In the real system, SU(2)$_+ \times$ SU(2)$_-$ is broken down to SU(2)$_s$. Keeping U(1)$_v$ as an exact symmetry, any spin-rotation invariant non-interacting Hamiltonian will automatically be invariant under SU(2)$_+ \times$ SU(2)$_-$ and, hence, the explicit breaking of this enhanced spin symmetry can only be captured by interaction terms. We refer to all such interaction terms collectively as ``intervalley Hund's interactions''. 

In order to associate a sign of the Hund's coupling with a member of a given Hund's pair of states, we use the simple explicit form of the Hund's interaction given in \equref{HundsCouplingInteraction} of the main text. By evaluation their respective Hartree and Fock energies, we can express the contribution of this interaction to each state in \tableref{ListOfOrderParameters}:

\begin{equation}
\begin{split}
    \Delta E_{\text{SP}}&=-\Delta E_{\text{SVP}}=\frac{2J_H}{N}\sum_{\vec{G}}\left(\left[\sum_{\vec{k}}\bar{F}_1^{bb}(\vec{k},\vec{G})\right]^2-4\left[\sum_{\vec{k}}\bar{F}_3^{bb}(\vec{k},\vec{G})\right]^2D_0^4\right)\\
    \Delta E_{\text{IVC}_-}&=-3\Delta E_{\text{SIVC}_-}=-\frac{3J_H}{2N}\sum_{\vec{k},\vec{q}}\left[(\bar{F}_{1}^{bb}(\vec{k},\vec{q}))^2+(\bar{F}_{2}^{bb}(\vec{k},\vec{q}))^2-4(\bar{F}_{3}^{bb}(\vec{k},\vec{q}))^2D_0^4-4(\bar{F}_{4}^{bb}(\vec{k},\vec{q}))^2 D_0^4 \right. \\& \left. +4(\bar{F}_{5}^{bb}(\vec{k},\vec{q}))^2w_0^2+4(\bar{F}_{6}^{bb}(\vec{k},\vec{q}))^2w_0^2-(\bar{F}_{7}^{bb}(\vec{k},\vec{q}))^2w_0^2D_0^4-(\bar{F}_{8}^{bb}(\vec{k},\vec{q}))^2w_0^2D_0^4\right]\\
    \Delta E_{\text{IVC}_+}&=-3\Delta E_{\text{SIVC}_+}=-\frac{3J_H}{2N}\sum_{\vec{k},\vec{q}}\left[(\bar{F}_{1}^{bb}(\vec{k},\vec{q}))^2+(\bar{F}_{2}^{bb}(\vec{k},\vec{q}))^2-4(\bar{F}_{3}^{bb}(\vec{k},\vec{q}))^2D_0^4-4(\bar{F}_{4}^{bb}(\vec{k},\vec{q}))^2 D_0^4 \right. \\& \left. -4(\bar{F}_{5}^{bb}(\vec{k},\vec{q}))^2w_0^2-4(\bar{F}_{6}^{bb}(\vec{k},\vec{q}))^2w_0^2+(\bar{F}_{7}^{bb}(\vec{k},\vec{q}))^2w_0^2D_0^4+(\bar{F}_{8}^{bb}(\vec{k},\vec{q}))^2w_0^2D_0^4\right]\\
    \Delta E_{\text{SSLP}_{\pm}}&=\mp\frac{2J_H}{N}\sum_{\vec{G}}\left(\left[\sum_{\vec{k}}\bar{F}_2^{bb}(\vec{k},\vec{G})\right]^2-4\left[\sum_{\vec{k}}\bar{F}_4^{bb}(\vec{k},\vec{G})\right]^2D_0^4\right)\\
    \Delta E_{\text{SLP}_{\pm}}&=0\\
    \Delta E_{\text{VP}}&=0
\end{split}
\end{equation}
The form of the energies in the chiral-decoupled limit, $w_0=D_0=0$, is indicated in the last two columns in \tableref{BehaviorOfEnergetics}. Their signs can be found in the last column of \tableref{ListOfOrderParameters}.

By construction of the order parameters in \tableref{ListOfOrderParameters}, we see that a state and its Hund's partner have an energy contributions from $J_H$ of opposite sign. Furthermore, all states that are their own Hund's partners, are not affected energetically to linear order in $J_H$.

\section{Superconducting instabilities}\label{AppSuperconductivity}

\subsection{Fluctuation-induced superconductivity}\label{AppFluctSC}
Finally, in this appendix we provide more details on how we determine the leading superconducting instabilities in \secref{NoSpinPolarization} that result from fluctuations of the candidate particle-hole orders of MSTG identified in this work. Since the applied formalism is closely connected to \refcite{PhysRevB.102.064501}, we will be brief and focus on the novel aspects relevant here---the spinless valley time-reversal symmetry $\widetilde{\Theta}$ of some of the states in \tableref{ListOfOrderParameters} and what happens in the absence of electron-phonon coupling---and illustrate the formalism explicitly with a minimal model for the IVC$_-$.

Integrating out the bosonic fields $\phi_q^j$ in the action $\mathcal{S}$ in \equref{FullActionForFluctuations} leads to an effective electron-electron interaction. Projecting this interaction to the electronic degrees of freedom, $f_{k;\eta,s}$, with band energies $\epsilon_{\eta}(\vect{k})$, that are closest to the chemical potential for each momentum $\vec{k}$, we perform a decoupling in the Cooper channel (focusing on inter-valley pairing which is expected to be dominant), with Hubbard-Stratonovich fields $\Delta^{\eta}_{s,s'}$, and obtain the saddle-point equations \cite{PhysRevB.102.064501}
\begin{equation}
    \Delta_{s,s'}^\eta(k) = - \sum_{\eta'}\int_{k'} \frac{\mathcal{V}^{s,s'}_{\bar{s},\bar{s}'}(k,\eta;k',\eta')}{\omega_{n'}^2+ \epsilon^2_{\eta'}(\vect{k}')} \Delta_{\bar{s}',\bar{s}}^{\eta'}(k'). \label{SaddlePointEquationGeneral}
\end{equation}
Here the interaction matrix elements have the form $\mathcal{V}^{s_1,s_2}_{s_3,s_4}(k,\eta;k',\eta') = \mathcal{W}^{s_1,s_2}_{s_3,s_4}(k,\eta;k',\eta') + \Delta\mathcal{V}^{s_1,s_2}_{s_3,s_4}(k,\eta;k',\eta')$, where the first part is the SU(2)$_+ \times$ SU(2)$_-$ symmetric contribution of the phonons from $S_{\text{phonon}}$ and the second is a result of the particle-hole fluctuations,
\begin{equation}
    \Delta\mathcal{V}^{s_1,s_2}_{s_3,s_4}(k,\eta;k',\eta') = - \frac{1}{2} \Lambda^j_{-\eta,s_2;-\eta',s_3}(-\vec{k},-\vec{k}')[\chi(k-k')]_{j,j'} \Lambda^{j'}_{\eta,s_1;\eta',s_4}(\vec{k},\vec{k}'),
\end{equation}
where $\Lambda^j_{\eta,s;\eta',s'}(\vec{k},\vec{k}') := \braket{u_{\eta,s}(\vec{k})|\lambda^j(\vec{k},\vec{k}')|u_{\eta',s'}(\vec{k}')}$ are the matrix elements of the coupling vertices $\lambda^j$ in \equref{phifcoupling} with the respect to the wavefunctions $u_{\eta,s}(\vec{k})$ of the low-energy fermions $f_{k;\eta,s}$.

Since the system has SU(2)$_s$ symmetry, we can always decompose pairing into singlet and triplet, which correspond to $\Delta_{s,s'}(k) = (i s_y)_{s,s'} \gamma_{k}^\eta$, $\gamma_{k}^\eta=\gamma_{-k}^{-\eta}$, and $\Delta_{s,s'}(k) = (i s_y \vec{d}\cdot \vec{s})_{s,s'} \Gamma_{k}^\eta$, $\Gamma_{k}^\eta=-\Gamma_{-k}^{-\eta}$, respectively. Their respective saddle-point equations immediately follow from \equref{SaddlePointEquationGeneral} and read as
\begin{subequations}\begin{equation}
    \gamma_{k}^+ = \int_{k'} \frac{M^s_{k,k'}}{\omega_{n'}^2+ \epsilon^2_{+}(\vect{k}')} \gamma_{k'}^+, \quad M^s_{k,k'} = -\frac{1}{4} \sum_{s,s',\eta,\eta'} ss'\, \mathcal{V}^{s,-s}_{-s',s'}(\eta k,\eta;\eta'k',\eta'), 
\end{equation}
and 
\begin{equation}
    \Gamma_{k}^+ = \int_{k'} \frac{M^t_{k,k'}}{\omega_{n'}^2+ \epsilon^2_{+}(\vect{k}')} \Gamma_{k'}^+, \quad M^t_{k,k'} = -\frac{1}{4} \sum_{s,s',\eta,\eta'} \eta\eta'\, \mathcal{V}^{s,-s}_{-s',s'}(\eta k,\eta;\eta'k',\eta'),
\end{equation}\label{SaddlePointEquations}\end{subequations}
respectively.

To make progress, we need to take into account time-reversal symmetry. To begin with $\widetilde{\Theta}$, we have $\widetilde{\Theta} \ket{u_{\eta,s}(\vec{k})} = \eta \ket{u_{-\eta,-s}(-\vec{k})}$ (technically, we are here using $\widetilde{\Theta}=s_x\eta_y \mathcal{K}$ but this can equally well be used as a definition of $\widetilde{\Theta}$ in \tableref{ListOfOrderParameters}). For any set of bosonic modes, $\phi_q^j$ that are all even ($\tilde{t}_\phi=+$) or odd ($\tilde{t}_\phi=-$) under $\widetilde{\Theta}$, we further have $\widetilde{\Theta} \lambda^j(\vec{k},\vec{k}') \widetilde{\Theta}^\dagger = \tilde{t}_\phi\,\lambda^j(-\vec{k},-\vec{k}')$ and, hence,
\begin{equation}
    \Lambda^j_{\eta,s;\eta',s'}(\vec{k},\vec{k}') =\eta \, \eta' \, \tilde{t}_\phi\, \left[\Lambda^j_{-\eta,-s;-\eta',-s'}(-\vec{k},-\vec{k}')\right]^*.
\end{equation}
Similarly, we have  $\Lambda^j_{\eta,s;\eta',s'}(\vec{k},\vec{k}') =\sigma \, \sigma' \, t_\phi\, \left[\Lambda^j_{-\eta,-s;-\eta',-s'}(-\vec{k},-\vec{k}')\right]^*$ for bosonic modes even ($t_\phi=+$) or odd ($t_\phi=-$) under $\Theta_s$ \cite{PhysRevB.93.174509,PhysRevB.102.064501}.
This allows us to rewrite the relevant SU(2)$_+ \times$ SU(2)$_-$-symmetry breaking part, $\Delta M^{s,t}$, in $M^{s,t} =  M_0^{s,t} +  \Delta M^{s,t}$ (with $M_0^{s,t}$ related to $\mathcal{W}$ and phonons) as
\begin{equation}
    \left(\Delta M^{s}\right)_{k,k'} = \frac{1}{8} \tilde{t}_\phi  \sum_{s,s',\eta,\eta'} ss'\eta\eta' \mathcal{F}_{k,\eta,s;k',\eta',s'}, \qquad  \left(\Delta M^{t}\right)_{k,k'} = \frac{1}{8} \tilde{t}_\phi  \sum_{s,s',\eta,\eta'} \mathcal{F}_{k,\eta,s;k',\eta',s'}, \label{FormEmergentTRS}
\end{equation}
when applying $\widetilde{\Theta}$. We here introduced
\begin{equation}
    \mathcal{F}_{k,\eta,s;k',\eta',s'} = \left[\Lambda^j_{\eta,s;\eta',s'}(\eta \vec{k},\eta'\vec{k}')\right]^*[\chi(\eta k-\eta' k')]_{j,j'} \Lambda^{j'}_{\eta,s;\eta',s'}(\eta \vec{k},\eta'\vec{k}') \geq 0,
\end{equation}
which is non-negative, since $\chi$ has to be positive definite due to stability \cite{PhysRevB.93.174509,PhysRevB.102.064501}. For conventional time-reversal symmetry, $\Theta_s$, singlet and triplet switch their role \cite{PhysRevB.102.064501}
\begin{equation}
     \left(\Delta M^{s}\right)_{k,k'} = \frac{1}{8} t_\phi  \sum_{s,s',\eta,\eta'}  \mathcal{F}_{k,\eta,s;k',\eta',s'}, \qquad \left(\Delta M^{t}\right)_{k,k'} = \frac{1}{8} t_\phi  \sum_{s,s',\eta,\eta'} ss'\eta\eta' \mathcal{F}_{k,\eta,s;k',\eta',s'}. \label{FormUsualTRS}
\end{equation}

Upon writing $\Delta_s(k) = \gamma_{k}^+/\sqrt{\omega_n^2 + \epsilon^2_+(\vec{k})}$ and $\Delta_t(k) = \Gamma_{k}^+/\sqrt{\omega_n^2 + \epsilon^2_+(\vec{k})}$, we can recast solving the saddle-point equations in \equref{SaddlePointEquations} as solving for the largest eigenvalue $\lambda_{t,s}(T)$ of the symmetric matrix $(\widetilde{M}^{s,t})_{k,k'} := M^{t,s}_{k,k'}/\sqrt{(\omega_n^2 + \epsilon^2_+(\vec{k}))(\omega_{n'}^2 + \epsilon^2_+(\vec{k}'))}$ as a function of temperature $T$. The critical temperature, $T_{c}^{s,t}$, in the two channels corresponds to $\lambda_{t,s}(T_{c}^{s,t}) = 1$.
Defining $(\Delta \widetilde{M}^{s,t})_{k,k'} := \Delta M^{t,s}_{k,k'}/\sqrt{(\omega_n^2 + \epsilon^2_+(\vec{k}))(\omega_{n'}^2 + \epsilon^2_+(\vec{k}'))}$, we immediately see from \equsref{FormEmergentTRS}{FormUsualTRS}, that $(\Delta\widetilde{M}^{s})_{k,k'} \geq (\Delta \widetilde{M}^{t})_{k,k'}$ if $t_\phi = +$ or $\tilde{t}_\phi = -$ and $(\Delta\widetilde{M}^{t})_{k,k'} \geq (\Delta \widetilde{M}^{t})_{k,k'}$ if $t_\phi = -$ or $\tilde{t}_\phi = +$.  

Let us first discuss the situation without electron-phonon coupling and, hence, $\widetilde{M}^{s,t} = \Delta \widetilde{M}^{s,t}$. As follows from the Perron-Frobenius theorem, given two square matrices $A$ and $B$ of equal size with $A_{ij} \geq |B_{ij}|$ and $A$ being irreducible \footnote{This means that its adjacency graph is strongly connected. In our case, it corresponds to the situation that one can scatter between any pair of momenta via some number of virtual states; at least for $D_0\neq 0$, this is generically expected to be the case.}, the largest eigenvalue of $A$ (the ``Perron root'') is larger than the magnitude of the largest eigenvalue of $B$. We then have $\lambda_{s}(T) \geq \lambda_{t}(T)$ and, hence, $T_c^s \geq T_c^t$ if $t_\phi = +$;
this means that singlet will generically dominate over triplet. Similarly, if 
$\tilde{t}_\phi = +$, we get $\lambda_{t}(T) \geq \lambda_{s}(T)$ leading to $T_c^t \geq T_c^s$, which means that triplet will dominate. Note that if the fluctuations have a well-defined behavior under both $\Theta_s$ and $\widetilde{\Theta}$ and we further have $t_\phi = \tilde{t}_\phi$, \equsref{FormEmergentTRS}{FormUsualTRS} imply that $\Delta M^s = \Delta M^t$, i.e., singlet and triplet remain degenerate.

In the scenario proposed in \refcite{PhysRevB.102.064501}, where electron-phonon coupling yields an important contribution in stabilizing superconductivity but predominantly in the SU(2)$_+ \times$ SU(2)$_-$-symmetric channel, we treat the symmetry-breaking $\Delta \widetilde{M}^{s,t}$ as a small perturbation to the SU(2)$_+ \times$ SU(2)$_-$-symmetric $\widetilde{M}^{s,t}_0 := \widetilde{M}^{s,t}-\Delta \widetilde{M}^{s,t}$ associated with the phonons. As follows from the analysis of \refcite{PhysRevB.93.174509}, we must have $\Delta^0_{t,s}(k) > 0$ for the unperturbed saddle-point solutions, i.e., leading eigenstates of $\widetilde{M}^{s,t}_0$. As such, it is clear that the leading (first-order) correction $\delta \lambda_{s,t} = \int_{k,k'} \Delta^0_{t,s}(k) (\Delta \widetilde{M}^{s,t})_{k,k'} \Delta^0_{t,s}(k')$ is larger for singlet (triplet), which then dominates, if $t_\phi = +$ ($t_\phi = -$) or $\tilde{t}_\phi = -$ ($\tilde{t}_\phi = +$).

We illustrate these statements more explicitly for the IVC$_-$ state, which is odd under $\Theta_s$ and even under $\widetilde{\Theta}$, i.e., characterized by $t_\phi=-$ and $\tilde{t}_{\phi}=+$. As already mentioned in \secref{NoSpinPolarization}, a minimal description is given by $[\chi(i\Omega_n,\vec{q})]_{j,j'} = \chi_0 \delta_{j,j'}/(\Omega_n^2 + c^2 \vec{q}^2 +\xi^{-2})$, $j=1,2$, and $\lambda^j(\vec{k},\vec{k}') = s_0\eta_j f_{\vec{k},\vec{k}'}$, where $f_{\vec{k},\vec{k}'}$ is a matrix in band space with  $f^\pdagger_{\vec{k},\vec{k}'} = f^\dagger_{\vec{k}',\vec{k}} = -f^T_{\vec{k}',\vec{k}}$. This leads to 
\begin{equation}
    \mathcal{F}_{k,\eta,s;k',\eta',s'} = \frac{2\chi_0 \delta_{s,s'}\delta_{\eta,-\eta'}}{\Omega_n^2 + c^2 \vec{q}^2 +\xi^{-2}} \left| \braket{u_{\eta,s}(\vec{k}) | f_{\vec{k},\vec{k}'} | u_{\eta,s}(\vec{k}') }\right|^2 \geq 0.
\end{equation}
So only matrix elements with $ss'\eta\eta' = -1$ are non-zero such that $\Delta M^{s,t}$ are identical in \equsref{FormEmergentTRS}{FormUsualTRS} for $t_\phi=-$ and $\tilde{t}_{\phi}=+$. We, thus, consistently see that IVC$_-$ fluctuations will favor the triplet state.

Finally, consider the SLP$_-$ state, which is odd under both $\Theta_s$ and $\widetilde{\Theta}$ and, thus, has $t_\phi=\tilde{t}_{\phi}=-$. To be consistent with the analysis above, fluctuations of this state must affect singlet and triplet in exactly the same way. This can also be easily seen explicitly by noting that its coupling vertex $\lambda$ in \equref{phifcoupling} cannot couple different valleys or spin species and, hence, $\mathcal{F}_{k,\eta,s;k',\eta',s'} \propto \delta_{s,s'}\delta_{\eta,\eta'}$. Consequently, we get $\Delta M^s = \Delta M^t$ in \equsref{FormEmergentTRS}{FormUsualTRS}.

\subsection{Order parameter in the polarized phase}\label{SCInPolarizedState}
Finally, we also illustrate the spin-structure of the superconducting order parameter in the presence of flavor polarization, relevant for $\nu$ with $2 \lesssim |\nu| \lesssim 3$. For both SP-based and SVP-based band resetting, the superconducting order parameter in the TBG-like bands has the form 
\begin{equation}
    H_{\text{SC,p}}^b = \sum_{\vec{k}} \Delta_{\vec{k}} \, \psi_{\vec{k};+}^\dagger \psi_{-\vec{k};-}^\dagger + \text{H.c.},
\end{equation}
where $\psi_{\vec{k};\pm}$ are the electronic operators of valley $\eta=\pm$ in the polarized TBG-like bands which are closest to the Fermi level for each $\vec{k}$. In principle, there are also finite band-off-diagonal matrix elements. However, since we are only interested in the predominant spin-structure of the superconductor, we do not need to take them into account. Note that, if the band resetting also involves intervalley coherent order, such as in \figref{Bands2}(c,d), the labels $\pm$ of $\psi_{\vec{k},\pm}$ have to be understood as combinations of the two valleys; the spin structure we discuss here will not be affected by it.

To transform back to the physical spin ($s=\uparrow,\downarrow$) basis, we write $\psi_{\vec{k};\eta} = \sum_{s} (u^*_{\vec{k};\eta})_s f_{\vec{k};\eta,s}$, where the two-component spinor, $u_{\vec{k};\eta}$, encodes the nature of the flavor polarization. The singlet, $\Delta_{\vec{k}}^s$, and triplet, $\vec{d}_{\vec{k}}$, component of the order parameter, coupling as $H_{\text{SC,p}}^b = \sum_{\vec{k}}\Delta_{\vec{k}}  f_{\vec{k};+,s}^\dagger \left( (s_0 \Delta_{\vec{k}}^s + \vec{s}\cdot\vec{d}_{\vec{k}}) i s_y \right))_{s,s'} f_{-\vec{k};-,s'}^\dagger + \text{H.c.}$, then follow from

\begin{equation}
    \left( (s_0 \Delta_{\vec{k}}^s + \vec{s}\cdot\vec{d}_{\vec{k}}) i s_y \right))_{s,s'} = (u_{\vec{k};+})_s (u_{-\vec{k};-})_{s'}.
\end{equation}
For the case of SP, where the spin polarization is in the same direction in the two valleys, say along the $s_z$ direction, we have
\begin{equation}
     (u_{\vec{k};\pm})_s  = \delta_{s,\uparrow}, \quad \Rightarrow \quad \Delta_{\vec{k}}^s = 0,\, \vec{d}_{\vec{k}} = -\Delta_{\vec{k}} (1,i,0)^T/2,
\end{equation}
corresponding to non-unitary triplet pairing. A Zeemann field, $\vec{B}_Z$, will not affect the polarization nor the superconductor (except for pinning the direction of their magnetizations).

For the SVP-type of polarization, we have instead (choosing the polarization along the $s_x$ axis)
\begin{equation}
    u_{\vec{k};\pm}  = \frac{1}{\sqrt{2}}\begin{pmatrix}\pm 1 \\ 1 \end{pmatrix} \quad \Rightarrow \quad \Delta_{\vec{k}}^s = \Delta_{\vec{k}}/2,\, \vec{d}_{\vec{k}} = \Delta_{\vec{k}} (1,0,0)^T/2,
\end{equation}
leading to an equal admixture of singlet and triplet pairing \cite{PhysRevResearch.2.033062}.
Applying $\vec{B}_Z$, will pin the orientation of the spin to be orthogonal to it and induce a canting of angle $\theta$ that increases with $|\vec{B}_Z|$. For $\vec{B}_Z$ along the $s_z$ direction, we thus get
\begin{equation}
    u_{\vec{k};\pm}  = \frac{1}{\sqrt{2(1+\sin\theta)}}\begin{pmatrix} \pm (1 + \sin\theta) \\ \cos\theta \end{pmatrix} \quad \Rightarrow \quad \Delta_{\vec{k}}^s = \cos\theta \Delta_{\vec{k}}/2,\, \vec{d}_{\vec{k}} = \Delta_{\vec{k}} (1,i \sin\theta,0)^T/2.
\end{equation}
So we see that this state continuously transforms into the non-unitary triplet when applying a magnetic field.
We note the behavior in magnetic field of these pairing states---the non-unitary triplet and singlet-triplet admixed phases---is the same in the absence of polarization, as shown in \refcite{PhysRevResearch.2.033062}.

\section{WZW terms in MSTG}\label{AppendixOnWZW}
In this appendix we will give more details on how we reached the conclusions in \secref{WZWTerms}. We will begin as in the main text with $|\nu|=0$. In this case, as in the main text, the low-energy physics is captured by
\begin{equation}
    H^{\text{D}}_0=\sum_{\vec{q}} \psi^\dagger_{\vec{q}}\left[\gamma_1 q_x+ \gamma_2 q_y\right] \psi^\pdagger_{\vec{q}}, \qquad \gamma_1 = \rho_x\mu_z s_0, \quad \gamma_2 = \rho_y \mu_0 s_0,   \label{AppendixDiracAction}
\end{equation} 
where we are using the notation for Pauli matrices defined in \tableref{paulinotation}. We will assume a pairing $\Delta$ with 2 real components and an insulating order parameter $m_j$ with three real components, incorporated as a Dirac masses,
\begin{equation}
    H^{\text{D}}_1=\sum_{\vec{q},j} \psi^\dagger_{\vec{q}} m_j \psi^\pdagger_{\vec{q}}+\sum_{\vec{q}} \left(\psi^\dagger_{\vec{q}} \Delta T \psi^\dagger_{-\vec{q}}+\text{H.c.}\right)
\end{equation}
with $T=\mu_x s_y$. We want to consider all possible pairings $\Delta$ and insulating states $m_j$ such that when the Dirac fermions are integrated out of the action corresponding to $H^{\text{D}}_0+H^{\text{D}}_1$, a WZW term will be produced. One way to do this would be to consider which of the options classified in \cite{Christos_2020} survive projection to a minivalley-valley locked subspace. Alternatively, we can consider which orders are compatible with the conditions derived in \cite{Christos_2020} for the projected form of the Dirac action. We will choose the second way in this appendix. We will first consider the compatible pairings $\Delta$ at $\nu=0$. At $\nu=0$, triplet pairing is excluded as we require SU(2)$_s$ spin symmetry to be intact before the bands acquire any ordering. A requirement for a pairing to be compatible with the existence of a WZW term is that it must satisfy the following conditions with the kinetic term's $\gamma$ matrices 
\begin{equation}\label{ConstraintDelta}
    \gamma_i\Delta T=-\Delta T\gamma_i^T\neq 0 \qquad i=1,2.
\end{equation}
We will additionally assume pairings which do not break moir\'e translational symmetry and intervalley pairing terms which preserve $U(1)_v$. We have also chosen a momentum independent $\Delta$. The only pairing compatible with all the above requirements is $\Delta=\mathbbm{1}$, transforming under the $A$ representation of the point group. As mentioned in the main text, this is the same pairing considered for TBG in \cite{Christos_2020} projected to the minivalley-valley locked subspace. For this pairing, we can then ask what are the compatible 3-component order parameters $m_j$, $j=1,2,3$. The conditions the insulating states must satisfy with the pairing and $\gamma$ matrices are:
\begin{equation}
    \text{Tr}[\gamma_{i_1}\gamma_{i_2}m_{j_1}m_{j_2}m_{j_3}]\propto \epsilon_{i_1 i_2 i_3 j_1 j_2 j_3}
\end{equation}
\begin{equation}
    m_j\Delta T=\Delta T m_j^T\neq 0 \qquad j=1,2,3
\end{equation}
We find the only two options are a 3-component spin-Hall state $\rho_z\mu_z(s_x,s_y,s_z)$ and a 2-component moir\'e density wave state $\rho_x(\mu_x,\mu_y)$ combined with a third component sublattice polarized state $\rho_z$. 

\begin{table}[tb]
    \centering
    \caption{We explicitly list the notation we use to denote the Pauli matrices which act on the sublattice, minivalley, valley, and spin degrees of freedom used throughout the text in the second column, and the notation used in \cite{Christos_2020} in the third column. }
    \begin{tabular}{ccc}
        \hline\hline
        & \,\, Notation in this work \,\, & \,\, Notation in \cite{Christos_2020} \,\, \\ \hline
        Sublattice & $\rho$ & $\rho$\\
        Minivalley & $\mu$ & $\mu$\\
        Valley & $\eta$ & $\tau$\\
        Spin & $s$ & $\sigma$ \\ \hline\hline
    \end{tabular}
    \label{paulinotation}
\end{table}

We can then consider the case at $|\nu|=2$. 
As argued in the main text, the low-energy Dirac theory takes the same form as at $\nu=0$, see \equref{AppendixDiracAction}. 

However, unlike the case at $\nu=0$, we no longer assume SU(2)$_s$ spin symmetry. It is simple to see the options which were compatible at $\nu=0$ for singlet pairing are still compatible at $\nu=2$. On the other hand, the breaking of spin rotation symmetry allows for additional options at $|\nu|=2$ which were not previously possible at $|\nu|=0$. It is easily verified that the triplet pairing $\Delta=\mu_z s_z$ satisfies \equref{ConstraintDelta}. We find two options which are compatible with this pairing. The first is the set of orders $\rho_z(s_x,s_y)$ and $\rho_z\mu_z s_z$, corresponding to our SSLP$_+$ and SSLP$_-$ respectively. The second is $\rho_x(\mu_x,\mu_y)$ and $\rho_z$, corresponding to a moir\'e density wave state and SLP$_+$ state respectively.

\end{appendix}

\end{document}